%% file: main.tex
\documentclass[a4paper,12pt,english,twoside,dvipsnames]{book}
\usepackage{fix-cm}
\usepackage[T1]{fontenc}
\usepackage[utf8]{inputenc}
\usepackage[a4paper,left=2.5cm,right=2.5cm,top=2.5cm,bottom=2.5cm,bindingoffset=0.5cm]{geometry}
\usepackage{graphicx}
\usepackage{import}
\usepackage{verbatim} 
\usepackage{amsmath}
\usepackage{mathtools}
\usepackage{titlesec}
\usepackage{pdfpages}
\usepackage{afterpage}
\usepackage{dashrule}
\usepackage{amsmath,amssymb,amsfonts}
\usepackage{algorithmic}
\usepackage[ruled,vlined]{algorithm2e}
\usepackage{multirow}
\usepackage{comment}
\usepackage{array}
\usepackage{ragged2e}
\usepackage{type1cm}
\usepackage{dirtytalk}
\usepackage{epigraph}
\usepackage{setspace}
\usepackage{indentfirst}
\usepackage[super]{nth}
\usepackage{enumitem}
\usepackage{caption}
\usepackage{lipsum}
\usepackage{upgreek}
\usepackage{fancyhdr}
\usepackage{url}
\usepackage{indentfirst}
\usepackage[hidelinks]{hyperref}
\hypersetup{colorlinks=false}
\usepackage{type1cm}
\usepackage{lettrine}
\usepackage{float}
\usepackage{subfigure}
\usepackage{multicol}
\graphicspath{{CHAPTER_IMAGES/}}
\newcolumntype{P}[1]{>{\centering\arraybackslash}p{#1}}
\def\blankpage{%
      \clearpage%
      \thispagestyle{empty}%
      \null%
      \clearpage}

\newcolumntype{R}[1]{>{\centering\let\newline\\\arraybackslash\hspace{0pt}}m{#1}}
\titleformat{\chapter}[display]{\normalfont\Large\bfseries}{\filleft \color{Plum}\fontsize{18}{35} \selectfont \MakeUppercase{\textbf{Chapter \thechapter}}}{0.01ex}{\vspace{1mm}\filleft \color{Black}\fontsize{30}{35}\selectfont{\textbf{}}}
\usepackage{babel,csquotes,xpatch}
\usepackage[backend=biber, bibstyle=ieee, citestyle=numeric-comp, mincitenames=1, maxcitenames=1, url=false, sorting=none, doi=true,
eprint=false]{biblatex}
\DeclareSourcemap{
  \maps[datatype=bibtex]{
    \map[overwrite]{
      \perdatasource{Thesis_Bibliography.bib}
       \step[fieldset=keywords, fieldvalue={,inbib}, append]
    }
  }
}
\begin{document}
\includepdf[pages=1, pagecommand={}, fitpaper=true]{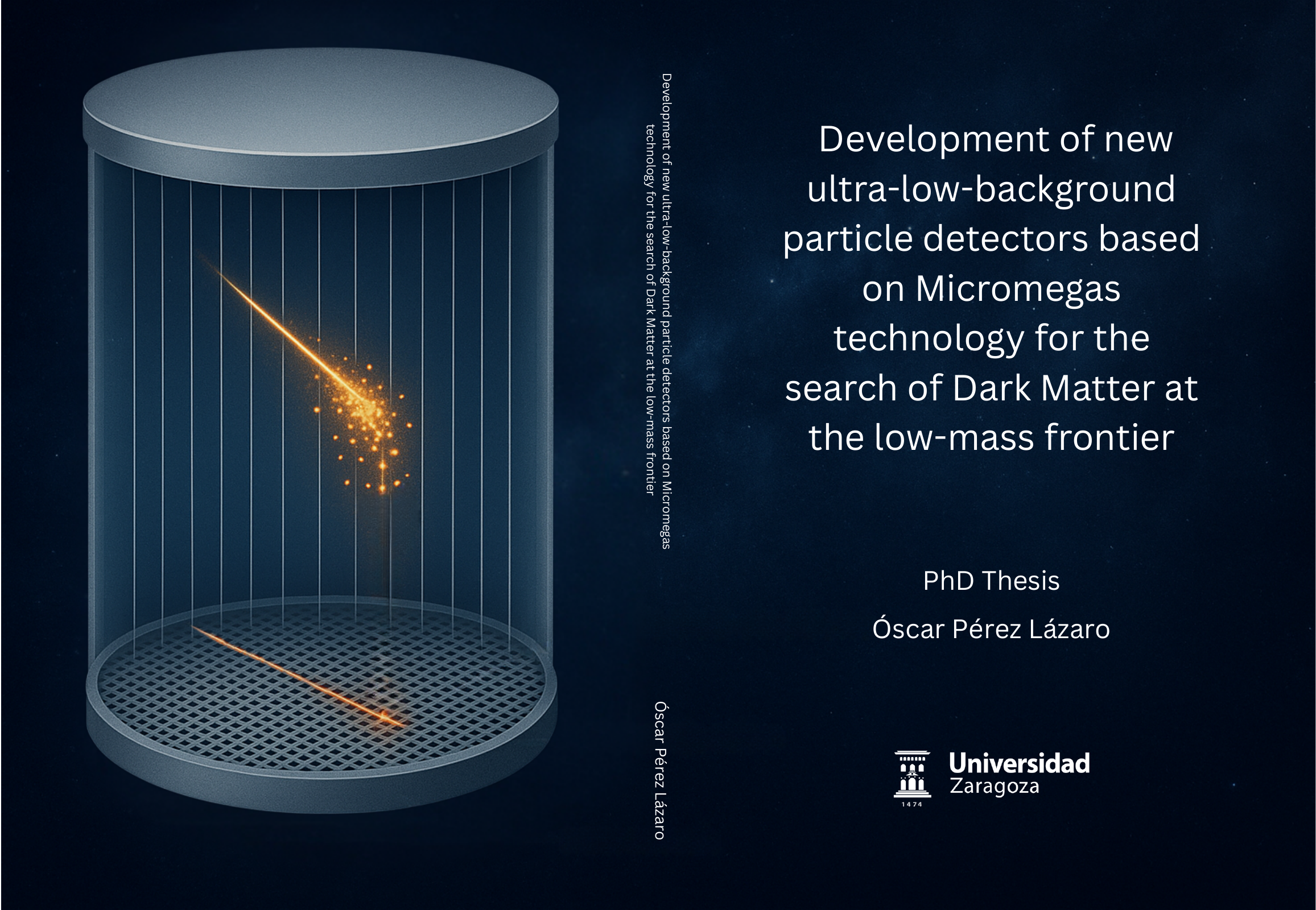}
\pagestyle{empty}

\input{Title_Page}
\blankpage
\newpage
\input{Acknowledgements}
\clearpage
\clearpage
\pagestyle{empty}                                
\addtocontents{toc}{\protect\thispagestyle{empty}}
\tableofcontents
\blankpage
\clearpage

\mainmatter
\pagenumbering{arabic}
\pagestyle{fancy}
\fancyhf{}
\fancyfoot[C]{\thepage}
\fancyhead[LE,RO]{Chapter 1: The Dark Matter Problem} 
\renewcommand{\headrulewidth}{1pt}
\onehalfspacing
\import{CHAPTERS/}{Chapter1}    
\clearpage
\pagestyle{fancy}
\fancyhf{}
\fancyfoot[C]{\thepage}
\fancyhead[LE,RO]{Chapter 2: WIMPs as Dark Matter Candidates} 
\renewcommand{\headrulewidth}{1pt}
\onehalfspacing
\import{CHAPTERS/}{Chapter2}   
\clearpage
\pagestyle{fancy}
\fancyhf{}
\fancyfoot[C]{\thepage}
\fancyhead[LE,RO]{Chapter 3: Fundamental Processes in Gaseous Detectors}    
\renewcommand{\headrulewidth}{1pt}
\onehalfspacing
\import{CHAPTERS/}{Chapter3}    
\clearpage
\pagestyle{fancy}
\fancyhf{}
\fancyfoot[C]{\thepage}
\fancyhead[LE,RO]{Chapter 4: Gaseous TPCs with MPGDs}     
\renewcommand{\headrulewidth}{1pt}
\onehalfspacing
\import{CHAPTERS/}{Chapter4}    
\clearpage
\pagestyle{fancy}
\fancyhf{}
\fancyfoot[C]{\thepage}
\fancyhead[LE,RO]{Chapter 5: TREX-DM Experiment}               
\renewcommand{\headrulewidth}{1pt}
\onehalfspacing
\import{CHAPTERS/}{Chapter5}    
\clearpage
\pagestyle{fancy}
\fancyhf{}
\fancyfoot[C]{\thepage}
\fancyhead[LE,RO]{Chapter 6: Radon and the Background Problem} 
\renewcommand{\headrulewidth}{1pt}
\onehalfspacing
\import{CHAPTERS/}{Chapter6}    
\clearpage
\pagestyle{fancy}
\fancyhf{}
\fancyfoot[C]{\thepage}
\fancyhead[LE,RO]{Chapter 7: GEM Preamplification Stage for Energy Threshold Reduction}               
\renewcommand{\headrulewidth}{1pt}
\onehalfspacing
\import{CHAPTERS/}{Chapter7}    
\clearpage
\pagestyle{fancy}
\fancyhf{}
\fancyfoot[C]{\thepage}
\fancyhead[LE,RO]{Chapter 8: \texorpdfstring{$^{37}$Ar}{37Ar} as a Low-Energy Calibration Source}               
\renewcommand{\headrulewidth}{1pt}
\onehalfspacing
\import{CHAPTERS/}{Chapter8}    
\clearpage
\pagestyle{fancy}
\fancyhf{}
\fancyfoot[C]{\thepage}
\fancyhead[LE,RO]{Chapter 9: Sensitivity}               
\renewcommand{\headrulewidth}{1pt}
\onehalfspacing
\import{CHAPTERS/}{Chapter9}    
\clearpage
\pagestyle{fancy}
\fancyhf{}
\fancyfoot[C]{\thepage}
\fancyhead[LE,RO]{Summary and Conclusions}               
\renewcommand{\headrulewidth}{1pt}
\onehalfspacing
\import{CHAPTERS/}{Chapter10}
\clearpage
\pagestyle{fancy}
\fancyhf{}
\fancyfoot[C]{\thepage}
\fancyhead[LE,RO]{Resumen y Conclusiones}               
\renewcommand{\headrulewidth}{1pt}
\onehalfspacing
\import{CHAPTERS/}{Chapter10_spa}    
\clearpage
\pagestyle{fancy}
\fancyhf{}
\fancyfoot[C]{\thepage}
\fancyhead[LE,RO]{Bibliography}
\renewcommand{\headrulewidth}{1pt}
\printbibliography[heading=bibintoc,title={Bibliography}]
\clearpage

\end{document}

%% file: Title_Page.tex
\begin{titlepage}
    \begin{center}
        \vspace*{\fill}
        \Huge
        \textbf{Development of new ultra-low-background particle detectors based on Micromegas technology for the search of Dark Matter at the low-mass frontier}
 
        \vspace{3.0cm}
        \normalsize
            Thesis submitted by\\  
        \Large
            \textbf{Óscar Pérez Lázaro}\\
        \vspace{1.0cm}
        \normalsize
            Supervised by\\  
        \large
            \textbf{Igor García Irastorza}
            
        \vspace{3.0cm}

        \Large
            \textbf{Doctor of Philosophy (Physics)} 
            
        \vspace{3.0cm}
            
        \Large
            \textbf{Área de Física Atómica, Molecular y Nuclear}\\
            \textbf{Departamento de Física Teórica}\\
            \textbf{Facultad de Ciencias}\\
            \textbf{Universidad de Zaragoza}\\
            \textbf{Zaragoza, España}\\
       \vspace{1.0cm}
            \textbf{2025}
         \vspace*{\fill}       
    \end{center}
\end{titlepage}

%% file: Acknowledgements.tex
\begin{center}
    {\color{Black}\fontsize{30}{35}\selectfont{\textbf{\textit{Agradecimientos}}}}
\end{center}
\vspace{0.5cm}
%
\justify
Muchas gracias a mi director de tesis, al grupo de investigación de acogida y a todos los que me han ayudado en el desarrollo de este trabajo. Gracias también a mi familia y a las personas que me han apoyado en estos años. 
\newline
\newline
\newline

%% file: CHAPTERS/Chapter1.tex
\chapter{The Dark Matter Problem} \label{Chapter1_Dark_Matter}
%
%
\section*{}
\parshape=0
\vspace{-13mm}
%
\section{Introduction} \label{Chapter1_Introduction}

The nature of Dark Matter is one of the most challenging open problems in modern physics. The term Dark Matter (\textit{dunkle Materie}) was coined by Fritz Zwicky~\cite{Zwicky_1933,Zwicky_1937} to describe the invisible matter that he used to account for the mass discrepancies he encountered in his studies of the Coma Cluster, and has been used since then to allude to the non-baryonic and cold (non-relativistic) matter that makes up around 26\% of the mass-energy balance in the universe.

While direct observations of Dark Matter are lacking, its presence can be inferred through the gravitational effects it has on ordinary matter. Understanding the characteristics of Dark Matter is crucial for a variety of factors:

\begin{itemize}
    \item It would revolutionise our comprehension of the fundamental components of the universe, potentially leading to what is usually called as Physics Beyond the Standard Model (BSM).
    \item Since Dark Matter constitutes a significant portion of the mass-energy balance of the universe (and an 85\% of the matter content), a detailed understanding is critical to have accurate cosmological models.
    \item It would provide insight into the early history of the universe and the mechanisms driving its evolution. Specifically, Cold Dark Matter could explain galaxy formation, galaxy clustering and the large-scale structures observed in the cosmos.
\end{itemize}

This chapter aims to provide an overview of the evidence we have of its existence, as well as a summary of the main Dark Matter candidates or proposed solutions to the problem.

\section{Evidence for Dark Matter} \label{Chapter1_Evidence}

\subsection{Galactic Rotation Curves} \label{Chapter1_Evidence_GalacticRotation}
\begin{figure}[htb]
\centering
\includegraphics[width=0.65\textwidth]{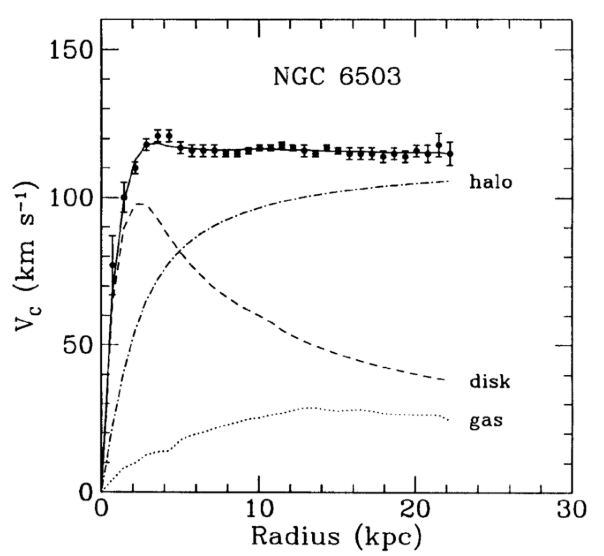}
\caption{Galactic rotation curve for NGC 6503. Shown are the experimental data (black dots with error bars) and the fit (black solid line), along with the different contributions to the fit: baryonic matter (disk and gas) and the Dark Matter (halo) needed to obtain a good agreement with the data. Extracted from~\cite{Freese}.}
\label{fig:chapter1_galactic_rotation_curve}
\end{figure}

The study of galaxy rotation curves is one of the earliest pieces of evidence for Dark Matter, dating back to the pioneering studies by Vera Rubin \textit{et al.}~\cite{Rubin}. Simply put, if only visible matter was present, galaxies would be composed of stars and gas, and their rotation speeds should decrease with increasing distance from the galactic center ($r^{-1/2}$ according to Newtonian dynamics). However, careful observations of velocity distribution of stars for spiral galaxies show an unusually high velocity for stars near the edge of the galaxy, displaying a flat shape that disagrees with the predictions from Newtonian gravity applied to the visible mass alone. This is commonly known as the \textit{galactic rotation problem}, and suggests the existence of invisible matter that contributes to the total mass distribution. See Figure~\ref{fig:chapter1_galactic_rotation_curve} for an example of a galactic rotation curve where baryonic matter alone is not enough to explain the experimental observations.

The Dark Matter contribution is usually modelled as a halo with decreasing density as one gets away from the centre of the galaxy. Navarro, Frenke and White found, performing N-body simulations, that the density profile for galaxies is very accurately described by the following distribution~\cite{navarro_frenk_white_model}:

\begin{equation}
    \rho (r) = \frac{\rho_s}{\left(\frac{r}{r_s}\right)\left(1+\frac{r}{r_s}\right)^2}
    \label{eq:chapter1_density_profile_NFW}
\end{equation}

where $r_s$ (scale radius) and $\rho_s$ (characteristic volume density) are the fitting parameters. This density profile (NFW model) works well for a wide variety of galaxy sizes. However, this distribution shows a rise (or cusp) as we approach the galactic centre, which is incompatible with some observations, such as those of dwarf galaxies, where the density profile in the central region is flat. This is what is known in small-scale cosmology as the core-cusp problem~\cite{core_cusp_problem}. For these cases, profiles such as the so-called pseudo-isothermal (pISO) model are found to better describe the observations:

\begin{equation}
    \rho (r) = \frac{\rho_s}{1+\left(\frac{r}{r_s}\right)^2}
    \label{eq:chapter1_density_profile_pISO}
\end{equation}

\begin{figure}[htb]
\centering
\includegraphics[width=0.95\textwidth]{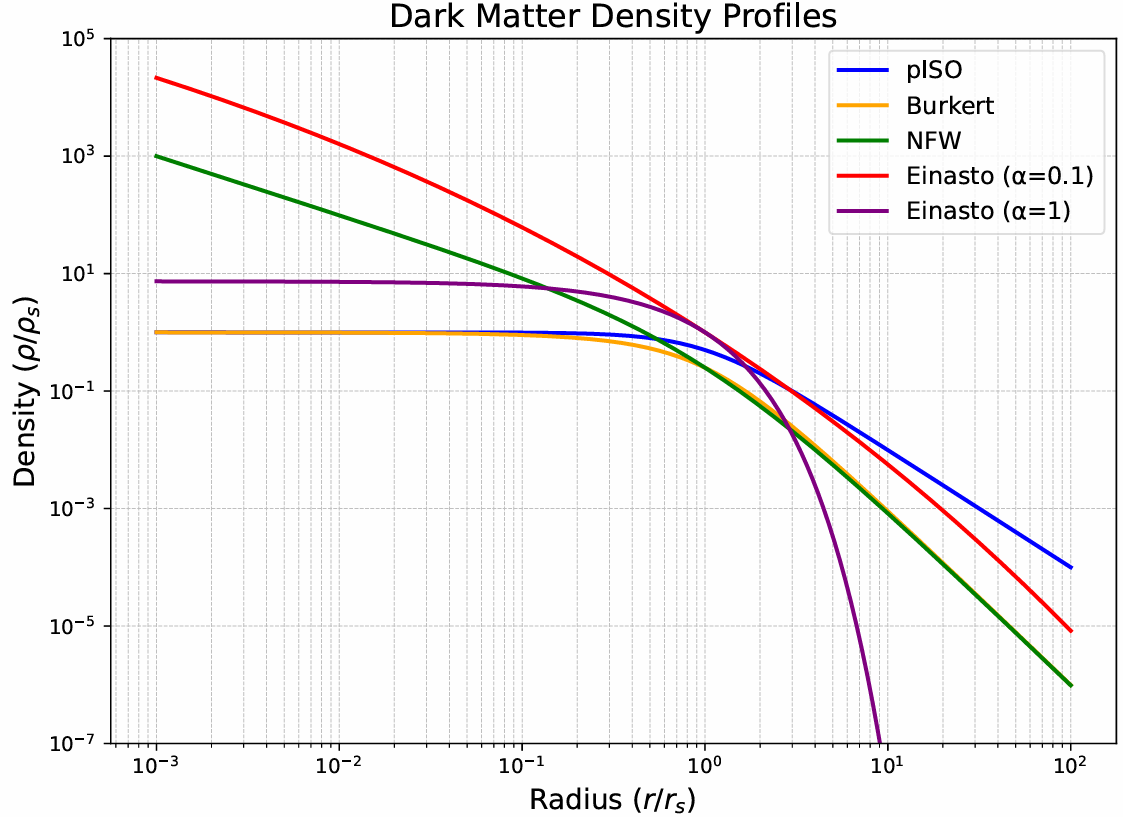}
\caption{Comparison of the density profiles (density as a function of radius) for different models of Dark Matter halo. Note the possibility of obtaining cuspy core and flat core profiles by varying $\alpha$ in the Einasto model. Source: own elaboration with formulas taken from~\cite{dark_matter_halo_models}.}
\label{fig:chapter1_dark_matter_halo_density_profiles}
\end{figure}

In general, there is a whole family of models to describe Dark Matter halo profiles. For example, the Burkert model is based on a modification of the pISO model, $\rho_{\mathrm{Burkert}}=\rho_{\mathrm{pISO}}(1+r/r_s)^{-1}$, so that the halo mass diverges more slowly~\cite{burkert_model}. Einasto model includes another controllable parameter, $\alpha$, which allows the density profile to be varied between cuspy core and flat core. Figure~\ref{fig:chapter1_dark_matter_halo_density_profiles} shows all the aforementioned Dark Matter halo models, illustrating the two types of profiles that can be achieved with the Einasto model.

\subsection{Velocities in Galaxy Clusters} \label{Chapter1_Evidence_Velocities}

When analysing the motion of galaxies within a cluster, it becomes apparent that their velocities are far too high to be explained by the visible mass alone. 
 
For example, in his studies, Zwicky applied the virial theorem to the Coma Cluster to obtain an estimate of the total mass. According to this theorem, the total kinetic energy in a gravitationally bound system in equilibrium is proportional to the total gravitational potential energy. By measuring the dispersion velocity of the galaxies, $v_{\mathrm{disp.}}$, the total mass of the cluster, $M$, can be calculated as:

\begin{equation}
    M \sim \frac{v^2_{\mathrm{disp.}}D}{G} \label{eq:chapter1_virial_theorem_mass}
\end{equation}

where $D$ is the size of the cluster and $G$ the gravitational constant. Zwicky found that, plugging in the observed dispersion velocities, the total mass was much larger than that estimated from cluster luminosity measurements (the \textit{missing mass problem}).

Subsequent studies in different clusters confirmed the universality of this discrepancy, estimating the ratio between total mass and visible or luminous mass (electromagnetically interacting matter) to be approximately 10:1.

Later, it was found that a halo density profile of Dark Matter, both for the galaxies and the clusters themselves, provides the additional gravitational pull necessary to maintain the observed cluster dynamics. In simulations, it has been observed that the density distributions used to model the Dark Matter halo at the cluster level are the same as those previously discussed in Section~\ref{Chapter1_Evidence_GalacticRotation}. This reveals that haloes are self-similar at different spatial scales.

\subsection{Gravitational Lensing} \label{Chapter1_Evidence_GravitationalLensing}

Gravitational lensing, a consequence of General Relativity, is the bending of light as it follows the curvature of spacetime surrounding a massive object. 

This phenomenon is a crucial evidence in favour of Dark Matter, because the observed distortion of background galaxies as their light passes through a galaxy cluster (the lens) demonstrates a greater lensing effect than expected based exclusively on the visible mass of the galaxy cluster, indicating the presence of additional invisible matter that contributes to the total gravitational influence. 

\begin{figure}[htb]
\centering
\includegraphics[width=0.7\textwidth]{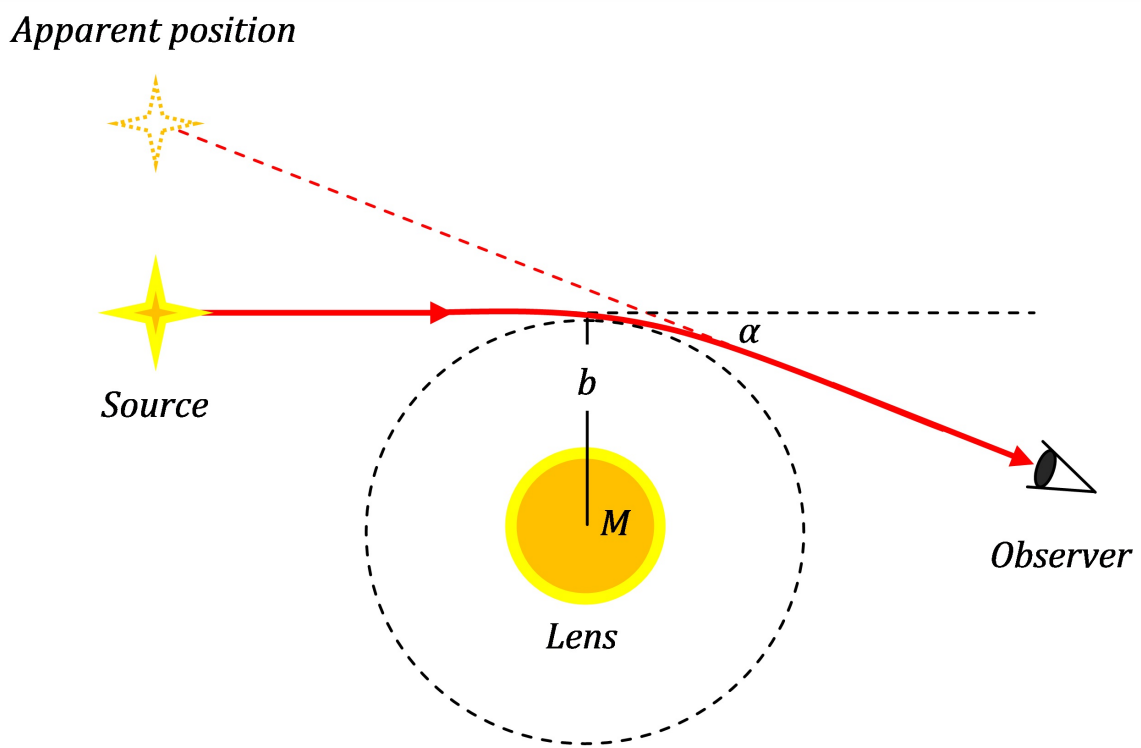}
\caption{Schematic of the deflection (angle $\alpha$) that a light source undergoes when it enters the gravitational potential of a source with mass $M$ at a distance $b$. Image extracted from~\cite{gravitational_lensing_einstein_ring}.}
\label{fig:chapter1_gravitational_lensing_deflection}
\end{figure}

The deflection of light by the gravity of an object is described by Einstein's field equations. For an object of mass $M$, the deflection angle is given by:

\begin{equation}
    \alpha = \frac{4GM}{c^2}\frac{1}{b}    \label{eq:chapter1_gravitational_lensing_deflection_angle}
\end{equation}

where $G$ is the gravitational constant, $c$ the speed of light and $b$ the impact parameter (the perpendicular distance from the path of the light beam to the centre of the object). This equation shows that, as one would expect, the more massive the object, the more it bends the light. On the other hand, if the light ray passes very far away from the gravitational zone of influence ($b$ is large), the angle is very small. Figure~\ref{fig:chapter1_gravitational_lensing_deflection} illustrates the different parameters of Equation~\ref{eq:chapter1_gravitational_lensing_deflection_angle}.

\begin{figure}[htb]
\centering
\includegraphics[width=0.8\textwidth]{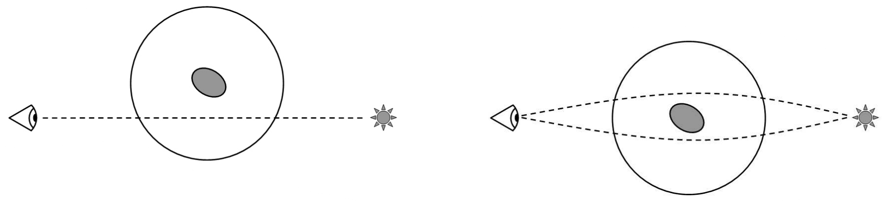}
\caption{Comparison between weak (left) and strong (right) gravitational lensing. Image extracted from~\cite{gravitational_lensing_weak_vs_strong}.}
\label{fig:chapter1_gravitational_lensing_weak_vs_strong}
\end{figure}

In gravitational lensing, two objects are defined: the lens (mass $M$ that deflects the light) and the source (the object whose light is deflected). Depending on the alignment between these two objects, a distinction is made between strong and weak gravitational lensing: in strong lensing, the lens and the source are aligned with the observer; in weak lensing, the lens is displaced with respect to the observer-source line. A schematic illustration of both cases is shown in Figure~\ref{fig:chapter1_gravitational_lensing_weak_vs_strong}.

\vspace{2mm}
\textbf{\normalsize Strong Lensing}
\vspace{0mm}

\begin{figure}[htb]
\centering
\includegraphics[width=0.55\textwidth]{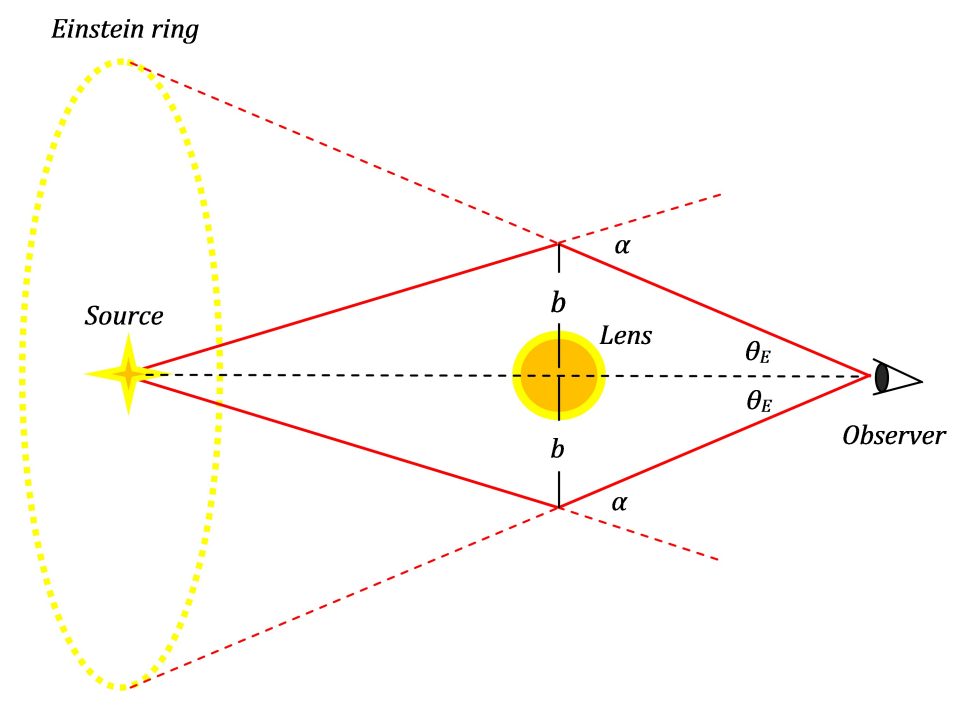}
\caption{Diagram of the creation of an Einstein ring in the observer plane due to ideal alignment of source-lens-observer and spherical symmetry of the lens. Image extracted from~\cite{gravitational_lensing_einstein_ring}.}
\label{fig:chapter1_gravitational_lensing_einstein_ring}
\end{figure}

In strong lensing, the very good alignment of lens and background object creates considerable visual distortions, including multiple images of the object, arcs, or even Einstein rings. An Einstein ring is a circular image of a background source that is formed as a consequence of the deflection of light by the gravitational field of the lens when the observer, the lens and the source are in ideal alignment conditions. A schematic of this phenomenon can be seen in Figure~\ref{fig:chapter1_gravitational_lensing_einstein_ring}. However, note that the radius of the ring is exaggerated in the image for the sake of clarity, since in reality the distances between source, lens and observer are very large with respect to the impact parameter $b$, so the small angle approximation works very well. Using this approximation, and expressing the angles in radians, it can be seen that $r_E = \theta_E D_S = \alpha D_{LS}$, where $r_E$ is the radius of the ring, $\theta_E$ the angular size of the ring, $D_S$ the observer-source distance, and $D_{LS}$ the lens-source distance. Substituting $\alpha$ from Equation~\ref{eq:chapter1_gravitational_lensing_deflection_angle}, and also taking into account that, in this approximation, $b = \theta_E D_L$, where $D_L$ is the observer-lens distance, we have that the angular size of the ring is:

\begin{equation}
    \theta_E = \sqrt{\frac{4GM}{c^2}\frac{D_{LS}}{D_L D_S}} \label{eq:chapter1_gravitational_lensing_einstein_ring}
\end{equation}

Normally, the lensing mass is more complex (like a cluster of galaxies), and lacks both circular symmetry and ideal alignment with the background object. In this case, rather than Einstein rings, we find partial arcs around the lens, as well as multiple images of the same object, depending on the shape of the object's gravitational potential. For example, strong lensing studies of the galaxy cluster Abell 1689~\cite{Halkola} show that visible mass is not enough to account for the observed multiple images and arcs, and the analysis of these distorted images allows for a mapping of the Dark Matter distribution in the cluster, which fundamentally agrees with a modification of the NFW density profile from Equation~\ref{eq:chapter1_density_profile_NFW}.

\vspace{2mm}
\textbf{\normalsize Weak Lensing}
\vspace{0mm}

In weak lensing, the object and lens alignment is not perfect. This is the most common case in our observations of the universe. As the lens is offset from the source-observer line, the distortions produced by the foreground lens are weaker. This makes the measurement of the lensing effect on a single background object impractical, so we resort to measurements of the lensing effect on an ensemble. A typical case of weak lensing would be the deformation that a cluster of galaxies produces in a background cluster of galaxies.

\begin{figure}[htb]
\centering
\includegraphics[width=0.65\textwidth]{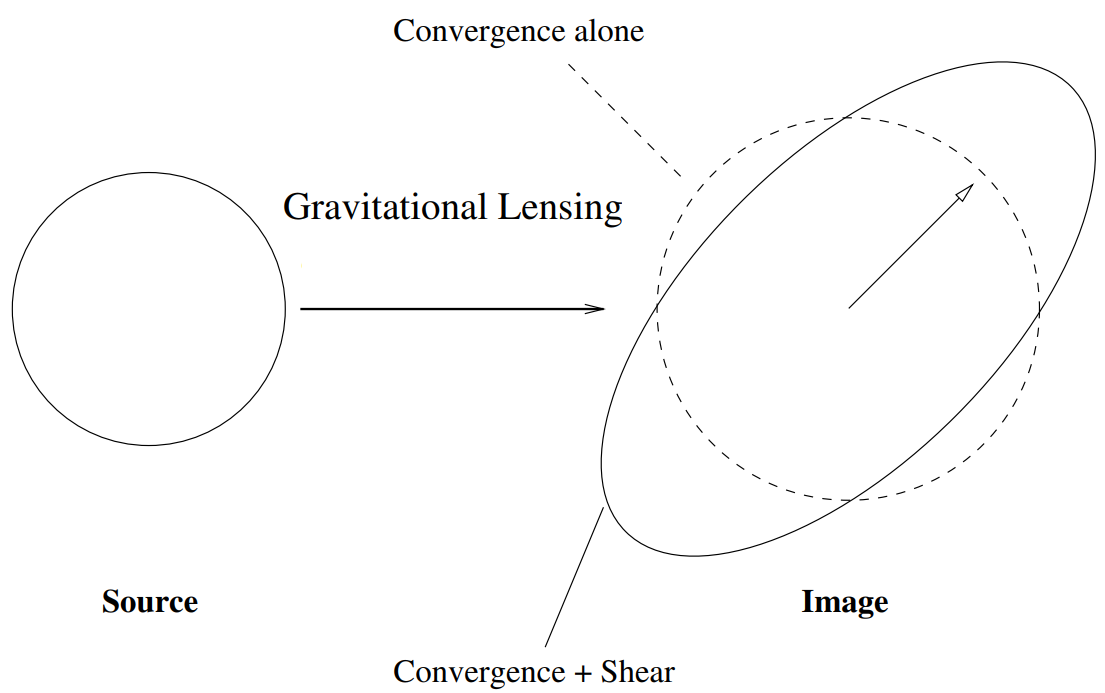}
\caption{Weak lensing effect of convergence and shear on a circular background object (left) as seen on the observer plane (right). Image extracted from~\cite{gravitational_lensing_weak}.}
\label{fig:chapter1_gravitational_lensing_weak_convergence_shear}
\end{figure}

Although not explicitly mentioned, Equations~\ref{eq:chapter1_gravitational_lensing_deflection_angle} and~\ref{eq:chapter1_gravitational_lensing_einstein_ring} are obtained assuming that the lens is a point mass $M$. For weak lensing, the lens is assumed to be extense, and depending on the distance to the centre of the lens, the enclosed mass will change, and with it the deflection angle. This makes the formalism somewhat more complicated, although the intuition is similar to that presented in strong lensing. For a detailed discussion of the weak lensing formalism, see~\cite{weak_gravitational_lensing_review}. For the purposes of this thesis, it is sufficient to highlight that in weak lensing, the transformation from the source plane to the observer plane is encoded in two quantities, known as convergence ($\kappa$) and shear ($\gamma$). Convergence describes the magnification of the background object, while shear describes the distortion (elongation or stretching) of the image. The effect of these two quantities can be seen in Figure~\ref{fig:chapter1_gravitational_lensing_weak_convergence_shear}. 

In particular, a very compelling piece of evidence in favour of Dark Matter comes from the dynamics of the merging of two galaxy clusters to form the so-called Bullet Cluster. Mapping the mass distribution of this merging through weak lensing studies on background galaxies show that the centre of the total mass is shifted with respect to the centre of the visible baryonic mass with a statistical significance of 8$\sigma$~\cite{Clowe}.

On the other hand, it is also worth mentioning the so-called cosmic shear, which refers to the weak lensing effects caused by the distribution of Dark Matter on the cosmological scale. The cosmic shear studies the distortion in galaxy shapes due to the cumulative effect of the large-scale structure of the universe. Analysis of distant galaxies through the cosmic shear provides a mapping of the Dark Matter of the universe~\cite{Kilbinger}.

\vspace{-3mm}
\subsection{Cosmic Microwave Background (CMB) and \texorpdfstring{\\}{ } Baryon Acoustic Oscillations (BAO)} \label{Chapter1_Evidence_CMB}

Both the CMB, microwave radiation that fills the entire observable universe, and BAO, ripples in the density of visible baryonic matter, are relics of the early universe and cornerstones of modern cosmology, and provide invaluable constraints on the composition of the universe and its evolution. Precise measurements of CMB anisotropies with NASA's Wilkinson Microwave Anisotropy Probe (WMAP) project~\cite{WMAP} and ESA's Planck satellite~\cite{planck_2006} have confirmed the so-called $\Lambda$CDM model of the universe, where Cold Dark Matter plays a fundamental role in the creation of observed structures and in the evolution of the universe. Similarly, sky surveys such as the Sloan Digital Sky Survey (SDSS)~\cite{SDSS} have provided observations and measurements of BAO, which are complementary to CMB measurements. 

\vspace{2mm}
\textbf{\normalsize Cosmology Review}
\vspace{0mm}

Assuming the cosmological principle (the universe is homogeneous and isotropic on the cosmological scale), the evolution of the universe is described by the Friedmann-Lemaître-Robertson-Walker (FLRW) metric:

\begin{equation}
    \mathrm{d}s^2 = - c^2 \mathrm{d}t^2 + a^2(t) \left[ \frac{\mathrm{d}r^2}{1-\kappa r^2} + r^2 \mathrm{d}\Omega^2 \right]
    \label{eq:chapter1_FLRW_metric}
\end{equation}

where $a(t)$ is the so-called scale factor, $\kappa$ represents the spatial curvature of the universe ($\kappa<1$ open universe, $\kappa=0$ flat universe, $\kappa>1$ closed universe), $r$ is the radial part and $\mathrm{d}\Omega^2 = \mathrm{d}\theta^2 + \sin^2\theta \mathrm{d}\varphi^2$ is the angular part. Applying the Einstein field equations, we obtain the Friedmann equations, which govern the evolution of the scale factor:

\vspace{-7mm}
\begin{equation}
\begin{split}
    H^2(t)&=\left(\frac{\dot{a}(t)}{a(t)}\right)^2 = \frac{8 \pi G }{3}\rho - \frac{\kappa c^2}{a^2(t)} + \frac{\Lambda c^2}{3} \\ \\
    & \frac{\ddot{a}(t)}{a(t)} =  -\frac{4 \pi G}{3}\left(\rho+\frac{3p}{c^2}\right) + \frac{\Lambda c^2}{3} \label{eq:chapter1_friedmann_equations}
\end{split}
\end{equation}

where $H(t)$ is the Hubble parameter\footnote{Many times, $H$ can be found expressed as $H=100h$~km/s/Mpc, where $h$ is dimensionless.}, $\rho(t)$ is the energy density, $\Lambda$ is the cosmological constant term and $p$ is the total pressure of the various components. It can be seen that by making the following change of variables:

\begin{equation}
    \rho \to \rho + \frac{\Lambda c^2}{8 \pi G} \qquad p \to p - \frac{\Lambda c^4}{8 \pi G} 
    \label{eq:chapter1_energy_density_variable_change}
\end{equation}

The cosmological constant term can be interpreted as a constant\footnote{Although in some cosmological models, $\Lambda$ depends on time.} energy density, $\rho_\Lambda = \Lambda c^2/8 \pi G$, that exerts negative pressure, $p=-c^2\rho_\Lambda $. This is the $\Lambda$ term in the $\Lambda$CDM model, and is used to model the dark energy of the universe. It is convenient to define the critical density, $\rho_c$, the energy density value for which the universe is flat ($\kappa=0$):

\begin{equation}
    \rho_c = \frac{3H^2}{8 \pi G}
    \label{eq:chapter1_critical_density}
\end{equation}

Defining the dimensionless parameter $\Omega = \rho/\rho_c$, and including the change of variables of Equation~\ref{eq:chapter1_energy_density_variable_change}, the first line of Equation~\ref{eq:chapter1_friedmann_equations} can be rewritten as:

\begin{equation}
    1 = \Omega (t) - \frac{\kappa c^2}{a^2(t)H^2(t)}
    \label{eq:chapter1_friedmann_equation_omega}
\end{equation}

where $\Omega (t)$ includes all forms of energy, $\Omega = \Omega_m + \Omega_r + \Omega_\Lambda$, with $\Omega_m$ the matter term, $\Omega_r$ radiation and $\Omega_\Lambda$ dark energy. Sometimes, the curvature term is also reabsorbed in the equation, defining a curvature density $\Omega_\kappa$.

\vspace{2mm}
\textbf{\normalsize Cosmic Microwave Background}
\vspace{0mm}

\begin{figure}[htb]
\centering
\includegraphics[width=0.8\textwidth]{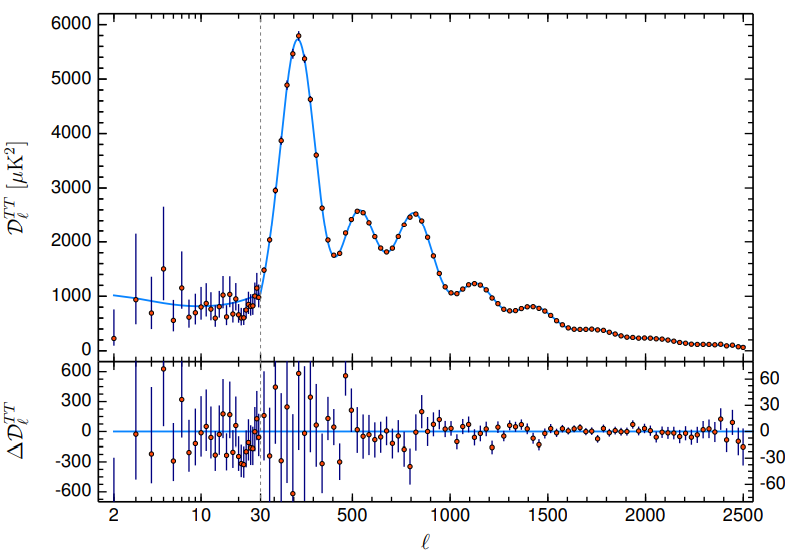}
\caption{Top: temperature power spectrum of the CMB, $D_\ell$, as a function of the multipole moment $\ell$. The red dots with error bars (68\% CL) are the experimental points extracted from Planck data, whereas the blue line is the best fit to the spectrum assuming a $\Lambda$CDM cosmological model. The grey line marks the transition from logarithmic to linear scale. Bottom: residuals of the best fit. Plot extracted from~\cite{planck_2020}.}
\label{fig:chapter1_planck_2018_power_spectrum}
\end{figure}

The CMB is the radiation left over from the early universe, dating back to about $t_r=$ 380000 years after the Big Bang. Before this time, radiation and matter (free protons and electrons) were in thermal equilibrium in a radiation-opaque plasma (Thomson scattering, explained in Section~\ref{Chapter3_Interactions_Photons}, was dominant). This plasma was fairly homogeneous, but it presented local overdensities of matter (in the inflationary picture, caused by primordial quantum fluctuations amplified by inflation). These overdensities led to the propagation of sound waves in the plasma due to the dynamics between gravitational attraction and pressure repulsion. At $t_r$, known as recombination, protons and electrons combined to form neutral atoms of hydrogen, leading to the decoupling of photons from matter (photons were allowed to travel freely). This remnant radiation is what we observe today as the CMB, redshifted due to the expansion of the universe. It is close to a perfect black body across all directions of the cosmos, with temperature $T_{\mathrm{CMB}}=2.7255\pm 0.0006$~K~\cite{CMB_temperature_2009}, but it presents some pretty small anisotropies on the order of $ \Delta T/T_{\mathrm{CMB}} \approx 10^{-5} $. These fluctuations vary from point to point, and are usually studied through the decomposition of the spatial map of the CMB into spherical harmonics $Y_{\ell m}$:

\begin{equation}
    \Delta T (\theta,\varphi)= T(\theta,\varphi)-T_{\mathrm{CMB}}  = \sum_{\ell,m} a_{\ell m}Y_{\ell m} (\theta,\varphi)
    \label{eq:chapter1_spherical_harmonics_decomposition}
\end{equation}

where $a_{\ell m}$ are the (complex) coefficients of the expansion. From this decomposition, and taking into account the principle of isotropy, the so-called power spectrum is defined as an average over $m$:

\begin{equation}
    C_\ell = \frac{1}{2\ell + 1}\sum_{m=-\ell}^{+\ell} \left|a_{\ell m}\right|^2
    \label{eq:chapter1_angular_power_spectrum}
\end{equation}

with $\ell$ the multipole moment of the spectrum, describing the angular size of the fluctuations ($\sim \pi/\ell$). This spectrum encodes the variance of the temperature fluctuations, and is usually presented as $D_\ell = \ell(\ell +1)C_\ell / (2\pi)$. In Figure~\ref{fig:chapter1_planck_2018_power_spectrum}, we can see the power spectrum $D_\ell$ of the CMB from the 2018 results~\cite{planck_2020} of the Planck mission.

It exhibits a series of peaks that provide information about the oscillations of the photon-baryon plasma prior to recombination. The position of the first peak ($\ell \approx 220$) agrees with a vision of a flat universe, which implies $\Omega \approx 1$ according to Equation~\ref{eq:chapter1_friedmann_equation_omega}. On the other hand, the relative height of the peaks depends on the total matter density, $\Omega_m$. It turns out that the gravitational potential necessary to produce the specific amplitude ratio computed from the data is larger than what baryonic matter ($\Omega_b$) alone could provide, which can be explained by including a form of matter that does not interact with light and clumps together, forming gravitational wells around matter overdensities. The best fit to the power spectrum data is provided by the $\Lambda$CDM cosmological model, in which most of the energy content of our flat universe is in the form of dark energy ($\Omega_\Lambda$), and most of the matter content is cold, non-baryonic matter that only interacts gravitationally ($\Omega_c$). Table~\ref{table:chapter1_cosmological_parameters_planck} shows some parameters extracted from the best fit to Planck 2018 data. Check Table 2 from~\cite{planck_2020} for the full set of parameters derived from the fit. The key conclusion for this thesis is that more than 2/3 of the universe is dark energy, whereas $\Omega_c \approx 0.26$ and $\Omega_b \approx 0.05$: not only do we need Dark Matter to explain our observations, on top of that it constitutes almost 85\% of all the matter content.

\begin{table}[htb]\centering
\begin{center}
\begin{tabular}{c|c}  \hline\hline
 \textbf{Parameter} &  \textbf{Value (68\% CL)} \\
 \hline
 $\Omega_\Lambda$ & $0.6847\pm 0.0073$ \\    
 \hline
 $\Omega_m$ & $0.3153\pm 0.0073$ \\ 
 \hline
 $\Omega_b h^2$ & $0.02237\pm 0.00015$ \\
 \hline
 $\Omega_c h^2$ & $0.1200\pm 0.0012$ \\
 \hline
 $H_0$ (km/s/Mpc) & $67.36\pm 0.34$ \\
 \hline
 r$_*$ (Mpc) & $144.43\pm 0.26$ \\
 \hline
 r$_{\mathrm{drag}}$ (Mpc) & $147.09\pm 0.26$ \\
 \hline\hline
\end{tabular}    
\end{center}
\caption{Present-day values of some cosmological parameters, derived from the fit to the data shown in Figure~\ref{fig:chapter1_planck_2018_power_spectrum}. Values extracted from~\cite{planck_2020}.}
\label{table:chapter1_cosmological_parameters_planck}
\end{table}

\vspace{0mm}
\textbf{\normalsize Baryon Acoustic Oscillations}
\vspace{0mm}

The same acoustic waves responsible for the CMB power spectrum peaks also left an imprint on the large-scale structure of the universe. In particular, BAO materialise as a preferential scale in galaxy clustering, and are used as a standard ruler in cosmological distance measurements. The BAO scale is defined by the distance travelled by the sound waves between the end of inflation and the decoupling of baryons from photons after recombination:

\begin{equation}
    r_{\mathrm{drag}} = \int_0^{t_{\mathrm{drag}}}\frac{c_s(t)}{a(t)}dt
    \label{eq:chapter1_BAO_scale_sound_horizon}
\end{equation}

where $c_s(t)$ is the speed of the sound waves and $t_{\mathrm{drag}}$ is the time of decoupling of baryons from photons, also called the drag epoch. Although similar, the drag epoch should not be confused with the time of last scattering $t_*$, defined as the time of decoupling of photons from baryons. Technically, $t_*$ is the time when the photon optical depth is 1, while $t_{\mathrm{drag}}$ is the time when the baryon optical depth is 1. Since there are so many more photons than baryons, the photons decoupled from baryons first, while the baryons continued to "see" the photons for a slightly longer time. This translates to $r_* < r_{\mathrm{drag}}$, with $r_*$ defined similarly to $r_{\mathrm{drag}}$. The Planck 2018 (present-day) values of these two quantities are given in Table~\ref{table:chapter1_cosmological_parameters_planck}. 

\begin{figure}[htb]
\centering
\includegraphics[width=0.8\textwidth]{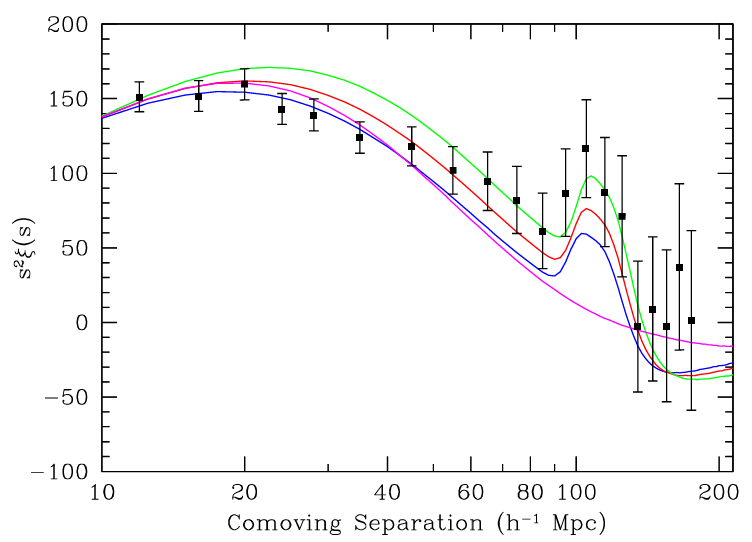}
\caption{Large-scale correlation function (in units of $s^2\xi(s)$) as a function of the comoving distance $s$. The black dots with error bars are the experimental points from SDSS data, the pink line represents an only-CDM model with $\Omega_m h^2=\Omega_c h^2=0.105$ and the green, red and blue lines are fits to the data using a $\Lambda$CDM model with $\Omega_m h^2 = 0.12$, $\Omega_m h^2 = 0.13$ and $\Omega_m h^2 = 0.14$, respectively. The three of them have $\Omega_b h^2 = 0.024$. The bump at around $s=100h^{-1}$~Mpc is the BAO scale, which cannot be explained by Dark Matter alone (pink line), but needs a $\Lambda$CDM model to have a good fit. Plot extracted from~\cite{Eisenstein_2005_BAO}.}
\label{fig:chapter1_BAO_large_scale_correlation_function}
\end{figure}

The way to search for the ripples of acoustic waves that froze in the drag epoch is to look at the clustering of galaxies in astronomical surveys such as SDSS. The formalism for describing the level of clustering is the so-called two-point correlation function $\xi(s)$. Without going into more detail, this function is defined as a measure of the excess probability, compared over a random occurrence, of finding a galaxy in a fixed volume element at a separation $s$ from another galaxy. The study of this function with SDSS data allowed the detection of the baryon acoustic peak for the first time in 2005~\cite{Eisenstein_2005_BAO}. Figure~\ref{fig:chapter1_BAO_large_scale_correlation_function} shows this result: the bump around $s=100h^{-1}$~Mpc is the preferential BAO scale, which taking $h\approx 0.68$ implies $s\approx 147$~Mpc, in line with the known value of $r_{\mathrm{drag}}$. The fit to these data requires a model containing both Dark Matter and baryonic matter (green, red and blue lines), reinforcing the idea of the $\Lambda$CDM model. However, a model with Dark Matter alone is not able to explain the excess (pink line).

In general, BAO studies in the latest generations of astronomical surveys complement the Planck 2018 CMB results (and results from other sources, such as supernovae data), helping to place more restrictive constraints on different cosmological parameters. For example, in~\cite{BAO_SDSS}, the combined results of the entire lineage of experiments that have used data from the completed SDSS (SDSS-I, SDSS-II, BOSS and eBOSS) are reported. Overall, all analyses point to a flat $\Lambda$CDM model. 

\vspace{2mm}
\textbf{\normalsize Shortcomings of $\Lambda$CDM}
\vspace{0mm}

Cold Dark Matter is a fundamental ingredient of the best cosmological model available to date, $\Lambda$CDM. However, this description is not perfect and has some discrepancies with observational evidence. Here we briefly describe some of the most relevant tensions in the CDM paradigm:

\begin{itemize}
    \item \textbf{Missing dwarf galaxies problem}: CDM simulations predict a much larger number of dwarf galaxies around large galaxies like the Milky Way than observed, indicating either an imperfect understanding of galaxy formation or a flaw in the CDM's explanation of small-scale structures.
    \item \textbf{Core-cusp problem}: already mentioned in Section~\ref{Chapter1_Evidence_GalacticRotation}, it refers to the problem of density distributions in galaxies: CDM simulations are in agreement with NFW-type density profiles that have a steep "cusp" around the origin, while the inferred distributions for galaxies do not always match this profile. In particular, dwarf galaxies have a flatter core than predicted by CDM.
    \item \textbf{Early formation problem}: according to the CDM paradigm, galaxy formation follows a hierarchical process whereby massive galaxies arise from successive mergers of Dark Matter halos of smaller galaxies. Recently, observations from the James Webb Space Telescope (JWST) revealed galaxies at $z>10$ redshifts. The presence of such massive structures in the early universe seems to contradict CDM simulations, according to which galaxy formation rates should be slower~\cite{JWST_tension_CDM_2023}. 
\end{itemize}

The presence of these and other challenges has led to refining the CDM model. In some cases, modifications such as the addition of a Warm Dark Matter component (such as sterile neutrinos, see Section~\ref{Chapter1_Explanations_sterile_neutrinos}) are enough to relieve some of these tensions.

\section{Explanations for the Dark Matter Problem} \label{Chapter1_Explanations_Dark_Matter}

If we refer to the Dark Matter problem more generally, meaning the discrepancies between observations and our current models, there are two main solutions to address the issue:

\begin{itemize}
    \item \textbf{The particle nature of Dark Matter}: this is the most widely accepted hypothesis, and consists in assuming that there really exists an invisible mass that we would call Dark Matter, composed of particles different from those that make up the Standard Model of particle physics. These particles would interact very weakly with ordinary matter (their main interaction mechanism would be gravity), but they could be detected through the weak interaction in dedicated experiments with sufficient sensitivity. A multitude of candidates have been proposed, including Weakly Interacting Massive Particles (WIMPs), axions or sterile neutrinos. There is also the possibility that Dark Matter is composed of Primordial Black Holes (PBHs). These are not particles \textit{per se}, but they are massive, non-luminous objects that fit the description of Dark Matter. The consensus is that, if this hypothesis is true, Dark Matter would not be monocomponent, but could be composed of a mix of PBHs and different particles, which could form part of a hypothetical dark sector yet to be discovered.
    \item \textbf{Modification of the Laws of Gravity}: in contrast to the particle hypothesis, alternative theories posit that discrepancies in observations attributed to Dark Matter are actually a consequence of an incomplete description of gravity on galactic and cosmological scales. In this framework, theories such as MOdified Newtonian Dynamics (MOND) and its relativistic generalisation Tensor-Vector-Scalar (TeVeS) gravity try to explain the various phenomena of Section~\ref{Chapter1_Evidence} without resorting to non-visible matter.
\end{itemize}

In this section, we will focus mainly on the different particles hypothesised to make up Dark Matter, both because one of them is the foundation of this thesis, and because it is the paradigm that best explains the phenomena discussed in Section~\ref{Chapter1_Evidence}. We will discuss the theoretical bases, evidence and experiments associated with some of the most compelling candidates. However, Subsection~\ref{Chapter1_Explanations_modified_gravity_alternative} is devoted to the hypothesis of gravity modification, pointing out some successes of alternative theories.

\subsection{Weakly Interacting Massive Particles} \label{Chapter1_Explanations_WIMPs}

WIMPs are among the most intensively studied Dark Matter candidates. They are hypothetical new particles appearing in extensions of the Standard Model that would interact only gravitationally and via the weak force, with masses typically in the range $10\mathrm{~GeV}-10\mathrm{~TeV}$ (though lower masses are currently under study). They are interesting from the point of view of cosmology because they could naturally explain the current observed abundance of Dark Matter through the freeze-out mechanism (in what is known as the \textit{WIMP miracle}). Their detection strategies are divided into direct searches (interactions with Standard Model particles, depositing part of their energy), indirect searches (WIMP-WIMP annihilations leading to detectable Standard Model products) and production in accelerators (reverse process to annihilation). Given their relevance to this thesis, their detailed discussion is left for Chapter~\ref{Chapter2_WIMPs}.

\subsection{Axions} \label{Chapter1_Explanations_axions}

Axions are particles that arise when solving the strong CP problem in Quantum ChromoDynamics (QCD) using the Peccei-Quinn mechanism. Apart from their theoretical motivation, they are good candidates for composing Cold Dark Matter due to their unique properties and compatibility with astrophysical and cosmological observations.

The Lagrangian of the Standard Model contains a CP-violating term:

\begin{equation}
    \mathcal{L}_\theta = -\theta\frac{g_s^2}{32\pi^2}G^a_{\mu\nu}\tilde{G}^{a,\mu\nu}
    \label{eq:chapter1_lagrangian_CP_term}
\end{equation}

where $G^a_{\mu\nu}$ is the gluon tensor field, $\tilde{G}^{a,\mu\nu}$ is its dual, $g_s$ the strong coupling constant and $\theta$ is a dimensionless parameter. This term implies a non-zero electric dipole moment for the neutron:

\begin{equation}
    d_n \sim \frac{e m_u m_d}{(m_u + m_d)m_n^2}\theta
    \label{eq:chapter1_electric_dipole_moment_neutron}
\end{equation}

where $e$ is the electric charge, and $m_u$, $m_d$ and $m_n$ are the masses of the up quark, down quark and neutron, respectively. Experimental measurements have found $d_n = (0.0\pm 1.1_{\mathrm{stat}}\pm 0.2_{\mathrm{sys}})\times 10^{-26}e\cdot \mathrm{cm}$~\cite{nEDM_2020}, which implies $\theta \lesssim 10^{-10}$. This fine-tuning of $\theta$ is what is known as the strong CP problem.

The Peccei-Quinn mechanism solves this problem dynamically. The solution consists in introducing a new field and a global anomalous chiral symmetry $U(1)_{PQ}$ which breaks spontaneously at an energy scale $f_a$. This symmetry breaking gives rise to a pseudo-Nambu-Goldstone boson, the axion. The axion field $a$ is coupled to the gluons through the term:

\begin{equation}
    \mathcal{L}_a = -\frac{a}{f_a}\frac{g_s^2}{32\pi^2}G^a_{\mu\nu}\tilde{G}^{a,\mu\nu}
    \label{eq:chapter1_lagrangian_axion_CP_term}
\end{equation}

The axion potential is minimized when the effective parameter $\theta_{\mathrm{eff}}=\theta + \langle a \rangle/f_a$ is zero. Thus, the vacuum expectation value of the axion, $\langle a \rangle$, takes the value $\langle a \rangle=-\theta f_a$, with axions being excitations of $a$ around this value of $\langle a \rangle$. It is therefore natural to redefine $a \rightarrow a + \langle a \rangle$, with the consequent cancellation of the CP-violating term in the Lagrangian.

Initially, it was thought that it was necessary for $f_a$ to be on the order of the energy of the electroweak scale for the theory to work, but such an axion would have a heavy mass which would have been discovered by now. Later, it was realised that this condition was not necessary, and much lighter axions were possible, such as those in the KSVZ~\cite{Kim_1979, SVZ_1980} and DFSZ~\cite{DFS_1981, Zhitnitsky_1980} models. These axions came to be called invisible axions, because they evaded the experimental and astrophysical constraints and were more weakly coupled than the original axions. The axion mass $m_a$ is related to $f_a$ and the QCD scale $\Lambda_{\mathrm{QCD}}\sim 200$~MeV as follows:

\begin{equation}
    m_a \sim \frac{\Lambda_{\mathrm{QCD}}^2}{f_a}
    \label{eq:chapter1_axion_mass}
\end{equation}

The axion mass can range from $\sim$ eV down to peV for high enough energy scales $f_a$, or even lower for what are called ultralight axions. Axions interact weakly with light through the coupling:

\begin{equation}
    \mathcal{L}_{a\gamma} = -\frac{1}{4}g_{a\gamma}aF_{\mu\nu}\tilde{F}^{\mu\nu} = g_{a\gamma}a\vec{E}\cdot \vec{B}
    \label{eq:chapter1_axion_photon_lagrangian}
\end{equation}

where $g_{a\gamma}\sim 1/f_a$ determines the strength of the interaction, which is very small for the allowed mass range. The conversion $a\longleftrightarrow \gamma $ inside a strong magnetic field $B$, often called Primakoff effect, is the main mechanism exploited to detect axions.

On top of constituting a good solution to the strong CP problem, axions are considered good Dark Matter candidates due to their low mass and weak interaction with matter. It is also worth noting that Equation~\ref{eq:chapter1_axion_mass} is valid for QCD axions, related to the Peccei-Quinn mechanism. However, there is a general class of particles, called Axion-Like Particles (ALPs), that are not tied to the QCD scale and, therefore, do not have a specific dependence between $m_a$ and $f_a$. ALPs are more flexible Dark Matter candidates, and arise in the context of String Theory or in the general context of models that extend the Standard Model.

Axions could be produced in the early universe mainly through two mechanisms:

\begin{itemize}
    \item \textbf{Vacuum Realignment}: it refers to the relaxation of the axion field towards the minimum during the Peccei-Quinn symmetry breaking.
    \item \textbf{Topological defects}: the decay of topological defects associated to the axion field, such as cosmic strings or domain walls, would give rise to a population of axions.
\end{itemize}

These production modes could yield the right amount of non-relativistic Dark Matter. An in-depth look into the production of axions is beyond the scope of this thesis, but~\cite{axion_cosmology_2016} can be checked for more information on the subject.

\begin{figure}[htb]
\centering
\includegraphics[width=0.85\textwidth]{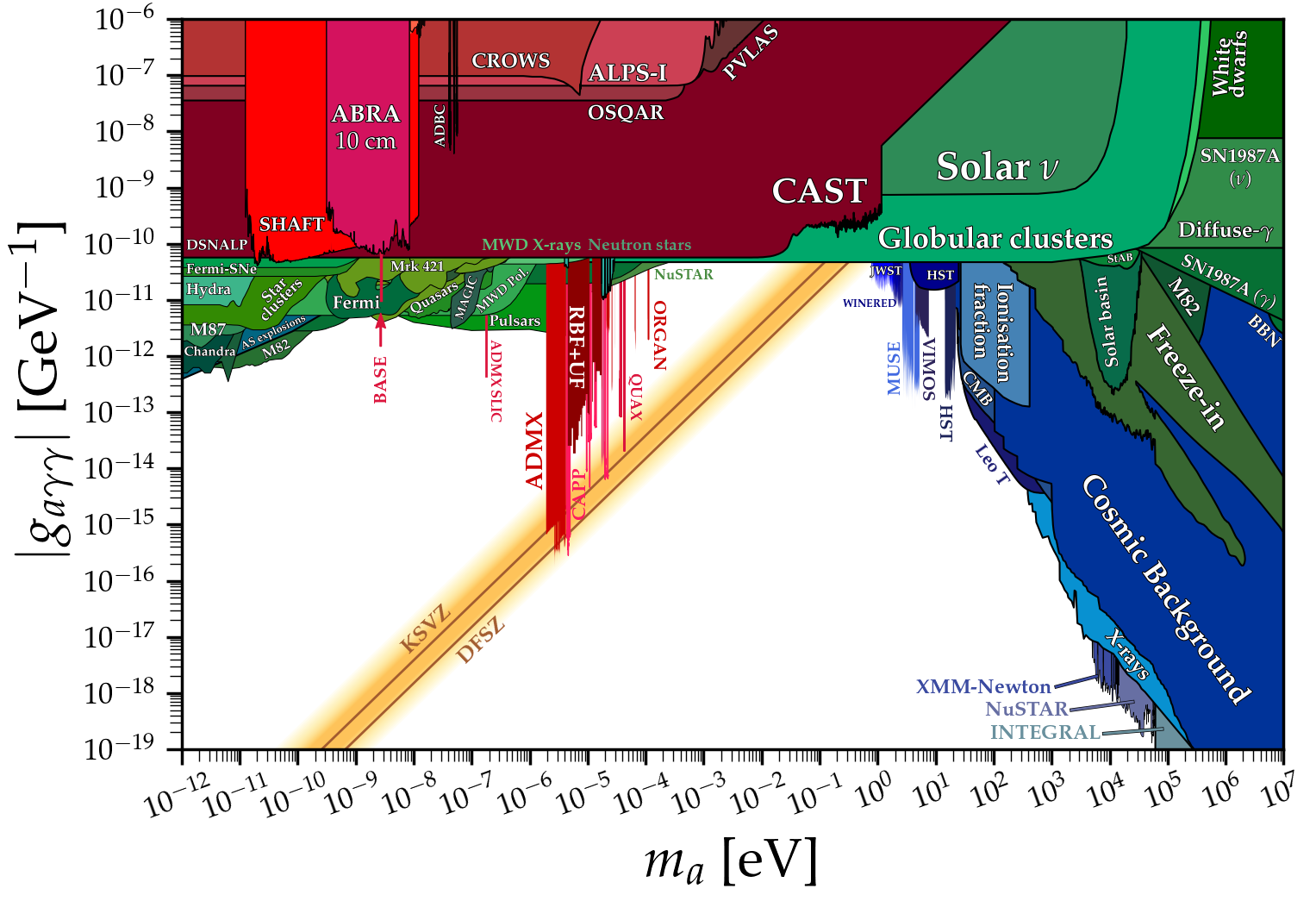}
\caption{Constraints to the axion-photon coupling as a function of the axion mass. The bounds are organised into three categories: astrophysical searches for axions \textit{not as Dark Matter} (green), cosmological/astrophysical bounds for axions \textit{as Dark Matter} (blue) and laboratory searches (red). This plot, together with all the references from which the various constraints have been drawned, can be found in~\cite{Axion_Limits}.}
\label{fig:chapter1_axion_photon_parameter_space}
\end{figure}

The search for axions and ALPs is mainly focused on the axion-photon interaction, and represented in parameter space $(m_a,g_{a\gamma})$, but other interactions such as axion-electron are also studied (see~\cite{axion_electron_2024} for a review on fermion-coupled axions). Here, we will limit ourselves to axion-photon couplings. The references for the most restrictive constraints from different sources can be found on~\cite{Axion_Limits}, and they are shown in Figure~\ref{fig:chapter1_axion_photon_parameter_space}. These can be broadly classified into astrophysical searches for axions \textit{not as Dark Matter} (green), cosmological/astrophysical bounds for axions \textit{as Dark Matter} (blue) and laboratory searches (red). Each of them is explained below.

\vspace{-2mm}
\textbf{\normalsize Astrophysical Constraints}
\vspace{0mm}

Astrophysical systems with intense magnetic fields and known stellar evolution are natural probes for axions produced by the Primakoff effect, providing constraints on $g_{a\gamma}$ (green in Figure~\ref{fig:chapter1_axion_photon_parameter_space}).

An example of such a system is globular clusters, dense stellar systems with well-known evolutionary processes. In this context, $R$ is usually defined as the number of horizontal branch stars divided by the number of red giants within the cluster. Energy losses via photon-axion conversion in stellar nuclei would mainly affect stars pertaining to the horizontal branch, thus shortening the lifetime of this phase and, therefore, reducing the $R$ ratio.

Another astrophysical system used is white dwarfs, the last stage of stellar evolution in low-mass stars. In white dwarfs, since the nuclear fuel has been exhausted, there is no heat source, and the only thing left is gradual cooling. The main process of energy loss in young white dwarfs is neutrino emission (Urca process). However, the existence of axions would add an extra cooling channel through photon-axion conversion. Observations of white dwarf luminosity functions suggest faster cooling rates than expected if only Standard Model processes are included, imposing a bound on $g_{a\gamma}$.

\vspace{2mm}
\textbf{\normalsize Cosmological Constraints}
\vspace{0mm}

If Dark Matter is composed of axions, they could have an impact on astrophysical systems or cosmological observations (blue in Figure~\ref{fig:chapter1_axion_photon_parameter_space}). An example would be birefringence, a phenomenon by which photons would rotate their plane of polarisation as they propagate through cosmic magnetic fields due to axion-photon mixing. This effect is studied in the CMB spectrum, where it would induce B-mode polarisation.

\vspace{2mm}
\textbf{\normalsize Laboratory Searches}
\vspace{0mm}

The direct search for axions via Primakoff effect in a laboratory environment (red in Figure~\ref{fig:chapter1_axion_photon_parameter_space}) can be grouped into three main categories: helioscopes, haloscopes and Light-Shining-through-Wall (LSW) experiments.

\begin{itemize}
    \item \textbf{Helioscopes}: they look for axions produced in the solar core via Primakoff effect by trying to convert them back into photons in the magnetic field of a magnet in a laboratory setting. Helioscopes base their analysis on the solar axion flux models. CAST (CERN Axion Solar Telescope), the most sensitive helioscope to date, used a 9-T, 10-m-long magnet to try to convert axions to photons by pointing at the sun during sunrise and sunset. CAST was active from 2003 to 2021, going through different phases and publishing different results. The most recent result of this extensive physics campaign is reported in~\cite{CAST_2024}, resulting in the most stringent experimental upper limit to $g_{a\gamma}$. The International Axion Observatory (IAXO), and its intermediate stage BabyIAXO (already under construction at DESY), seek to improve the sensitivity to the axion-photon coupling, with projections of $g_{a\gamma}\sim 1.5\times 10^{-11}$~GeV$^{-1}$ for BabyIAXO and $g_{a\gamma}\sim 10^{-12}$~GeV$^{-1}$ for IAXO~\cite{BabyIAXO_2021}, which means an improvement of about an order of magnitude with respect to CAST.
    \item \textbf{Haloscopes}: they are designed to detect axions from the galactic halo by converting them into photons in a resonant cavity immersed in a strong magnetic field. Analyses in haloscopes rely on axion flux models of the galactic Dark Matter halo. The main haloscope is ADMX (Axion Dark Matter eXperiment), which uses resonant cavities in the microwave range to convert axions into photons. It is most sensitive to axion masses in the $10^{-6}$~eV to $10^{-4}$~eV range.
    \item \textbf{LSW experiments}: they try to detect axions by a process of photon-axion-photon conversion: in essence, a laser immersed in a magnetic field produces axions, which pass through a wall that is opaque to the laser photons; on the other side of the wall, these axions are reconverted into photons, which are detected as an excess over the detector background. These are the purest laboratory experiments, relying neither on models of axion production in the sun nor on models of Dark Matter. Examples of LSW experiments are ALPS (Any Light Particle Search) at DESY or OSQAR (Optical Search of QED vacuum magnetic birefringence, Axion and photon Regeneration) at CERN.
\end{itemize}

\vspace{-4mm}
\subsection{Sterile Neutrinos} \label{Chapter1_Explanations_sterile_neutrinos}

Sterile neutrinos are hypothetical particles that appear in extensions to the Standard Model. Unlike active neutrinos (which have weak isospin $\pm 1/2$), sterile neutrinos do not interact with the gauge bosons of the Standard Model. Their existence is motivated by the neutrino mass problem, and they appear by providing a solution to it through the seesaw mechanism\footnote{In a very simplified form, the neutrino mass problem arises because neutrinos were assumed to be massless according to the Standard Model, but experimental evidence from neutrino oscillations shows that they have small but non-zero masses. The seesaw mechanism is a solution to this problem, and consists in the introduction of heavy, right-handed neutrinos (sterile neutrinos), which interact with the active neutrinos. This interaction produces two mass eigenstates, a heavy one (associated with the sterile neutrino) and a light one (associated with the active neutrino).}.

Therefore, sterile neutrinos would only interact gravitationally with Standard Model particles, and through weak mixing with active neutrinos, with an interaction strength encoded by the mixing angle $\theta$. This makes them good candidates for Dark Matter. In the different models proposed, sterile neutrinos have masses ranging from the keV scale to the GeV order. However, the typical Dark Matter candidate has a mass in the $1\mathrm{~keV}\lesssim m_s\lesssim 50\mathrm{~keV}$ range, and falls into the category of what is known as Warm Dark Matter. Unlike Cold Dark Matter, which includes non-relativistic particles at the time of early universe structure formation, Warm Dark Matter refers to lighter, non-relativistic particles but with sufficient velocity to suppress the formation of smaller-scale structures such as dwarf galaxies.

In order to be Dark Matter, sterile neutrinos must be stable on cosmological scales. Since their main decay channel is the radiative decay $\nu_s \rightarrow ~ \nu_a + \gamma$, it must have a half-life on the order of the age of the universe. This property is equivalent to imposing a very small mixing angle ($\sin^2(2\theta)\ll 1$), fulfilled for sterile neutrinos with $m_s\sim$~keV.

The thermal\footnote{There are other mechanisms that do not involve the thermal production of sterile neutrinos through the weak interaction (possible thanks to the mixing $\theta$ with active neutrinos), such as out-of-equilibrium decay of heavier particles. However, we focus on thermal production as it is the paradigmatic and most studied option in the sterile neutrino field.} production of sterile neutrinos in the early universe is hypothesised to be driven by two main mechanisms:

\begin{itemize}
    \item \textbf{Non-resonant production}: also called the Dodelson-Widrow (DW) mechanism, it consists in the production via mixing with active neutrinos by oscillations. This mechanism can occur without incurring a large lepton asymmetry, and the efficiency of sterile neutrino production depends on the mixing angle $\theta$ and the mass $m_s$. As we will see below, this option is very constrained by the current state of the parameter space.
    \item \textbf{Resonant production}: also called the Shi-Fuller mechanism, it requires a large lepton asymmetry to enhance the DW process by resonant oscillations. The resonance is due to the modification of the neutrino potential by the lepton asymmetry, which enhances the conversion efficiency of active neutrinos into sterile neutrinos. It is motivated by the fact that the lepton asymmetry $n_L$ is not as constrained as the baryon asymmetry $n_B$, although a fine-tuning of $n_L$ is necessary for the resonant mechanism to be relevant.
\end{itemize}

The main sources of observational evidence for sterile neutrinos come from indirect detection methods or astrophysical/cosmological constraints:

\begin{itemize}
    \item \textbf{X-ray observations}: the idea is to detect the photon produced by radiative decay via $\nu_s \rightarrow ~ \nu_a + \gamma$. Considering $m_s\gg m_a$, conservation of energy and momentum imposes that the photon produced is monoenergetic with energy $E_\gamma = m_s/2$, which is in the X-ray range for typical keV-scale sterile neutrinos. This rare decay is trying to be captured with X-ray telescopes such as Chandra~\cite{Chandra_2012}, XMM-Newton~\cite{XMM-Newton_2017} or NuSTAR~\cite{NuSTAR_2017}, among others. In fact, a line was recently detected at 3.5~keV that has been interpreted as a sterile neutrino with mass $m_s \sim 7$~keV. This line has been confirmed independently with XMM-Newton and Chandra in galaxy clusters~\cite{Galaxy_clusters_3.5keV_sterile_neutrino_2014}, as well as in the Andromeda~\cite{Andromeda_3.5keV_sterile_neutrino_2014} and Milky Way~\cite{Milky_Way_3.5keV_sterile_neutrino_2015} galaxies. However, the signal is not clear-cut, because it is at the resolution limit of some telescopes, and because it admits alternative explanations. New missions such as Athena~\cite{Athena_2016} are expected to shed light on this matter.
    \item \textbf{Phase-space arguments}: the value of the phase-space density of Dark Matter in galaxies cannot be higher than that of a degenerate Fermi gas due to the Pauli exclusion principle~\cite{sterile_neutrinos_review_2018}. This imposes a lower limit on the mass of the sterile neutrino, which is model-independent. If, in addition, a particular primordial distribution function (dependent on the production mechanism) is assumed, the mass range can be further constrained to $m_s>1.7$~keV~\cite{Boyarsky_phase_space_sterile_neutrinos_2009}.
    \item \textbf{Structure formation}: the study of the Lyman-$\alpha$ forest\footnote{The Lyman-$\alpha$ forest is the set of absorption lines that appear between the Lyman-$\alpha$ transition of neutral hydrogen and the redshifted wavelength of this transition in the spectrum of a quasar or a distant galaxy. These absorption lines are produced by clouds of neutral hydrogen between the observer and the object. Different redshifted Lyman-$\alpha$ lines are seen, corresponding to clouds at different distances.} is useful to probe the properties of the intergalactic medium. Different regions of this medium are at different distances, and thus provide information about different stages in the evolution of the universe. Comparison of the Lyman-$\alpha$ forest with hydrodynamical simulations of large-scale structure formation (which assume a particular matter distribution) allows different models of Dark Matter to be studied. For example, experiments like eBOSS of SDSS have put a limit on the mass of resonantly produced sterile neutrinos~\cite{eBOSS_Lyalpha_sterile_neutrinos_2017}.
    \item \textbf{Big Bang Nucleosynthesis}: in the case of resonantly-enhanced production, a significant lepton asymmetry would result in an increase of the effective neutrino density during nucleosynthesis, which would alter the known abundances of the light elements~\cite{BBN_constraints_sterile_neutrinos_2005}.
    \item \textbf{Laboratory constraints}: in beta-decay experiments, sterile neutrinos could alter the energy spectrum of the emitted electrons. Experiments like KATRIN~\cite{KATRIN_2005} are focused on studying the beta decay of tritium, one of the least energetic beta decays (with $Q_\beta = 18.6$~keV), and among their goals is to look for distortions in the spectrum that could point to the existence of sterile neutrinos. Projections of mixing angle $\theta$ sensitivity in experiments such as KATRIN's TRISTAN upgrade have been studied in, for example,~\cite{TRISTAN_sensitivity_2015}.
\end{itemize}

\begin{figure}[htb]
\centering
\includegraphics[width=0.9\textwidth]{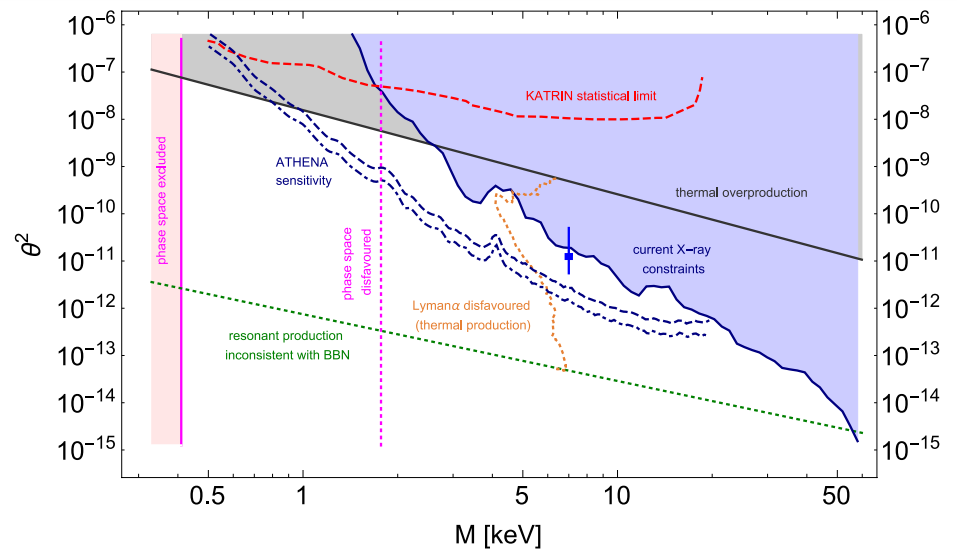}
\caption{Constraints of the Dark Matter window for sterile neutrinos in the parameter space $(m_s, \theta^2)$. The different bounds come from phase-space considerations (model-independent, solid pink line, and model-dependent, dashed pink line), X-ray bounds (solid blue line), X-ray projections (dashed blue line), BBN constraints (resonant production, dashed green line), structure formation constraints (resonant production, dashed orange line) and experimental projections from KATRIN (dashed red line). The black solid line represents the non-resonant production that produces the right amount of Dark Matter consistent with our observations (above that line there is an overproduction). The blue marker with error bars is the claim $m_s \approx 7$~keV from X-ray observations, at the limit of current sensitivity. Plot taken from~\cite{sterile_neutrinos_review_2018}.}
\label{fig:chapter1_sterile_neutrinos_parameter_space_constraints}
\end{figure}

Figure~\ref{fig:chapter1_sterile_neutrinos_parameter_space_constraints} shows the constraints accumulated using the methods described here. As can be seen, the parameter space for sterile neutrinos is quite restricted, which means that it is not currently considered as the most likely Dark Matter candidate. In fact, the non-resonant production mechanism is practically ruled out: the solid black line represents the production by this mechanism to give the correct Dark Matter density (above this line it would produce more Dark Matter than observed), while the dashed pink line gives a lower bound to $m_s$ due to phase-space considerations. Considering the X-ray constraints, this mechanism is practically limited to a window of $2\mathrm{~keV}\lesssim m_s \lesssim 3\mathrm{~keV}$. Therefore, the most favoured model at present is the resonant production, as it extends the parameter space up to the green dashed line (constraint due to BBN). However, it is also rather limited, as the Lyman-$\alpha$ forest bounds (dashed orange line, calculated with resonant production) are more restrictive than the phase-space considerations. The blue solid dot shows the already discussed claim of $m_s \approx 7$~keV due to X-ray observations, which is at the sensitivity limits, and will be clarified by ATHENA.

\subsection{Primordial Black Holes} \label{Chapter1_Explanations_PBHs}

PBHs are a type of black hole hypothesised to have formed in the early universe, due to the collapse of large density fluctuations resulting from perturbations in the inflationary epoch. This differentiates them, for example, from black holes resulting from stellar evolution. To obtain PBHs, the density contrast $\delta = (\rho-\bar{\rho})/\bar{\rho}$, where $\rho$ is the local overdensity and $\bar{\rho}$ the average background density, has to exceed a certain threshold $\delta_c$, which is estimated to be $\delta_c \approx 0.45$~\cite{PBHs_lecture_notes}. This implies that $\rho \sim \bar{\rho}$. On the other hand, the mass $M$ to form a black hole has to be concentrated within the Schwarzschild radius $R_S$, $\rho = M/R_S^3 \sim c^6/(G^3M^2)$. Combining this with the observation that the average background density coincides with the cosmological density at a time $t$ after the Big Bang, $\bar{\rho}\sim 1/(Gt^2)$ (valid for a flat, radiation-dominated universe like the early universe), we have:

\begin{equation}
    M\sim \frac{c^3 t}{G} \sim 10^{15} \left( \frac{t}{10^{-23}\mathrm{~s}} \right)\mathrm{~g} \sim M_\odot \left( \frac{t}{10^{-5}\mathrm{~s}} \right)
    \label{eq:chapter1_PBHs_mass}
\end{equation}

where $M_\odot =10^{33}$~g is the solar mass. The range of possible masses for PBHs is very wide, spanning masses from the Planck scale, $M=M_P \sim 10^{-5}$~g (corresponding to the Planck time $t_P \sim 10^{-43}$~s), to $M\sim 10^{51}$~g (corresponding to the recombination epoch, $10^{13}$~s), or even larger.

PBHs are well motivated from a theoretical point of view, as they could be the seed for supermassive black holes ($10^6-10^{10}$~$M_\odot$) in galactic centres~\cite{PBHs_lecture_notes}. Also, they are a solid Dark Matter candidate, as they possess the properties of being massive, non-baryonic\footnote{Normal stellar black holes cannot compose the totality of Dark Matter, since their formation comes from baryonic matter, which is subject to the constraint $\Omega_b \approx 0.05$.}, and they interact mainly via gravity.

There are different strategies to search for PBHs:

\begin{itemize}
    \item \textbf{Evaporation}: if PBHs were Dark Matter, they should be stable on scales larger than the age of the universe. It is known that PBHs with masses $M<10^{15}$ g would have evaporated via Hawking radiation by now~\cite{Hawking_1975}, and those within the range $10^{15}$~g $\lesssim M\lesssim 10^{17}$~g would be evaporating today. The latter would emit gamma rays that should be visible over the background. This excess has been investigated in data from telescopes such as FermiLAT~\cite{FermiLAT_2018_PBHs}.
    \item \textbf{Microlensing}: this effect has the same basis as the strong and weak gravitational lensing explained in Section~\ref{Chapter1_Evidence_GravitationalLensing}. However, unlike these, where the lens is a sufficiently massive object to observe changes in the direction of the light from the source, in microlensing the lens is a low-mass, compact object, with mass on the planetary or stellar order, such as PBHs. Detecting the deflection of light due to such objects is difficult, but the change in source brightness can be tracked, as the small size of the lens makes the source-lens alignment dynamic on human scales (from seconds to years rather than millions of years). This transient event is rare, and is therefore sought after in large astronomical surveys such as OGLE~\cite{OGLE_2019}.
    \item \textbf{Cosmic Microwave Background}: if PBHs were present in the early universe, mass accretion would alter the statistical properties of the CMB temperature and polarisation map, which would place limits on their abundance. A review on these bounds and how to derive them can be found in~\cite{PBHs_CMB_constraints_2024}.
    \item \textbf{Gravitational Waves}: the merger of PBHs would produce gravitational waves detectable in interferometers such as LIGO/Virgo/KAGRA. An update with the latest results of the search for PBHs with data from the O3 run of the LIGO-Virgo-KAGRA network can be found in~\cite{LIGO_Virgo_KAGRA_PBHs_2024}.
    \item \textbf{Dynamical effects}: PBHs would have dynamical effects on stars, wide binaries or dwarf galaxies. Many of these effects would involve the heating or destruction of astronomical systems by the passage of nearby PBHs~\cite{PBHs_observational_evidence_review_2024}.
\end{itemize}

\begin{figure}[htb]
\centering
\includegraphics[width=0.85\textwidth]{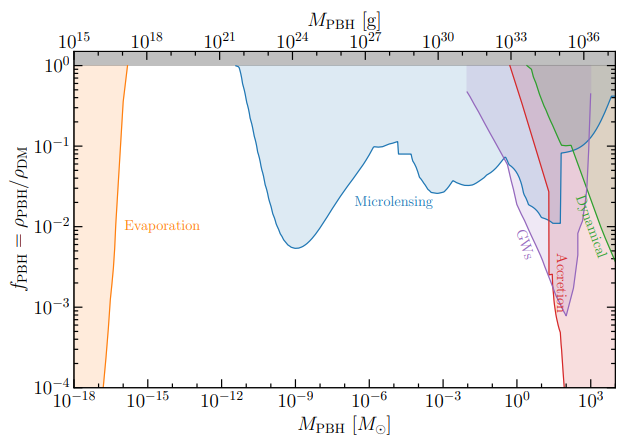}
\caption{Constraints on the abundance of PBHs as a Dark Matter component in the parameter space $(M_{\mathrm{PBH}}, f_{\mathrm{PBH}})$. The different bounds come from evaporation (orange), microlensing (blue), gravitational waves (purple), accretion in the CMB (red) and dynamical effects (green). Plot extracted from~\cite{Green_2024_PBHs}.}
\label{fig:chapter1_PBHs_parameter_space_constraints}
\end{figure}

Although there are observations that could be explained by PBHs (see~\cite{PBHs_observational_evidence_review_2024} for an extensive review of them), there is currently no unambiguous evidence for their existence. However, experiments dedicated to their search help to put constraints on their abundance today. This abundance is usually given as a quotient:

\begin{equation}
    f(M)=\frac{\Omega_{\mathrm{PBH}}}{\Omega_c}
    \label{eq:chapter1_PBHs_density_fraction}
\end{equation}

where $\Omega_{\mathrm{PBH}}$ is the density parameter for PBHs today (defined in a similar fashion to $\Omega_c$ in Section~\ref{Chapter1_Evidence_CMB}). With this definition, $0\leq f(M)\leq 1$, depending on the fraction of the total Dark Matter constituted by PBHs. Figure~\ref{fig:chapter1_PBHs_parameter_space_constraints} shows the constraints on the abundance of PBHs in the parameter space $(M,f(M))$, stemming from the different strategies explained above. As can be seen, in almost all the mass range, the possibility that PBHs are the major component of Dark Matter has been ruled out, but there is still an interesting window to explore, known as the \textit{asteroid mass window}, in the interval $10^{17}\mathrm{~g}\lesssim M \lesssim 10^{22}\mathrm{~g}$.

\vspace{-3mm}

Although PBHs are well motivated, the possibility of them being part of the Dark Matter of the universe is not without criticism, due to some problems of parameter fine-tuning. In particular, the density parameter $\Omega_{\mathrm{PBH}}$ can be expressed as~\cite{PBHs_lecture_notes}:

\begin{equation}
    \Omega_{\mathrm{PBH}} = 10^{18} \beta(M) \left( \frac{M}{10^{15}\mathrm{~g}} \right)^{-1/2}
    \label{eq:chapter1_PBHs_density_parameter}
\end{equation}

where $\beta(M)$ is the fraction of the mass of the universe in the form of PBHs at the time of their formation. Therefore, placing bounds on $f(M)$ imposes bounds on $\beta(M)$. In order for $\Omega_{\mathrm{PBH}}\sim \Omega_c$, $\beta$ must be tiny: in the mass range considered in Figure~\ref{fig:chapter1_PBHs_parameter_space_constraints}, $10^{15}\mathrm{~g}<M<10^{37}\mathrm{~g}$, one obtains $10^{-18}\lesssim\beta\lesssim 10^{-7}$, which is some serious fine-tuning for $\beta$. There are attempts to propose scenarios in which these extremely low values could arise naturally, check~\cite{PBHs_lecture_notes} for one possible explanation.

\subsection{Other Particle Approaches} \label{Chapter1_Explanations_other_particle_approaches}

In the quest to determine the nature of Dark Matter, numerous candidates have been proposed, each coming from different theoretical frameworks. Although the focus is usually on well-established candidates such as WIMPs, axions, sterile neutrinos or PBHs, there is a variety of less conventional possibilities. For the sake of completeness, some of them are briefly discussed in this section.

One example is Fuzzy Dark Matter (FDM), which posits that Dark Matter is made up of ultra-light bosons (with masses on the order of 10$^{-20}$~eV) that could exhibit wavelike properties on cosmological scales, having an impact on the formation process of large-scale structures.

Dark photons are gauge bosons associated with a $U(1)$ symmetry of the hypothetical dark sector. They could interact weakly with Standard Model photons through kinetic mixing, which would open channels for their detection. If dark photons had a small mass, they would be viable Dark Matter candidates.

Another candidate arises from extra dimensions in Kaluza-Klein models. In particular, Dark Matter could be made up of Kaluza-Klein particles, excited states of particles that appear in the compactification of extra spatial dimensions and which may have a very weak coupling with conventional matter.

Self-Interacting Dark Matter (SIDM) proposes that Dark Matter is not collisionless as in the CDM paradigm, and its constituents would have interactions with each other apart from gravitation. These interactions would be strong enough to alter the behaviour of Dark Matter on astrophysical scales, affecting the formation of structures in the universe.

The absence of a signal in the search for Dark Matter has also given rise to more pessimistic perspectives. One option is the so-called nightmare scenario, which refers to the possibility that Dark Matter is composed of particles that are only gravitationally coupled, so their gravitational pull is ultimately the only thing we can detect. An equivalent option is a particle that interacts so weakly with Standard Model particles that we will never be able to detect it by conventional means.

\subsection{The Modified Gravity Alternative} \label{Chapter1_Explanations_modified_gravity_alternative}

One of the most notable alternatives to explain the anomalies associated with the \textit{missing mass problem} without invoking non-visible matter is MOdified Newtonian Dynamics (MOND), introduced by Milgrom in 1983~\cite{Milgrom_1983_MOND}. MOND is based on a modification of Newton's second law for accelerations below a certain threshold $a_0$:

\begin{equation}
    ma \longrightarrow m\mu\left(\frac{a}{a_0}\right)a
    \label{eq:chapter1_MOND}
\end{equation}

where $\mu(x)$ is a function satisfying $\mu(x)\approx x$ for $x\ll 1$ (low acceleration regime) and $\mu(x)\approx 1$ for $x\gg 1$ (restoring Newtonian dynamics). In this framework, phenomena such as asymptotic flatness in galactic rotation curves are naturally explained by the low acceleration regime at galactic distances:

\begin{equation}
    \frac{a^2}{a_0} = G\frac{M}{r^2} \xRightarrow{a=v^2/r} v = (GMa_0)^{1/4}
    \label{eq:chapter1_MOND_flat_curve}
\end{equation}

where $M$ is \textit{only baryonic} mass, and the acceleration scale $a_0$ has to be calibrated by fitting the rotation curves observed for different galaxies. 

In fact, recent measurements point to extended galactic rotation curves that remain flat up to $r\sim 1$~Mpc~\cite{Mistele_2024}, well beyond the expected virial radii of Dark Matter halos. This is difficult to understand with CDM simulations, because rotation curves should have an asymptotic decline. However, MOND predicts indefinitely flat rotation curves, in agreement with these results.

Another success of MOND is naturally predicting the observed Baryonic Tully-Fisher Relation, $v^4\propto M$ (see Equation~\ref{eq:chapter1_MOND_flat_curve}).

Very recently, MOND has also been used to explain JWST's data on early galaxy formation~\cite{JWST_McGaugh_2024}, which is \textit{a priori} inconsistent with the slow hierarchical build-up from CDM predictions. However, CDM simulations rely on numerous inputs such as the stellar initial mass function, which may not be constant across cosmic times, and thus may differ substantially for high redshifts~\cite{JWST_tension_CDM_2023}.

Nevertheless, it should be noted that despite some predictive successes of MOND, a cosmological model based on this theory is not available, so that observations such as the acoustic peaks of the CMB are hardly explainable with MOND alone. Furthermore, MOND is insufficient to explain the dynamics observed in galactic clusters, requiring also a certain amount of Dark Matter. One of the models most advocated by MOND proponents is the addition of a sterile neutrino with mass $m_\nu \approx 11$~eV/c$^2$ as a candidate for Hot Dark Matter ($\nu$HDM model)~\cite{Banik_MOND_review_2022}.

Likewise, MOND is not a relativistic theory, so it is not able to explain phenomena such as gravitational lensing, which is explained by General Relativity together with Dark Matter. Relativistic extensions such as Tensor-Vector-Scalar gravity (TeVeS)~\cite{TeVeS_Bekenstein_2004} have been proposed, but they also come with their own problems. The various efforts to embed MOND in a relativistic framework are reviewed in detail in~\cite{MOND_relativistic_extensions_review_2011}.

For an extensive review about the MOND paradigm, with its successes and pitfalls, and a comparison with $\Lambda$CDM, check~\cite{Banik_MOND_review_2022}.

In conclusion, modified gravity approaches are able to explain with very few parameters several galactic-scale phenomena, notably rotation curves and Baryonic Tully-Fisher Relation, but fail in cluster and cosmological observations, for which they have to resort to a Dark Matter component anyway. This complicates matters further, requiring both Dark Matter and a modification of General Relativity, one of our most extensively tested theories. For this reason, the CDM paradigm is still the most accepted description in the scientific community.

%% file: CHAPTERS/Chapter2.tex
\chapter{WIMPs as Dark Matter Candidates} \label{Chapter2_WIMPs}
%
{
\lettrine[loversize=0.15]{A}{mong} the proposed particle candidates for Dark Matter, Weakly Interacting Massive Particles (WIMPs) have gained significant attention during the last decades due to their compatibility with current cosmological observations and several Beyond the Standard Model frameworks. WIMPs are hypothesised to be a new class of particles with masses ranging from a few~GeV to a few~TeV that only interact via weak and gravitational forces with ordinary matter, making their direct detection challenging.
%
\section*{}
\parshape=0
\vspace{-20.5mm}
}
This chapter provides a brief updated review of the WIMP landscape, explaining the motivation for why WIMPs are good Dark Matter candidates, and delving into the different strategies proposed for their detection, along with the main experiments that help to explore the WIMP parameter space.

\section{Motivation} \label{Chapter2_Motivation}

Let us assume the existence of a new particle $\chi$ that is neutral, massive (with mass $m_\chi$), stable and weakly interacting with the known particles of the Standard Model. Theoretical calculations suggest that such a particle could explain the observed abundance and distribution of Dark Matter in the universe. This section outlines how.

In the early universe, the temperature of the universe $T$ is high and all particles are in thermal equilibrium. In these conditions, $\chi$ has a number density $n_{\chi, \mathrm{eq}}$ which is defined by the Fermi-Dirac (if it is a fermion) or Bose-Einstein (if it is a boson) statistics. For $T\gg m_\chi$, $n_{\chi, \mathrm{eq}}\propto T^3$, the same as for photons in thermal equilibrium. On the other hand, for $T\ll m_\chi$, the regime is classical and the energy distribution is defined by Maxwell-Boltzmann statistics. Therefore, $n_{\chi, \mathrm{eq}}\propto (m_\chi T)^{3/2}\exp\left(-m_\chi /T\right)$, which reflects that $\chi$'s are exponentially suppressed.

In a static picture of the universe, thermal equilibrium would persist indefinitely, leading to the exponential extinction of 
$\chi$ particles. However, the universe is expanding, so this reasoning is not correct.

For high temperatures, $\chi$ particles are abundant and in equilibrium with their annihilation products according to $\chi + \chi \leftrightarrow \mathrm{SM}+\mathrm{SM}$. As the temperature drops, $n_{\chi, \mathrm{eq}}$ approaches an exponential, but there comes a point where the expansion of the universe reduces the efficiency of the annihilation reaction by pushing the particles $\chi$ away from each other faster than they interact. In technical terms, the interaction rate $\Gamma =\langle \sigma_A v \rangle n_\chi$, with $\langle \sigma_A v \rangle$ the thermally-averaged annihilation cross-section times the relative velocity, falls below the expansion rate of the universe, encoded by the Hubble parameter $H=\dot{a}/a$. The point at which $\Gamma = H$ is known as the \textit{freeze-out} time, and thereafter the remaining $\chi$ particles exist as a relic abundance.

What we have just discussed qualitatively is described formally by the Boltzmann equation, which determines the time evolution of $n$~\cite{Jungman_1996}:

\begin{equation}
    \dot{n_\chi} + 3Hn_\chi = -\langle \sigma_A v \rangle \left(n^2_\chi-n^2_{\chi,\mathrm{eq}} \right)
    \label{eq:chapter2_boltzmann_equation}
\end{equation}

The term on the left that includes $H$ is responsible for the expansion of the universe, while the right-hand side accounts for the balance between annihilation (term $n^2_\chi$) and production (term $n^2_{\chi,\mathrm{eq}}$) of $\chi$'s. Equation~\ref{eq:chapter2_boltzmann_equation} is valid both for Dirac ($\chi \neq \bar{\chi}$) and Majorana particles ($\chi = \bar{\chi}$).

\begin{figure}[htb]
\centering
\includegraphics[width=0.75\textwidth]{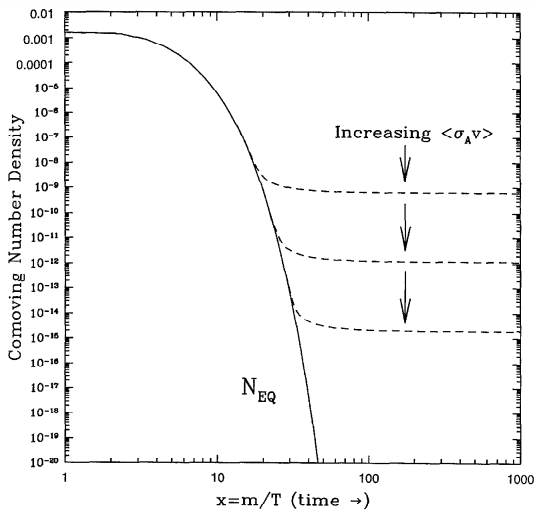}
\caption{Solutions to the Boltzmann equation for the comoving number density for $\chi$'s as a function of the parameter $m_\chi/T$. The black solid line represents the thermal equilibrium situation $n_{\chi,\mathrm{eq}}$, while the dashed lines are the actual solutions that include the expansion of the universe for different values of $\langle \sigma_A v \rangle$. It can be seen how larger cross-sections lead to later freeze-out times and smaller relic densities. Taken from~\cite{Jungman_1996}.}
\label{fig:chapter2_Boltzmann_equation}
\end{figure}

The Boltzmann equation has no analytical solution and is usually solved by numerical methods. Figure~\ref{fig:chapter2_Boltzmann_equation} shows solutions to this equation for different values of $\langle \sigma_A v \rangle$, together with the equilibrium value $n_{\chi,\mathrm{eq}}$ if the universe were static. It is observed that, as $\langle \sigma_A v \rangle$ increases, the freeze-out time increases, and the relic density becomes smaller. This is expected, because as the annihilations become more efficient, more expansion time is needed to reach the freeze-out point, and since the suppression is exponential around that point, we expect noticeable differences in the surviving number density.

\begin{figure}[htbp]
\centering
\includegraphics[width=0.8\textwidth]{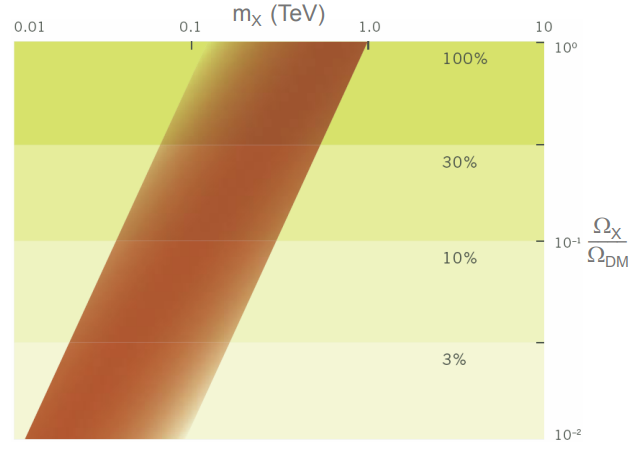}
\caption{Parameter space $(m_\chi, \Omega_\chi/\Omega_\mathrm{DM})$. The brown band represents the proportion of Dark Matter accounted for by $\chi$'s that have coupling strength $0.5<k<2$, $g^4_\chi = k g^4_{\mathrm{weak}}$, values motivated by the WIMP miracle. Extracted from~\cite{WIMPs_lecture_notes_Feng_2022}.}
\label{fig:chapter2_WIMP_miracle}
\end{figure}

Approximate solutions also exist for Equation~\ref{eq:chapter2_boltzmann_equation}, see~\cite{Jungman_1996} for a detailed argumentation and derivation. For the purposes of this manuscript, we will simply point out that, by imposing the freeze-out condition $\Gamma (T_f) = H(T_f)$, the freeze-out temperature is found to be around $T_f \sim m_\chi/20\ll m_\chi$ (so particles are cold when they freeze out), and that:

\begin{equation}
    \left(\Omega_\chi h^2\right)_0 \approx \frac{3\times 10^{-27}\mathrm{~cm^3~s^{-1}}}{\langle \sigma_A v \rangle}
    \label{eq:chapter2_WIMP_miracle_1}
\end{equation}

where $H=100h$~km/s/Mpc, $\Omega_\chi=\rho_\chi/\rho_c$ is the relic density of $\chi$ particles and the subscript 0 indicates present-day values. This result is, in principle, independent of the mass $m_\chi$. Taking into account that $\left(\Omega_\mathrm{DM} h^2\right)_0 \approx 0.12$ (Table~\ref{table:chapter1_cosmological_parameters_planck}), an interaction cross-section of $\langle \sigma_A v \rangle \approx 3\times 10^{-26}\mathrm{~cm^3~s^{-1}}$ is needed to identify the $\chi$ particles with the main component of Dark Matter. Interestingly, this is a typical value for a weak-scale interaction cross-section. On top of that, dimensionally~\cite{WIMPs_lecture_notes_Feng_2022}:

\begin{equation}
    \langle \sigma_A v \rangle \propto \frac{g^4_\chi}{16\pi^2 m^2_\chi}
    \label{eq:chapter2_WIMP_miracle_2}
\end{equation}

where $g_\chi$ is the characteristic coupling entering the annihilation cross-section. Parameterising $g^4_\chi = k g^4_{\mathrm{weak}}$, with $g_{\mathrm{weak}} \approx 0.65$ the weak-scale coupling, Equation~\ref{eq:chapter2_WIMP_miracle_1} can be linked to the mass $m_\chi$, $\Omega_\chi \propto m^2_\chi/(k g^4_{\mathrm{weak}})$. Figure~\ref{fig:chapter2_WIMP_miracle} shows the parameter space $(m_\chi, \Omega_\chi/\Omega_\mathrm{DM})$, for values of $k$ compatible with the weak scale, $0.5<k<2$ (brown band). It can be concluded that masses in the range $(100\mathrm{~GeV},1\mathrm{~TeV})$ would account for all Dark Matter (top of the band), $\Omega_\chi \approx \Omega_\mathrm{DM}$. This is the so-called \textit{WIMP miracle}: neutral, stable\footnote{To ensure that their density has not changed with respect to the freeze-out time.} $\chi$ particles with mass and interactions on the order of the weak scale, $(m_\chi,g_\chi)\sim (m_{\mathrm{weak}},g_{\mathrm{weak}})$, would be produced by the freeze-out mechanism with a non-relativistic relic density equal to the observed density of Dark Matter. This result is striking because it links Dark Matter (cosmology) and new physics at the weak scale (particle physics), two seemingly unrelated fields. 

$\chi$ particles produced through the freeze-out mechanism are called WIMPs\footnote{Or thermal WIMPs to stress the freeze-out mechanism, because the name WIMP is an umbrella term that can also include other production mechanisms such as \textit{freeze-in}, and masses and interaction strengths below the weak scale~\cite{Billard_2022}.}. Among the best theoretically motivated WIMP candidates is the neutralino, the Lightest Supersymmetric Particle (LSP) in many Minimal Supersymmetric Standard Model (MSSM) scenarios. Supersymmetry (SUSY) is an extension to the Standard Model proposed to solve issues such as the hierarchy problem. It posits that there is symmetry between fermions and bosons, so that a new particle is added for each particle in the Standard Model. The neutralino is a stable fermion resulting from a linear combination of the photino and higgsino (supersymmetric partners of the photon and Higgs boson, respectively), and is typically a Majorana particle. It has masses and interactions on the weak scale, which makes it an interesting candidate. Theoretical studies of the neutralino have shown that, in large regions of MSSM parameter space, its predicted relic abundance is compatible with the cosmological abundance of Dark Matter. See~\cite{Jungman_1996} or ~\cite{WIMPs_lecture_notes_Feng_2022} for an extensive review of the subject.

The WIMP miracle, coupled with the independent motivation to search for SUSY at the weak scale, has led to decades of experimental efforts to detect WIMPs. However, the absence of both SUSY hints at the LHC and WIMP signals in direct detection experiments have severely constrained the parameter space of WIMPs, practically ruling out the masses and interaction ranges motivated by the WIMP miracle. Nevertheless, it would be incorrect to assert that the thermal WIMP paradigm is dead, since beyond the coincidence of the annihilation cross-section with weak interactions, Equation~\ref{eq:chapter2_WIMP_miracle_2} shows that the value of $\sigma_A$ depends on both $g_\chi$ and $m_\chi$, and thus the correct relic density can be achieved with interactions and masses outside the weak scale~\cite{Feng_2008}. Indeed, if the $g_\chi \sim g_{\mathrm{weak}}$ condition is relaxed, the mass is free to move over a wider range of values. Currently, searches have shifted towards low-mass WIMPs, $m_\chi < 10$~GeV, as it is the least explored region of the parameter space, and because of its increasing accessibility due to improvements in the energy threshold of the various experiments.

\section{WIMP Searches} \label{Chapter2_WIMP_Searches}

The possibility of WIMPs as Dark Matter has motivated extensive research and experimental efforts. To this end, experiments have been designed to directly detect WIMPs by capturing their interactions with atomic nuclei or electrons ($\chi + \mathrm{SM}\rightarrow \chi + \mathrm{SM}$). These experiments normally operate deep underground to remove the cosmic background, and on top of that they have several layers of shielding to isolate the detectors as much as possible. These interactions would mainly occur through elastic scattering (see Section~\ref{Chapter3_Interactions_Neutral_Particles}) between WIMPs and a target nucleus, although other channels such as inelastic scattering or WIMP-electron scattering (for very light WIMPs) are also being explored. The elastic scattering cross-section depends on the mass and specific properties of both the WIMP and the target nucleus, and can be either Spin Independent or Spin Dependent.

On the other hand, indirect detection methods focus on searching for the secondary products of WIMP annihilation ($\chi + \chi \rightarrow \mathrm{SM}+\mathrm{SM}$) or decay ($\chi \rightarrow \mathrm{SM}$), such as high-energy photons or cosmic rays. Lastly, colliders such as the LHC at CERN seek to produce WIMPs directly ($\mathrm{SM}+\mathrm{SM} \rightarrow \chi + \chi$), as a by-product of high-energy collisions.

\begin{figure}[htbp]
\centering
\includegraphics[width=0.8\textwidth]{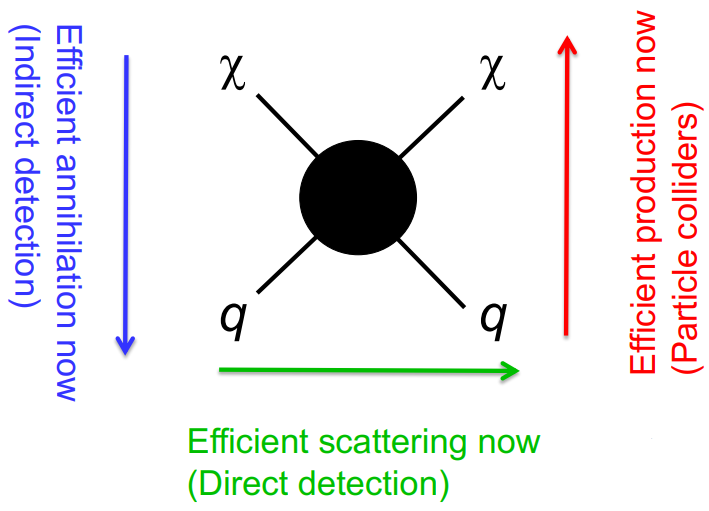}
\caption{Illustration summarising the different WIMP detection strategies. Image extracted from~\cite{WIMPs_lecture_notes_Feng_2022}.}
\label{fig:chapter2_WIMPs_searches}
\end{figure}

Figure~\ref{fig:chapter2_WIMPs_searches} encapsulates the different detection strategies.

\subsection{Direct Searches} \label{Chapter2_Direct_Searches}

As mentioned above, direct detection experiments aim to record the interactions of WIMPs with atomic nuclei or electrons. Given that WIMPs are neutral, interactions with atomic electrons are rare and are not expected to be the dominant channel. Therefore, we will focus on nuclear recoils induced by elastic interactions of WIMPs with target nuclei, as most of the searches (and, in particular, TREX-DM, the experiment on which this thesis is based) are focused on this channel. The expected elastic scattering rate per unit detector mass is given by~\cite{Lewin_1996}:

\begin{equation}
    \frac{dR}{dE_{\mathrm{nr}}}=\frac{\rho_\chi}{m_\chi m_N}\int_{v_{\mathrm{min}}}^{v_{\mathrm{max}}}vf(v)\frac{d\sigma}{dE_{\mathrm{nr}}}dv 
    \label{eq:chapter2_differential_scattering_rate}
\end{equation}

where $\frac{dR}{dE_{\mathrm{nr}}}$ is the differential scattering rate, $\rho_\chi$ is the local WIMP density (typically assumed to be 
0.3~GeV/cm$^3$), $m_\chi$ and $m_N$ are the WIMP and target nucleus masses, $v$ is the WIMP velocity in the Earth-based detector frame, $f(v)$ the velocity distribution relative to the detector, $v_{\mathrm{min}}$ is the minimum velocity necessary to produce a nuclear recoil of energy $E_{\mathrm{nr}}$, $v_{\mathrm{max}}$ is the maximum WIMP velocity, namely the escape velocity in the detector frame (assuming they are gravitationally bound to the Milky Way) and $\frac{d\sigma}{dE_{\mathrm{nr}}}$ is the differential scattering cross-section.

\vspace{2mm}
\textbf{\normalsize Interaction Cross-Section}
\vspace{0mm}

In general, the differential cross-section can be expressed as~\cite{Jungman_1996}:

\begin{equation}
    \frac{d\sigma}{dE_{\mathrm{nr}}} = \frac{m_N}{2v^2 \mu^2} \sigma_0F^2(E_{\mathrm{nr}})
    \label{eq:chapter2_differential_cross_section_decomposition}
\end{equation}

where $\mu = m_\chi m_N/(m_\chi + m_N)$ is the reduced mass of the WIMP-nucleus system, $\sigma_0$ is the cross-section at zero momentum transfer and $F(E_{\mathrm{nr}})$ the form factor, which accounts for the loss of coherence at high momentum transfers. The total cross-section is determined by the sum of the individual contributions of all nucleons in the nucleus, and has two terms: Spin Independent (SI) and Spin Dependent (SD). Applying Equation~\ref{eq:chapter2_differential_cross_section_decomposition} to the interaction between the WIMP and the nucleus:

\begin{equation}
    \frac{d\sigma}{dE_{\mathrm{nr}}} = \frac{m_N}{2v^2 \mu^2}\left[ \sigma_{0,\mathrm{SI}}F^2_{\mathrm{SI}}(E_{\mathrm{nr}}) + \sigma_{0,\mathrm{SD}}F^2_{\mathrm{SD}}(E_{\mathrm{nr}}) \right]
    \label{eq:chapter2_differential_cross_section_SI_SD}
\end{equation}

In the SI interaction, the individual amplitudes do not depend on the orientation of the spins, and therefore the sum of the contributions is equivalent to a coherent interaction with the nucleus as a whole. It can be proved that~\cite{Jungman_1996}:

\begin{equation}
    \sigma_{0,\mathrm{SI}} = \frac{4}{\pi} \mu^2 \left( Zf_p + (A-Z)f_n \right)^2
    \label{eq:chapter2_spin_independent_cross_section}
\end{equation}

where $Z$ and $A$ are the atomic and mass numbers of the nucleus, and $f_n$ and $f_p$ are the fundamental couplings to neutrons and protons. In general, $f_n \approx f_p$, which results in:

\begin{equation}
    \sigma_{0,\mathrm{SI}} = \frac{4}{\pi}\mu^2 f^2_n A^2 = \sigma_n \frac{\mu^2}{\mu_n^2}A^2
    \label{eq:chapter2_spin_independent_cross_section_simplified}
\end{equation}

where $\mu_n$ the reduced mass of the WIMP-nucleon system, and $\sigma_n$ is the WIMP-nucleon cross-section, defined as $\sigma_n =\frac{4}{\pi}\mu_n^2 f^2_n$. This is usual in order to be able to directly compare cross-sections in different materials. Equation~\ref{eq:chapter2_spin_independent_cross_section_simplified} shows that heavy nuclei are preferred to enhance sensitivity.

In the SD interaction, the amplitude WIMP-nucleon changes sign depending on the spin of the nucleon. As nucleons usually have alternating spin, their contributions to the total amplitude eventually cancel out, except for those nucleons that are unpaired. Consequently, the total SD cross-section is a function of the net nuclear spin, $J$:

\begin{equation}
    \sigma_{0,\mathrm{SD}} = \frac{32}{\pi} G_F^2 \mu^2 \left[ a_p\langle S_p \rangle + a_n\langle S_n \rangle \right]^2 \frac{J+1}{J}
    \label{eq:chapter2_spin_dependent_cross_section}
\end{equation}

with $G_F$ the Fermi constant, $a_n$ and $a_p$ the effective couplings to neutrons and protons, and $\langle S_n \rangle$ and $\langle S_p \rangle$ the expectation values of the neutron and proton spin operators in the nucleus. SD interactions are only relevant in isotopes with non-zero spin, and it is typically assumed that WIMPs couple exclusively either to protons ($a_n=0$) or neutrons ($a_p=0$). Therefore, nuclei with an odd proton number (such as $^{19}$F or $^{127}$I) are sensitive to WIMP-proton interactions, whereas those with an odd number of neutrons (such as $^{17}$O or $^{131}$Xe) could probe WIMP-neutron interactions. 

In most SUSY models, the SI cross-section dominates over the SD one for nuclei with $A\gtrsim 30$~\cite{Shan_2011} due to the $A^2$ dependence from Equation~\ref{eq:chapter2_spin_independent_cross_section}. This fact, coupled with the limited number of isotopes suitable for SD interactions (which, moreover, often require enrichment due to their low natural abundance) makes the SI interaction the most exploited direct detection channel for WIMPs. Therefore, we will refer purely to SI interactions in what follows.

\vspace{2mm}
\textbf{\normalsize Quenching Factor}
\vspace{0mm}

When detecting nuclear recoils such as those produced in elastic scattering of WIMPs or neutrons, it is crucial to take into account the quenching factor (QF), a parameter that allows to compare them with electron recoils.

Most background sources, such as photons or electrons, interact mainly with the electrons of the target material (see Chapter~\ref{Chapter3_Interactions}), giving rise to a ionisation/scintillation signal that can be calibrated in energy and is measured in units of electron-equivalent energy, e.g. keV$_{\mathrm{ee}}$. However, in nuclear recoils of WIMPs or neutrons, not all of the energy transferred in the collision (measured in nuclear-recoil energy units, e.g. keV$_{\mathrm{nr}}$) is subsequently converted into a measurable signal in the form of ionisation/scintillation. Part may be lost as thermal energy or through other processes. The quenching factor, $QF(E_\mathrm{nr})$, quantifies this difference:

\begin{equation}
    E_{\mathrm{meas.}}(\mathrm{keV}_\mathrm{ee}) = QF(E_\mathrm{nr}) \times E_\mathrm{nr}(\mathrm{keV}_\mathrm{nr})
\end{equation}

In other words, $QF(E_\mathrm{nr})$ allows to obtain the total deposited recoil energy, $E_\mathrm{nr}$, from the measured ionisation/scintillation energy, $E_{\mathrm{meas.}}$. It is important to note that the QF depends on the recoil energy and the detector material. In any WIMPs search experiment, calibrating the QF is essential to correctly interpret the measured signals. This is done combining:

\begin{itemize}
    \item X-ray/gamma sources (such as $^{55}$Fe, $^{137}$Cs, $^{109}$Cd): to set the scale (keV$_{\mathrm{ee}}$) of electron recoils.
    \item Neutron sources (such as AmBe, $^{252}$Cf, or neutron generators): to calibrate the response to nuclear recoils.
\end{itemize}

In this thesis, for simplicity of notation, we will omit the electron-equivalent when reporting measured energies $E_{\mathrm{meas.}}$; for example, keV$_{\mathrm{ee}}\equiv$ keV.

\vspace{2mm}
\textbf{\normalsize Velocity Distribution}
\vspace{0mm}

The distribution of velocities is usually assumed to be Maxwellian around the galactic centre:

\begin{equation}
    f(v')\propto \exp\left(-v'^2/v_0^2\right)
    \label{eq:chapter2_velocity_distribution}
\end{equation}

where $\vec{v}' = \vec{v} + \vec{v}'_E$ is the WIMP velocity in the galactic frame, with $\vec{v}$ the WIMP velocity in the Earth frame and $\vec{v}'_E$ the velocity of Earth in the galactic frame. On the other hand, $v_0 \approx $ 220~km/s is the characteristic velocity of WIMPs. This distribution is truncated at the escape velocity $v'_{\mathrm{esc}}\approx 544$~km/s, that is, $f(v'>v'_{\mathrm{esc}})=0$, since WIMPs with higher velocities are not bound to the galaxy. 

\vspace{2mm}
\textbf{\normalsize Annual Modulation}
\vspace{0mm}

A special type of direct detection involves the search for an annually modulated WIMP signal due to the motion of the Earth around the Sun, which alters the relative velocity of the detector with respect to the galactic Dark Matter halo. Given the Earth's orbit around the Sun, an annual periodic variation in the WIMP interaction rate is expected.

The Sun moves around the galactic centre with a velocity $\vec{v}'_\odot \approx 232$~km/s, while the Earth moves around the Sun with a velocity $\vec{v}'_\oplus \approx 30$~km/s in an orbit tilted at an angle of $\theta \approx 60^\circ$ with respect to the galactic plane. Therefore, the velocity of a detector located on the Earth is given by:

\begin{equation}
    v'_E = v'_\odot + v'_\oplus\cos(\theta)\cos{\left[\omega(t-t_0)\right]} 
    \label{eq:chapter2_velocity_earth_frame}
\end{equation}

where $w=2\pi/T$ with $T=1$~y, and the phase $t_0$ is set to June 2, when the combined velocity reaches its maximum. Correspondingly, the minimum is recorded in December. 

This time dependence propagates to the interaction rate (Equation~\ref{eq:chapter2_differential_scattering_rate}), as $v'_E$ affects $f(v)$ (Equation~\ref{eq:chapter2_velocity_distribution}) and $v_{\mathrm{min}}$. It can be proved that~\cite{Billard_2022}:

\begin{equation}
    R(t) = B(t) + S_0 + S_m\cos{\left[\omega(t-t_0)\right]}
    \label{eq:chapter2_annual_modulation}
\end{equation}

where $R(t)$ is the total rate, $B(t)$ is the (potentially) time-dependent background rate, $S_0$ is the unmodulated signal and $S_m$ the amplitude of the modulated part. The main problems with this method are that the background rate can be time-dependent, so a very exhaustive control is required in order not to find spurious time dependencies, and that the amplitude $S_m$ is very small, $S_m/S_0 \sim 3$\%~\cite{Lewin_1996}, hence very sensitive detectors are needed.

\subsubsection{Key Factors Affecting Sensitivity} \label{Chapter2_Direct_Searches_Sensitivity}

Putting together Equations~\ref{eq:chapter2_differential_cross_section_decomposition} and~\ref{eq:chapter2_spin_independent_cross_section_simplified}, plugging it into Equation~\ref{eq:chapter2_differential_scattering_rate} and rearranging:

\begin{equation}
    \frac{dR}{dE_{\mathrm{nr}}}=\frac{\rho_\chi}{m_\chi} \frac{\sigma_n}{2\mu_n^2}A^2 F^2_{\mathrm{SI}}(E_{\mathrm{nr}})\int_{v_{\mathrm{min}}}^{v_{\mathrm{max}}}\frac{f(v)}{v}dv \label{eq:chapter2_differential_scattering_rate_simplified}
\end{equation}

Due to the exponential form of $f(v)$, it is possible to express the differential scattering rate as a decaying exponential~\cite{Lewin_1996}: 

\begin{equation}
    \frac{dR}{dE_{\mathrm{nr}}} \propto \exp\left(-E_{\mathrm{nr}}/E_0 r\right) F^2_{\mathrm{SI}}(E_{\mathrm{nr}}) \label{eq:chapter2_differential_scattering_rate_exponential}
\end{equation}

with $E_0$ the characteristic energy associated to $v_0$ and $r = 4m_\chi m_N /(m_\chi + m_N)^2$ a kinematic factor.

Based on these two equations, we will now review the main ingredients playing a role in the sensitivity of an experiment, understood as the minimum interaction cross-section it can detect or exclude for a given range of masses at a given confidence level (usually 90\%). It is a measure of the potential to distinguish signal over background. We will focus on how to access the low-mass parameter space motivated in Section~\ref{Chapter2_Motivation}.

First of all, we will define signal, $S$, as the number of WIMP events that an experiment with detection efficiency $\varepsilon (E_\mathrm{nr})$, duration $T$ and mass $M$ accumulates in the energy range $(E_\mathrm{thr}, E_\mathrm{max})$, with $E_\mathrm{thr}$ the energy threshold of the experiment. Similarly, the background $B$ is defined as the number of non-WIMP events that the experiment registers. These come from varied sources such as material contamination or cosmic rays. The calculation of the expected signal is given by:

\begin{equation}
    S = M\times T \int_{E_\mathrm{thr}}^{E_\mathrm{max}} \frac{dR}{dE_{\mathrm{nr}}} \varepsilon(E_\mathrm{nr}) dE_{\mathrm{nr}}
    \label{eq:chapter2_expected_signal_experiment}
\end{equation}

where the detector efficiency is related to the quenching factor. Therefore, in an experiment that records $N$ counts, the background-only hypothesis ($B$) is compared with the signal-plus-background hypothesis ($B+S$) to see if discovery can be claimed. If it cannot, an upper bound is placed on the largest cross-section compatible with $N$. The factors that most affect $S$, and thus sensitivity, are:

\begin{itemize}
    \item \textbf{Astrophysical considerations}: the local density $\rho_\chi$ and velocity distribution $f(v)$ depend on the model chosen to describe the galactic Dark Matter halo (see Section~\ref{Chapter1_Evidence_GalacticRotation}).
    \item \textbf{Exposure}: it is defined as $\mathcal{E}=M\times T$, and is typically presented in~kg$\cdot$y or~ton$\cdot$y. The more mass or data-taking time, the better the sensitivity.
    \item \textbf{Energy threshold}: the exponential form of $\frac{dR}{dE_{\mathrm{nr}}}$ highlighted in Equation~\ref{eq:chapter2_differential_scattering_rate_exponential} indicates that most of the interactions will be at low energy, so the energy threshold $E_\mathrm{thr}$ of the experiment is crucial to accumulate signal, making the upper limit $E_\mathrm{max}$ less relevant. This is all the more important if we want to access the low-mass WIMPs range.
    \item \textbf{Background level}: a low background level allows a higher sensitivity for the same amount of signal. It is one of the most difficult parameters to keep under control, and is usually the limiting factor in many experiments (see Chapter~\ref{Chapter6_Radon_problem} for its study in an experiment such as TREX-DM).
    \item \textbf{Target material}: as previously mentioned, materials with high mass numbers are preferred due to the $A^2$ dependence of Equation~\ref{eq:chapter2_differential_scattering_rate_simplified}. Despite this, as in all elastic collisons, there is another parameter that comes into play: the mass ratio $m_\chi/m_N$. This can be seen in the kinematic parameter $r$ that appears in the exponential of Equation~\ref{eq:chapter2_differential_scattering_rate_exponential}: if $m_N \gg m_\chi$ or $m_\chi \gg m_N$, $r \rightarrow 0$ so the mass mismatch will cause the exponential to suppress interactions more quickly. For standard WIMPs with $m_N \sim O(100)\mathrm{~GeV}$, heavy nuclei such as Xe ($\approx 131.3$) are used. However, in the low-mass range, lighter nuclei such as Ne or Ar are preferred to optimise sensitivity.
\end{itemize}

\begin{figure}[htbp]
\centering
\includegraphics[width=0.48\textwidth]{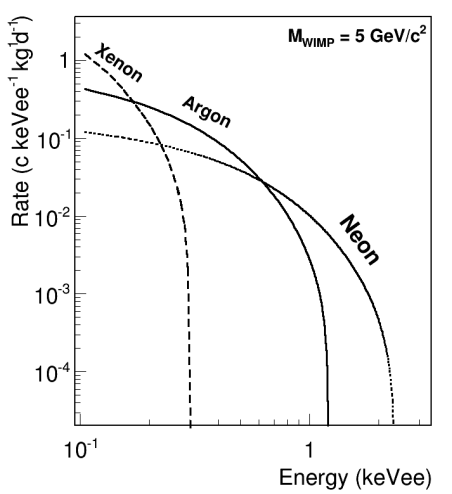}
\includegraphics[width=0.505\textwidth]{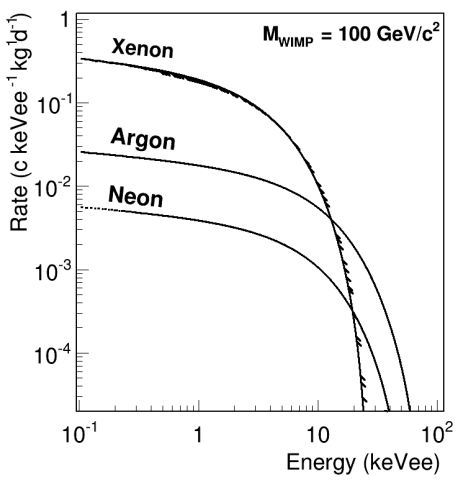}
\caption{Differential scattering rates $\frac{dR}{dE}$ (in ~c/keV$_{\mathrm{ee}}$/kg/day) as a function of energy (in keV$_{\mathrm{ee}}$) for two WIMP masses: 5~GeV (left) and 100~GeV (right). A quenching factor has been applied to transform from keV$_{\mathrm{nr}}$ to keV$_{\mathrm{ee}}$. Extracted from~\cite{tesis_javi_gracia}.}
\label{fig:chapter2_WIMP_scattering_rate}
\end{figure}

The last four points can be easily observed in the comparison given in Figure~\ref{fig:chapter2_WIMP_scattering_rate}: high exposures and low backgrounds are necessary to be sensitive to low scattering rates, on the order of $\frac{dR}{dE}\sim 10^{-2}-1$~c/keV/kg/day. On the other hand, the low energy threshold plus low $A$ combination is optimal for low-mass WIMPs (here, $m_\chi = 5$~GeV): Ar and Ne present a wider spectrum than Xe, and a low threshold (typically < 1 keV) is necessary as the spectra are concentrated in the low energy region. For higher masses (here, $m_\chi = 100$~GeV), the $A^2$ factor of rate enhancement for Xe makes it more suitable, assuming the energy threshold is at least within the range $1-10$~keV.

\subsubsection{Detection Technologies and Leading Experiments}
\label{Chapter2_Direct_Searches_Experiments}

\begin{figure}[htbp]
\centering
\includegraphics[width=1.0\textwidth]{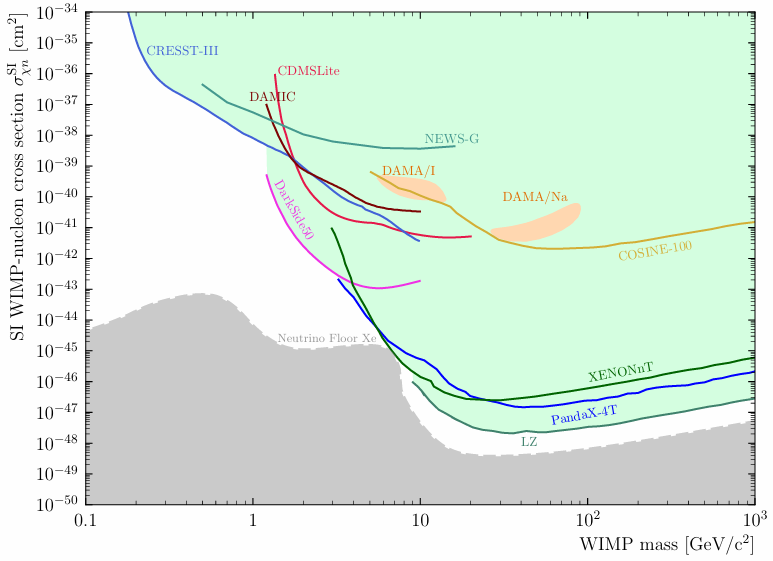}
\caption{Exclusion plot for the SI WIMP-nucleon cross-section vs. WIMP mass parameter space, with bounds from CRESST-III~\cite{CREST-III_2019}, NEWS-G~\cite{NEWS-G_2018}, DAMIC~\cite{DAMIC_2020}, CDMSLite~\cite{CDMSLite_2018}, DarkSide50~\cite{DarkSide50_2022}, COSINE-100~\cite{COSINE-100_2021}, XENONnT~\cite{XENONnT_2023, XENONnT_2025}, PandaX-4T~\cite{PandaX-4T_2025, PandaX-4T_2023_1, PandaX-4T_2023_2} and LZ~\cite{LZ_2025}, together with neutrino fog region for Xe~\cite{neutrino_floor_Xe_2021} (grey) and DAMA/LIBRA~\cite{Savage_2009} claimed discovery (orange shaded region). Plot prepared using~\cite{wimp_limits_plotting_tool}.}
\label{fig:chapter2_WIMP_exclusion_plot}
\end{figure}

In this section, a brief overview of the main technologies employed in the direct search for WIMPs is given, together with the main experiments in each category. Figure~\ref{fig:chapter2_WIMP_exclusion_plot} shows some of the most competitive bounds (as of early 2025) in the parameter space $(m_\chi,\sigma_n)$, with $\sigma_n$ the SI WIMP-nucleon cross-section.

\vspace{2mm}
\textbf{\normalsize Bolometers}
\vspace{0mm}

Bolometers measure the temperature rise induced by nuclear recoils in crystals held at cryogenic temperatures. The low temperatures, on the order of mK, allow very good energy resolution and low energy threshold, which makes this technology particularly suitable for low-mass searches, even in the $m_\chi < 1$~GeV range. Some of them also detect ionisation or scintillation as a resource for background rejection. Main experiments include:

\begin{itemize}
    \item EDELWEISS (France, Modane Underground Laboratory): it employs Ge bolometers with ionisation readout.
    \item CRESST (Italy, Laboratori Nazionali del Gran Sasso): it uses CaWO$_4$ bolometers, discriminating via phonon plus light signals. It has reached sub-100-eV energy thresholds, and is the leading experiment in the sub-GeV mass area, as showcased in Figure~\ref{fig:chapter2_WIMP_exclusion_plot}.
    \item SuperCDMS (Canada, SNOLAB): it uses Ge and Si bolometers with a readout based on phonon plus charge.
\end{itemize}

\vspace{2mm}
\textbf{\normalsize Scintillators}
\vspace{0mm}

Scintillators measure the light emitted upon excitation of a scintillator crystal after interaction with a WIMP. Some of them use Pulse Shape Discrimination (PSD) to differentiate between electronic and nuclear recoils.

The DAMA/LIBRA experiment, located at the LNGS in Italy, has claimed an annual modulation signal with its NaI scintillators, accumulating more than 22 independent annual cycles over the different phases and upgrades of the experiment, starting in 1995. This signal coincides in period and phase with that expected from Equation~\ref{eq:chapter2_annual_modulation}, and has an amplitude in the range [2-6]~keV of (0.01014 $\pm$ 0.00074)~c/d/kg/keV with a statistical significance of 13.7$\sigma$~\cite{Bernabei_2023}.

Given the extraordinary robustness and high confidence level of the signal, multiple independent experiments have attempted to reproduce these results, using different techniques and target materials. All of them have failed to confirm the annual modulation signal, or the detection of a WIMP in the range $(m_\chi ,\sigma_{n})\sim (10-100\mathrm{~GeV},10^{-40}-10^{-42}\mathrm{~cm}^2)$ predicted by DAMA/LIBRA. In fact, that region of the parameter space is completely discarded by leading experiments, as can be seen in Figure~\ref{fig:chapter2_WIMP_exclusion_plot}. Among all the experiments, ANAIS (Annual Modulation with NaI Scintillators) and COSINE-100 stand out. ANAIS is located at the Canfranc Underground Laboratory (under the Spanish Pyrenees), while COSINE-100 can be found at Yangyang Underground Laboratory in South Korea. They both look for an annual modulation signal using the same target material as DAMA/LIBRA, in order to compare results as cleanly as possible. After 3 years of data taking, ANAIS reports results incompatible with the DAMA/LIBRA observation at almost 3$\sigma$ confidence level~\cite{ANAIS_2024}, while COSINE challenges the claim of DAMA/LIBRA with more than 3$\sigma$ after 6.4 years of data taking~\cite{COSINE_2024}. This further calls into question the already controversial detection claim, pointing to possible errors in the analysis or to unaccounted background sources that fluctuate over time.

\vspace{2mm}
\textbf{\normalsize Dual-Phase TPCs}
\vspace{0mm}

This technology is based on the use of liquid noble gases (Xe or Ar) in equilibrium with a small gaseous phase to detect nuclear recoils by a combination of scintillation (signal 1 or S1) and ionisation (signal 2 or S2). The S1 signal is detected by an array of PhotoMultiplier Tubes (PMTs), and serves as the event trigger. The S2 signal drifts into a gas phase, where it is amplified and detected by the same array of PMTs. The combination of the two signals with the use of a TPC allows the 3D reconstruction of the events, a crucial tool for background discrimination. The leading experiments are:

\begin{itemize}
    \item XENONnT (Italy, LNGS): successor of XENON1T, it uses $\sim$ 6 tons of liquid xenon (LXe). It has recently entered the neutrino fog for the first time~\cite{XENONnT_2024}, the area of parameter space in which solar and atmospheric neutrinos produce an indistinguishable signal from the scattering of light WIMPs with a Xe nucleus.
    \item LUX-ZEPLIN (LZ, located at Sanford Underground Research Facility, USA): TPC with 7 tons of LXe, one of the main competitors of XENONnT, currently offering the most stringent limits in the $10-1000$~GeV mass range (see Figure~\ref{fig:chapter2_WIMP_exclusion_plot}).
    \item PandaX-4T (China Jinping Underground Laboratory, China): another LXe-based experiment. Its latest results are shown in Figure~\ref{fig:chapter2_WIMP_exclusion_plot}, and support the null results obtained by XENONnT and LZ.
    \item DarkSide-50 (Italy, LNGS): employing liquid argon (LAr), it currently has the most competitive bounds in the $m_\chi \sim (1,3)$~GeV range.
\end{itemize}

\vspace{2mm}
\textbf{\normalsize CCD Detectors}
\vspace{0mm}

Charge-Coupled Devices (CCDs) are semiconductor detectors that measure the ionisation induced by WIMP recoils with nuclei or electrons. They present very low energy thresholds, on the eV scale, and they also provide very good spatial resolution. Their main advantage is the sensitivity to sub-GeV WIMPs, a range in which WIMP-electron interaction can already be probed. The main experiment is DAMIC (SNOLAB, Canada, currently being upgraded at LSM, France), which uses Si CCDs and has the best limits to date for the WIMP-electron scattering cross-section on the MeV scale~\cite{DAMIC_electron_2019}.

\vspace{2mm}
\textbf{\normalsize Gaseous Detectors}
\vspace{0mm}

Gaseous detectors are based on identifying nuclear recoils by ionisation of a gas (usually a noble gas) contained in a pressurised chamber. Gas TPCs allow for track reconstruction of interactions, crucial in background discrimination. On the other hand, they have the potential to reach low energy thresholds, which makes them ideal candidates for sub-GeV WIMPs and directional searches. Directional detection seeks to observe the angular distribution of nuclear recoils, with the aim of detecting the daily modulation of the WIMP wind due to the Earth's rotation around its axis.

NEWS-G (New Experiments With Spheres-Gas, located in SNOLAB, Canada) is one of the main gaseous detector experiments, and it employs a Spherical Proportional Counter (SPC) filled with a light noble gas to maximise sensitivity down to the sub-GeV range, where it has set exclusion limits only surpassed by CRESST (see Figure~\ref{fig:chapter2_WIMP_exclusion_plot}). DRIFT (Directional Recoil Identification From Tracks, located at the Boulby Underground Laboratory, UK), uses a negative-ion TPC with gases such as CS$_2$ or SF$_6$ to track directions of nuclear recoils.

TREX-DM, explained in detail in Chapter~\ref{Chapter5_TREXDM}, is a low-background, non-directional, high-pressure gas TPC focused on searching for low-mass WIMPs, potentially in the sub-GeV range. Sensitivity projections are discussed in Chapter~\ref{Chapter9_Sensitivity}.

\subsection{Indirect and Collider Searches} \label{Chapter2_Indirect_Collider_Searches}

The principle of indirect detection lies in the assumption that WIMPs can interact and annihilate each other, leaving other Standard Model particles (such as positrons, antiprotons, photons or neutrinos) as a by-product of the annihilation, whereas collider searches are based on the opposite process, the production of WIMPs as a result of the collision of known particles. This can be expressed by the process:

\begin{equation}
    \chi + \chi \leftrightarrow \mathrm{SM}+\mathrm{SM}
    \label{eq:chapter2_indirect_collider_searches}
\end{equation}

In indirect detection (left to right), the annihilation cross-section $\sigma_{\chi + \chi \rightarrow \mathrm{SM}+\mathrm{SM}}$ is probed, while in collider production (right to left), the production cross-section $\sigma_{\mathrm{SM}+\mathrm{SM} \rightarrow \chi + \chi}$ is constrained. Both processes are complementary to the direct searches, as they study different aspects of the interplay between Dark Matter and the Standard Model.

\vspace{2mm}
\textbf{\normalsize Indirect Detection}
\vspace{0mm}

Indirect searches look for excess particles in areas with a high Dark Matter density, namely galactic centers and galaxy clusters. The main indirect searches focus on:

\begin{itemize}
    \item \textbf{Gamma ray studies}: gamma rays could either be a direct product of the annihilation process, or be produced via inverse Compton scattering of charged particles. Some particularly interesting astronomical objects for Dark Matter detection are dwarf spheroidal galaxies (dSphs). These are galaxies with a very high dark-to-luminous mass ratios, no gas and few intrinsic sources of high-energy photons, which leads to high signal-to-background ratios. These characteristics, coupled with the fact that many of them are close to us, make them the perfect target for gamma ray searches~\cite{dSphs}. Many experiments have placed constraints on $\langle \sigma_{\chi + \chi \rightarrow \mathrm{SM}+\mathrm{SM}} v \rangle$ during the last decade. Among them, the most prominent come from the Fermi Large Array Telescope (Fermi-LAT)~\cite{Fermi-LAT}, the High-Altitude Water Cherenkov Observatory (HAWC)~\cite{HAWC} and the three main Imaging Atmospheric Cherenkov Telescopes (IACTs) experiments: the VERITAS gamma-ray observatory (Very Energetic Radiation Imaging Telescope Array System)~\cite{VERITAS}, MAGIC~\cite{MAGIC} (Major Atmospheric Gamma Imaging Cherenkov Telescopes) and HESS~\cite{HESS} (High Energy Stereoscopic System). These searches have focused on the paradigmatic WIMP mass range (100~GeV $\lesssim m_X \lesssim $ 100~TeV), though alternative regions such as the UltraHeavy Dark Matter range (UHDM, $m_X \gtrsim $ 100~TeV) are increasingly being explored, for example by the VERITAS collaboration~\cite{VERITAS_UHDM}.
    \item \textbf{Cosmic ray excess}: the annihilation or decay of WIMPs can produce high-energy charged particles such as electrons, positrons, protons and antiprotons. Unlike gamma rays, these particles are deflected by interstellar magnetic fields, so directional information is lost. Therefore, the strategy is to look for anomalies in cosmic-ray spectra that do not fit the astrophysical predictions. One of the best-known observations is the positron excess in the high-energy range detected by the PAMELA satellite~\cite{PAMELA_2009}, and later confirmed by AMS-02~\cite{AMS-02_2014} and Fermi-LAT~\cite{Fermi-LAT_2017}. It was initially interpreted as a potential Dark Matter signal. However, for this to be the case, the annihilation cross-section would have to be two orders of magnitude larger than that obtained by thermal production. On the other hand, there are model-dependent uncertainties (such as the density profile used) and other alternative explanations with astrophysical origin such as pulsars~\cite{Rocamora_2024}, so it is not easy to determine the source of the emission and to draw conclusions.
    \item \textbf{Neutrino searches}: neutrinos could be a direct product of the annihilation of WIMPs, or an indirect product through secondary decays of heavy particles. Mainly, neutrino excesses are probed by looking at the galactic center (similar to gamma-ray searches), the Sun and the Earth. In these two cases, it is assumed that WIMPs could accumulate in the solar/Earth's core due to scattering and gravitational trapping, which would increase their density and favour their annihilation, hence neutrino excesses could be searched for. The most restrictive limits in searches of the galactic center currently belong to Fermi-LAT and HESS~\cite{Zornoza_2021}, while the Earth core searches are not competitive with the direct detection experiments. However, the results coming from solar WIMPs by experiments such as IceCube~\cite{IceCube_2016}, ANTARES~\cite{ANTARES_2016} or SuperK~\cite{SuperKamiokande_2015} are the most sensitive for the spin-dependent WIMP-proton cross-section.
\end{itemize}

\vspace{2mm}
\textbf{\normalsize Collider Searches}
\vspace{0mm}

Colliders like the LHC try to produce Dark Matter candidates as a by-product of high-energy proton-proton collisions. Due to the weak interaction of WIMPs, if produced, they escape the detector without leaving energy, so their presence is inferred by missing transverse momentum $p_T$ in events with Standard Model particles involved.

A common method is the so-called mono-X searches, in which WIMPs are produced together with visible particles such as a jet, photon or weak boson. Schematically, $p+p \rightarrow \chi + \chi + j/\gamma/Z/W$. The constraints they provide are model-independent, but there is a background challenge due to neutrinos, which also leave no energy. Another approach is the search for new resonances: in some models, the DM-Standard Model interaction is mediated by a new $Z'$ particle from the dark sector, which can decay to visible (such as quarks or lepton pairs) or invisible (involving WIMPs) final states. Finally, in SUSY and other scenarios, WIMPs are usually identified with the new lightest particle, typically the neutralino. In these models, SUSY production of particles, such as squarks or gluinos, is predicted to decay into dijets or leptons plus neutralinos. The lack of evidence for SUSY in the LHC data has placed strong constraints and ruled out many scenarios involving light SUSY partners.

For a detailed review of collider searches, see~\cite{DM_colliders_2018}.

%% file: CHAPTERS/Chapter3.tex
\chapter{Fundamental Processes in Gaseous Detectors} \label{Chapter3_Gaseous_Detectors}

{
\lettrine[loversize=0.15]{T}{he} main idea behind a particle detector is to study particles based on the analysis of the signal generated in the interaction of an incident particle with a target material. The objectives of these detectors vary from simple radiation counting (such as Geiger-Müller detectors) to the complex identification and characterisation of the different properties of the particle (particle type, energy, and even direction and reconstruction of the interaction like in Time Projection Chambers). Within the classification of detectors according to the possible interactions, some of the most important are ionisation detectors (creation of ion-electron pairs), scintillation detectors (emission of light in the excitation/de-excitation process) and bolometers (production of heat in the detection material).
%
\section*{}
\parshape=0
\vspace{-20.5mm}
}

In this chapter, and because they are central to the development of this thesis, we focus on the study of detectors whose detection material is gaseous, used in ionisation mode. Gaseous detectors have played a fundamental role in experimental particle physics since the beginning of the 20th century, starting with the cloud chambers that led to the discovery of particles such as the positron or the muon. In a gaseous detector in ionisation mode, the gas is ionised when a particle passes through it, producing ion-electron pairs. These charges drift towards the detector electrodes when a suitable electric field is applied. Near the electrode, the charges undergo a multiplication process due to an intense electric field, resulting in the generation of an electrical signal that can be detected, and the charge can be duly processed and analysed.

%
\section{Interaction of Particles in a Gaseous Detector} \label{Chapter3_Interactions}
The first step to detect a particle in a gaseous detector is to ionise the medium. This is done differently depending on the type of particle. This section is devoted to the study of the interaction mechanisms that different particles exhibit in a detector.

\subsection{Photons} \label{Chapter3_Interactions_Photons}

Photons interact in a gaseous detector by one of four interactions: photoionisation, elastic scattering, inelastic scattering and pair production. All these interactions are processes governed by probabilities (described by the cross-sections of each process), in which a photon loses all or part of its energy. Therefore, the most intuitive quantity to get an idea of the frequency of interaction is the mean free path, $\lambda$, defined as the average distance a photon travels before interacting in a material. This quantity comes from the application of the Beer-Lambert law to a monoenergetic photon beam of intensity $I_0$:

\begin{equation}
    I(x) = I_{0}\exp\left(-n\sigma x\right) = I_{0}\exp\left(-x/\lambda\right)
\end{equation}

where $I(x)$ is the intensity of the beam after travelling a distance $x$, $\sigma$ is the total cross-section (sum of the four individual cross-sections), $n$ is the number density of the medium and we define $\mu = \left( n\sigma\right)$ as the linear attenuation coefficient. The mean free path is therefore defined as $\lambda = \mu^{-1}$, and is the distance at which the beam intensity is reduced by a factor $e$. Sometimes, the mass attenuation coefficient is used, $\mu_{m}=\mu/\rho$, where $\rho$ is the density of the material. This is the quantity that is often tabulated in sources such as~\cite{NIST_XCOM}.

\begin{figure}[htb]
\centering
\includegraphics[width=0.85\textwidth]{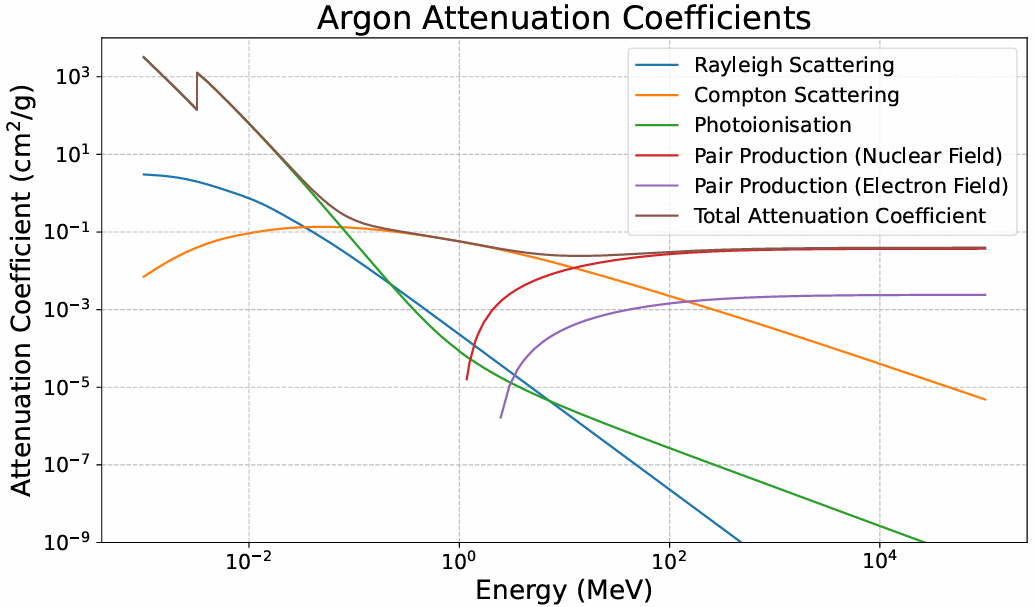}
\includegraphics[width=0.85\textwidth]{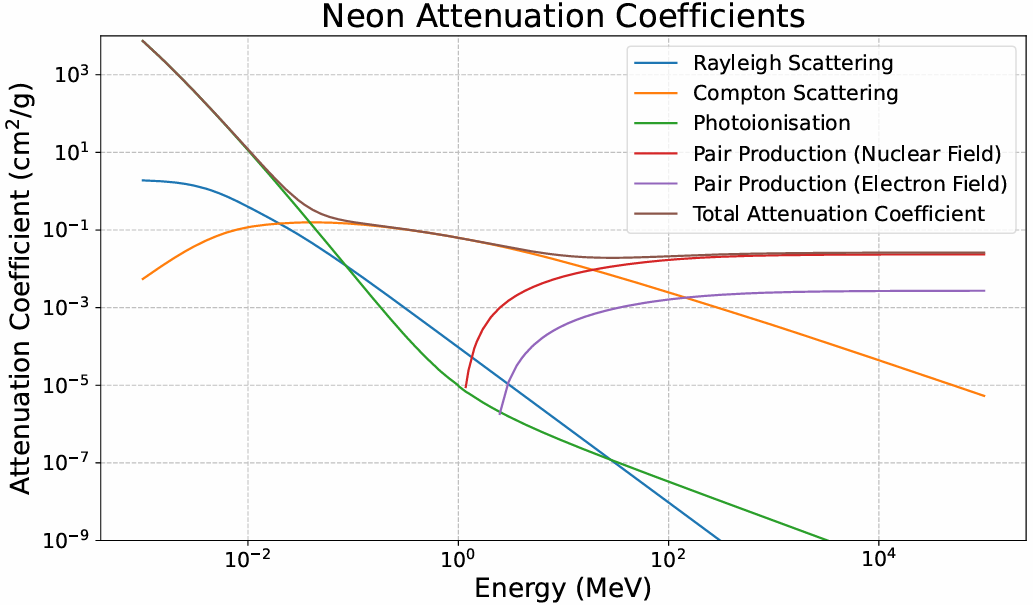}
\caption{Mass attenuation coefficients for argon (top) and neon (bottom). All the individual interaction processes are included. Elaborated with data extracted from~\cite{NIST_XCOM}.}
\label{fig:chapter3_attenuation_coefficients}
\end{figure}

In Figure~\ref{fig:chapter3_attenuation_coefficients}, the energy dependence of total and individual mass attenuation coefficients for argon and neon is shown. Each of the four processes is discussed below.

\vspace{2mm}
\textbf{\normalsize Photoionisation}
\vspace{0mm}

In photoionisation, also known as the photoelectric effect (although this name comes from the study of this effect in metals), a photon is completely absorbed by an atom in the gas, resulting in the ejection of an electron (photoelectron) from one of the atomic shells. This process requires that the energy of the photon, $E_{\gamma}$, exceeds the binding energy of the electron in the corresponding shell, $E_{\mathrm{shell}}$. The kinetic energy of the ejected electron, $E$, is given by the difference between the energy of the photon and the binding energy of the shell:

\begin{equation}
    E = E_{\gamma} - E_{\mathrm{shell}} \label{eq:photoelectric}
\end{equation}

The cross-section of the photoelectric effect depends mainly on the energy of the incident photon ($E_{\gamma}$) and the atomic number ($Z$) of the material. The following expression, valid up to energies of $\sim$ 500~keV (the electron mass), reflects the dependence on these parameters:

\begin{equation}
    \sigma_{\mathrm{pe}}\propto \frac{Z^5}{E_{\gamma}^{7/2}} \label{eq:photoelectric_sigma}
\end{equation}

Photoionisation has the highest probability at low energies (typically of the order of $\sim$ 10~keV), due to the inverse energy dependence. At higher energies, the probability of other interactions (Compton scattering or pair production) becomes dominant. On the other hand, there is a very strong dependence on the atomic number, which makes gases with high $Z$ more efficient at absorbing photons. This is a factor that influences the detector design phase of experiments, and explains why many of them use gases such as argon and xenon.

When the energy of the incident photons is equal to or slightly higher than the binding energy of the electrons in the innermost shells (such as the K-shell), there is an abrupt jump in the photoelectric cross-section (and thus in the mass attenuation coefficient) known as the shell edge effect. This increase in the photoelectric probability is explained by the availability of more electrons to interact with (increase in effective $Z$). This phenomenon can be observed for argon in Figure~\ref{fig:chapter3_attenuation_coefficients} around 3.20~keV, which is the binding energy of the K-shell. In the case of neon, the K-shell energy is at 0.87~keV, so the shell edge is not visible in Figure~\ref{fig:chapter3_attenuation_coefficients}.

In photoionisation, the photoelectron leaves a vacancy when it is expelled, which makes the atom unstable. This vacancy can be filled by means of two mechanisms:

\begin{itemize}
    \item \textbf{Fluorescence}: in this atomic de-excitation mechanism, the vacancy left by the photoelectron in the shell $i$ is occupied by another electron from an outer shell $j$, emitting a photon of energy $E_i-E_j$. These transitions are usually called characteristic X-rays, because the energy is in the X-ray range and they are different for each atom. It is usual to find these transitions referred to in Siegbahn notation (where $\mathrm{K = 1s_{1/2}}$, $\mathrm{L1 = 2s_{1/2}}$, $\mathrm{L2 = 2p_{1/2}}$, $\mathrm{L3 = 2p_{3/2}}$, $\mathrm{M1 = 3s_{1/2}}$, and so on). Figure~\ref{fig:chapter3_characteristic_xrays_transitions} shows the various possible transitions between energy levels. An approximation to the real values of these transitions can be obtained by applying the Bohr atomic model. Thus, for example, for the $\mathrm{K\alpha}$ lines one obtains:

    \begin{equation}
        E_{\mathrm{K\alpha}}= E_{i=1}-E_{j=2}=\frac{m_\text{e} q_\text{e}^4}{8 h^2 \varepsilon_{0}^2} \left( \frac{1}{1^2} - \frac{1}{2^2} \right)(Z-1)^2 \approx \frac{3}{4}(Z-1)^2 \times 13.6~\mathrm{eV}
    \end{equation}

    For argon ($Z=18$), this gives $E_{\mathrm{K\alpha}}=2.95~\mathrm{keV}$, which matches the experimental value~\cite{xray_data_booklet}. This formula works well for low $Z$ values. This fluorescence X-ray can either interact in the gas volume, or it can leave it without interacting. In the first case, it contributes to the reconstruction of the full energy of the initial photon, while in the second case, the energy reconstruction is incomplete, giving rise to the so-called escape peak.
    
    \begin{figure}[htbp]
    \centering
    \includegraphics[width=0.6\textwidth]{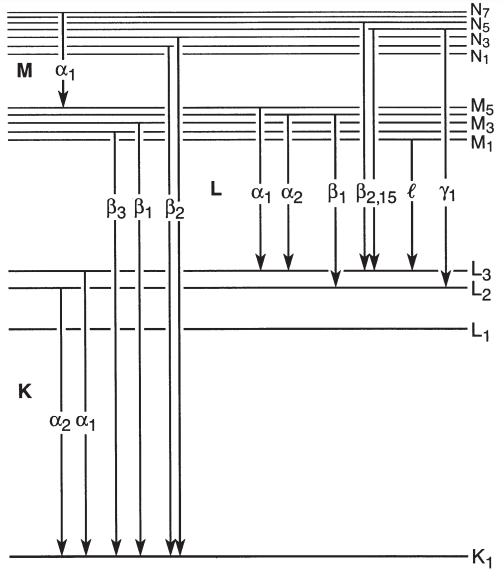}
    \caption{Characteristic X-ray transitions using Siegbahn notation. Extracted from~\cite{xray_data_booklet}.}   \label{fig:chapter3_characteristic_xrays_transitions}
    \end{figure}
    
    \item \textbf{Auger electron}: in this case, the vacancy in the shell $i$ is also occupied by an electron in a higher shell $j$, but the energy difference $E_i-E_j$ is transmitted to another electron in a shell $a$. This is a non-radiative process, in which the $a$-shell electron, known as the Auger electron, is stripped from the atom if $E_i-E_j > E_a$, with $E_a$ the binding energy of that electron. Auger transitions are also often denoted by X-ray spectroscopy notation; thus, for example, a KLL transition would be one in which a vacancy in the K shell is occupied by an electron in the L shell, transferring the energy difference to another electron in the L shell.
\end{itemize}

\begin{figure}[htbp]
\centering
\includegraphics[width=0.6\textwidth]{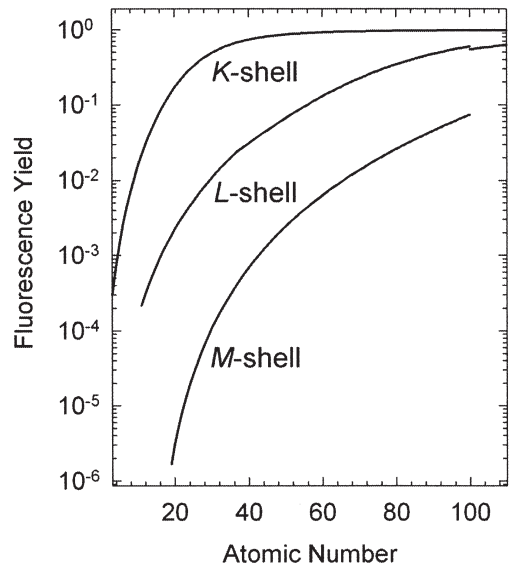}
\caption{Fluorescence yield as a function of $Z$ for K, L and M shells. L and M shells are averages over subshells. Extracted from~\cite{xray_data_booklet}.}   \label{fig:chapter3_fluorescence_yield}
\end{figure}

Both mechanisms compete to fill the vacancy left by the photoelectron. The probability that the atom is stabilised by one method or the other is measured by the fluorescence yield, $\omega$. This quantity indicates the number of times a fluorescence event occurs per photon absorbed, and $1-\omega$ gives the probability of Auger electron emission. In general, fluorescence is more likely as $Z$ increases, and is much more likely to occur in the innermost layers ($\omega_\mathrm{K}>\omega_\mathrm{L}>\omega_\mathrm{M}$). This dependence can be clearly seen in Figure~\ref{fig:chapter3_fluorescence_yield}. In particular, $\omega_\mathrm{K}$ for neon ($Z=10$) is $< 2$\%, while for argon ($Z=18$) it is $\approx 13$\%. In both cases, $\omega_\mathrm{L}$ is negligible (of the order of 0.01\% for neon, and 0.1\% for argon).

In both processes, occupying the vacancy generated by the photoelectron creates another vacancy in a higher shell (or two vacancies in the case of Auger transitions), so that several successive transitions are necessary to reach a situation of atomic stability. In the following, we will illustrate how these mechanisms come into play in an X-ray calibration of a Micromegas detector using $\mathrm{Ar-5\%iC_{4}H_{10}}$ at 1~bar as gas (the principle of operation of these detectors is explained in Section~\ref{Chapter4_Micromegas}, whereas the reason for the inclusion of isobutane is discussed in Section~\ref{Chapter3_Charge_Generation}).

\begin{figure}[htbp]
\centering
\includegraphics[width=1.0\textwidth]{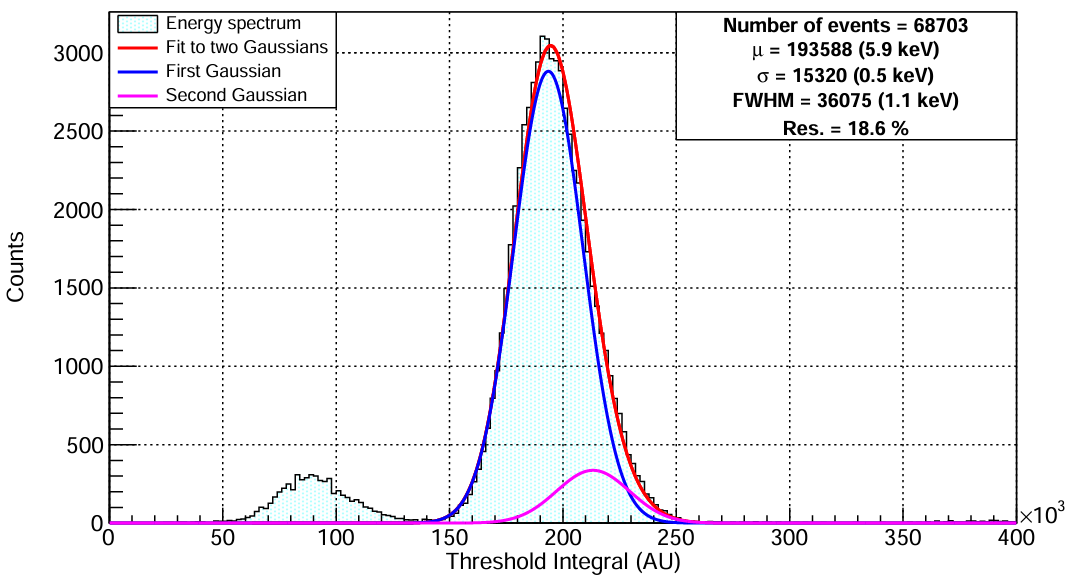}
\caption{$^{55}$Fe calibration spectrum using a microbulk Micromegas detector filled with $\mathrm{Ar-5\%iC_{4}H_{10}}$ at 1~bar. Fit to the combined photopeak around 6~keV (red) is included, along with the individual gaussians resulting from the combined fit: the 5.9~keV (blue) and 6.5~keV (pink) lines (the intensity quotient 24/3 between them has been imposed to perform the fit). The escape peak around 3~keV constitutes $\sim$ 10\% of the events. Source: own elaboration.}   \label{fig:chapter3_fe55_spectrum}
\end{figure}

In Figure~\ref{fig:chapter3_fe55_spectrum}, a $^{55}$Fe calibration spectrum is shown. $^{55}$Fe is an isotope ($T_{1/2}=2.76$~y \cite{lnhb_table_radionucleides}) that decays through electron capture:

\begin{equation}
    ^{55}\mathrm{Fe} + e^- \rightarrow ~ ^{55}\mathrm{Mn} + \nu_e
\end{equation}

With almost 90\% probability~\cite{lnhb_table_radionucleides}, the captured electron belongs to the K shell. As can be seen in Figure~\ref{fig:chapter3_fluorescence_yield}, the fluorescence yield of the manganese ($Z=25$) K-layer is $\approx$ 30\% (so $\approx$ 27\% of the total decays). Therefore, 70\% (around 63\% of the total decays) of K-shell de-excitations occur via Auger emissions. However, these electrons are absorbed in the source encapsulation, so only fluorescence X-rays are of interest. $^{55}$Mn has two main K-shell characteristic X-rays: 5.9~keV and 6.5~keV, with intensities of roughly 24\% and 3\%~\cite{lnhb_table_radionucleides} (referred to the total 27\%). The first comes from the combination of the $\mathrm{K\alpha_1}$ and $\mathrm{K\alpha_2}$ lines (with intensities of 16\% and 8\%, respectively), and the second comes mainly from the $\mathrm{K\beta_1}$ line.

These X-rays interact with argon ($\mathrm{iC_{4}H_{10}}$ will be ignored as it is the minority component) mainly by photoelectric effect (see Figure~\ref{fig:chapter3_attenuation_coefficients}), producing photopeaks at 5.9~keV and 6.5~keV, which are blended into one due to the resolution of the detector and the difference in intensities. When electrons are stripped from the K shell (binding energy 3.2~keV), 87\% of the time de-excitation will occur via Auger emission, so all the energy will be registered. However, 13\% of the time, fluorescence photons will be produced, mostly from $\mathrm{K\alpha}$ emission, with energy 2.95~keV (the remaining energy up to 3.2~keV will almost always be reabsorbed as Auger electrons, since $\omega_\mathrm{L}$ is negligible). These $\mathrm{K\alpha}$ photons can either be reabsorbed, contributing to the full photopeak (but leaving an energy deposit that is spatially distinguishable from the main deposit), or escape from the detector without interacting, producing an escape peak around 3~keV. The proportion of fluorescence photons escaping from the detector depends strongly on the particular detector geometry and source positioning, as well as the gas and pressures used. From the calibration in Figure~\ref{fig:chapter3_fe55_spectrum}, it can be seen that the escape peak contains about 10\% of total events, indicating that a large part of the Ar fluorescence photons are escaping from the detector.

\vspace{2mm}
\textbf{\normalsize Elastic Scattering}
\vspace{0mm}

Elastic scattering, also known as Rayleigh scattering, occurs when there is no change in energy in the incident photon, only a change of direction. This interaction is most relevant when the wavelength of the incident photon is much larger than the size of the atom or molecule (in other words, when the energy of the photon is much smaller than the atomic energy scale). In this case, the photon interacts with the atom or molecule as a whole, inducing a dipole moment in the electronic cloud. The photon is scattered in a direction other than the incident direction, and the effective cross-section of this scattering follows the law of dipole radiation:

\begin{equation}
    \sigma_{\mathrm{R}}\propto \frac{\alpha^2}{\lambda_\gamma^4} \propto \alpha^2 E_\gamma^4 \label{eq:rayleigh_sigma_visible}
\end{equation}

where $\alpha$ is the polarizability of the atom or molecule, and $\lambda_\gamma$ and $E_\gamma$ are the wavelength and energy of the impinging photon. This formula is typically valid for visible light and the near UV, where $\lambda_\gamma >$ atom size. However, the energies of interest in this thesis are in the X-ray range, where $\lambda_\gamma <$ atom size. In this case, the incident photon can interact individually with the bound electrons making up the electron cloud, and Rayleigh scattering is usually modelled as Thomson scattering (elastic scattering of photons by free electrons) modified by the atomic form factor (to account for the internal atomic structure):

\begin{equation}
    \frac{d\sigma_\mathrm{R}}{d\Omega} = r_e^2 \left(\frac{1+\cos^2\theta}{2}\right) F^2(q,Z) \label{eq:rayleigh_sigma_xray}
\end{equation}

where $\frac{d\sigma_R}{d\Omega}$ is the differential cross-section, $r_e$ is the classical electron radius, $\theta$ is the scattering angle and $F(q,Z)$ is the atomic form factor, with $q$ the momentum transfer. Obtaining $\sigma_R$ is usually done by numerical integration of Equation~\ref{eq:rayleigh_sigma_xray}, using tabulated atomic form factors~\cite{NIST_XCOM}. This is a complicated task beyond the scope of this thesis, and it suffices to note that $\sigma_R$ increases with $Z$ (due to a larger number of electrons to interact with) and decreases with $E_\gamma$ (as $E_\gamma$ increases, $q$ increases, and $F(q,Z)$ decreases). This behaviour can be seen in Figure~\ref{fig:chapter3_attenuation_coefficients}.

\vspace{2mm}
\textbf{\normalsize Inelastic Scattering}
\vspace{0mm}

In inelastic scattering, more commonly known as Compton scattering, a photon (typically X-ray) of energy $E_\gamma$ interacts with a free or weakly bound electron. An energy and momentum transfer from the photon to the electron occurs, resulting in a scattered photon with energy $E_{\gamma^\prime}$ and an electron with energy $E_e$. For gases, the binding energies of the electrons in the last shells are of the order of eV, so they can be considered free electrons for all practical purposes.

Following the laws of conservation of energy and momentum, the wavelength shift for the incident photon can be derived:

\begin{equation}
    \Delta\lambda = \frac{h}{m_e c}(1 - \cos{\theta}) \label{eq:compton_wavelength_shift}
\end{equation}

where $\Delta\lambda = \lambda' - \lambda$ is the wavelength shift, $h$ is Planck's constant, $m_e$ is the rest mass of the electron, $c$ is the speed of light and $\theta$ is the scattering angle. This shows that the shift only depends on the scattering angle and is independent of the initial energy of the photon. Using the relation $E_\gamma = hc/\lambda$, Equation~\ref{eq:compton_wavelength_shift} can be expressed in terms of energy:

\begin{equation}
    E_{\gamma^\prime} = \frac{E_\gamma}{1 + (E_\gamma/m_e c^2)(1-\cos\theta)} \label{eq:compton_energy_shift}
\end{equation}

The higher the scattering angle (with maximum at $\theta =180^\circ$), the higher the energy transferred to the electron ($E_e = E_\gamma - E_{\gamma^\prime}$).

\begin{figure}[htbp]
\centering
\includegraphics[width=0.8\textwidth]{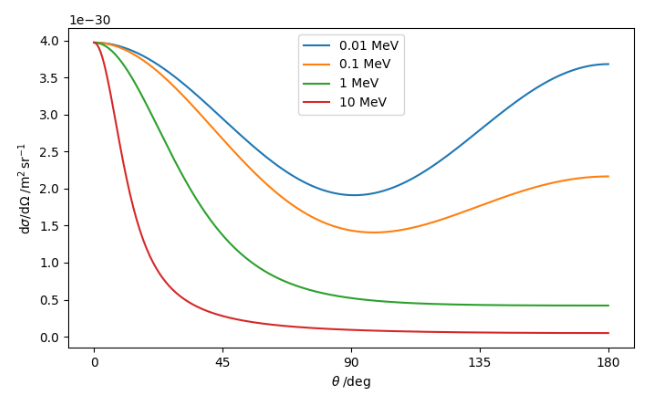}
\caption{Klein-Nishina differential cross-section $\frac{d\sigma}{d\Omega}$ as a function of the scattering angle $\theta$ for different incident energies $E_\gamma$. Plot extracted from~\cite{klein_nishina}.}  
\label{fig:chapter3_klein_nishina}
\end{figure}

The differential cross-section of the Compton effect is described by the Klein-Nishina formula:

\begin{equation}
    \frac{d\sigma}{d\Omega} = \frac{1}{2} r_e^2 \left(\frac{E_{\gamma^\prime}}{E_\gamma}\right)^{2} \left[\frac{E_{\gamma^\prime}}{E_\gamma} + \frac{E_\gamma}{E_{\gamma^\prime}} - \sin^2(\theta)\right] \label{eq:compton_klein_nishina}
\end{equation}

where $\frac{d\sigma}{d\Omega}$ is the differential cross-section and $r_e$ is the classical electron radius. The dependence of the cross-section on the scattering angle for different energies can be seen in Figure~\ref{fig:chapter3_klein_nishina}. It can be seen how backscattering ($\theta >90^\circ$) is equiprobable at low energies ($E_\gamma =$ 0.01~MeV), but is suppressed as the energy increases.

\vspace{2mm}
\textbf{\normalsize Pair Production}
\vspace{0mm}

Pair production is a process in which a high-energy photon interacts with a strong electromagnetic field, resulting in the creation of an electron-positron pair. For this process to occur, $E_\gamma >$ 1.022~MeV = $2m_e$, with $m_e$ the rest mass of the electron. Pair production plays a fundamental role in the attenuation of gamma rays in matter, and can occur in the presence of the electronic or nuclear Coulomb field:

\begin{equation}
    \gamma + A \rightarrow e^- + e^+ + A \label{eq:pair_production}
\end{equation}

where $\gamma$ is the incident photon and $A$ represents the atomic nucleus or an electron.

In the case of the nuclear field, the nucleus absorbs some of the recoil momentum to conserve momentum, but it does not significantly affect its state because $m_N > m_e$. The differential cross-section of the process is described by the Bethe-Heitler formula. For the purposes of this thesis, it is sufficient to note that:

\begin{equation}
    \frac{d\sigma}{dE_e} \propto Z^2 \label{eq:pair_production_nuclear_cross_section}
\end{equation}

where $E_e$ is the energy of the electron produced, and $Z$ is the atomic number. This shows that this process is more likely in materials with high atomic number, and the dependence can be understood by noting that $\frac{d\sigma}{dE_e} \propto E^2 \propto (Ze)^2$, where $E$ is the nuclear electric field.

In the case of the electron field, also called triplet production, the electron plays an active role: by absorbing part of the recoil momentum, it will alter its final state and acquire a certain velocity, since its mass is equal to that of the pair produced. Triplet production has a higher energy threshold, and can only occur when $E_\gamma >$ 2.044~MeV = $4m_e$~\cite{pair_production}, because the energy needed to produce the pair and respect momentum conservation is higher.

The cross-section of this process is more complex than for nuclear production, because the recoil electron shares momentum and energy along with the produced pair, leading to a more complicated energy distribution among the three particles in the final state. Suffice it to note that this cross-section is much smaller than in nuclear production, and scales with $Z$ instead of $Z^2$, because the electromagnetic field is produced by an electron (the field is $E \propto e$ rather than $E \propto Ze$), not by the atomic nucleus. Thus, the more electrons an atom has, the more likely the interaction, hence the $Z$ dependence.

The energy thresholds 1.022~MeV (for nuclear production) and 2.044~MeV (for electron production) can be seen in Figure~\ref{fig:chapter3_attenuation_coefficients}. Nuclear production dominates over electronic production, and starts to be the main interaction for photons with $E_\gamma\sim$ 10~MeV.

\subsection{Charged Particles} \label{Chapter3_Interactions_Charged_Particles}

This section has been prepared following~\cite{stopping_power} as a main reference. 

Charged particles, as they pass through a gas, lose energy and change their direction continuously. There are two main modes of energy loss: 

\begin{itemize}
    \item Inelastic collisions, also called Coulomb interactions. The particle transfers energy to the medium through ionisations and atomic/molecular excitations.
    \item Bremsstrahlung or braking radiation: emission of photons due to the deceleration of particles by nuclear electric fields.
\end{itemize}

As the energy loss is continuous, it is convenient to define the energy loss per unit length, also known as stopping power:

\begin{equation}
    S(E) = -\frac{dE}{dx} \label{eq:stopping_power}
\end{equation}

As with photons (see Section~\ref{Chapter3_Interactions_Photons}), the definition in Equation~\ref{eq:stopping_power} is also known as linear stopping power, while mass stopping power is defined as $S_m(E) = S(E)/\rho$, and is the quantity typically found tabulated in sources such as~\cite{ESTAR_PSTAR_ASTAR}. The study of $S(E)$ varies as a function of the particle mass: heavy particles (such as protons, alpha particles and ions) mainly ionise the medium and deviate little from their initial trajectory, while light particles (such as electrons and positrons) undergo multiple deflections, and have radiative losses at relatively low energies. In both cases, the range of a particle with initial energy $E_i$ and final energy $E_f$ is defined as:

\begin{equation}
    R = \int_{E_f}^{E_i} S(E)^{-1} dE \label{eq:range_charged_particles}
\end{equation}

The range represents the average pathlength travelled by a particle that is slowed down from $E_i$ to $E_f$. To calculate this expression, the Continuous-Slowing-Down Approximation (CSDA) is followed, which means that energy losses are assumed to be continuous along the path, with an average energy loss per unit length described by $S(E)$. This approximation starts to fail as we approach energies of the order of outer atomic electrons, $O(10-100)$~eV, especially for the case of light particles due to their low mass. This is because the energy losses in a single collision start to represent a significant fraction of the total remaining energy, and thus are no longer continuous.

It is important to distinguish between range and penetration depth or projected range: while the range represents the total length travelled by the particle, penetration depth is a measure of the amount of matter traversed by the particle, measured along the incident direction of the particle. For heavy particles, such as alphas and protons, the two definitions are almost identical. However, in the case of light particles, it is common for the range to be larger than the penetration depth, because their trajectories are wiggly due to multiple scattering rather than straight lines.

We now describe in detail the stopping power for both types of particles.

\vspace{2mm}
\textbf{\normalsize Heavy Particles}
\vspace{0mm}

Heavy particles are characterised by a mass $M$ that is very large compared to the electron mass, $m_e$. The stopping power is mainly described by inelastic scattering energy transfers from the incident particle to the atomic electrons, and is expressed as:

\begin{equation}
    S(E)= n \int W\frac{d\sigma}{dW}dW \label{eq:stopping_power_inelastic}
\end{equation}

where $W$ is the magnitude of the energy transfer, $\frac{d\sigma}{dW}$ is the differential cross-section of the inelastic processes and $n=\rho N_A Z/A$ is the number density of electrons in the gas, with $\rho$, $Z$ and $A$ the density, atomic number and atomic weight of the gas atoms and $N_A$ Avogadro's number. The calculation of this integral is usually separated into two components: soft collisions (the incident particle sees the atomic electrons as bound, but its energy is much higher than the atomic binding energies) and hard collisions (the incident particle sees the atomic electrons as free and at rest).

A relevant quantity in the study of stopping power is the maximum energy transfer from a particle of mass $M$ to an electron of mass $m_e$, $W_\mathrm{max}$:

\begin{equation}
    W_\mathrm{max} = \frac{2\tau (\tau +2)m_e c^2}{1+2(\tau +1)(m_e/M) + (m_e/M)^2} \label{eq:max_energy_transfer}
\end{equation}

with $\tau$ the kinetic energy of the incident particle relative to its rest mass, i.e., $T=\tau Mc^2$. Given the heavy mass situation, $M\gg m_e$, Equation~\ref{eq:max_energy_transfer} takes the form:

\begin{equation}
    W_\mathrm{max} \approx 2\tau (\tau +2)m_e c^2 =  2m_e c^2 \beta^2 \gamma^2 \label{eq:max_energy_transfer_2}
\end{equation}

where $\beta = v/c$ for the incident particle, $\gamma = (1-\beta^2)^{-1/2}$ is the Lorentz factor, and the relativistic relation $\tau = \gamma -1$ was used.

\begin{figure}[htb]
\centering
\includegraphics[width=1.0\textwidth]{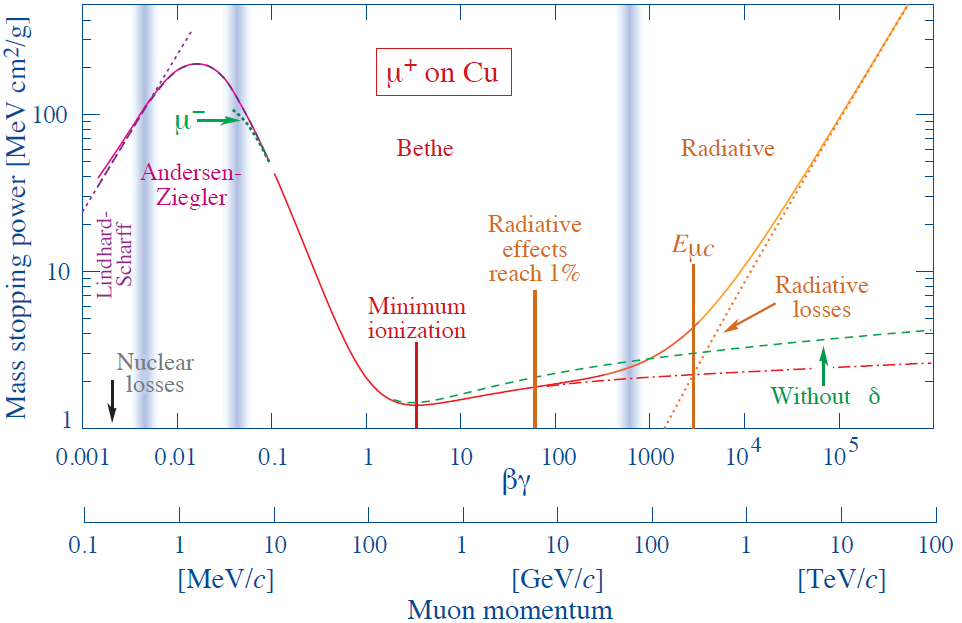}
\caption{Mass stopping power of antimuons ($\mu^+$) in copper as a function of $\beta\gamma$. Solid lines correspond to the total mass stopping power. Legend from left to right: solid purple: low-energy region where Bethe-Bloch equation does not work; dashed purple: Lindhard proportional approach at ultra low energies; dashed green: so-called Barkas effect to correct for electric charge at very low energies; solid red transforming into solid orange: Bethe-Bloch + radiative stopping power; dashed red: only Bethe-Bloch ionisation stopping power (dashed green without density correction $\delta$); dashed orange: only radiative stopping power. Plot extracted from~\cite{particle_data_group}.}   \label{fig:chapter3_stopping_power_muons}
\end{figure}

It can be shown that, at first Born approximation, Equation~\ref{eq:stopping_power_inelastic} for heavy particles can be expressed in terms of $W_\mathrm{max}$ as follows:

\begin{equation}
    S(E) = 4 \pi r_e^2 m_e c^2 \rho N_A\frac{Z}{A}\frac{z^2}{\beta^2} \left[\ln \left(\frac{W_\mathrm{max}}{I}\right) - \beta^2 - \frac{\delta}{2} - \frac{C}{Z}\right] \label{eq:bethe_bloch}
\end{equation}

where $z$ is the charge of the incident particle, $r_e$ is the classical electron radius, $I$ is the mean excitation energy of the gas and $\delta$ and $C/Z$ are the density and shell correction terms. This is the so-called Bethe-Bloch equation for the stopping power, and its form for antimuons ($\mu^+$) in copper as a function of $\beta\gamma$ can be seen in Figure~\ref{fig:chapter3_stopping_power_muons}.

The main ranges of interest of Equation~\ref{eq:bethe_bloch} are reviewed below:

\begin{itemize}
    \item At ultra low energies, the Bethe-Bloch formula is not valid and other descriptions must be used. For $\beta\gamma < 0.01$, the particle velocity is comparable to the orbital velocities of bound electrons ($v \approx 0.01$~c), and the stopping power can be described as proportional to $\beta$ (Lindhard approach). For $0.01<\beta\gamma < 0.05$, phenomenological formulae such as those developed by Andersen and Ziegler are usually used.
    \item At very low energies, $\beta\gamma\approx 0.1$, the particle velocity is greater than the orbital velocity of electrons, but not enough to see them as free particles. In this range, shell corrections have to be applied to take into account the atomic structure of the gas. This is done by including the $C/Z$ term in the formula, which contributes to a higher stopping power, $S(E)\propto\beta^{-2}(\ln(\beta^2\gamma^2)-C/Z)$.
    \item At low energies, $0.1<\beta\gamma <1$, $S(E)$ is dominated by the $\beta^{-2}$ term. In this regime, the stopping power increases rapidly as the particle velocity decreases, leading to a large energy loss at low velocities. This is particularly relevant for heavy particles such as protons or alpha particles, which exhibit higher energy loss towards the end of their path, forming the so-called Bragg peak.
    \item As we approach relativistic energies, $1<\beta\gamma <10$ ($\beta\gamma\sim O(1)$), a minimum is reached in the stopping power, and $S(E)$ begins to rise slightly due to the logarithmic term, $S(E)\propto \ln(\gamma^2)$. The energy loss in this range is practically constant, $S(E)/\rho\approx 1-2$~MeV$\cdot$cm$^2\cdot$g$^{-1}$, and particles in this energy range (such as cosmic muons) are called Minimally Ionising Particles or MIPs.
    \item Well within the relativistic energy range, $\beta\gamma >10$, $S(E)$ follows a logarithmic rise. As velocity increases and we enter the ultra relativistic range, $\beta\gamma >1000$, the medium starts becoming polarised, limiting the field extension and effectively reducing the energy loss. This is taken into account through the density effect correction ($\delta$), and $S(E)\propto \ln(\gamma^2)-\delta$. However, in this ultra relativistic range, the stopping power is dominated by radiative losses (not ionisation) and is practically linear with energy, $S(E)\approx [S(E)]_{\mathrm{rad}}\propto E$.
\end{itemize}

\vspace{2mm}
\textbf{\normalsize Light Particles}
\vspace{0mm}

For light particles (in particular, electrons and positrons), unlike for heavy particles, radiative losses due to bremsstrahlung are relevant at lower energies, and start to dominate for $E\sim O(10)$~MeV. It is therefore convenient to define from the outset:

\begin{equation}
    S(E) = [S(E)]_{\mathrm{col}}+[S(E)]_{\mathrm{rad}} \label{eq:stopping_power_light_particles}
\end{equation}

Although the exact form is more complex, it is a good enough approximation to assume $[S(E)]_{\mathrm{rad}}\propto E$.
For collisional losses, dominant at low energies, relativistic and spin effects have to be included in the kinematics of the scattering process. In addition, in the case of electrons, exchange effects associated with the indistinguishability of the incident electron and the target have to be taken into account. All in all, $[S(E)]_{\mathrm{col}}$ can be expressed as:

\begin{equation}
    [S(E)]_{\mathrm{col}}=2 \pi r_e^2 m_e c^2 \rho N_A\frac{Z}{A}\frac{1}{\beta^2} \left[\ln \left(\frac{T}{I}\right)^2 + \ln \left(1+\frac{\tau}{2}\right) + F^\pm (\tau) - \delta\right] \label{eq:collision_stopping_power_light_particles}
\end{equation}

where, as before, $T=\tau m_e c^2$. On the other hand, 

\begin{equation}
\begin{split}
    & \mathrm{Electrons}\rightarrow F^- (\tau)= (1-\beta^2)\left[1+\frac{\tau^2}{8}-(2\tau +1)\ln 2\right] \\ \\
    & \mathrm{Positrons}\rightarrow F^+ (\tau)= 2\ln 2 - \left(\frac{\beta^2}{12}\right)\left[23+\frac{14}{\tau +2}+\frac{10}{(\tau +2)^2}+\frac{4}{(\tau +2)^3}\right] \label{eq:f_electrons_positrons}
\end{split}
\end{equation}

As opposed to Equation~\ref{eq:bethe_bloch} for heavy charged particles, the shell correction term $C/Z$ is not included in Equation~\ref{eq:stopping_power_light_particles}. A full discussion on this topic can be found in~\cite{stopping_power} or in~\cite{particle_data_group}.

\begin{figure}[htbp]
\centering
\includegraphics[width=1.0\textwidth]{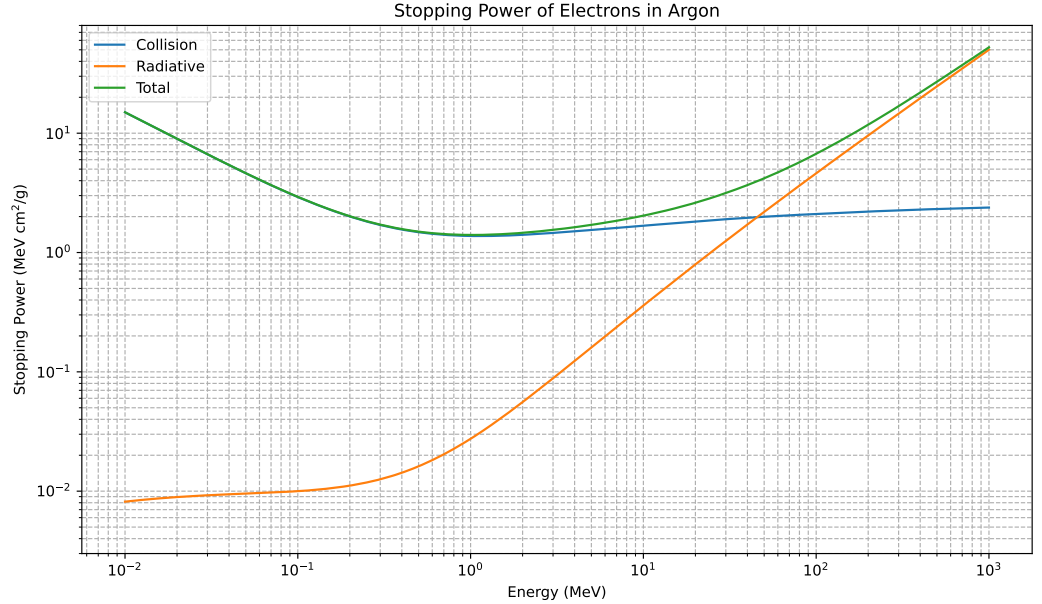}
\caption{Mass stopping power of electrons in argon. Both collision and radiative losses are included. Plot elaborated with data extracted from~\cite{ESTAR_PSTAR_ASTAR}.}   \label{fig:chapter3_stopping_power_electrons_argon}
\end{figure}

\begin{figure}[htbp]
\centering
\includegraphics[width=1.0\textwidth]{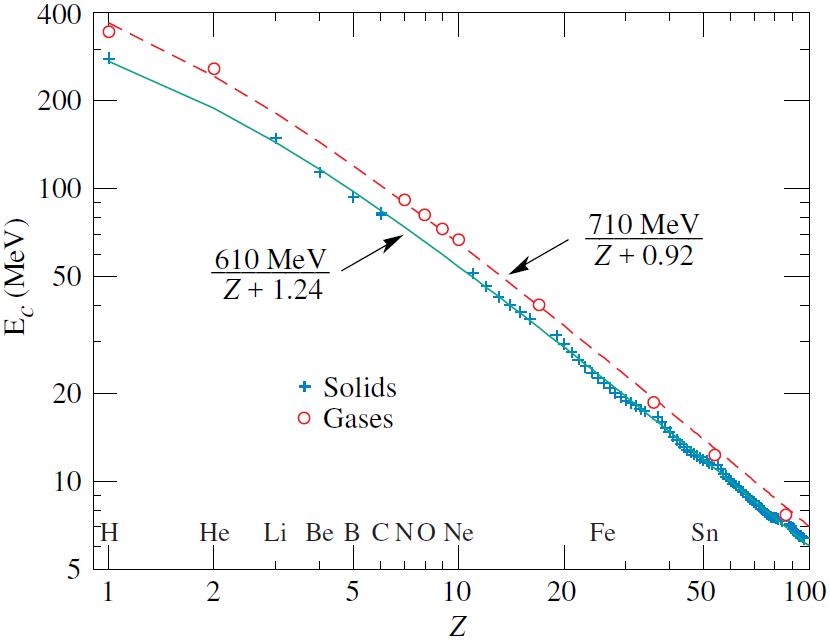}
\caption{Critical energy for electrons as a function of $Z$. Both experimental data and fits are shown. The blue markers are for solid and liquid elements, while red markers are for gases. Plot extracted from~\cite{particle_data_group}.}   \label{fig:chapter3_critical_energy_electrons}
\end{figure}

Stopping power for electrons and positrons is described by Equation~\ref{eq:collision_stopping_power_light_particles} up to $O(10)$~MeV, and is dominated by radiative losses at higher energies. This can be seen in Figure~\ref{fig:chapter3_stopping_power_electrons_argon}, where total, radiative and collision stopping powers of electrons in argon are shown. It can be intuited that the approximation $[S(E)]_{\mathrm{rad}}\propto E$ works well for $E\gtrsim 500$~keV. On the other hand, it can also be seen that there is a point at which the two curves intersect: the energy losses due to collision are equal to the losses due to radiation. This point is called the critical energy, $E_c$, and can be parameterised by the following equation:

\begin{equation}
   E_c = \frac{a}{Z+b} \label{eq:critical_energy}
\end{equation}

where $a=610$~MeV and $b=1.24$ for solids, and $a=710$~MeV and $b=0.92$ for gases. From Figure~\ref{fig:chapter3_critical_energy_electrons}, it can be seen that the agreement of this formula with experimental data $(Z,E_c)$ is very good.

\subsection{Neutral Particles} \label{Chapter3_Interactions_Neutral_Particles}
Uncharged particles, such as neutrons or WIMPs (discussed in Chapter~\ref{Chapter2_WIMPs}), do not ionise the gas directly as they lack electric charge. These particles are more penetrating than photons or charged particles, and are detected through indirect ionisations via nuclear interactions. The main interaction mechanisms for neutrons are described below:

\begin{itemize}
    \item Elastic scattering, $(n,n)$. The particle collides with a nucleus, transferring some of its kinetic energy to the nucleus but without changing its internal state. The nucleus recoils as part of the collision, with enough energy to ionise the gas. The amount of ionisation (and thus signal) produced depends on the energy of the recoil nucleus and the ionisation potential of the gas. This is one of the main interaction modes for low-energy neutrons (thermal or epithermal), as well as WIMPs, where the interaction can be Spin Independent or Spin Dependent (see Section~\ref{Chapter2_Direct_Searches}).
    \item Inelastic scattering, $(n,n\gamma)$. Upon collision, the neutron is temporarily absorbed by the nucleus, being re-emitted with energy lower than the initial energy. In this process, the nucleus is left in an unstable excited state, returning to the ground state through $\gamma$ emissions. This process is relevant for fast neutrons, but not for WIMPs.
    \item Absorption reactions. This includes reactions in which other particles are produced as a consequence of neutron absorption. Some of the most important are radiative capture, $(n,\gamma)$ (the neutron is absorbed by the nucleus, which is left in an excited state that decays by photon emission), transmutation, $(n,p)$ or $(n,\alpha)$ (here, the excited state decays by emitting a proton or an alpha particle) and fission (the nucleus splits into two or more smaller nuclei).
\end{itemize}

Normally, the amount of ionisation for the same incident particle energy is smaller in neutron or WIMP nuclear recoils than in electron recoils (direct ionisation interactions such as photons or charged particles). In this sense, the quenching factor (QF) is defined as the ratio of ionisation yield produced by a nuclear recoil to that produced by an electron recoil of the same energy. This is essential to calibrate the energy response of the detector in the search for WIMPs.

%
\section{Charge Generation} \label{Chapter3_Charge_Generation}
Charge generation in gaseous detectors is initiated by the interaction of an incident particle with atoms or molecules in the medium. These interactions have been covered in Section~\ref{Chapter3_Interactions}, and lead to the production of electron-ion pairs that form the basis of the signals that are subsequently detected. For each interaction, an electron-ion pair is produced from the initial energy deposit. This is known as primary ionisation. Secondary ionisation is due to additional ionisations by the electrons released in the primary ionisation. In general, it might be thought that a cascade of secondary ionisations occurs until the last electrons have energies below the ionisation potential of the gas, $I_{\mathrm{ion}}$. However, in practice, the average energy needed to produce an electron-ion pair, $W$, is somewhat higher, because some of the energy is lost in inelastic processes other than ionisation, such as atomic excitations or rotational/vibrational excitations (in molecules), with energies $I_{exc}$. These values (together with other properties) for different atoms and molecules are given in Table~\ref{table:chapter3_ionisation_energy_fano_factors}. It has been experimentally proved that the mean number of electron-ion pairs, $\langle N_0\rangle$, produced by an incident particle of energy $E$ is proportional to $W$:

\begin{equation}
    \langle N_0\rangle = \frac{E}{W} \label{eq:number_total_electron_ion_pairs}
\end{equation}

In reality, this number is an average, and therefore the total number of pairs produced, $N_0$, fluctuates around this value. These deviations are governed by the Fano factor, which quantifies the fluctuations with respect to a perfectly Poissonian process. Thus, the variance $\sigma_{N_0}$ is given by:

\begin{equation}
    \sigma^2_{N_0} = FN_0 \label{eq:Fano_factor}
\end{equation}

In the case $F<1$, the statistical fluctuations are smaller than the variance of a purely Poissonian distribution. This is because the ionisation events are not entirely independent of each other, leading to a narrower distribution of $N_0$. This is typical for noble gases or mixtures of noble gases with hydrocarbons such as isobutane, where the Fano factor is around 0.2-0.3. Specific values for different gases can be found in Table~\ref{table:chapter3_ionisation_energy_fano_factors}. The Fano factor determines the intrinsic resolution of the detector:

\begin{equation}
    R(\%\mathrm{FWHM})=2.35\frac{\sigma_{N_0}}{\langle N_0\rangle}=2.35\sqrt{\frac{F}{\langle N_0\rangle}}=2.35\sqrt{\frac{FW}{E}}\times 100\% \label{eq:resolution_fwhm_fano_factor}
\end{equation}

As might be expected, the statistical resolution is better the higher the energy ($R\propto 1/\sqrt{E}$), the smaller the Fano factor and the lower the average ionisation energy ($R\propto \sqrt{FW}$). For example, photons of 5.9~keV like the ones in Figure~\ref{fig:chapter3_fe55_spectrum} in a detector with argon have an intrinsic statistical resolution of $R(E=5.9~\mathrm{keV})=7.5$\%. Also, 22~keV photons from a $^{109}$Cd source like the one used in TREX-DM and 8~keV photons from copper fluorescence have intrinsic resolutions of $R(E=22~\mathrm{keV})=4$\% and $R(E=8~\mathrm{keV})=6.5$\% in argon and $R(E=22~\mathrm{keV})=3.5$\% and $R(E=8~\mathrm{keV})=6$\% in neon.

\begin{table}[htbp]\centering
\begin{center}
\begin{tabular}{c|c|c|c|c}  \hline\hline
 \textbf{Gas} &  \textbf{$I_{\mathrm{exc}}$ (eV)} & \textbf{$I_{\mathrm{ion}}$ (eV)} & \textbf{$W$ (eV)} & \textbf{$F$} \\
 \hline
 He & 19.8 & 24.5 & 45 & 0.17 \\    
 \hline
 Ne & 16.7 & 21.6 & 30 & 0.17 \\ 
 \hline
 Ar & 11.6 & 15.7 & 26 & 0.23 \\
 \hline
 Xe & 8.4 & 12.1 & 22 & 0.17 \\
 \hline
 CH$_4$ & 8.8 & 12.6 & 30 & 0.26 \\
 \hline
 iC$_{4}$H$_{10}$ & 6.5 & 10.6 & 26 & 0.26 \\
 \hline\hline
\end{tabular}    
\end{center}
\caption{Lowest excitation energy ($I_{\mathrm{exc}}$), lowest ionisation energy ($I_{\mathrm{ion}}$), mean energy for electron-ion pair production ($W$) and measured Fano factor ($F$) for different gases. Values extracted from~\cite{tesis_javi_gracia}.}
\label{table:chapter3_ionisation_energy_fano_factors}
\end{table}

In gaseous detectors, it is common to use noble gases such as argon, neon or xenon as the primary ionisation medium, due to their high ionisation efficiency and low electron attachment (see Section~\ref{Chapter3_Charge_Drift}). However, the use of a pure noble gas poses different problems:

\begin{itemize}
    \item Secondary ionisations: as mentioned above, the primary charge is capable of generating secondary ionisations. However, unwanted secondary ionisations can occur due to excitation-deexcitation processes in which energy is released in the form of photons, typically in the UV range. The contribution of these photons can lead to uncontrolled avalanches (instability) or to the appearance of charge at places other than the original interaction point (worse spatial resolution). 
    \item Instability at high fields: noble gases tend to produce electric discharges when subjected to high electric fields, such as those used to amplify primary charges. This leads to detector instability.
\end{itemize}

To alleviate these problems, it is common to add a secondary gas in small proportions (typically 1-10\%). This gas is known as a quencher, and is usually a carbon-containing molecule, such as CO$_2$ or the hydrocarbons CH$_4$ (methane) or iC$_{4}$H$_{10}$ (isobutane). Quenchers have large absorption cross-sections for UV photons, and are also effective at dissipating energy through vibrational and rotational modes. This improves both detector stability by minimising unwanted discharges, and spatial resolution by locating the charge close to the primary interaction points.

%
\section{Charge Drift} \label{Chapter3_Charge_Drift}

Once the electron-ion pair production phenomenon has taken place, the electron-ion pairs are transported under the influence of an electric field. This movement of charges towards the amplification region, called drift, is fundamental in signal detection, and is responsible for parameters such as detection efficiency and spatial and temporal resolutions. The effect of drift on these parameters is explained below.

\vspace{2mm}
\textbf{\normalsize Drift Velocity}
\vspace{0mm}

When an electric field $\vec{E}$ is applied, the generated electrons and ions move towards the electrodes (ions towards the cathode, electrons towards the anode). The movement of charges is the result of a balance between electric attractive forces and braking by collisions with gas molecules. This is encapsulated in the formula for the drift velocity, which is directly dependent on the applied electric field:

\begin{equation}
    \vec{v}_{e,i} = \mu_{e,i}(E)\vec{E} \label{eq:drift_velocity_ions_electrons}
\end{equation}

where $\vec{v}_{e,i}$ is the electron/ion drift velocity and $\mu_{e,i}(E)=q_{e,i}/(\nu(E)m_{e,i})$ is the so-called electron/ion mobility, with $q_{e,i}$ and $m_{e,i}$ the electric charge and mass of electron/ion, and $\nu$ the mean collision frequency. Electrons, being much lighter than ions, have a higher mobility, and therefore drift much faster towards the anode than ions towards the cathode. On the other hand, mobility depends on the specific gas mixture and its pressure through $\nu$: the heavier the gas and the higher the pressure, the more frequent the collisions of electrons/ions with the gas, and therefore the lower the drift velocity.

The drift velocity affects the time resolution of the detector: fast mixtures allow for short drift times and narrower pulses, which makes it possible to resolve events very close to each other. These mixtures are ideal for high-energy physics applications in which there is a high event rate and pile-up is a limiting factor. For example, the Micromegas detectors from the tracking system of the COMPASS experiment~\cite{COMPASS_experiment} use a mixture of Ne/C$_{2}$H$_{6}$/CF$_{4}$ (80/10/10), where ethane is used as a quencher, and carbon tetrafluoride is used to increase the drift velocity of the mixture.

\vspace{2mm}
\textbf{\normalsize Diffusion of Charges}
\vspace{0mm}

As the charges drift due to the electric field, a random spread occurs that causes the charge cloud to deviate from the direction of the field lines. This is a result of thermal motion and collisions with the gas, and is known as diffusion. Diffusion affects both spatial ($x-y$) and temporal ($z$) resolution through its two components:

\begin{itemize}
\item \textbf{Transverse diffusion}: it occurs perpendicularly to the electric field, deforming the charge cloud laterally. This blurs the location of the event in the $x-y$ plane, worsening the spatial resolution of the detector.
\item \textbf{Longitudinal diffusion}: it happens parallel to the direction of the electric field. For electrons, this results in a variance in arrival times at the readout plane, broadening the signals, and thus worsening the temporal resolution.
\end{itemize}

Mathematically, the diffusion of the electronic charge cloud after a time $t$ is described by a Brownian motion. In the $x-y$ plane, the origin is the initial position of the interaction, while in the $z$ axis the spread is centred around $v_e t$. Thus, the charge cloud has a spatial distribution described by a (normalised) Gaussian:

\begin{equation}
    f(x,y,z,t)= \left(\frac{1}{\sqrt{4\pi D_T t}}\right)^2 \left(\frac{1}{\sqrt{4\pi D_L t}}\right)\exp{\left(-\frac{(x^2+y^2)}{4D_T t}-\frac{(z-v_e t)^2}{4D_L t}\right)} \label{eq:diffusion_gaussian_density_function}
\end{equation}

where $D_T$ and $D_L$ are the transverse and longitudinal diffusion coefficients. The standard deviations of this distribution are given by:

\begin{equation}
    \sigma_T = \sqrt{2D_T t} ~~~~~~ \sigma_L = \sqrt{2D_L t} \label{eq:diffusion_length}
\end{equation}

\begin{figure}[htbp]
\centering
\includegraphics[width=0.4960\textwidth]{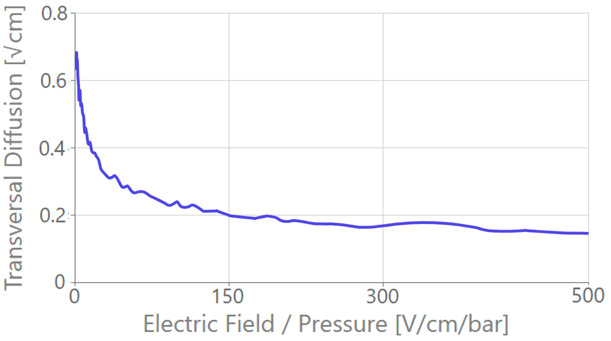}
\includegraphics[width=0.4960\textwidth]{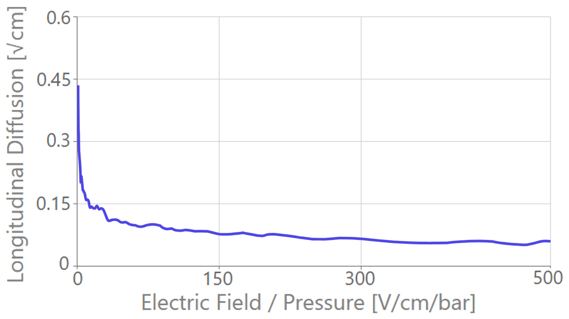}
\caption{Transverse and longitudinal diffusion coefficients ($D'$) for pure argon as a function of the reduced electric field. Plot extracted from~\cite{gas_graphs}, where more gas mixtures can be plotted interactively.}
\label{fig:chapter3_transverse_longitudinal_diffusion_pure_argon}
\end{figure}

This quantity is also called the diffusion length. The diffusion coefficients are related to the mobility $\mu_e$ through the Einstein formula:

\begin{equation}
    D=\frac{\mu_e k_B T}{e} \label{eq:Einstein_formula}
\end{equation}

where $k_B$ is the Boltzmann constant and $T$ is the temperature. Since $D$ depends on $\mu_e$, it also depends on the particular gas mixture, the pressure $P$ and the electric field $E$. $D_T$ and $D_L$ are different due to the asymmetry introduced by $E$: in the transverse component, there is no electric field, and therefore the diffusion is only due to thermal motion, whereas in the longitudinal direction there is an electric field, and thus the diffusion is not only governed by thermal energy, but also by the energy acquired during the time between collisions. $D$ is usually presented in units of cm$^2$/s. However, defining the drift length $L$ travelled after a time $t$ with the relation $L=v_e t$, Equation~\ref{eq:diffusion_length} can be expressed as:

\begin{equation}
   \sigma = \sqrt{\frac{2DL}{v_e}}=D'\sqrt{L}\label{eq:diffusion_length_redefinition}
\end{equation}

where $D'=\sqrt{2D/v_e}$ is a redefinition of the diffusion coefficient. This is the quantity used to simulate drift in Garfield++~\cite{garfield++}, and is typically presented in units of $\mathrm{\sqrt{cm}}$. In Figure~\ref{fig:chapter3_transverse_longitudinal_diffusion_pure_argon}, the transverse and longitudinal diffusion coefficients ($D'$) for argon are shown. In general, transverse coefficients are larger than longitudinal coefficients.

To mitigate the loss of spatial and time resolution due to diffusion, two strategies are followed: 

\begin{itemize}
    \item Using high electric fields: at low reduced fields ($E/P$, with $P$ the pressure), diffusion coefficients are higher because they are dominated by thermal collisions; for higher values of $E/P$, the energy acquired between collisions is greater than thermal energy, so the effect of thermal diffusion is minimised and the diffusion coefficients get smaller, as shown in Figure~\ref{fig:chapter3_transverse_longitudinal_diffusion_pure_argon}.
    \item Choosing an appropriate gas mixture: pure noble gases possess relatively high diffusion coefficients because most of the collisions are elastic (the only inelastic mechanism is excitation), whereas the addition of a quencher increases the ways to effectively dissipate energy inelastically (excitation, vibrational or rotational modes), reducing diffusion.
\end{itemize}

\vspace{2mm}
\textbf{\normalsize Loss of Charge}
\vspace{0mm}

The loss of charge in the drift process can be mainly due to two mechanisms: recombination and electron attachment. These phenomena reduce the number of free charges available for amplification. Both are briefly discussed below.

Recombination occurs when a free electron, $e^-$, is captured by a positive ion, $M^+$, to form a neutral atom or molecule:

\begin{equation}
   e^- + M^+ \rightarrow M \label{eq:charge_loss_recombination}
\end{equation}

Two types of recombination can be distinguished:

\begin{itemize}
    \item \textbf{Volumetric}: it occurs in the gas volume, especially when the electron and ion concentrations are high. The recombination rate is proportional to the densities of electrons ($n_e$) and ions ($n_i$), $R=\alpha n_e n_i$, where $\alpha$ is the recombination coefficient.
    \item \textbf{Surface}: it occurs on the chamber walls, when electrons diffuse and lose their charge at the surface. It is most common for events whose interaction occurs near the edges. 
\end{itemize}

Recombination can be relevant in situations of high ionisation density (such as the end of an alpha particle track), and can be mitigated by applying strong electric fields to reduce the time electrons and ions spend close to each other.

\textbf{Electron attachment}

Electron attachment is a critical phenomenon in gaseous detectors whereby free electrons in the gas are captured by electronegative molecules to form negative ions. The most representative case is the O$_2$ molecule, which has a high electroaffinity (0.43 $\pm$ 0.2~eV) and can be present in the sensitive volume as an impurity. This can result in significant charge loss, affecting detection efficiency.

The most usual electron attachment mechanism was proposed by Bloch and Bradbury in 1935, and revised by Herzenberg. According to this model, called the BBH model, the attachment process begins with the capture of an electron by an electronegative molecule, forming a negative ion in an excited state:

\begin{equation}
    A + e^- \rightarrow A^{-*} \label{eq:bbh_model_first_step}
\end{equation}

De-excitation can occur with the consequent release of the free electron, which would not contribute to the overall attachment:

\begin{equation}
    A^{-*} \rightarrow A + e^- ~~~~~~ A^{-*} + M \rightarrow A + M + e^- \label{eq:bbh_model_detachment}
\end{equation}

The first process is the opposite of the one described in Equation~\ref{eq:bbh_model_attachment}, while in the second the electron release occurs through the collision of the excited ion with a molecule $M$ of the buffer gas (without energy transfer to $M$).

However, there are two competing processes with Equation~\ref{eq:bbh_model_detachment}, in which de-excitation leads to a stabilisation of the negative ion:

\begin{equation}
    A^{-*} \rightarrow A^- + \gamma ~~~~~~ A^{-*} + M \rightarrow A^- + M^* \label{eq:bbh_model_attachment}
\end{equation}

The first process is a stabilisation by emission of a photon, while the second involves an energy transfer to a molecule of the gas $M$, leaving it in an excited state $M^*$.

The branching ratios of Equation~\ref{eq:bbh_model_detachment} vs. Equation~\ref{eq:bbh_model_attachment} depend on the nature of the gas used: for pure noble gases there are no vibrational or rotational modes, so processes in Equation~\ref{eq:bbh_model_detachment} are dominant. On the other hand, they are inert from the point of view of electronegativity. However, complex molecules such as the typical quenchers (CH$_4$, iC$_{4}$H$_{10}$, etc.) have a dense vibrational spectrum, which favours stabilisation by the second process in Equation~\ref{eq:bbh_model_attachment}.

There is another, less common type of attachment, based on the breaking of a molecule into two parts due to the collision with the free electron, which is immediately captured by the positive ion resulting from the dissociation process. This is known as dissociative attachment:

\begin{equation}
    AB + e^- \rightarrow (AB)^{-*} \rightarrow \begin{cases}
        AB + e^-\\A + B^-
    \end{cases} \label{eq:dissociative_attachment}
\end{equation}

In this process, the excited ion $(AB)^{-*}$ is formed as an intermediate state, and can either go back to the ground state by re-emitting the electron (top process) or break into two, absorbing the electron (bottom process). However, for this process to take place, the energy of the free electrons has to be in the vicinity of the threshold for molecular dissociation (4.6~eV for O$_2$ and 5.5~eV for H$_2$O). In practice, in typical gaseous detectors, this contribution is practically negligible, as the average energy per electron is usually less than 1~eV~\cite{Huk_attachment}.

The presence of attachment-producing contaminants, such as O$_2$, is mainly caused by:

\begin{itemize}
\item Leaks in the detector or in the volumes in direct connection with the sensitive volume: poor sealing of some detector components (such as junctions between vacuum tubes or feedthroughs) contributes to the presence of critical points through which air, containing O$_2$ (among other things), can leak.
\item Outgassing of materials: outgassing is the dynamic process by which material surfaces release adsorbed gas molecules. When these surfaces are in contact with the gas, outgassing introduces contaminants that can contribute to attachment.
\end{itemize}

To mitigate these problems, several strategies are followed: performing leak tests to monitor the different elements of the detector and defining acceptable leak levels for the experiment, heating the high-outgassing materials that are going to be in contact with the gas to remove the adsorbed compounts (\textit{bake-out} procedure), or the inclusion of filtering material in the gas system to regulate the proportions of the different contaminants.

On the other hand, as already mentioned, the inclusion of quenchers such as iC$_{4}$H$_{10}$ makes the attachment process more effective in the presence of contaminants, so it is also crucial to keep the quencher ratio at low percentages (1-10\%, typically), providing stability to the detector without degrading its detection efficiency.

A more thorough and systematic study of attachment in different gas mixtures can be found in~\cite{Huk_attachment}.

%
\section{Charge Amplification} \label{Chapter3_Charge_Amplification}

Charge amplification is perhaps one of the most critical processes in gaseous detectors, since the generation of a readable signal that can be processed by an acquisition system depends on it. 

By applying a sufficiently high electric field in a small region of the detector, the free electrons produced in the primary interactions that have drifted towards the amplification region acquire sufficient energy between collisions to produce new ionisations, which in turn cause new ionisations, in a process known as avalanche multiplication. This phenomenon is characterised by the gain $G$, which determines how much the primary charge is amplified. This process is described in more detail below.

The avalanche process begins when the primary electron cloud enters a region with a high electric field $E$ (typically E $\sim$ O(10)~kV/cm). This field provides energy to the electrons, accelerating them between collisions. The energy $K$ acquired by an electron between collisions is given by:

\begin{equation}
    K = eE\lambda \label{eq:avalanche_energy_acquired_between_collisions}
\end{equation}

where $\lambda$ is the electron mean free path, which is determined by the type of gas and the chamber pressure. If the electric field is strong, and the energy acquired is high enough ($K > W$), the electrons released from impact ionisation will be able to produce subsequent ionisations. This initiates a chain reaction.

\begin{figure}[htb]
\centering
\includegraphics[width=1.0\textwidth]{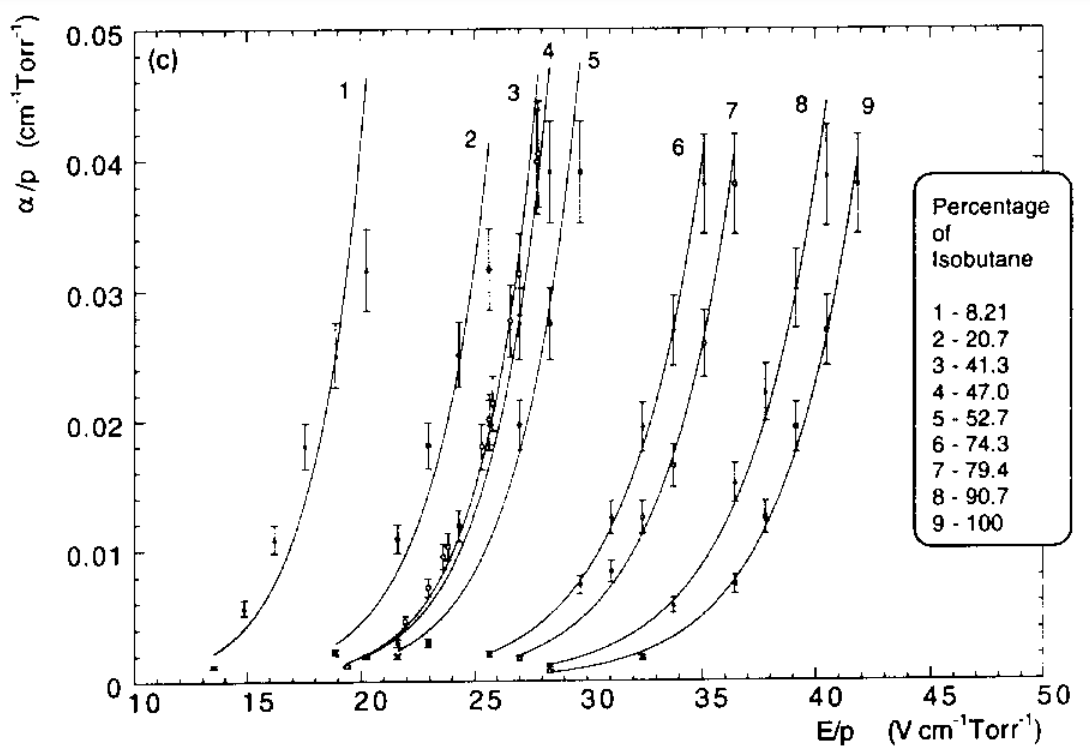}
\caption{First Townsend coefficient ($\alpha/P$) as a function of the reduced electric field ($E/P$) for different mixtures of Ar-iC$_{4}$H$_{10}$. Both experimental data and fits to Equation~\ref{eq:first_townsend_coefficient_exponential_dependence} are shown. Plot extracted from~\cite{first_townsend_coefficient_sauli}.}   \label{fig:chapter3_first_townsend_coefficient_argon_isobutane}
\end{figure}

The ionisation cascade is governed by a parameter called the first Townsend coefficient, $\alpha$, which encodes the average number of ionisations per unit distance travelled per electron (therefore, $\alpha = \lambda^{-1}$). This coefficient depends on pressure $P$ and electric field $E$, and is often modelled from empirical data. In some sources, such as~\cite{first_townsend_coefficient_sauli}, they parameterise it with an exponential dependence:

\begin{equation}
    \frac{\alpha}{P}= A\exp\left(-B\frac{P}{E}\right) \label{eq:first_townsend_coefficient_exponential_dependence}
\end{equation}

where $A$ and $B$ are constants specific to the gas mixture, obtained from a fit to the data. Measurements of $\alpha$ for argon mixtures with different percentages of isobutane, together with their fit to Equation~\ref{eq:first_townsend_coefficient_exponential_dependence}, can be found in Figure~\ref{fig:chapter3_first_townsend_coefficient_argon_isobutane}.

From the definition of $\alpha$, it can be deduced that the number of charges $dn$ produced by a cloud of $n$ charges at a distance $dx$ from the origin of the amplification region is proportional to $\alpha$: 

\begin{equation}
    dn = n\alpha dx \label{eq:townsend_avalanche_amplification_differential}
\end{equation}

Assuming a uniform electric field over the entire amplification region ($E(x)=E \Rightarrow \alpha(E(x))=\alpha(E)=\alpha$), the integration of Equation~\ref{eq:townsend_avalanche_amplification_differential} gives the absolute gain $G$ of the detector, defined as:

\begin{equation}
    G = \frac{n_f}{n_0} = \exp\left(\alpha d\right) \label{eq:gain_avalanche_amplification}
\end{equation}

where $n_f=n(d)$ is the final charge after traversing the entire amplification gap $d$ and $n_0$ is the number of primary charges entering the amplification region. Typical values of $G$ lie within the $10^3-10^6$ range.

As with the generation of primary charge, the avalanche process for each individual electron is inherently statistical, and therefore we expect fluctuations in gain around the value given in Equation~\ref{eq:gain_avalanche_amplification}. Unlike primary charges, the process is not quasi-Poissonian, but follows a so-called Polya distribution~\cite{polya_single_electron}:

\begin{equation}
    P(g,\theta) = \frac{(1+\theta)^{(1+\theta)}}{\Gamma(1+\theta)}\frac{g^\theta}{G^{(1+\theta)}}\exp\left(-(1+\theta)\frac{g}{G}\right) \label{eq:polya_distribution_gain}
\end{equation}

where $g$ represents the values that the gain can take, $\Gamma$ is the gamma function, $\langle g\rangle=G$ is the mean gain of the distribution and $\theta$ is the shape parameter, which quantifies the amplitude of the fluctuations: if $\theta \rightarrow 0$, $P(g,\theta) = G^{-1}\exp(-g/G)$ (exponential decay); if $\theta \rightarrow \infty$, $P(g,\theta)$ approaches a delta function (minimal fluctuations). It can also be shown that the variance of the distribution is:

\begin{equation}
    \sigma^2_g =\frac{G^2}{1+\theta} \label{eq:polya_distribution_variance}
\end{equation}

It is important to note that this distribution describes the amplification undergone by a single primary electron, the so-called single-electron response. However, assuming that the electrons amplify independently of each other, by virtue of the Central Limit Theorem, we can ensure that the sum of $\langle N_0\rangle$ Polya distributions with mean $G$ and variance $\sigma_g$ will follow a Gaussian distribution with mean $\langle N\rangle= \langle N_0\rangle G$ and variance $\sigma_N^2 = \langle N_0\rangle\sigma_g^2$. 

With these considerations, the intrinsic energy resolution of the detector (Equation~\ref{eq:resolution_fwhm_fano_factor}) can be updated by quadratically combining the statistical uncertainties of the primary ionisation and amplification:

\begin{equation}
    R(\%\mathrm{FWHM})=2.35\sqrt{\frac{\sigma^2_{N_0}}{\langle N_0\rangle^2} + \frac{\sigma^2_{N}}{\langle N\rangle^2}}=2.35\sqrt{\left(F+\frac{1}{1+\theta}\right)\frac{W}{E}}\times 100\% \label{eq:resolution_fwhm_fano_factor_plus_amplification}
\end{equation}

There is another factor to take into account that reduces the effective gain, the so-called space charge effect. This is the field distortion effect produced by the excessive accumulation of positive ions, and leads to a loss of gain. This effect can occur as a consequence of too strong electric fields (too many ions are generated in the amplification process) or in high-rate environments, where the positive ions have not yet finished drifting when there is already a new event being amplified.

As was the case in primary charge generation, the use of a quencher with the main gas helps to control the avalanche process, both by avoiding unwanted ionisation of UV photons and by dissipating the energy of the positive ions (to minimise space charge effects). Also, in practical realisations of detectors, the avalanche process is confined to a small region with a high electric field, which provides controlled amplification and minimises space charge effects. This is what happens near the anode wire in proportional counters or inside the microholes in GEMs or Micromegas, as we will see in the following chapter. 

%% file: CHAPTERS/Chapter4.tex
\chapter{Gaseous TPCs with MPGDs} \label{Chapter4_TPCs_Micromegas}

{
\lettrine[loversize=0.15]{T}{he} Time Projection Chamber (TPC) is an essential instrument in experimental particle physics. It is known for its ability to track ionising particles in three dimensions and to measure energy deposits with high precision. 
%
\section*{}
\parshape=0
\vspace{-20.5mm}
}

The development of TPCs starts with the first wire-based detectors, and goes up to the modern MicroPattern Gaseous Detectors (MPGDs), of special interest in this thesis, among which the Micromegas detector (Micro-Mesh Gaseous Structure) and the GEM (Gas Electron Multiplier) stand out. These technologies have achieved extraordinary levels of resolution and sensitivity, allow for particle discrimination, and have the advantage of scalability, allowing for larger detectors and, thus, larger quantities of target gas. For all these reasons, they are widely used in experimental particle physics, and more recently, in rare-event experiments such as Dark Matter searches.

\section{TPCs} \label{Chapter4_TPCs}

The foundations of TPC technology began with the invention of the Multi-Wire Proportional Chamber (MWPC) by Georges Charpak in 1968~\cite{Charpak_1968}. In this detector, a grid of tightly-spaced, parallel wires is used as anode in a gas-filled chamber. As particles pass through the chamber, they ionise the gas, releasing electrons that drift to the wires under an electric field. When close to the wires, the Coulomb interaction $\sim r^{-2}$ produces an electric field capable of creating an avalanche amplification phenomenon, yielding a readable signal. Since each wire acts as an independent proportional counter down to 1-mm wire spacing, a 2D image of the interaction can be obtained. The development of MWPCs was a revolution in particle physics, resulting in a qualitative and quantitative leap over bubble and cloud chambers, and ultimately earned Charpak the Nobel Prize in Physics in 1992.

\begin{figure}[htbp]
\centering
\includegraphics[width=1.0\textwidth]{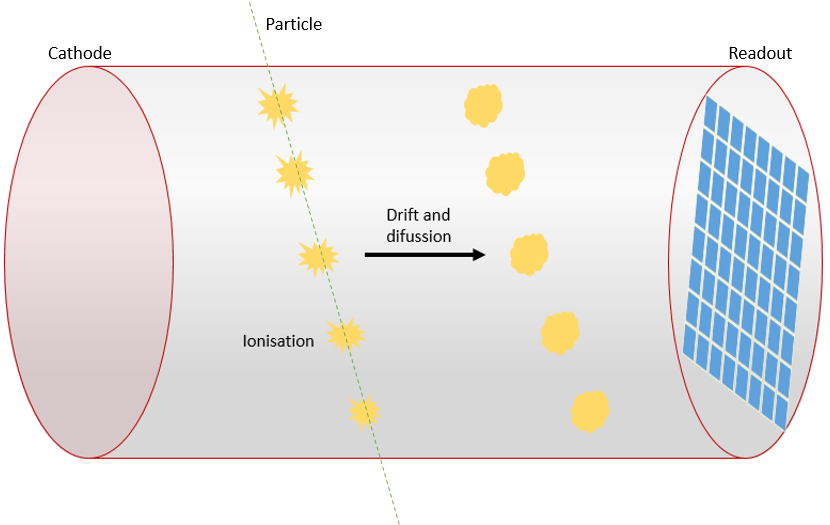}
\caption{Illustration of a cylindrical TPC composed of a cathode, a drift volume filled with gas where ionisations take place and a segmented readout towards which charges drift. Source: own elaboration.}
\label{fig:chapter4_TPC_illustration}
\end{figure}

In 1974, David Nygren of Lawrence Berkeley National Laboratory (LBNL) proposed the concept of the TPC~\cite{Nygren_2018}, which is based on the principles of MWPCs but also registers the timing of drift in the Z direction. The key is to use a uniform electric field to drift the electrons produced in the primary ionisations to a readout plane. Figure~\ref{fig:chapter4_TPC_illustration} shows a schematic representation of a cylindrical TPC with a segmented square readout.

The segmentation of the readout plane (implemented as a series of wires, strips, or pads) provides two-dimensional (X-Y) spatial information, while the time delay between the initial trigger and the last electron cloud of the event allows to infer the Z coordinate, knowing the drift velocity. The trigger can be given by the detection of scintillation photons (absolute Z) or ionisation electrons (relative Z). Thus, each ionisation cluster is mapped into a three-dimensional coordinate, allowing the full track of the incident particle to be reconstructed.

A key advantage of the TPC lies in its ability to combine these spatial coordinates with additional information such as energy deposition. This enables the identification of the particle type based on the specific ionisation profile ($dE/dx$).

Several design elements are critical for the optimal performance of a TPC:

\begin{itemize}
    \item \textbf{Drift Field}: its value has to be high enough to avoid recombination, but not so high as to produce secondary ionisations. This is achieved with electric fields in the range of $10^2 - 10^3$~V/cm. On the other hand, the uniformity of the electric field along the drift region is essential for maintaining a constant drift velocity, which ensures Z is accurately calculated. This is typically accomplished by using field cages or shapers with a series of resistive voltage dividers.
    \item \textbf{Gas Mixture}: this affects not only the drift velocity but also the diffusion and attachment properties. Fast electron transport ensures good timing resolution, ideal in high-event-rate environments, while minimal diffusion favours high spatial resolution. Optimised mixtures are selected to balance these parameters according to needs.
    \item \textbf{Readout Segmentation}: the granularity of the readout plane (typically $\sim$ 1~mm for wire grids and sub-mm for MPGDs) ultimately determines the resolution in the X-Y plane.
\end{itemize}

The first major TPC application can be found in the PEP-4 experiment at SLAC in the early 1980s~\cite{Hilke_2010}, and since then its presence has extended to branches of physics as diverse as high-energy physics, nuclear physics, astroparticle physics or medical imaging.

\section{MPGDs} \label{Chapter4_MPGDs}

The use of wires as sensing anodes in the first TPCs was a huge success, but it did not come without problems:

\begin{itemize}
    \item Spatial resolution is constrained by wire spacing, which comes with the challenge of dealing with high electrostatic forces when wires are too close to each other.
    \item The rate is limited by ion backflow: ions created near the wires move into the drift region and travel very slowly. This can create space charge effects that distort the electric field and, consequently, the particle track. This was a problem in PEP-4 and was addressed by placing a gating grid on top of the wire grid to avoid the passage of ions~\cite{Hilke_2010}.
\end{itemize}

These shortcomings led to the development of new technologies in the 1990s, among which MPGDs stand out. MPGDs are a broad family of gaseous detectors that leverage modern microfabrication techniques to replace traditional wire grids with finely patterned electrode structures, achieving sub-mm separations between electrodes. They offer better spatial resolution, higher rates and improved longevity.

\subsection{Micromegas Detector} \label{Chapter4_Micromegas}

\begin{figure}[htbp]
\centering
\includegraphics[width=0.85\textwidth]{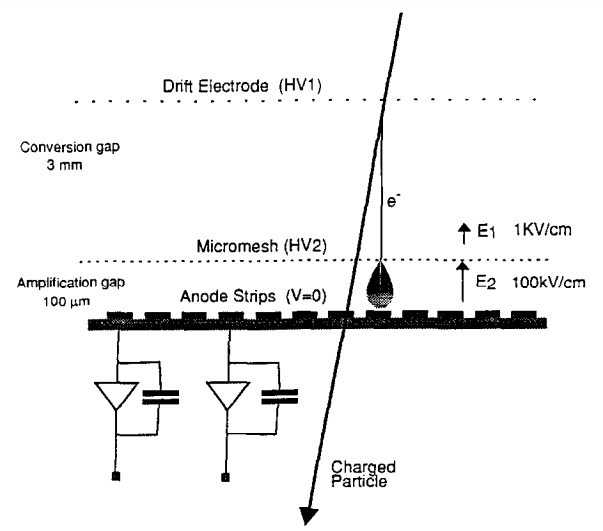}
\caption{Schematic image of a Micromegas detector, taken from the original paper~\cite{Giomataris_1995}.}
\label{fig:chapter4_Micromegas_illustration}
\end{figure}

The Micromegas detector, devised by Ioannis Giomataris and his collaborators in 1996~\cite{Giomataris_1995}, belongs to the MPGD family. Fundamentally, a Micromegas detector is a two-stage parallel-plate avalanche chamber with a narrow amplification gap~\cite{Attie_2021}. A thin metallic micromesh divides the sensitive volume of the TPC into a drift (or ionisation) region, where the primary ionisations are created, and an amplification region, where an intense electric field produces an avalanche effect on the electrons that pass through the mesh. An illustration of the Micromegas detector can be seen in Figure~\ref{fig:chapter4_Micromegas_illustration}, while Figure~\ref{fig:chapter4_microbulk_micromegas} shows a small Micromegas detector and a zoomed image of the micromesh.

The length of the drift region can vary, from a few millimeters in compact detectors to several meters in large-scale TPCs, and the drift field is in the range of $10^2 - 10^3$~V/cm, as explained above.

As for the amplification gap, its thickness is typically between 50 and 200~$\upmu$m, and is precisely maintained by insulating pillars. Once the primary electrons pass through the micromesh (facilitated by the "funnel effect" of the field lines), they encounter a strong electric field on the order of tens of kV/cm.

The main advantages of Micromegas readouts include:

\begin{itemize}
    \item \textbf{High Spatial Resolution}: the finely segmented readout, combined with the uniform and narrow amplification gap, has enabled spatial resolutions as fine as 12~$\upmu$m (RMS) in optimised configurations~\cite{Derre_2001}. This allows for extremely precise track reconstruction.
    \item \textbf{Fast Response}: the short distance over which charge multiplication occurs results in rapid signal formation, with time resolutions as low as $\sim 25$~ps in R\&D set-ups dedicated to precise timing such as PICOSEC~\cite{PICOSEC_2018}.
    \item \textbf{Excellent Energy Resolution}: The homogeneity of the amplification gap reduces fluctuations in the avalanche process, achieving outstanding energy resolutions. In experiments like CAST, this can be as low as around 12\% (FWHM) at 5.9~keV in argon~\cite{microbulk_cast_2011}, close to the theoretical limit of 7.5\% imposed by statistical fluctuations (calculated in Section~\ref{Chapter3_Charge_Generation}).
    \item \textbf{Gain}: dedicated test set-ups have proved that very high gains, even higher than 10$^6$, are attainable~\cite{Derre_2000}. Although these results are obtained in optimal conditions, under real experimental conditions it is possible to achieve gains of $10^3 - 10^4$, which means that energy thresholds < 1~keV are reasonable.
    \item \textbf{Radiopurity}: Micromegas detectors are built with radiopure materials (see Section~\ref{Chapter4_Micromegas_Radiopurity}), making them ideally suited for low-background experiments.
    \item \textbf{Scalability}: fabrication technologies allow the construction of large detectors, which is vital in experiments using large TPCs to increase detector mass.
\end{itemize}

These characteristics position Micromegas as a natural detector choice in astroparticle physics, and in particular in the search for Dark Matter. Micromegas are used in axion experiments such as BabyIAXO~\cite{IAXO_FerrerRibas_2023}, the successor of the successful CAST experiment at CERN~\cite{MicrobulkCAST_2010}, or in WIMP experiments such as TREX-DM, discussed in depth in Chapter~\ref{Chapter5_TREXDM}.

\vspace{2mm}
\subsubsection{Fabrication Processes} \label{Chapter4_Micromegas_Fabrication_Processes}

The traditional Micromegas involves suspending a metallic mesh above the anode plane by means of insulating spacers attached to the anode. Early designs employed fishing lines as spacers, but modern fabrication techniques now use photolithography to produce uniform, robust pillars that ensure a stable gap across large areas.

The fabrication techniques have evolved significantly, yielding two prominent methods: bulk Micromegas and microbulk Micromegas. Each of them has been refined to address specific experimental demands, such as large-area coverage, mechanical robustness, high energy resolution, and ultra-low radiopurity. In the following subsections, we detail the fabrication processes of both bulk and microbulk Micromegas. For an extensive review of all manufacturing techniques, see~\cite{Attie_2021}.

\vspace{10mm}
\textbf{\normalsize Bulk Micromegas}
\vspace{0mm}

\begin{figure}[htbp]
\centering
\includegraphics[width=1.0\textwidth]{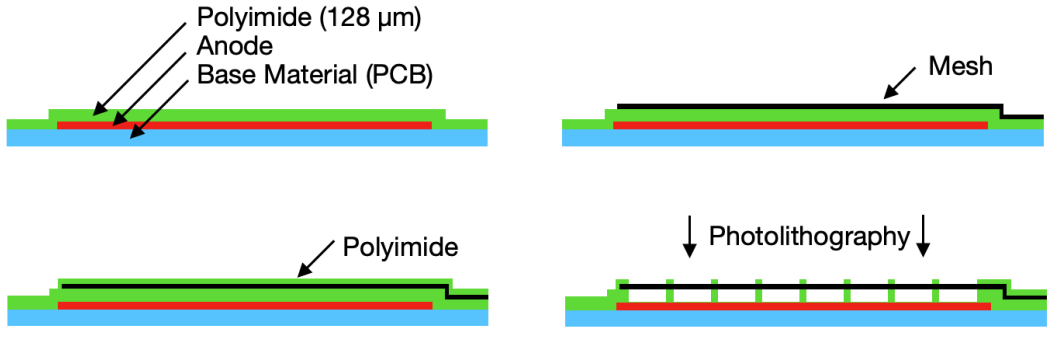}
\caption{Schematic view of the bulk Micromegas manufacturing process. Image extracted from~\cite{Attie_2021}.}
\label{fig:chapter4_bulk_fabrication}
\end{figure}

A key development in Micromegas technology came in the mid 2000s with bulk Micromegas~\cite{Giomataris_2004}, a manufacturing technique in which both mesh and anode are integrated into a single, monolithic structure. The process begins with a Printed Circuit Board (PCB) that carries a conductive anode pattern. This PCB serves as the substrate upon which the detector is built. As shown in Figure~\ref{fig:chapter4_bulk_fabrication}, the manufacturing process has four steps:

\begin{enumerate}
    \item \textbf{Lamination of the PCB}: the first step involves laminating the PCB with an insulating, photo-imageable polyimide material. The thickness of this polyimide layer is critical, as it defines the amplification gap of the detector. For example, to achieve a typical amplification gap of 128~$\upmu$m, two layers of 64~$\upmu$m polyimide are laminated sequentially.
    \item \textbf{Mesh placement}: following lamination, a metallic woven mesh (usually made from stainless steel) is carefully placed atop the polyimide layer. This mesh acts as the divider between the drift and amplification regions, and its precision is paramount to ensure uniform performance across the active area.
    \item \textbf{Second lamination}: an additional layer of the same photo-imageable polyimide is then laminated on top of the mesh. This step encapsulates the mesh within the insulating material, integrating it with the PCB.
    \item \textbf{Photolithographic etching}: a black mask featuring a defined hole pattern is applied, and the assembly is exposed to UV light. The areas under the holes become resistant to subsequent chemical treatments. An acid bath then removes the surrounding polyimide, leaving behind an array of cylindrical pillars and a supporting frame that defines the amplification gap.
\end{enumerate}

The resulting structure offers several advantages: by embedding the mesh directly into the PCB via the insulating pillars, detectors become robust against mechanical stress. In addition, the process can be integrated within industrial production techniques, which is essential for scaling up and building large-area detectors. However, one of their drawbacks is that the woven mesh itself, with a typical thickness of around 30~$\upmu$m, can introduce variations in the amplification gap (sometimes exceeding 10\%~\cite{Attie_2021}), potentially affecting energy resolution. In some cases, using an electroformed mesh can mitigate this issue, though it may come at the cost of reduced mechanical robustness.

\vspace{2mm}
\textbf{\normalsize Microbulk Micromegas}
\vspace{0mm}

\begin{figure}[htb]
\centering
\includegraphics[width=0.45\textwidth]{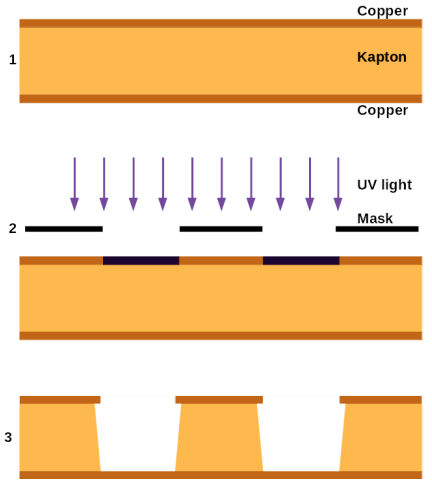}
\includegraphics[width=0.5\textwidth]{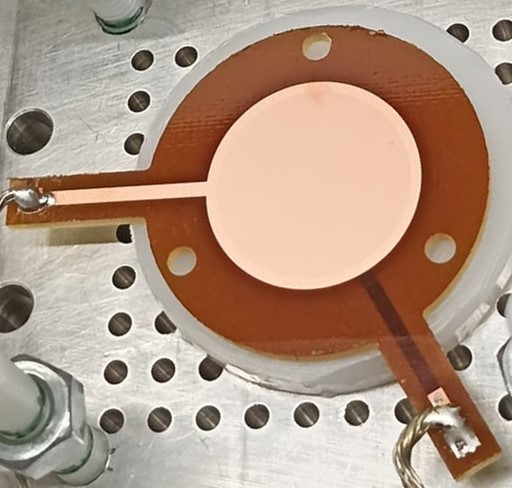}
\includegraphics[width=0.95\textwidth]{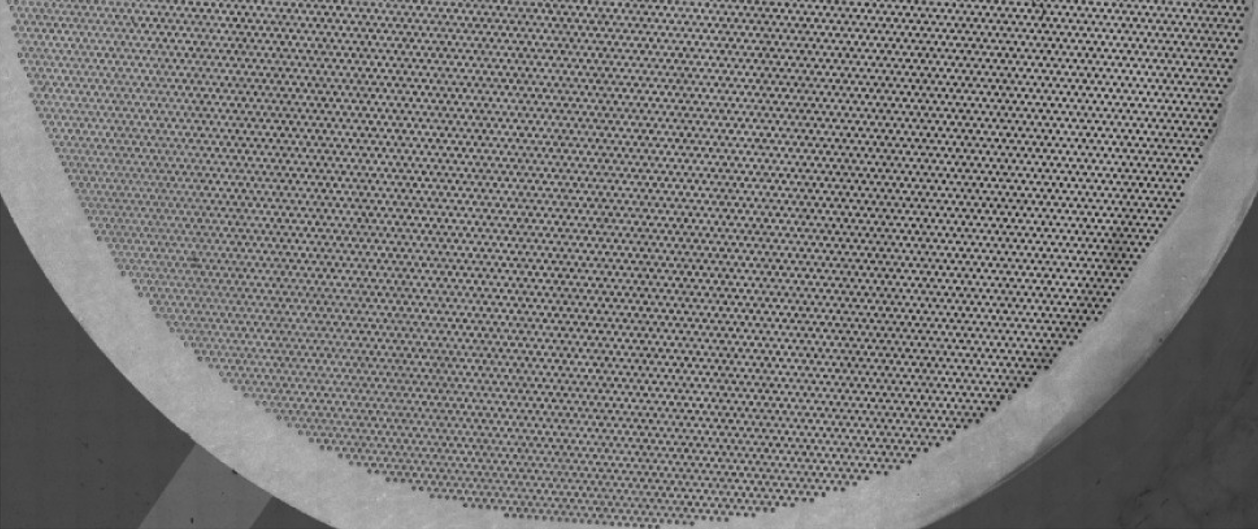}
\caption{Top left: illustration of the fabrication process in the microbulk technique. Image extracted from~\cite{Attie_2021}. Top right: a small microbulk Micromegas like the one used in Chapter~\ref{Chapter7_GEM-MM}. Bottom: zoomed view of the lower half of a microbulk Micromegas like the one shown in the figure, taken at CEA Saclay using a microscope.}
\label{fig:chapter4_microbulk_micromegas}
\end{figure}

In the late 2000s, the microbulk technique was introduced~\cite{Andriamonje_2010}, further improving the bulk technology. In this process, the mesh, supporting pillars, and readout structures are integrated into a single entity. The process, depicted in Figure~\ref{fig:chapter4_microbulk_micromegas}, has the following steps:

\begin{enumerate}
    \item \textbf{Double-sided copper-clad kapton foil}: the starting point is a thin kapton foil that is metallised on both sides with copper. One copper layer serves as the readout anode, while the opposite layer will eventually form the micromesh. 
    \item \textbf{Photochemical etching to define the mesh}: a photolithographic technique is used to etch circular holes in the copper layer that will become the mesh. This patterning is inspired by GEM production techniques and is critical for achieving the uniformity required for a consistent amplification gap.
    \item \textbf{Kapton etching to form the amplification gap}: following the definition of the mesh, the underlying kapton is chemically etched away in the regions corresponding to the holes. This etching process proceeds until the kapton is completely removed down to the anode layer. The remaining kapton around the etched holes acts as supporting pillars, giving mechanical stability to the thin structure. The gap is typically determined by the original thickness of the kapton, with standard values being around 50~$\upmu$m; however, gaps as small as 12.5–25~$\upmu$m have been fabricated for applications involving heavy gas mixtures at high pressures~\cite{Attie_2021}.
\end{enumerate}

Detectors produced with the microbulk technique offer several benefits: excellent radiopurity due to the low material budget (no additional materials apart from kapton and copper are used); very good energy resolution, often achieving values better than 12-15\% FWHM at 5.9~keV with argon mixtures. This improvement is attributed to the uniformity of the amplification gap, which reduces statistical fluctuations during the avalanche multiplication process; the integration of the mesh and readout into a single structure from the outset helps mitigate issues related to mechanical assembly and alignment that are more prevalent in the bulk method.

\subsubsection{Radiopurity} \label{Chapter4_Micromegas_Radiopurity}

Radiopurity refers to the intrinsic low levels of radioactive contaminants in the materials and components used to construct a detector. This property is essential in rare event searches, such as Dark Matter detection or neutrino-less double beta decay experiments, where small traces of radioactivity can generate background signals that mimic signal events.

In the context of Micromegas detectors, achieving high radiopurity is a critical design criterion. Both the bulk and microbulk fabrication methods strive to minimise the radioactive background, but there are intrinsic differences between the two techniques. Bulk Micromegas are typically manufactured on PCBs using photo-imageable polyimide materials and involve the use of a woven or electroformed mesh attached via insulating pillars. Although these materials are generally selected for their favourable electrical properties and mechanical stability, the multi-component assembly (including adhesives and additional insulating layers) can introduce higher radioactivity levels than those achievable with the microbulk technique~\cite{Cebrian_2010}. Indeed, microbulk Micromegas have a double-sided copper-clad kapton laminate as base material, which means that copper and kapton are the main materials used. 

A significant challenge regarding radiopurity arises during the chemical etching and cleaning stages in the fabrication process. Traditional processes often employ tap water and potassium-based chemical baths (such as KOH and KMnO$_4$) to etch the kapton and clean the foils. These procedures, however, carry the risk of depositing trace amounts of uranium and thorium (from tap water) and potassium onto the detector surfaces, which could compromise the sensitivity of the experiment making use of the detectors. Recent ideas, such as the use of deionised water, have shown promising results in terms of improved radiopurity.

In addition to controlling the chemical baths, monitoring using High-Purity Germanium (HPGe) detectors and specialised screening instruments such as the BiPo-3 detector has been essential to validate the radiopurity of the final readouts. Assays of raw materials and complete microbulk readouts have shown contamination levels for the natural decay chains of $^{238}$U and $^{232}$Th of less than a few tens of~$\upmu$Bq/cm$^2$~\cite{Cebrian_2010}, with similar values for $^{40}$K. See~\cite{Cebrian_2010} for a detailed study of the radiopurity of Micromegas readouts.

\subsection{GEM Detector} \label{Chapter4_GEM}

\begin{figure}[htbp]
\centering
\includegraphics[width=1.0\textwidth]{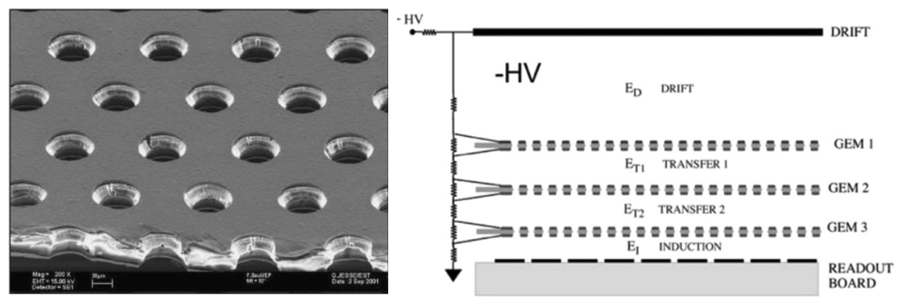}
\caption{Left: view under microscope of a GEM foil with a 50-$\upmu$m gap, 70-$\upmu$m hole diameter and 140-$\upmu$m hole pitch. Right: schematic of a triple GEM detector. Both images were taken from~\cite{Sauli_2016}.}
\label{fig:chapter4_GEM}
\end{figure}

In 1997, Fabio Sauli created the GEM~\cite{Sauli_1997}, another cornerstone in the field of MPGDs. Like Micromegas, GEMs utilise the principle of electron multiplication in a confined geometry, and they are made from a thin copper-kapton-copper foil, across which a regular pattern of microscopic holes is perforated through a precise photochemical etching process. Figure~\ref{fig:chapter4_GEM} shows a microscope image of a GEM foil.

GEMs divide the detector volume into three regions: 

\begin{enumerate}
    \item \textbf{Drift region}: it is the space between the cathode and the top copper layer of the GEM, where primary ionisations take place. The drift field has values on the order of $10^2 - 10^3$~V/cm. The specific value is chosen to optimise the collection efficiency of electrons in the holes.
    \item \textbf{Amplification gap}: it is the narrow region between the two copper electrodes, defined by the thickness of the kapton layer, typically 50~$\upmu$m. The applied amplification field is on the order of tens of kV/cm.
    \item \textbf{Transfer or induction region}: it is the volume between the bottom copper layer of the GEM and the anode, where the amplified charge is drifted towards the readout. The transfer field is in the same order of magnitude as the drift field, and the value is selected to optimise the extraction efficiency of the electrons coming out of the GEM holes.
\end{enumerate}

There is a clear conceptual link between GEMs and microbulk Micromegas, because the photochemical etching techniques used to define the micromesh and readout segmentation in microbulk Micromegas are inspired by those employed in GEM production. In principle, GEMs are also as radiopure as microbulk detectors.

GEMs are often used in combination with other GEMs, creating stacked structures in order to spread the amplification voltage across different stages, which reduces the individual voltages at each detector, improving the robustness of the overall set-up. In these cases, several transfer regions are defined by the configuration of cascaded GEMs (see Figure~\ref{fig:chapter4_GEM}). In this thesis, the combination of a GEM foil with a microbulk Micromegas is explored in Chapter~\ref{Chapter7_GEM-MM}, but with the aim of improving the overall gain of the system and reducing the energy threshold. In this case, the GEM functions as a pre-amplification stage.

Both technologies have found widespread application in high-energy physics experiments such as ALICE at CERN's LHC, where GEMs and Micromegas are deployed to achieve fine spatial resolutions and high particle rates.

\section{Spherical Proportional Counter} \label{Chapter4_SPC}

\begin{figure}[htbp]
\centering
\includegraphics[width=1.0\textwidth]{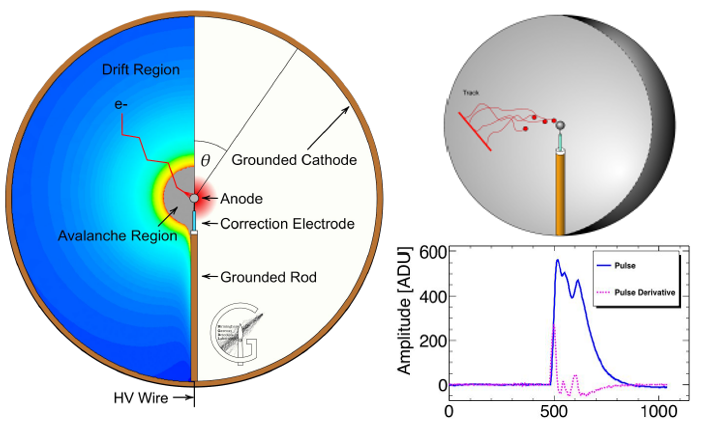}
\caption{Sketch of an SPC (left), together with a muon event inside the SPC and its signal shape (right). Illustrations taken from~\cite{Katsioulas_2021}.}
\label{fig:chapter4_SPC_sketch}
\end{figure}

Finally, Spherical Proportional Counters (SPCs), also developed by I. Giomataris and colleagues in the 2000s~\cite{Giomataris_2008}, feature a novel spherical geometry, distinct from the planar geometries of other detectors. In its classic configuration, the SPC consists of a large, hollow spherical vessel filled with a detection gas, with a single, small anode positioned at its centre. When ionising radiation interacts with the gas, it produces primary electrons that drift radially towards the anode, near which the electric field increases due to the inverse-square law, $\sim r^{-2}$, producing a multiplication phenomenon that results in a readable signal. This configuration effectively splits the sphere into two regions, drift and amplification. Figure~\ref{fig:chapter4_SPC_sketch} depicts the operating principle of an SPC, together with the typical signal produced by a track-like event such as a muon.

\begin{figure}[htbp]
\centering
\includegraphics[width=1.0\textwidth]{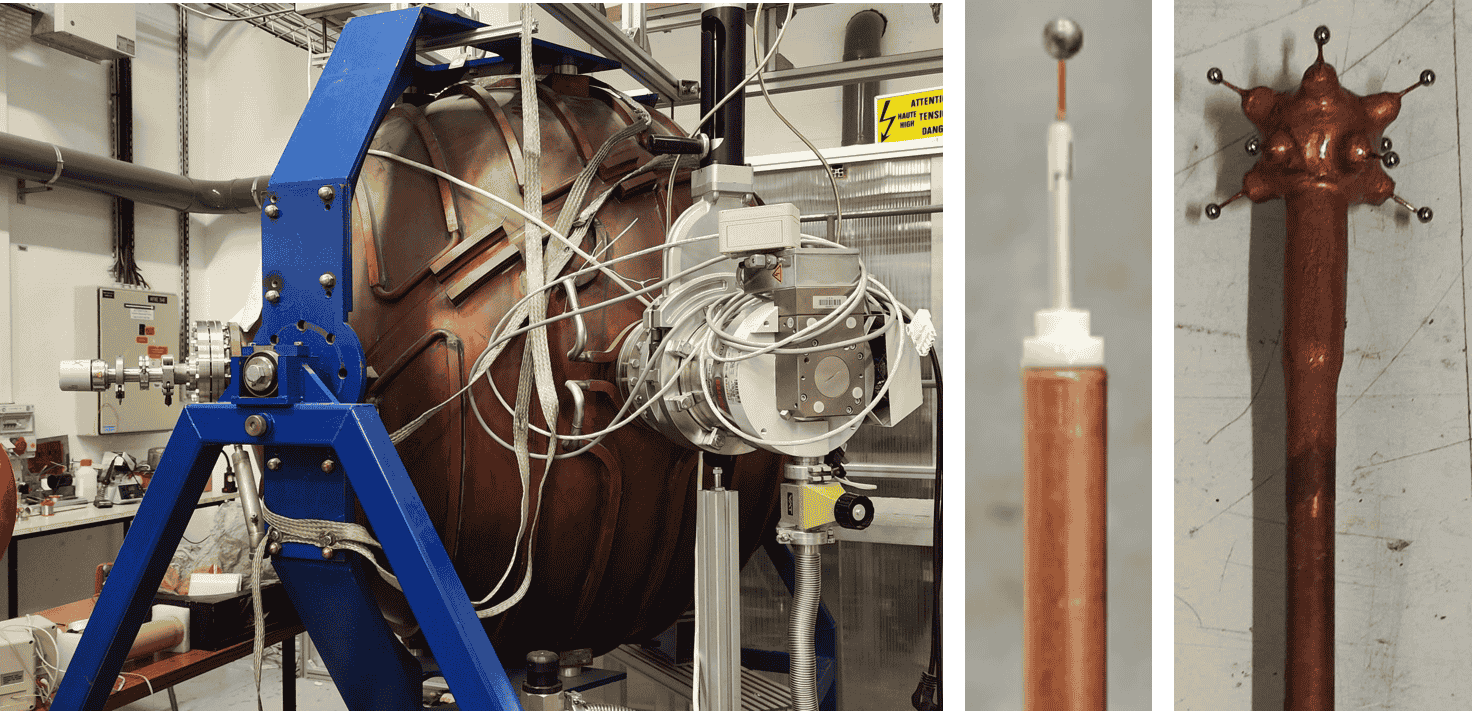}
\caption{Pictures of a copper SPC (left), a single anode (centre) and an ACHINOS multi anode (right). Photos taken at CEA Saclay.}
\label{fig:chapter4_copper_SPC_sensors}
\end{figure}

Recent developments have expanded the capabilities of the SPC by introducing a multi-anode readout, often referred to as the ACHINOS (due to its sea urchin shape). By segmenting the central electrode into multiple anodes, it becomes possible not only to collect the overall signal but also to extract spatial information regarding the point of origin of the ionisation event. This effectively transforms the SPC into a spherical TPC. In this hybrid configuration, the multi-anode arrangement allows for the reconstruction of the event topology in three dimensions. This added granularity enhances the ability of the detector to discriminate between different types of interactions, thereby improving background rejection and overall performance. Figure~\ref{fig:chapter4_copper_SPC_sensors} shows a copper SPC, together with a comparison between a single anode and an ACHINOS anode.

This simple design offers several advantages: the inherently low material budget (sphere and anode, which can be produced with radiopure materials such as copper) minimises sources of radioactive contamination; the spherical symmetry ensures a uniform response over the entire volume, providing excellent energy resolution even in a single-channel readout configuration.

An example of the use of SPCs in the search for low-mass Dark Matter is the NEWS-G experiment at SNOLAB (see Section~\ref{Chapter2_Direct_Searches_Experiments}). During this thesis, an SPC was operated at CEA Saclay to test a low-energy calibration technique (see Section~\ref{Chapter8_Tests_Saclay}).

%% file: CHAPTERS/Chapter5.tex
\chapter{TREX-DM Experiment} \label{Chapter5_TREXDM}
\vspace{-4mm}
{
\lettrine[loversize=0.15]{T}{REX-DM} (TPC for Rare Event eXperiments - Dark Matter)~\cite{TREX-DM_2016, TREX-DM_2020} is a novel, direct-search experiment conceived to look for low-mass Weakly Interacting Massive Particles (WIMPs). As explained in Section~\ref{Chapter2_Motivation}, traditional WIMP experiments have focused on the mass range motivated by the WIMP miracle ($m_\chi \sim 100\mathrm{~GeV} - 1\mathrm{~TeV}$), but recent experimental efforts have shifted the focus towards the low-mass region (typically below 10~GeV). Recording these interactions requires a detector with an exceptionally low energy threshold, well below 1~keV, and an ultra-low-background environment. In addition, sufficient exposure (detector mass $\times$ data collection time) is needed to accumulate enough statistics. Check Section~\ref{Chapter2_Direct_Searches_Sensitivity} for a deeper analysis of these requirements.
%
\section*{}
\parshape=0
\vspace{-20.5mm}
}

TREX-DM achieves this through a high-pressure gaseous TPC that operates with argon- or neon-based mixtures. The vessel has a central cathode, effectively separating it into two sensitive volumes. The design of the detector leverages light gases, which are particularly sensitive to low-energy nuclear recoils from low-mass WIMPs, while high pressure increases the target mass (up to $\sim$ 0.3~kg for argon or $\sim$ 0.16~kg for neon). The heart of the system is its two Micromegas detectors (one on each side), a technology well suited for rare event searches due to its low intrinsic radioactivity and potential for low energy thresholds.

\begin{figure}[htbp]
\centering
\includegraphics[width=1.0\textwidth]{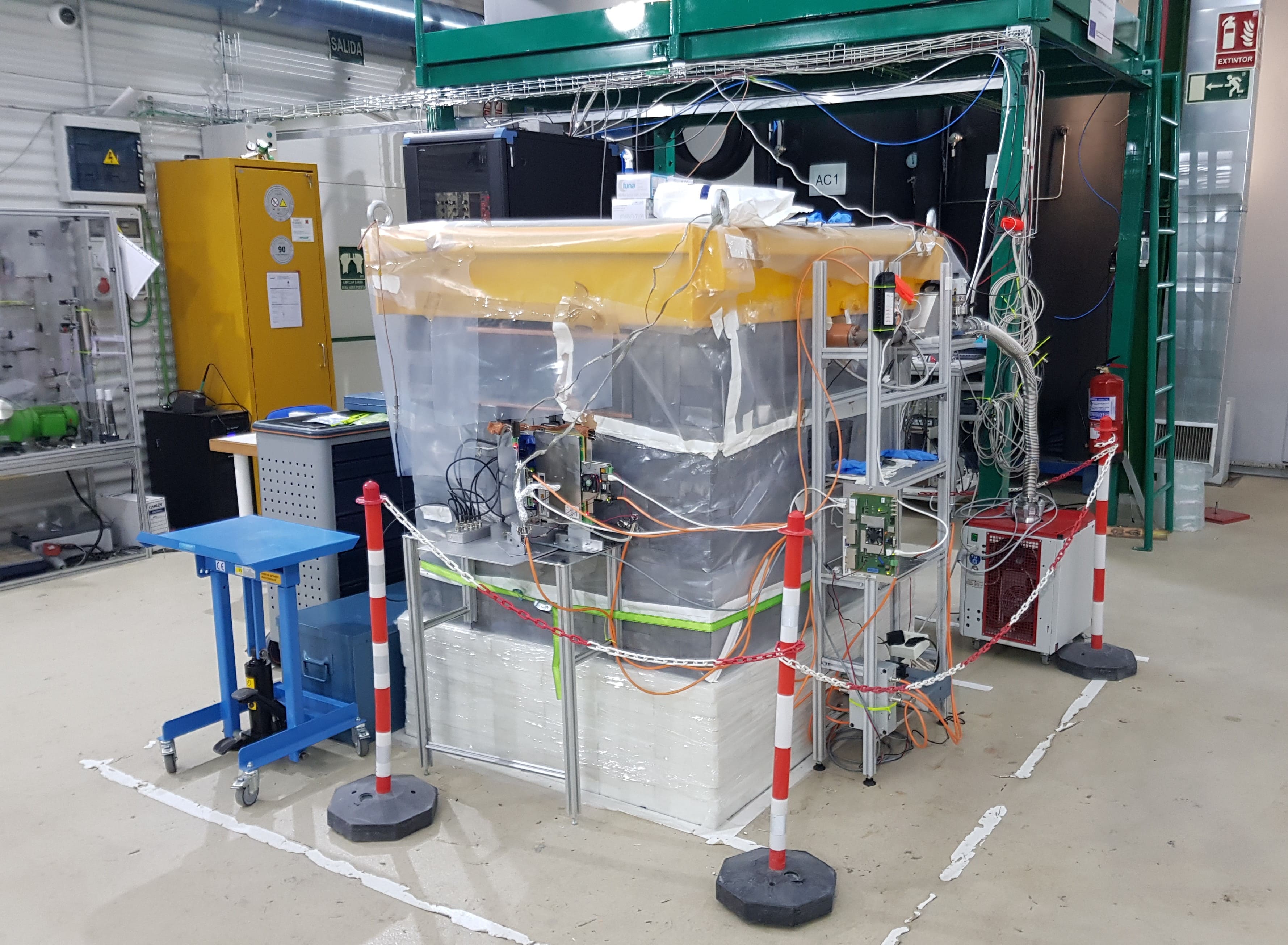}
\caption{TREX-DM operating at Lab2400 in the LSC. The detector is surrounded by its complete copper and lead shielding (including the ceiling), with the DAQ system connected and active. Note that the polyethylene top layer and water tanks (neutron moderator) are not yet installed.}
\label{fig:chapter5_trex-dm_lab2400}
\end{figure}

Initially commissioned and tested at Universidad de Zaragoza, TREX-DM was subsequently moved to the Laboratorio Subterráneo de Canfranc (see Figure~\ref{fig:chapter5_trex-dm_lab2400}). The underground location (at a depth of $\sim$ 2450~m.w.e.) is critical for mitigating cosmic ray-induced backgrounds. Since its installation in 2018, the detector has been undergoing commissioning, with full physics data runs planned to start in the near future.

In this chapter, we will describe the various elements that comprise the TREX-DM experiment, present an overview of its background model and the associated challenges encountered, and outline the experimental timeline from its initial installation at LSC, through to the current status. Finally, we include a dedicated section about drift velocity measurements, which were performed in order to validate the simulation results using TREX-DM directly.

\section{Description of the Experiment} \label{Chapter5_Description}

\begin{figure}[htbp]
\centering
\includegraphics[width=1.0\textwidth]{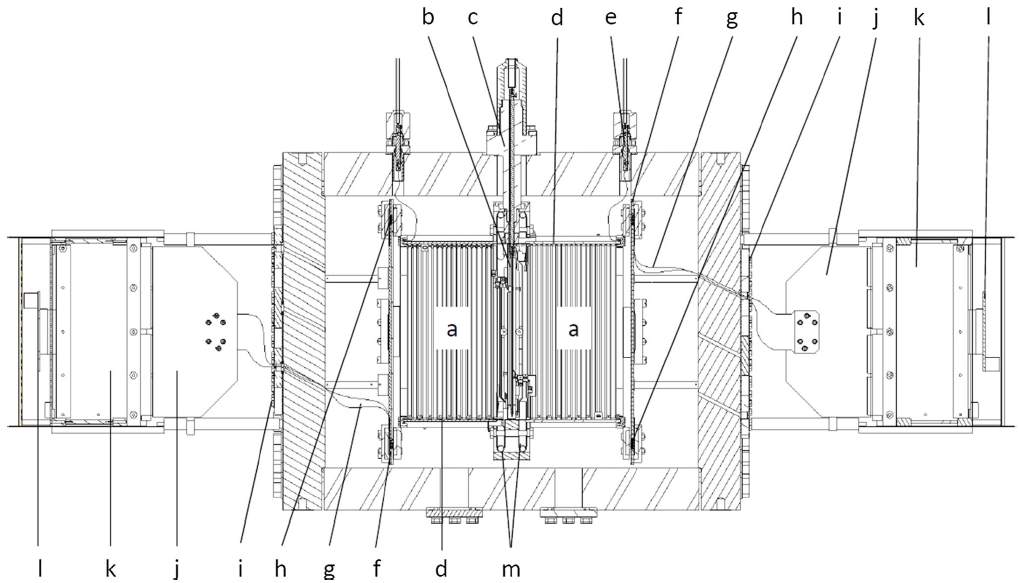}
\caption{TREX-DM layout with all its elements: active volumes (a), central cathode (b), high-voltage feedthrough (c), field cage (d), feedthrough for the last ring of the field cage (e), Micromegas detectors (f), flat cables to extract the signals (g), Micromegas-to-flat-cable connectors (h), feedthroughs for the flat cables (i), DAQ electronics (j, k, l) and calibration feedthroughs (m). Taken from~\cite{TREX-DM_2016}.}
\label{fig:chapter5_trex-dm_layout}
\end{figure}

The performance of the detector depends on the integration of several interrelated subsystems: the vessel, the shielding assembly, the detectors and calibration system, the gas system, the data acquisition electronics, the analysis software, and the slow control network. In the following subsections, each component is briefly described. Figure~\ref{fig:chapter5_trex-dm_layout} shows a schematic layout of TREX-DM with all its ancillary parts to help contextualise the elements described below. 

\subsection{Vessel} \label{Chapter5_Description_Vessel}

The chamber that constitutes the TPC is a cylindrical, $\sim 60$~L copper vessel with an inner diameter of 0.5~m and a length of 0.5~m. The vessel walls are 6~cm thick, allowing operation at pressures up to 10 bar (certified at 12 bar), while constituting the innermost part of the shielding at the same time. It is made of Electrolytic Tough Pitch copper (C11000, 99.9\% purity) from the company Sanmetal, and its screening using GDMS has verified its radiopurity, showing a content of $^{238}$U and $^{232}$Th below 0.062~mBq/kg and 0.020~mBq/kg, respectively~\cite{TREXDM_Bckg_Assessment_2018}.

\begin{figure}[htbp]
\centering
\includegraphics[width=1.0\textwidth]{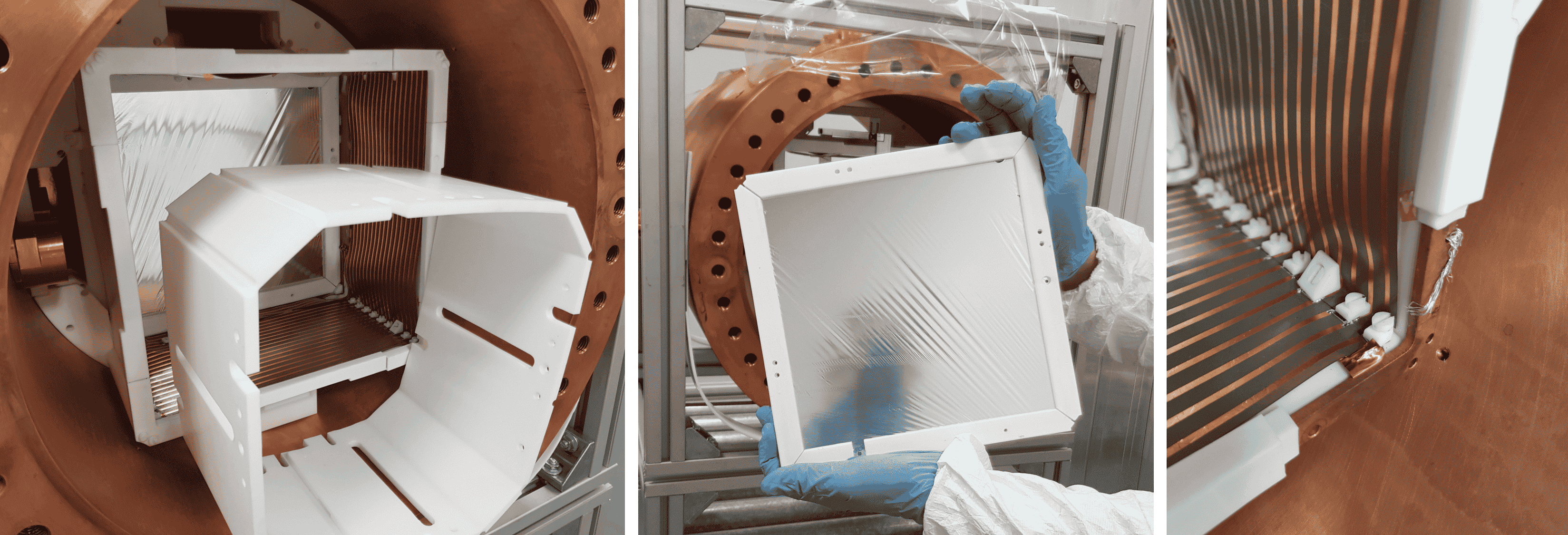}
\caption{Vessel open in the clean room of LSC. Left: PTFE piece to be installed within the field cage. Centre: mylar cathode mounted inside the PTFE frame. Right: detailed view of the PCB circuit and resistors of the field cage, together with the high-voltage connection at the last ring.}
\label{fig:chapter5_cathode_field_cage_last_ring}
\end{figure}

The original central cathode was constructed from aluminised mylar foil (see Figure~\ref{fig:chapter5_cathode_field_cage_last_ring}) with a thickness of 30–50~$\upmu$m. However, due to surface contamination affecting the background levels (see Section~\ref{Chapter6_Surface_Contamination}), this was replaced in March 2024 by a copper-clad (on both sides), 50-$\upmu$m-thick kapton laminate. The choice of a copper-kapton-copper cathode is motivated by the known radiopurity of copper and kapton (see Section~\ref{Chapter4_Micromegas_Radiopurity}).

The cathode is connected to high voltage by a tailor-made feedthrough, and divides the vessel into two active volumes. Their combined sensitive volume is $\sim$ 20~L, which corresponds to 0.32~kg of Argon or 0.16~kg of Neon mass at 10~bar. At each side there is a 16-cm-long field cage built from copper strips imprinted on a kapton Printed Circuit Board (PCB) mounted on PTFE supports. The copper strips are interconnected through a series of resistors (see Figure~\ref{fig:chapter5_cathode_field_cage_last_ring}), which distribute the voltage gradually from the cathode to the last strip, ensuring a uniform electric field over the whole drift distance.

\subsection{Shielding} \label{Chapter5_Description_Shielding}

The shielding system was designed following extensive material assays and Monte Carlo simulations (using, for example, Geant4~\cite{Geant4}). Its purpose is to reduce ambient backgrounds from gamma rays, neutrons, and electrons to levels that permit the identification of low-energy events.

\vspace{2mm}
\textbf{\normalsize Inner Copper Shielding}
\vspace{0mm}

The inner shielding consists of a 5-cm-thick layer of Oxygen Free Electronic copper (C10100, 99.99\% purity) from the company Luvata. Radiopurity measurements using HPGe detectors and Glow Discharge Mass Spectrometry (GDMS) have confirmed that the levels of $^{238}$U and $^{232}$Th are below 0.012~mBq/kg and 0.0041~mBq/kg, respectively~\cite{TREXDM_Bckg_Assessment_2018}.

\vspace{2mm}
\textbf{\normalsize Lead Shielding}
\vspace{0mm}

Surrounding the copper is a 20-cm-thick lead castle. The lead bricks have been selected for low-activity, with GDMS-measured contaminations for $^{238}$U and $^{232}$Th of 0.33~mBq/kg and 0.10~mBq/kg, respectively~\cite{TREXDM_Bckg_Assessment_2018}. The lead bricks are arranged in overlapping, staggered layers to avoid direct paths to the detector volume.

\vspace{2mm}
\textbf{\normalsize Neutron Shielding}
\vspace{0mm}

The outermost layer consists of 40 cm of moderator material, comprising polyethylene blocks for the floor and the ceiling, and water tanks for the sides. Currently, only the polyethylene floor is installed, with plans to place the rest of the shielding underway.

\begin{figure}[htbp]
\centering
\includegraphics[width=1.0\textwidth]{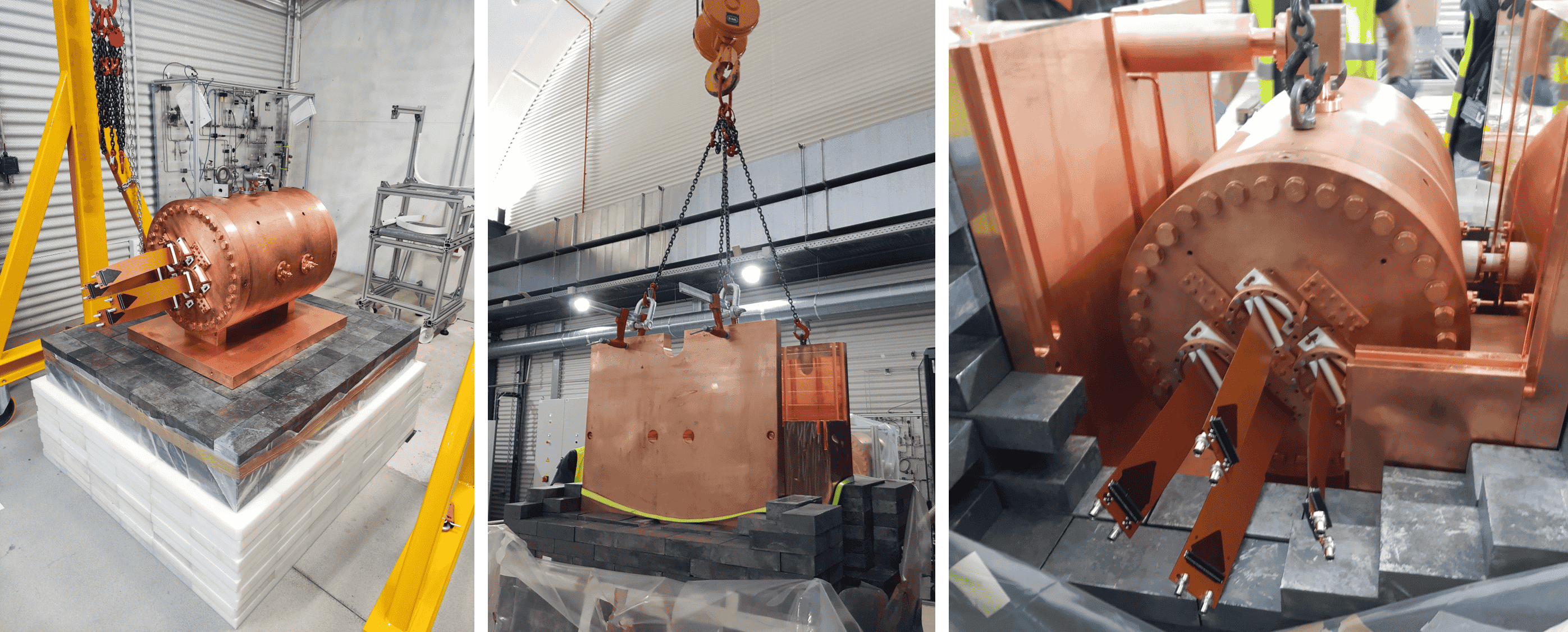}
\caption{Left: the vessel resting on its copper base during the relocation of TREX-DM to Lab2500, with minimal shielding in place, showing a polyethylene floor and the lead and copper bottom layers. Centre: installation of the copper shielding inside the lead castle. Right: the vessel placed inside a partially closed shielding. Note the characteristic staggered arrangement of the lead bricks is visible in all images.}
\label{fig:chapter5_vessel_shielding}
\end{figure}

\subsection{Background Model} \label{Chapter5_Description_Background_Model}

A background model is essential for estimating the projected sensitivity of TREX-DM to low-mass WIMPs. The model is constructed by combining radiopurity measurements of detector components with Monte Carlo simulations of the detector response. These simulations utilise the full detector geometry, including the shielding, vessel, field cage, and Micromegas readouts, and incorporate the measured environmental fluxes of gamma rays, neutrons, and muons at LSC. The output is a prediction of the energy spectrum of background events. Background rates are usually expressed in differential rate units (see Equation~\ref{eq:chapter2_differential_scattering_rate_simplified}), dru for short, with $1\mathrm{~dru}=1$~c/keV/kg/d.

\vspace{2mm}
\textbf{\normalsize Material Screening and Radiopurity}
\vspace{0mm}

A systematic material screening campaign was undertaken over several years using HPGe spectrometry, GDMS, and ICP-MS. The campaign provided activity levels for key isotopes in the $^{238}$U and $^{232}$Th decay chains as well as for $^{40}$K, $^{60}$Co, and other radionuclides present in the detector components. For instance, the $^{40}$K present on the Micromegas readout planes was identified as one of the dominant internal contributors, with a measured value of $3.45\pm 0.40$ $\upmu$Bq/cm$^2$. However, measurements of the new detectors installed in 2022 indicate that the $^{40}$K activity in the new Micromegas is $1.07\pm 0.23$ $\upmu$Bq/cm$^2$, a factor of 3 improvement. For argon-based mixtures, the $^{39}$Ar beta-emitter ($Q_\beta =565$~keV, $t  = 268$~y) has been specifically evaluated. Cosmogenically produced at the surface, atmospheric argon typically exhibits an activity of $^{39}$Ar of around 1~Bq/kg. However, underground-sourced argon (as demonstrated by the DarkSide-50 collaboration, which reported an activity of 0.73~mBq/kg~\cite{darkside_argon}) yields a reduction factor of about 1400. Using atmospheric argon, the background in TREX-DM would be dominated by $^{39}$Ar, reaching rates up to 219~dru in the region of interest~\cite{TREXDM_Bckg_Assessment_2018}. Therefore, the use of argon free of $^{39}$Ar (also called depleted argon) is critical for minimising this contribution. The other way to avoid it altogether is the use of neon.

\vspace{2mm}
\textbf{\normalsize Simulation}
\vspace{0mm}

The simulation chain is based on the Geant4 toolkit to assess both the interactions of environmental radiation (gamma rays, neutrons, and muons) with the TREX-DM detector and the activity of most of the components used in the detector, obtained from the material screening campaign. The complete geometry of the detector, including the shielding (copper, lead, and neutron moderator), the vessel, the field cage, and the Micromegas readouts, is implemented in the simulation.

\vspace{2mm}
\textbf{\normalsize Estimation of Background Contributions}
\vspace{0mm}

The overall background rate is derived by scaling the simulated contributions by the measured environmental fluxes/activities from the material screening programme. The main sources of background have been identified and quantified as follows:

\begin{itemize}
    \item \textbf{Micromegas Readout Planes}: the contribution from $^{40}$K in the TREXDM.v1 Micromegas was estimated at $<2.68$ dru. However, with the measured reduction in $^{40}$K activity in TREXDM.v2 Micromegas, the theoretical contribution from the readout planes is expected to be below 1~dru.
    \item \textbf{Copper Vessel}: cosmogenic activation in the copper vessel, primarily resulting in $^{60}$Co, contributes around 1.6 dru. The vessel’s large mass and surface area mean that even low activity levels can have a notable effect on the overall background.
    \item \textbf{Field Cage PCB and Resistors}: the PCB has been measured to contribute < 1.54~dru, while the resistors add < 0.48~dru.
    \item \textbf{Muon-Induced Events}: although the experiment is located deep underground, residual muons induce events at a rate of approximately 0.33~dru.
    \item \textbf{Intrinsic Gas Radioactivity}: for argon-based mixtures, $^{39}$Ar contributes a substantial background if atmospheric argon were used. With underground-sourced argon, this contribution is reduced dramatically, as discussed above.
\end{itemize}

\begin{figure}[htb]
\centering
\includegraphics[width=1.0\textwidth]{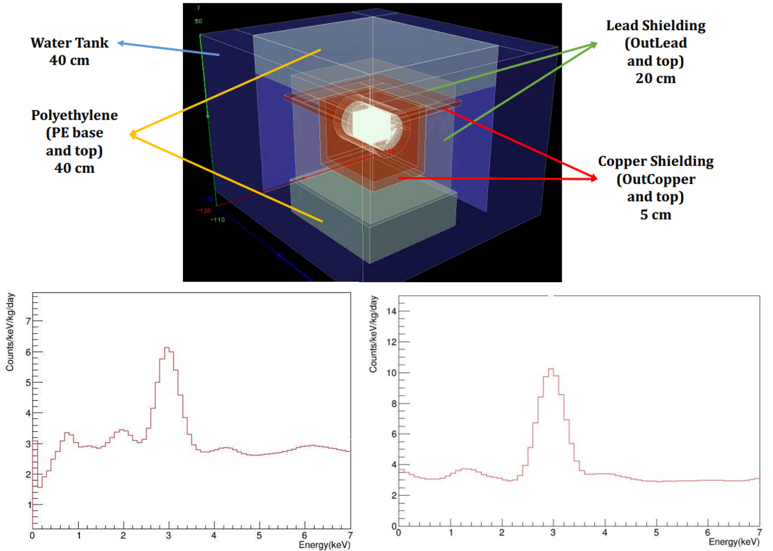}
\caption{Top: full geometry of TREX-DM as implemented in Geant4 simulations. Bottom: simulated differential rate spectrum in the low-energy region for argon (left) and neon (right) mixtures. The peak around 3~keV is due to $^{40}$K. The range stops at 7~keV to avoid interference of copper fluorescence at 8~keV. All plots extracted from~\cite{TREXDM_Bckg_Assessment_2018}.}
\label{fig:chapter5_background_model}
\end{figure}

When accounting for all these contributions, the expected overall background level in the region of interest (typically 2-7 keV) was estimated to be in the range of 1-10 dru, as Figure~\ref{fig:chapter5_background_model} indicates.

\subsection{Readout Planes} \label{Chapter5_Description_Detectors}

TREX-DM uses two 50-$\upmu$m-gap microbulk Micromegas detectors, each of them placed on the inner side of the endcaps that close the chamber. Characterisation of detectors before installation involves electrical continuity tests and capacitance measurements to check how many (if any) defective channels there are, calibrations to test energy resolution and gain uniformity across the detector, and radiopurity assays of samples. The best-performing detectors are selected for installation.

Throughout this thesis, the two Micromegas planes will be referred to as North and South. This nomenclature is given in terms of the original location of the experiment in Lab2400 of the LSC: North detector is ‘looking’ at the French side, while the South detector points to the Spanish side. When relocating to Lab2500, this nomenclature became meaningless as the orientations changed, so nowadays Left is used interchangeably for the North detector, and Right for the South detector. This reference is taken from the position where you insert the calibration sources into the vessel.

The evolution of the detectors has been marked by two main design iterations. The initial TREXDM.v1 design (2017) established a robust baseline based on the microbulk technology, while the TREXDM.v2 upgrade (tested in 2021, installed in 2022) introduced improvements in readout design, connectivity, and radiopurity, ultimately enhancing the performance.

\subsubsection{TREXDM.v1 Design} \label{Chapter5_Description_Detectors_v1}

Two Micromegas readout planes were manufactured at CERN and first installed in September 2017 at the Zaragoza set-up. Each plane provided an effective surface of $\sim 25\times 25$~cm$^2$, the largest microbulk detectors ever built. The readout was segmented into 256 channels in the X direction and 256 in the Y direction, with a strip pitch of roughly 1~mm.

Flat cables extracted signals from the strips and sent them to interface cards located outside the vessel. The interface cards are subsequently connected to the DAQ electronics by means of cables with ERNI connectors. The connections on both ends of the flat cables were made by special silicone-based connectors (Zebra Gold 8000C from Fujipoly) chosen for their radiopurity compared to other options~\cite{TREXDM_Bckg_Assessment_2018}.

\begin{figure}[htbp]
\centering
\includegraphics[width=0.91\textwidth]{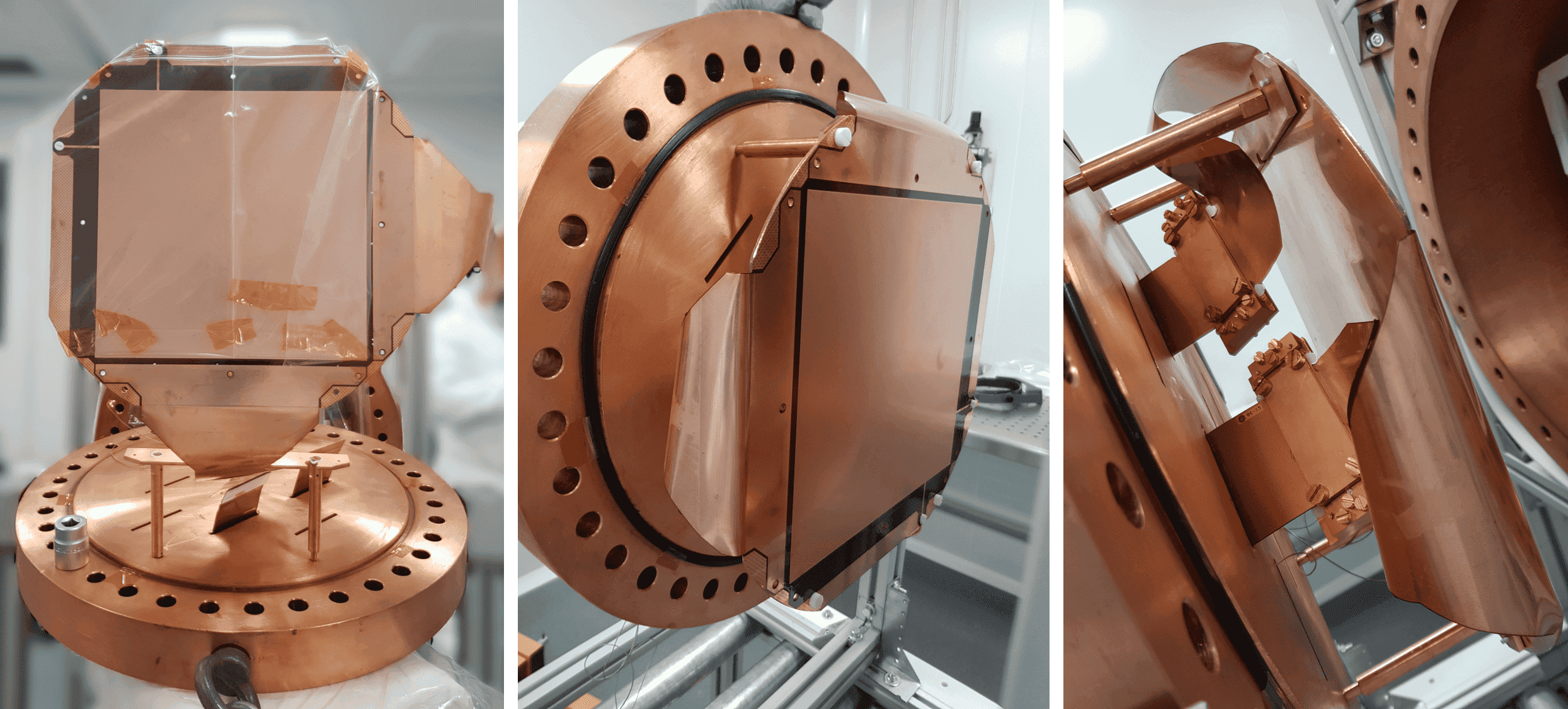}
\caption{A Micromegas from the TREXDM.v1 design in the clean room. Left: front view, with active area protected with plastic, one flat cable disconnected. Centre: Micromegas mounted on the endcap. Right: detailed view of the connections between the flaps of the detector and the flat cables.}
\label{fig:chapter5_micromegas_v1}
\end{figure}

See Figure~\ref{fig:chapter5_micromegas_v1} for images of a TREXDM.v1 detector installed in TREX-DM during an intervention in the clean room.

\subsubsection{TREXDM.v2 Design} \label{Chapter5_Description_Detectors_v2}

The upgraded TREXDM.v2 design introduced several changes with respect to TREXDM.v1. Here, we briefly review the most significant. For a thorough technical description and the reasons behind the new design, check~\cite{tesis_hector_2024}.

\vspace{2mm}
\textbf{\normalsize Readout Geometry}
\vspace{0mm}

\begin{figure}[htb]
\centering
\includegraphics[width=1.0\textwidth]{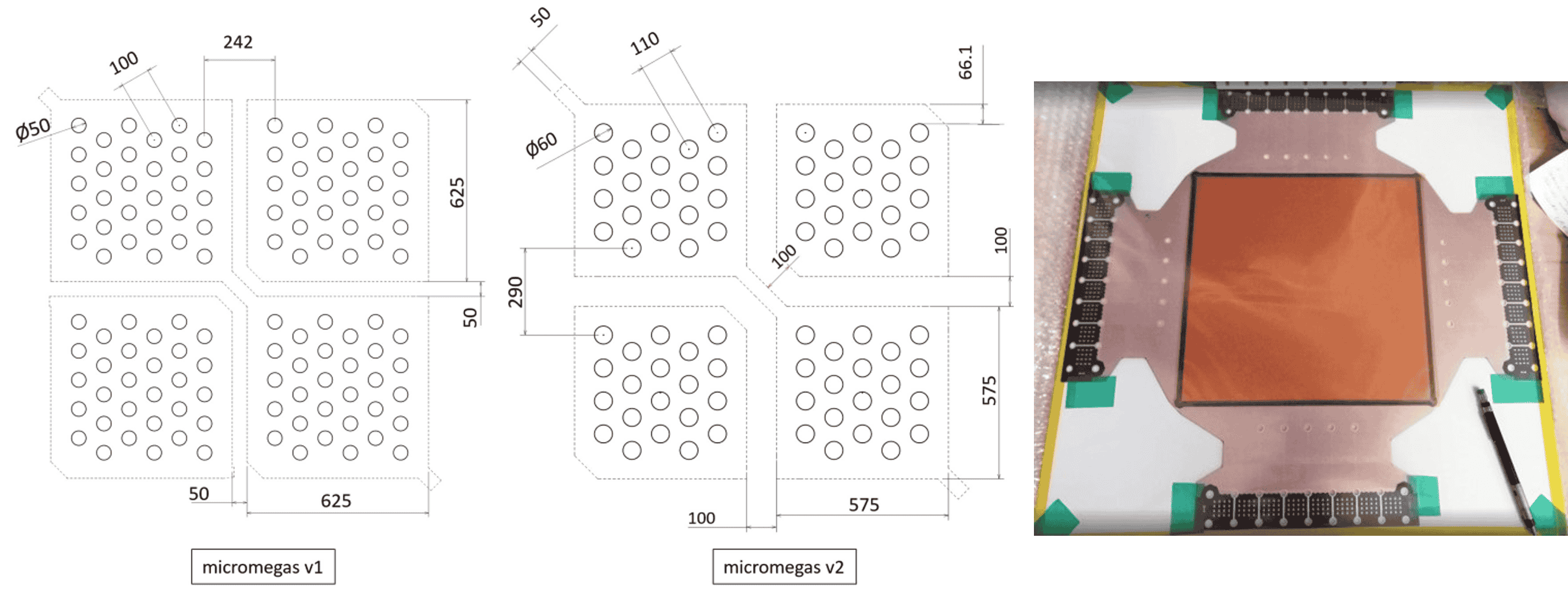}
\caption{Left: different pattern parameters between TREXDM.v1 and TREXDM.v2 designs. Taken from~\cite{tesis_hector_2024}. Right: a Micromegas plane from the TREXDM.v2 design.}
\label{fig:chapter5_micromegas_v2}
\end{figure}

Although the active area remains practically identical ($\sim 24.6\times 24.6$~cm$^2$) and the readout maintains 256 channels per direction (X and Y), the new design features a modified hole pattern (see Figure~\ref{fig:chapter5_micromegas_v2}). In contrast to the previous D50P100 pattern (50~$\upmu$m diameter holes with a 100~$\upmu$m spacing between the centres of the holes, also called pitch) where the inter-pixel separation is only 50~$\upmu$m, the TREXDM.v2 detectors utilise a D60P110 pattern. Here, the mesh holes have a diameter of 60~$\upmu$m and are arranged with a pitch of 110~$\upmu$m, which results allows for inter-pixel separation of 100~$\upmu$m. This change was made for two key reasons:

\begin{itemize}
    \item The D60P110 pattern is simpler to manufacture, yielding higher-quality holes. This goes in the direction of optimising gain and energy resolution.
    \item Although the electron collection efficiency is marginally lower than with D50P100 \cite{tesis_hector_2024}, the increased inter-pixel distance reduces the probability of leakage currents between pixels, particularly if a pixel is short-circuited with the mesh.
\end{itemize}

\vspace{5mm}
\textbf{\normalsize FaceToFace Connection}
\vspace{0mm}

The new Micromegas incorporate a FaceToFace connection, where the circuit with the pads and the flat cables are in direct contact. This configuration, combined with a revised pad layout (the channel pads are distributed in four sides instead of two), increases the separation between adjacent channels, reducing the likelihood of leakage currents if a pixel is short-circuited with the mesh.

\begin{figure}[htb]
\centering
\includegraphics[width=1.0\textwidth]{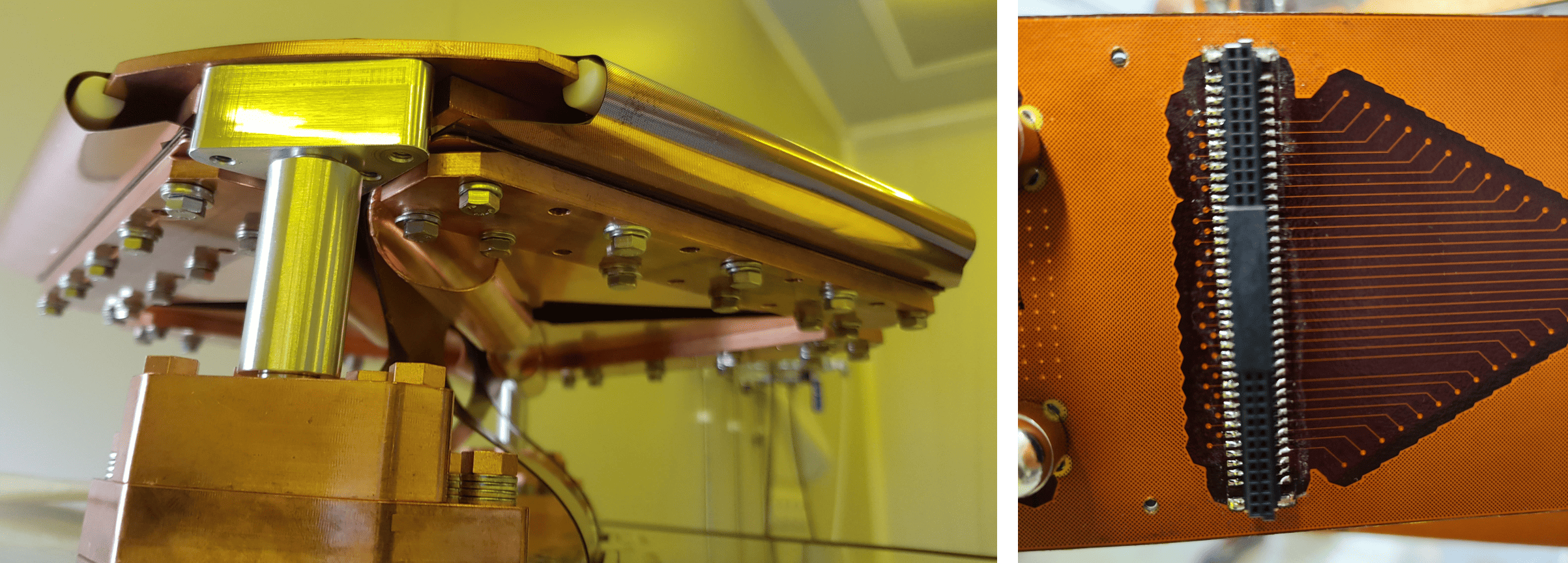}
\caption{Left: detailed view of the FaceToFace connection between pads and flat cable. Right: ERNI connector soldered at the end of a flat cable.}
\label{fig:chapter5_micromegas_v2_connections}
\end{figure}

The other end of the flat cables has a soldered ERNI connector (see Figure~\ref{fig:chapter5_micromegas_v2_connections}) which is then connected to the DAQ electronics by means of a cable, so that an interface card is not necessary.

\vspace{2mm}
\textbf{\normalsize Radiopurity}
\vspace{0mm}

Also, this new design was measured to have a lower $^{40}$K activity (by a factor 3), one of the main isotopes in the background model (see Section~\ref{Chapter5_Description_Background_Model}). One of the hypotheses is that the change of pattern in the active area affected the manufacturing process, reducing by 25\% the area exposed to the KOH bath, the most aggressive of all, but above all favouring the subsequent cleaning with KMnO$_4$ with larger diameter holes.

\vspace{1mm}
\textbf{\normalsize Performance Comparison}
\vspace{0mm}

The upgrade from TREXDM.v1 to TREXDM.v2 brought two major improvements. Firstly, TREX-DM.v2 finally allowed the simultaneous operation of both detectors: previously, the South detector in TREX-DM.v1 was inoperative due to persistent leakage current issues. Secondly, TREX-DM.v1 had a lot of dead strips (linked to strips shorted to mesh), which significantly worsened event reconstruction and energy resolution. Switching to TREX-DM.v2 resolved these issues.

\begin{figure}[htb]
\centering
\includegraphics[width=1.0\textwidth]{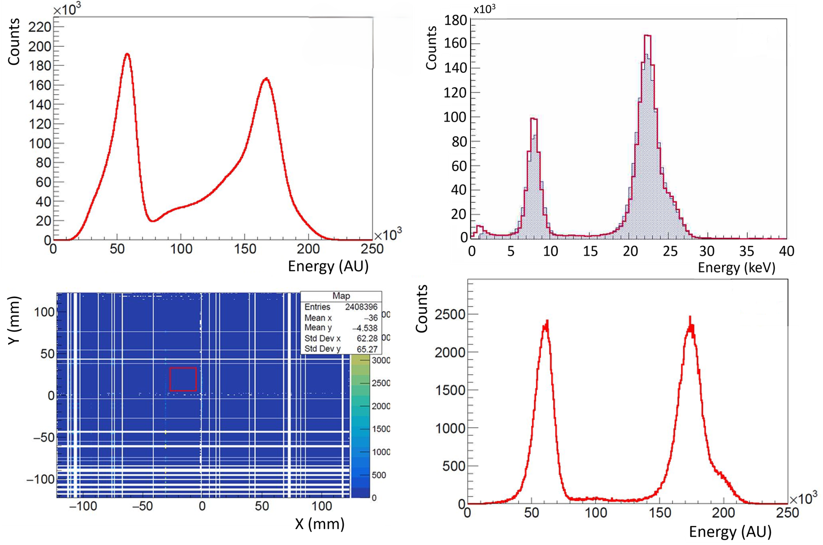}
\caption{$^{109}$Cd calibrations (22~keV and 8~keV from copper fluorescence are visible). Top left: 24-hour calibration taken with the North detector from the TREXDM.v1 design, $V_{\mathrm{mesh}}=365$~V. Top right: calibrations for the North detector from the TREXDM.v2 design at $V_{\mathrm{mesh}}=365$~V (blue, 15 hours) and $V_{\mathrm{mesh}}=375$~V (red, 16 hours), a gain map has been applied to correct for gain inhomogeneities. The energy threshold improves from $\approx$ 1.5~keV (blue) to below 1~keV (red). Bottom left: hitmap of the North detector from the TREXDM.v1 design for the 24-h calibration above, showcasing all the dead strips that worsen energy and spatial resolution. Bottom right: energy spectrum for the dead-strip-free region of the hitmap enclosed in red.}
\label{fig:chapter5_performance_comparison_detectors}
\end{figure}

As illustrated in Figure~\ref{fig:chapter5_performance_comparison_detectors}, the improvement in energy resolution is remarkable. Using Ne-2\%iC$_{4}$H$_{10}$ at 4 bar, TREX-DM.v1 typically achieved around 20\% FWHM at 22 keV, whereas TREX-DM.v2 routinely reaches approximately 15\% FWHM at the same energy. In TREX-DM.v1, a small area free of dead strips was needed to achieve comparable resolution (see Figure~\ref{fig:chapter5_performance_comparison_detectors}). Additionally, TREX-DM.v2 offers sub-keV energy thresholds in stable operation, as demonstrated in Figure~\ref{fig:chapter5_performance_comparison_detectors}.

\vspace{-3mm}
\subsection{Calibration System} \label{Chapter5_Description_Calibration}

\begin{figure}[htb]
\centering
\includegraphics[width=1.0\textwidth]{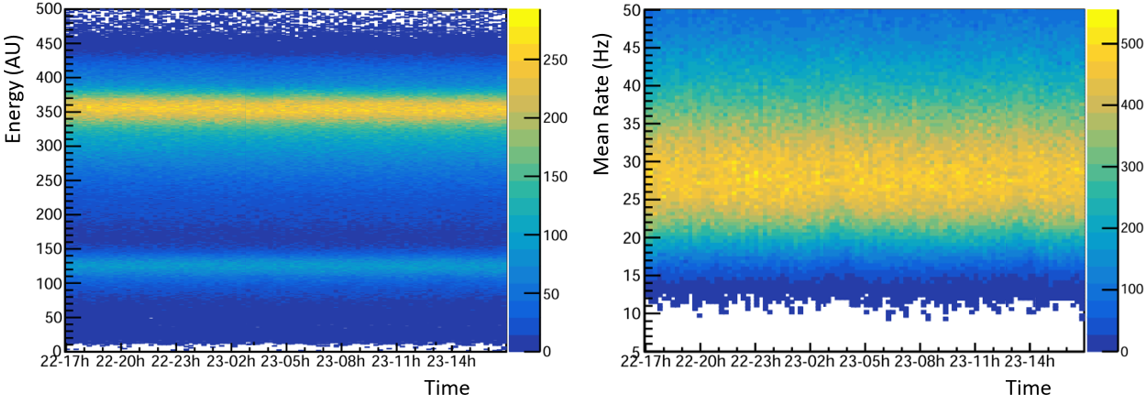}
\caption{Monitoring the gain (left) and rate (right) stability of TREX-DM North detector during a 24-h $^{109}$Cd calibration run.}
\label{fig:chapter5_trexdm_gain_rate_stability}
\end{figure}

TREX-DM uses a robust calibration system designed to ensure that the energy scale and gain of the detector remain stable over time. In routine operation, a movable $^{109}$Cd source is deployed on a weekly basis. Figure~\ref{fig:chapter5_trexdm_gain_rate_stability} illustrates the stability in gain and rate achieved under normal operating conditions.

In addition, TREX-DM has started to incorporate a low-energy calibration source based on $^{37}$Ar into the calibration protocol (see Chapter~\ref{Chapter8_Ar37}), though with much lower frequency (every few months). Neutron calibrations are also planned to characterise the detector's response to nuclear recoils. 

On top of that, a $^{83m}$Kr source is currently under development to provide periodic homogeneous calibrations across the detectors. $^{83m}$Kr is particularly useful for calibration purposes due to its short half-life of 1.83 hours and its two-step decay process, which emits conversion electrons, characteristic X-rays and Auger electrons that give overall energies of 32.1 keV and 9.4 keV~\cite{lnhb_table_radionucleides}. The gaseous nature of this metastable isotope makes it especially suitable for gaseous detectors like TREX-DM, as it can be introduced and removed from the system without contamination concerns. 

\vspace{1mm}
\textbf{\normalsize Sources}
\vspace{0mm}

The system uses two exempt $^{109}$Cd sources that emit X-rays at $\sim$ 3, 22, 25, and a gamma at 88~keV. These sources are mounted on the side of the vessel so that each source faces one of the active volumes, guaranteeing that both of them are calibrated under similar conditions.

In addition to the primary calibration peaks from $^{109}$Cd, the system consistently records an 8~keV copper fluorescence peak, mainly due to the copper present on the readout planes (and also the cathode, after replacing the mylar one with a copper-kapton-copper foil).

\vspace{1mm}
\textbf{\normalsize Mechanism}
\vspace{0mm}

\begin{figure}[htbp]
\centering
\includegraphics[width=0.77\textwidth]{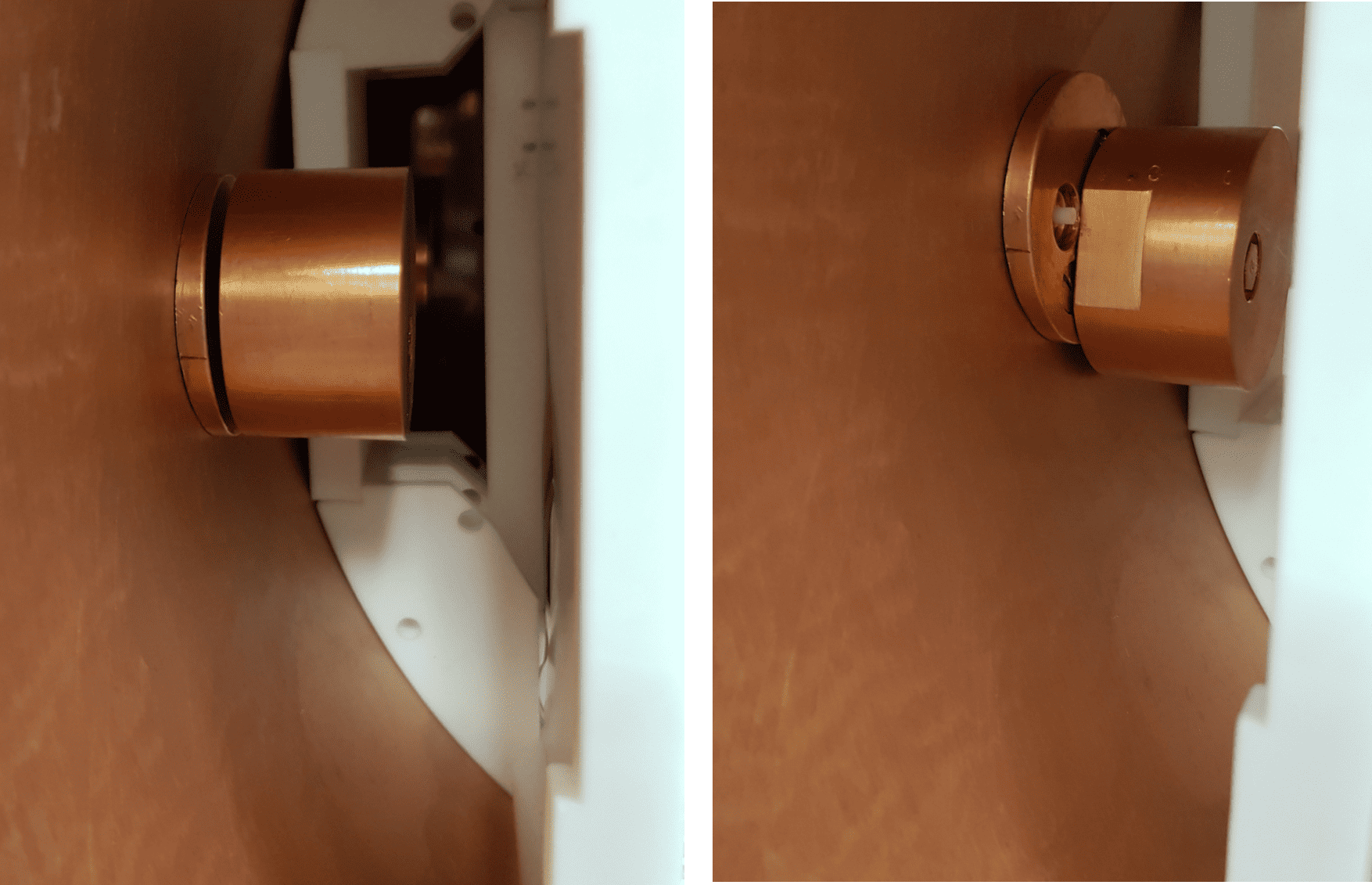}
\caption{Copper block inside the vessel. Left: closed with source out (parking position). Right: open with source in (calibration position).}
\label{fig:chapter5_calibration_system}
\end{figure}

The sources are mounted on a rod inserted via a leak-tight port. A lever mechanism is employed to move a 6-cm-thick copper piece, which can be interposed to completely block the source when calibration is not desired (see Figure~\ref{fig:chapter5_calibration_system}). This design not only provides an effective means of controlling the calibration exposure but also has been validated by a leak-test to ensure that the integrity of the vessel is maintained during operation.

\subsection{Data Acquisition System} \label{Chapter5_Description_DAQ}

The Data Acquisition (DAQ) system is responsible for digitising, processing, and storing the signals from the Micromegas readouts. It has been engineered to handle the high channel density of TREX-DM. It is made up of the following elements:

\begin{enumerate}
    \item Four FEC (Front-End Card)-Feminos cards with chip AGET (ASIC for General Electronic readout of TPCs). Two cards for each detector.
    \item A Trigger Control Module (TCM) card to synchronise the trigger and the clock for all the FEC-Feminos cards.
    \item A computer using a custom-made software to handle the acquisition.
    \item A switch to connect all the elements.
\end{enumerate}

\begin{figure}[htbp]
\centering
\includegraphics[width=1.0\textwidth]{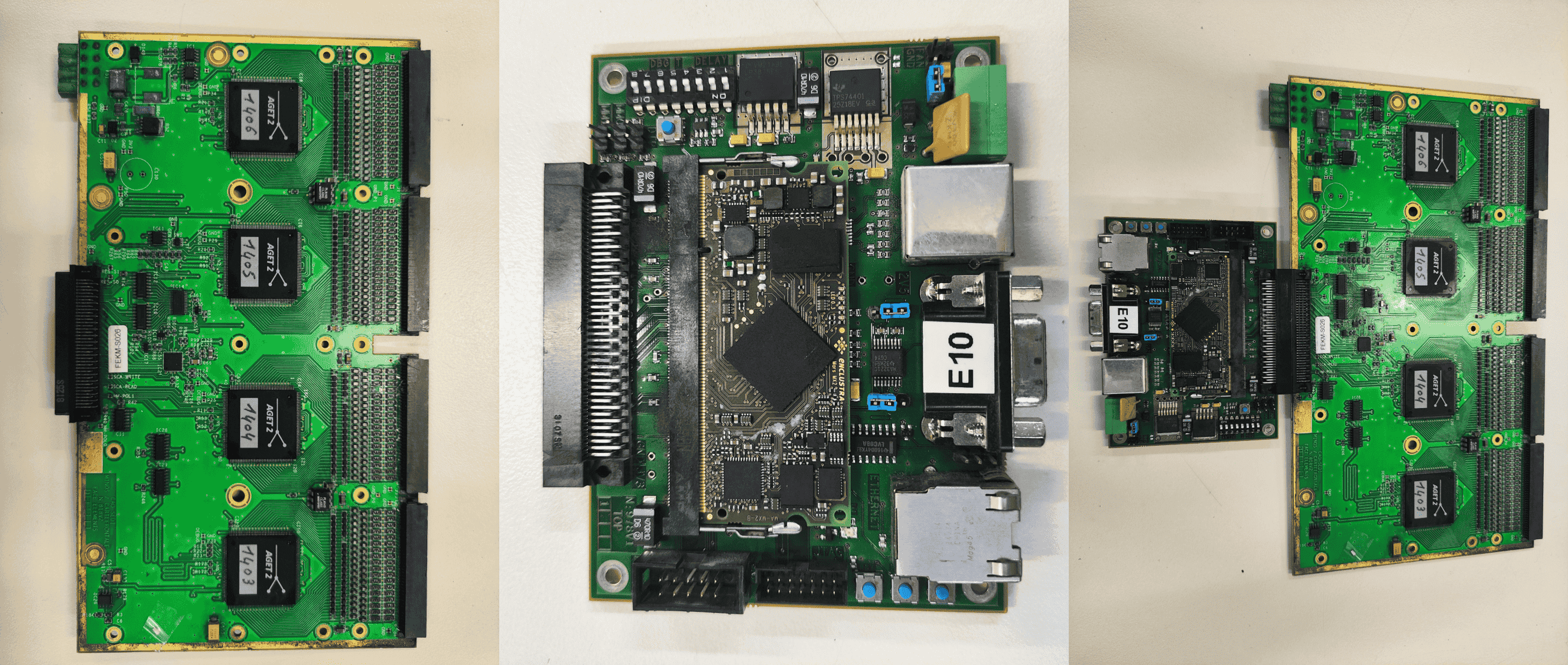}
\caption{Photo of the Front-End Card (FEC) with four AGET chips (left), Feminos card (centre) and the combination FEC-Feminos (right).}
\label{fig:chapter5_fec_agets_feminos}
\end{figure}

\begin{figure}[htbp]
\centering
\includegraphics[width=0.8\textwidth]{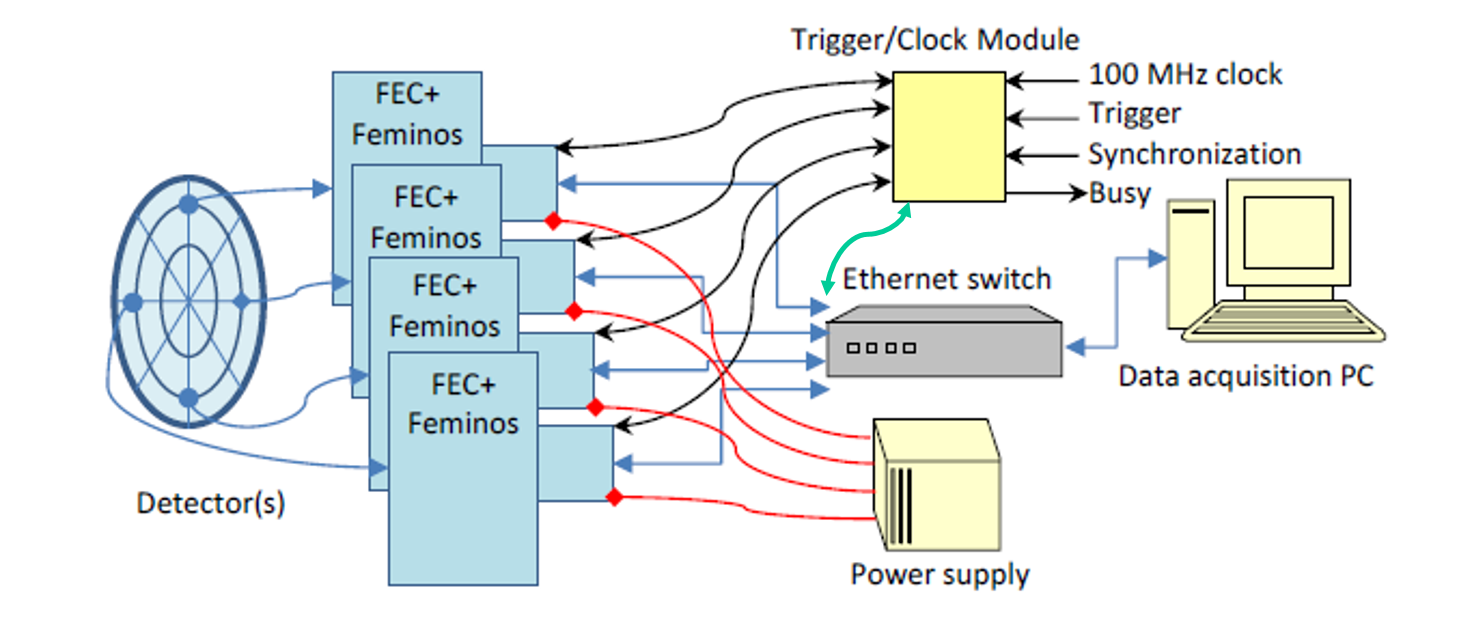}
\caption{Illustration showing the architecture of the DAQ system of TREX-DM. Image adapted from~\cite{Feminos_2014}.}
\label{fig:chapter5_daq_config}
\end{figure}

Both the FEC-Feminos (shown in Figure~\ref{fig:chapter5_fec_agets_feminos}) and the TCM are custom-made electronics cards developed at CEA Saclay as a solution for data acquisition in nuclear and high-energy physics experiments. The connection between the different elements is made with Ethernet cables, and the schematic set-up can be seen in Figure~\ref{fig:chapter5_daq_config}. Note that, for simplicity, there is only one detector in the image, but it represents the two sides of TREX-DM. Each of the elements is explained in more detail below.

\vspace{2mm}
\textbf{\normalsize Back-End DAQ PC}
\vspace{0mm}

A dedicated PC runs the DAQ software under a Linux operating system. A custom application in C++ manages data collection, assembly of the digitised data into coherent events, and data storage. Data are recorded in a raw format (.aqs) later converted to ROOT~\cite{ROOT_1997} format using REST-for-Physics (see Section~\ref{Chapter5_Description_REST}). The computer is connected to the front-end electronics through a common switch.

\vspace{2mm}
\textbf{\normalsize Front-End Cards (FECs) with AGET Chips}
\vspace{0mm}

Each custom-made FEC incorporates 4 AGET chips, and each chip processes 64 channels, yielding 256 channels per board. The total number of FECs (2 for each side) is matched to the number of readout channels in the Micromegas system (512 for each side).

The AGETs can generate the trigger based on an adjustable threshold applied to the signal (it can be fine-tuned to trigger on single-channel pulses), and they provide the amplification, shaping and storage of the analog signals.

It has 512 time bins and the minimum sampling rate is 1 MHz (1 $\upmu$s/bin), while the maximum is 100 MHz (10 ns/bin). Therefore, the temporal size of the window for each event ranges from $\sim$ 5 $\upmu$s to 500 $\upmu$s.

\vspace{2mm}
\textbf{\normalsize Feminos Cards}
\vspace{0mm}

A total of 4 Feminos cards (2 per Micromegas detector) are deployed. Their role is to interface with the AGETs to perform pedestal subtraction and to digitise the analog signals (with 12-bit precision), which are then transmitted to the central DAQ PC.

\vspace{2mm}
\textbf{\normalsize TCM Board}
\vspace{0mm}

The TCM board synchronises data from both halves of the TPC distributing a 100 MHz clock, and provides trigger signals to the Feminos cards.

The TCM has 24 ports, so it can sustain systems with up to 24 Feminos cards.

\vspace{-10mm}
\subsection{Software: REST-for-Physics} \label{Chapter5_Description_REST}

The REST-for-Physics~\cite{REST_2022} (Rare Event Searches Toolkit for Physics) framework is the primary platform for processing and analysing TREX-DM data. Developed within the research group during the past $\sim$ 10 years, it is a ROOT-based software suite and has been adapted for the specific needs of rare event searches. REST-for-Physics provides a unified, modular environment to process both experimental and Monte Carlo data through a well-defined event processing chain. This chain encompasses the transformation of raw acquisition data into a standard ROOT format, followed by successive stages of signal analysis and event reconstruction.

\vspace{2mm}
\textbf{\normalsize Data Conversion and Raw Signal Analysis}
\vspace{0mm}

The first stage in the REST-for-Physics workflow involves converting the raw data generated by the acquisition system into a ROOT-based format. Since the acquisition system outputs data in a format specific to the electronics cards used, a dedicated conversion process is implemented for each system. For example, TREX-DM uses the \textit{MultiFEMINOSToSignalProcess} to convert data from the FEC-Feminos system with AGET chips into \textit{RawSignalEvents}. This process ensures that the raw signals, recorded as 4096 ADC values per channel (12-bit precision), are properly formatted for further analysis.

Once in ROOT format, the data undergo the initial \textit{RawSignalAnalysisProcess}. In this process, REST-for-Physics applies algorithms to extract key observables from the digitised waveforms, such as pulse amplitude, integrated charge, peak timing or number of signals in an event.

\vspace{2mm}
\textbf{\normalsize From Raw Signal to Detector Signal}
\vspace{0mm}

At this point, the processed \textit{RawSignalEvents} are transformed into what is termed \textit{DetectorSignalEvents}. Here, the raw time bins are calibrated using the known sampling time of the electronics, converting the 512 time bins to the time domain. This stage also incorporates zero suppression: channels or time bins that do not exceed a defined noise threshold above the baseline are suppressed to reduce data volume and to focus the analysis on regions containing significant signal.

\vspace{2mm}
\textbf{\normalsize From Detector Signal to Detector Hits}
\vspace{0mm}

Following the \textit{RawToDetectorSignalProcess}, the processed \textit{DetectorSignalEvents} are transformed into so-called \textit{DetectorHitsEvents}. This step involves associating signals with physical locations on the Micromegas readout, taking into account the known geometry and segmentation of the detector (e.g. a strip pitch of $\sim$ 1mm). This is achieved with \textit{DetectorSignalToHitsProcess}. This process uses what is called the \textit{decoding}, a mapping between the channel numbers of the AGET chip and the corresponding Micromegas strips in the XY plane. The \textit{decoding} is incorporated in the \textit{DetectorReadout}, a class that generates/stores a readout description.

Moreover, the drift velocity is used in this process to translate the time differences recorded in the detector signal into spatial $Z$ coordinates along the drift direction of the TPC. In the absence of a trigger that provides a $t_0$ (such as S1 in dual-phase TPCs, see Section~\ref{Chapter2_Direct_Searches_Experiments}), these $Z$ coordinates are not absolute, but they are useful to know the length of the pulse in $Z$, $\Delta Z$. This is why drift velocity measurements such as the ones presented in Section~\ref{Chapter5_Drift_Velocity_Measurements} are important to correctly calibrate $\Delta Z$. This is crucial for well-defined linear tracks such as the ones left by alpha particles.

After converting each detector signal into one or several (depending on the method used) detector hits, \textit{DetectorHitsAnalysisProcess} calculates observables that describe the topology of each event (such as the mean $X\mathrm{/}Y$ positions, mean energy per hit, etc.)

\vspace{2mm}
\textbf{\normalsize From Detector Hits to Track}
\vspace{0mm}

The \textit{DetectorHitsToTrackProcess} transforms a \textit{DetectorHitsEvent} into a \textit{TrackEvent}. Tracks are defined as clusters of hits that have a spatial proximity relation. If the distance from a group of hits to another group of hits is larger than the \textit{clusterDistance} parameter, then those tracks will be considered independent inside the \textit{TrackEvent}. Tracks can be used for event reconstruction purposes, which is useful in background discrimination. For instance, the reconstructed event length in alpha particle interactions can be compared with expectations from models such as those provided by NIST ASTAR~\cite{ESTAR_PSTAR_ASTAR}, thus determining the identity of the particle.

\vspace{2mm}
\textbf{\normalsize Modular Architecture and Traceability}
\vspace{0mm}

REST-for-Physics is structured into a series of modular processes, each responsible for a distinct step in the data processing chain. This modular design facilitates the consistent treatment of both experimental and simulated data. It also ensures traceability: every analysis run is tagged with the corresponding version of the software. REST-for-Physics is an open-source project and it is distributed under a GNU public license at GitHub~\cite{REST_github}.

\subsection{Gas System} \label{Chapter5_Description_Gas_System}

A gaseous TPC relies on a gas mixture as the interaction medium for the particles being detected. In a TPC with charge readout using Micromegas, the gas composition must be optimised to ensure efficient performance across the processes explained in Chapter~\ref{Chapter3_Gaseous_Detectors}: ionisation of the gas, drift of the primary electrons, and electron avalanche amplification. The gas system is responsible for delivering and maintaining the appropriate gas mixture. This is vital for stable, controlled operation and detector performance in terms of gain and resolution.

\vspace{2mm}
\textbf{\normalsize Gas}
\vspace{0mm}

TREX-DM operates at a maximum pressure of 10 bar with two primary gas mixtures: an argon-based mixture with 1\%iC$_{4}$H$_{10}$ and a neon-based mixture with 2\%iC$_{4}$H$_{10}$. Argon and neon are selected for their high ionisation yield, and flammability requirements limit the percentage of iC$_{4}$H$_{10}$ used.

\vspace{2mm}
\textbf{\normalsize Elements of the System}
\vspace{0mm}

The gas system incorporates a network of valves and sensors/actuators to enable different modes of operation, including vacuum pumping, gas filling, recirculation, and pressure adjustment. A mass flow controller at the chamber inlet and a back-pressure regulator at the outlet allow for automated control of both pressure and flow rate. 

Also, given that both electron drift and amplification are highly sensitive to gas impurities, a recirculation system composed of a pump plus filters continuously circulates the gas through the TPC. Originally, oxygen and moisture filters were installed in the loop to remove impurities. However, systematic studies indicated that these filters were introducing $^{222}$Rn into the system, significantly affecting the background levels. Operation in semi-open loop showed it was possible to remove them from the gas system without compromising gas purity. See Chapter \ref{Chapter6_Radon_problem} for more information.

All these elements are integrated into the slow control infrastructure (see Section \ref{Chapter5_Description_Slow_Control}), allowing continuous logging of pressure, temperature, and flow rate.

\vspace{2mm}
\textbf{\normalsize Operation}
\vspace{0mm}

Firstly, the TPC is evacuated using a turbo pump to remove residual contaminants. Once a good enough vacuum level is achieved ($10^{-3}-10^{-4}$ mbar in $1-2$ days), the selected gas mixture is introduced from a pre-mixed bottle until the target pressure is reached. At this stage, the recirculation system is activated (closed loop), using a pump and filters to maintain gas purity over extended operation. There is also the option of operating in open loop, bypassing the recirculation pump and the filters, using a continuous gas flow that goes directly through the exhaust valve after leaving the vessel.

Unlike other experiments where gas recovery is necessary due to high costs or environmental reasons (such as SF$_6$ in the DRIFT experiment), TREX-DM does not employ a recovery system. When a change is required (whether for a new gas mixture, maintenance, or pressure reduction), the existing gas is vented to the atmosphere before restarting the operation.

\subsection{Slow Control} \label{Chapter5_Description_Slow_Control}

The Slow Control (SC) system is a critical component of the TREX-DM experiment, designed to continuously monitor the state of its various subsystems (including the gas system and all the power systems affecting the detectors) and act on them in real-time based either on pre-programmed conditions or at the discretion of the operator.

\begin{figure}[htbp]
\centering
\includegraphics[width=0.7\textwidth]{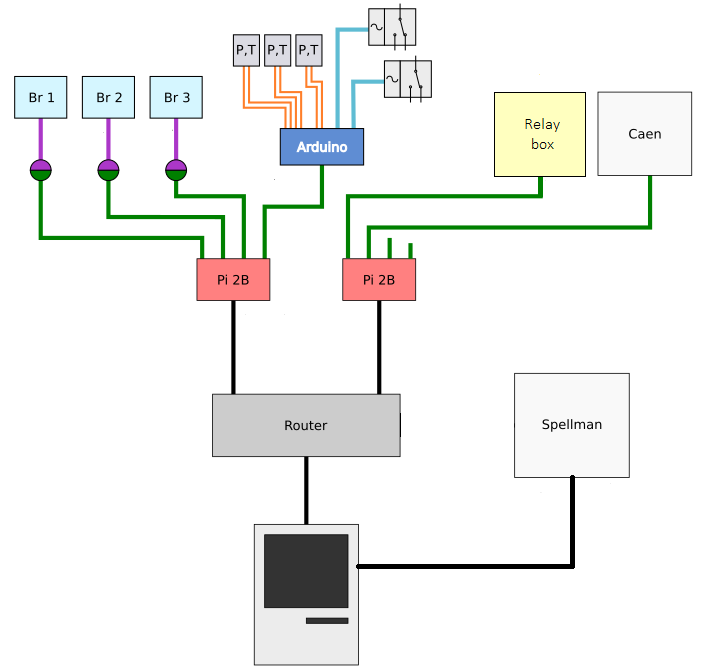}
\caption{Architecture of the SC system of TREX-DM. Source: own elaboration.}
\label{fig:chapter5_slow_control}
\end{figure}

The SC of TREX-DM is designed to manage key parameters such as gas flow, pressure, temperature, and power supplies. The schematic of the SC can be seen in Figure~\ref{fig:chapter5_slow_control}, and it shows the different elements of the system and their interconnection. We describe them in detail below.

\vspace{2mm}
\textbf{\normalsize Gas Panel}
\vspace{0mm}

The gas panel includes various components controlled via the SC. Three pressure and temperature sensors, distributed at different points, monitor conditions across the gas panel. These sensors, along with the recirculation pump and the two electrovalves (one at the inlet, one at the outlet) responsible for regulating the opening and closing of the vessel (thus isolating it when necessary), are connected to an Arduino board. The Arduino acts as an intermediary, transmitting data to a Raspberry Pi.

The gas flow and pressure regulation involve two flow controllers and a back-pressure regulator. The flow controller at the inlet manages the gas entering the TPC, while the back-pressure regulator controls the chamber pressure. The flow controller at the outlet regulates the exiting gas. These three devices are controlled via the Raspberry Pi connected to the gas panel.

\vspace{2mm}
\textbf{\normalsize Power System}
\vspace{0mm}

In the power system, the CAEN power supply provides high voltage to the Micromegas planes and is linked to a separate Raspberry Pi for control. The Spellman power supply, which provides the drift voltage, is directly linked to the SC computer. For example, the SC can reduce the voltage applied to the Micromegas if pressure drops or sparks are detected, or decrease the drift field intensity if sparks occur at the cathode.

Additionally, a relay box with four relays is connected to the same Raspberry Pi, allowing independent remote switching of the low-voltage power supplies for the readout electronics. This configuration enables the operator to restart electronic cards remotely, providing flexibility during data taking or troubleshooting.

\vspace{2mm}
\textbf{\normalsize Central PC}
\vspace{0mm}

Both Raspberry Pi units (one for the gas system and the other for the power systems) are connected to a network switch, which routes data to a central SC computer. This computer has a Python-based GUI to monitor and control all aspects of the SC. All the sensor readings are logged in real time to a database.

\section{Technical Challenges} \label{Chapter5_Technical_Challenges}

In order to provide a realistic overview of the day-to-day operation of TREX-DM, we summarise the main technical challenges encountered during routine operations and interventions. These challenges affect data taking, maintenance in the clean room, and overall system reliability. The following subsections detail the issues of electronic noise, leakage currents in the Micromegas, gas leaks, and gas quality, drawing on records from recent intervention logs and daily operational data.

\subsection{Electronic Noise} \label{Chapter5_Technical_Challenges_Noise}

One of the recurring challenges in operating TREX-DM is the management of electronic noise. Ambient interference and internal grounding issues have repeatedly degraded performance, worsening the effective energy threshold with respect to the intrinsic potential of the detector. The problem is observed as baseline fluctuations in the readout electronics, with noise levels being influenced by both the noise sources and the capacitance of
the detector. Typical capacitance values between strips and mesh are around 80-110 pF (at the microbulk detector level) and 15-20 pF (at the flat cable level that extracts the signal). Noise levels can vary considerably, sometimes even from one day to the next, forcing us to take corrective measures that delay operations and require extensive troubleshooting.

\vspace{2mm}
\textbf{\normalsize Optimisation of Grounding}
\vspace{0mm}

A key step in reducing noise involves optimisation of the grounding scheme. Copper braid cables are employed to join different grounding points. However, these connections are sometimes made between unexpected elements (for example, a metallic gas pipe might become linked to the ground of the TCM board). As a result, the grounding improvement process often becomes a trial-and-error procedure. In several interventions, iterative adjustments (including repositioning the braid cables, testing alternative grounding points, and verifying continuity with a multimeter) have been necessary, sometimes taking several days before noise levels drop to acceptable limits. In some cases, interference from electronics rack fans has also required to switch off these fans to prevent additional noise pick-up.

\vspace{3mm}
\textbf{\normalsize Flat Cables}
\vspace{0mm}

\begin{figure}[htbp]
\centering
\includegraphics[width=1.0\textwidth]{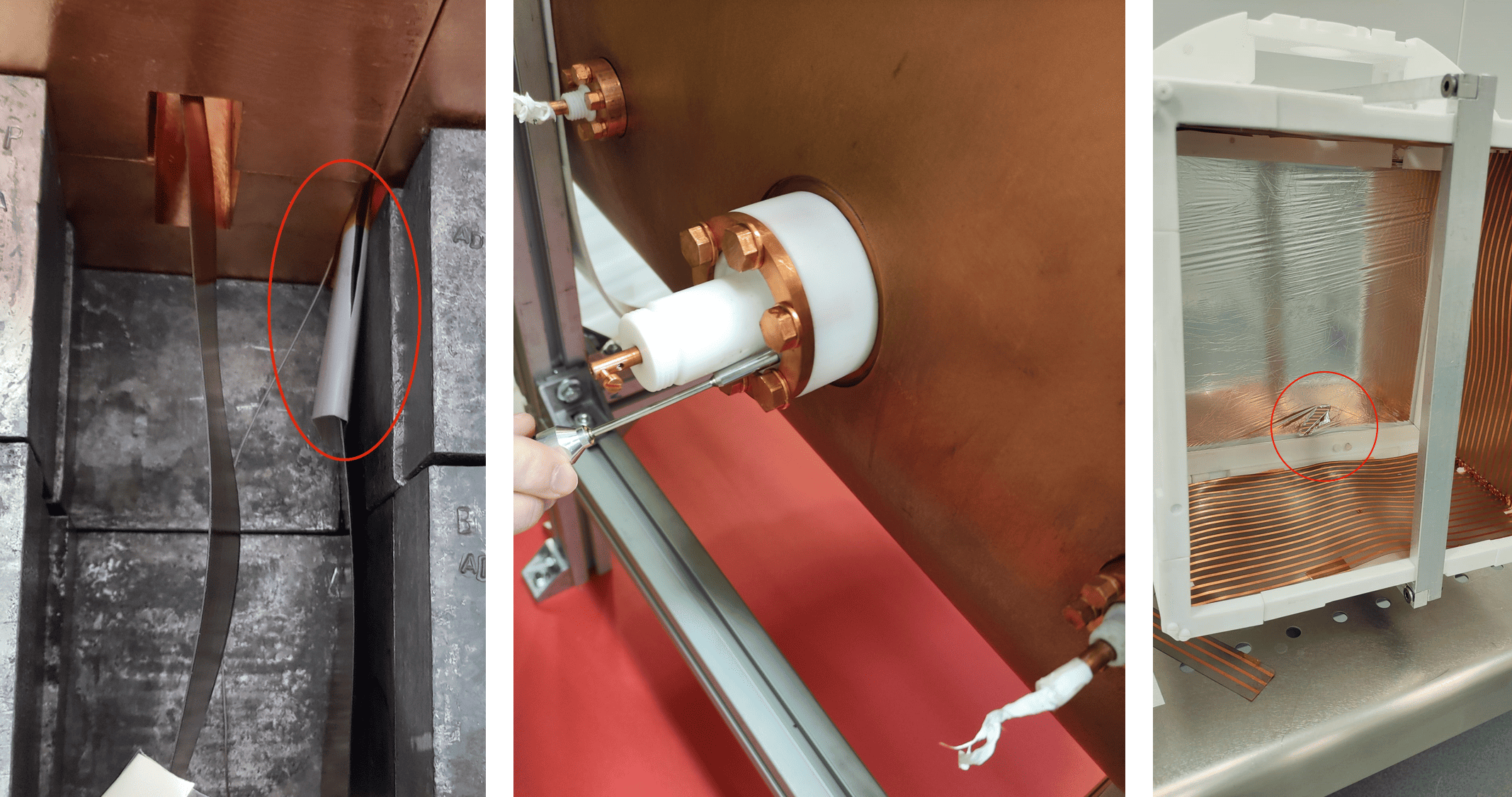}
\caption{Left: flat cables from TREX-DM.v1 emerging from the shielding; the red circle highlights the section where a cable was found to be in contact with the lead shielding, causing increased noise. Centre: leak test of the central high-voltage feedthrough performed after a significant leak was detected. Right: tearing of the mylar cathode (circled in red) immediately after its replacement, leading to a delay as the cathode had to be reinstalled.}
\label{fig:chapter5_technical_issues}
\end{figure}

Special attention is given to the flat cable connections that transport signals from the Micromegas. These cables are sensitive to their placement: if they touch other components such as the shielding, or if they are not properly positioned, they can pick up noisy signals. For example, with TREXDM.v1 detectors, data-taking was halted a couple of weeks because noise levels were too high (energy threshold $\sim$ several keV). After extensive trial and error, it was found that one flat cable was in contact with the lead shielding, and the outer kapton insulation had worn off. Wrapping that cable in insulating material (see Figure~\ref{fig:chapter5_technical_issues}) led to a significant improvement in noise performance, allowing to resume data-taking.

\vspace{2mm}
\textbf{\normalsize Impact of Interventions}
\vspace{0mm}

Every time the system is disconnected for maintenance or intervention, the subsequent reinstallation process proves to be time-consuming and unpredictable. Reconnecting all the components and re-optimising the grounding often introduces additional noise issues. In some instances, the only solution is to raise the threshold settings on the AGET chips to suppress the noisy channels, which elevates the energy threshold.

\subsection{Leakage Currents} \label{Chapter5_Technical_Challenges_Leakage_Currents}

Leakage currents in the Micromegas readout constitute a significant technical challenge. These currents arise when individual channels become short-circuited with the mesh, thereby creating unwanted conduction paths. Such shorts limit the maximum attainable mesh voltage and, in some cases, force us to disconnect the problematic channel (leaving it floating rather than referenced to ground). For reference, currents during normal operation are in the order of nA, while leakage currents reach values of $\upmu$A.

When a channel consistently exhibits leakage or tripping behaviour, it is isolated or desoldered to prevent interference with adjacent channels. However, a persistent leakage can sometimes induce sparks or propagate leakage to neighbouring channels. In these cases, several channels may need to be disconnected or removed, affecting a larger region of the detector. In practice, if a leakage current is detected during a voltage ramp-up, the process of locating and disconnecting the problematic channel and then working on the noise again can be time-consuming, sometimes taking days.

To mitigate these issues, we perform systematic channel-by-channel testing during interventions or while testing new detectors. By executing controlled voltage sweeps on individual channels, abnormal current responses can be identified early. Once detected, the problematic channels are either disconnected or set to float during the reinstallation of the electronics cards.

The situation is even more critical for the GEM component. A spark within the GEM that creates a conduction path between the top and bottom layers cannot be remedied by disconnecting a single channel. For instance, during the March 2024 intervention to install a GEM stage, continuity tests revealed a short circuit between the top and bottom layers. This discovery delayed the operation on that side for several months and ultimately led to the decision to remove and replace the GEM during a subsequent intervention in July 2024.

Potential causes for leakage currents include occasional sparks during voltage ramp-ups, mechanical stress on the connections, and contamination of contact surfaces. To minimise these issues, we have set a protocol that involves slow, incremental voltage increases and limits on the maximum current per channel from the power supply. Despite these precautions, ensuring that all channels operate reliably remains an ongoing operational task, and the process of identifying and isolating leakage currents often adds significant time to intervention schedules.

\subsection{Gas Leaks} \label{Chapter5_Technical_Challenges_Gas_Leaks}

Gas leaks represent a significant operational challenge in TREX-DM, as even minor leaks can compromise gas purity and, consequently, the detector’s gain and resolution. Two main issues arise from leaks: first, when the chamber is below atmospheric pressure during the filling protocol, external air can infiltrate the system, requiring multiple recirculation cycles to purge impurities once the target pressure has been achieved; and second, even if TREX-DM operates at overpressure, retrodiffusion can allow air to enter if the leak is substantial. Acceptable leak levels in TREX-DM are around $10^{-8}$~mbar$\cdot$l/s.

Leaks may occur at various points in the gas system. Notably, feedthroughs and calibration ports are especially vulnerable. During routine maintenance or interventions, the disconnection of tubes and repositioning of the chamber often results in the appearance of leaks. This happened, for example, when transporting the vessel to the new site at Lab2500 in 2023. Figure~\ref{fig:chapter5_technical_issues} shows a helium leak test being performed at the central high-voltage feedthrough. 

In addition, the gas panel itself is a critical area where leaks have been observed. In some cases, leaks in valves or tube unions have been attributed to over-tightening of nuts, which deforms the joint. Routine leak testing, using a combination of helium and nitrogen at different pressures, is essential to identify these problems early and to mitigate their impact before they affect detector performance.


\section{Timeline of the Experiment} \label{Chapter5_Timeline}

The development of TREX-DM has proceeded through multiple phases, with each period characterised by specific operating conditions, interventions, and upgrades. The following timeline summarises the major milestones and key interventions from late 2018 until October 2024, highlighting the technical challenges encountered during each phase. The idea is to provide a general view of the evolution of the experiment.

\vspace{2mm}
\textbf{\normalsize Late 2018 - September 2019: Commissioning with TREXDM.v1 Detectors}
\vspace{0mm}

First commissioning runs were conducted underground at Hall A of Lab2400 using the original TREXDM.v1 detectors with Ar-1\%iC$_{4}$H$_{10}$/Ne-2\%iC$_{4}$H$_{10}$. Data were taken with both the North and South detectors, though most runs utilised only the South detector because of leakage current issues. Pressures ranged from 1.2 to 8~bar. Analysis of early background runs revealed that the overall background level was too high, prompting investigations into unaccounted-for components contributing to background.

\vspace{2mm}
\textbf{\normalsize January 2020 - October 2020: Ne Runs and Initial Interventions}
\vspace{0mm}

Following a long 4-month shutdown (Sept 2019 - Jan 2020) aimed at addressing leakage current and noise issues, the experiment resumed with neon runs using only the North detector. 4~bar was adopted as the operating pressure for the time being, nominal mesh voltage was set to 365~V and drift field to 160~V/cm/bar. First low-gain runs highlighted problems with alpha contamination that also affected the background levels through low-energy emissions. Surface contamination on the cathode was thought to be the main problem, which led to the decision to replace the original mylar cathode in Oct 2020 with a new (mylar) one. However, the field cage PCB was also a suspect, so it was decided to insert two PTFE pieces (see Figure~\ref{fig:chapter5_cathode_field_cage_last_ring}), one in each side, to cover the field cage with the aim to avoid its direct contact with the sensitive volume.

\vspace{2mm}
\textbf{\normalsize November 2020 - February 2021: Cathode Replacement, $^{222}$Rn and Surface Contamination}
\vspace{0mm}

Neon runs after the intervention to replace the cathode showed the background levels (both at high and low energy) remained the same. During this period, some runs were taken with both detectors because part of the intervention was dedicated to trying to bring the South detector back to operation. Runs in sealed mode indicated the background problem was mainly due to active $^{222}$Rn in the system. Subsequent runs switching the filters on and off provided evidence about the source of $^{222}$Rn, pointing to the moisture filter as the main culprit. 

Even though most of the rate was due to $^{222}$Rn, data from these runs showed there was also a surface contamination component that could be traced to the PTFE pieces inside the field cage. This information prompted an intervention to machine a thin layer off these surfaces, to check if the surface contamination component could be reduced. Operating conditions remained at 4 bar, with mesh voltages alternating between 365~V in nominal-gain runs (energy range $\sim 1-100$~keV) and 210~V in low-gain runs (energy range $\sim 0.5-10$~MeV), and a drift field of 160~V/cm/bar.

\vspace{2mm}
\textbf{\normalsize February 2021 - May 2022: Extended Ne Runs and Purifier Studies}
\vspace{0mm}

In this phase, the experiment conducted extended neon runs (with some tests at 6 bar) after removing a thin PTFE layer from the piece covering the field cage, correlating the intervention with a reduction in surface contamination (see Section~\ref{Chapter6_Reduction_Attempt}).

During this period, several methods were tested to reduce radon contamination in the system, including the use of different moisture and oxygen filters (with the aim of introducing less radon), testing molecular sieves and activated charcoal (with the idea of trapping radon) and operating in a semi-sealed open-loop mode to bypass the purification and recirculation branches altogether. All this is covered in Section~\ref{Chapter6_Active_Rn}.

These runs were terminated prior to the intervention to install new detectors. Conditions were maintained at 4 bar, with mesh voltages around 210-365~V and a drift field of 160~V/cm/bar.

\vspace{2mm}
\textbf{\normalsize End of 2021 / Beginning of 2022: Detector Characterisation}
\vspace{0mm}

A dedicated test bench in Zaragoza was used to characterise and evaluate the five TREXDM.v2 detectors. Performance comparisons in terms of gain, energy resolution and number of active channels led to the selection of the best two detectors to be installed in TREX-DM. 

\vspace{2mm}
\textbf{\normalsize June 2022 – October 2022: Installation and Commissioning of New\\ TREXDM.v2 Detectors}
\vspace{0mm}

New TREXDM.v2 microbulk Micromegas detectors were installed in June-July 2022 to replace the old detectors that exhibited numerous dead strips (compromising energy resolution and track reconstruction, see Figure~\ref{fig:chapter5_performance_comparison_detectors}) and leakage currents issues (to the point that one of the detectors was not operational).

Stable-operation runs were carried out to study the upgrades introduced during the intervention. In these runs, low-gain (at 210~V) and high-gain (around 365-375~V) modes were tested at 4 bar, with the drift field adjusted to 100~V/cm/bar according to the new electron transmission curves.

\vspace{2mm}
\textbf{\normalsize October 2022: Dismantling of the Experiment}
\vspace{0mm}

By orders from the LSC, TREX-DM was required to be relocated from Hall A in Lab2400 to Lab2500, despite the latter not being fully conditioned. The experiment was dismantled in October 2022 without a clear timeline for reinstallation.

\vspace{2mm}
\textbf{\normalsize October 2022 - December 2022: Study of a Calibration Technique}
\vspace{0mm}

During this period, a study was conducted at CEA Saclay on a low-energy X-ray calibration technique based on $^{37}$Ar production via neutron irradiation of calcium powder. This work aimed to develop a calibration technique that could be applied to TREX-DM to extend its calibration to low energies. Chapter~\ref{Chapter8_Ar37} is devoted to this topic.

\vspace{2mm}
\textbf{\normalsize January 2023 - May 2023: Importing the Calibration Technique and Testing of the GEM-Micromegas System}
\vspace{0mm}

Efforts were made in Zaragoza to import the $^{37}$Ar calibration technique by acquiring the necessary components (such as the CaO powder) and preparing the system (design of the vessel containing the powder, the valves, etc.). Once everything was ready, the tests were halted due to the unavailability of a proper neutron source.

In parallel, extensive tests were conducted on the GEM-MM setup. Both small test Micromegas units and one of the spare, full-scale TREX-DM Micromegas detectors (replicas of the installed detectors) were used to evaluate the feasibility of adding a GEM preamplification stage above the Micromegas. The tests focused on assessing the stability of the system in terms of voltages and quantifying the potential improvement in energy threshold. See Chapter~\ref{Chapter7_GEM-MM} for more information.

\vspace{2mm}
\textbf{\normalsize May 2023 - October 2023: Recommissioning at Lab2500}
\vspace{0mm}

TREX-DM was reinstalled in Lab2500, a process that required significant site conditioning including new electrical wiring, gas tubing, and complete reinstallation of the shielding from scratch. This phase also involved restarting the slow control system, re-optimising noise levels, and re-testing the detectors to ensure proper operation after transport. 

The relocation to Lab2500 resulted in an accumulated delay of approximately one year, during which various technical difficulties, such as potential gas leaks and leakage currents arising from the transport process, had to be addressed.

Data-taking at Lab2500 started with Ar-1\%-iC$_{4}$H$_{10}$ at 1 bar, due to the difficulty to acquire neon and the persistent background problem, which makes unnecessary to raise the pressure further.

\vspace{2mm}
\textbf{\normalsize October 2023 - January 2024: Alpha Directionality Studies}
\vspace{0mm}

With the experiment recommissioned in Lab2500, efforts shifted towards a detailed analysis of the background to refine the alpha directionality analysis. These studies were critical to quantify the surface contamination levels of cathode and Micromegas.

\vspace{2mm}
\textbf{\normalsize January 2024 - March 2024: Drift Velocity Measurements}
\vspace{0mm}

A dedicated campaign was undertaken to measure the electron drift velocity at different drift fields. These measurements were essential for validating the TPC’s time-to-space conversion parameters, ensuring accurate track reconstruction.

\vspace{2mm}
\textbf{\normalsize March 2024: GEM and Cathode Intervention}
\vspace{0mm}

An intervention was carried out to install a GEM foil on top of the Micromegas on one side of the detector and to replace the mylar cathode with a copper-kapton-copper composite with the aim of improving surface contamination levels on the cathode.

After this process, continuity tests revealed a short circuit between the top and bottom layers of the GEM that rendered that side non-operational, resulting in a temporary delay of several months.

\vspace{2mm}
\textbf{\normalsize March 2024 - July 2024: Data-Taking and $^{37}$Ar Production and Testing}
\vspace{0mm}

During this period, data-taking with the working side indicated that the new cathode was not cleaner than the old mylar one in terms of surface contamination, leading to the hypothesis that handling issues and brief exposures to atmospheric radon might be contributing to persistent surface contamination.

Meanwhile, the $^{37}$Ar calibration source was finally produced at Centro Nacional de Aceleradores (CNA) in Sevilla in April 2024 by irradiating calcium powder with 5~MeV neutrons. This had been halted due to uncertainties regarding where to irradiate the calcium powder with neutrons. An alternative of using the AmBe neutron source previously used at CEA Saclay was considered, but this option was ultimately ruled out in favour of conducting the irradiation in Spain, where logistics and transport are considerably simpler. 

The $^{37}$Ar source was tested in May 2024 in TREX-DM. Since the energy threshold was around 1~keV, the entire energy spectrum could not be seen, but the calibration process worked as expected.

\vspace{2mm}
\textbf{\normalsize July 2024: Another GEM Intervention}
\vspace{0mm}

Another intervention focused on substituting the problematic GEM foil was performed. It was replaced with another working GEM, and electrical tests were performed after installation to ensure proper operation. Detailed work was undertaken to reinstall electronics and re-optimise electronic noise, with the idea of having the GEM side working in optimal conditions.

\vspace{2mm}
\textbf{\normalsize July 2024 - October 2024: Data-Taking with GEM Side and New $^{37}$Ar Runs}
\vspace{0mm}

These months were devoted to fine-tuning the GEM side, with the aim of lowering the energy threshold in order to be able to see the whole $^{37}$Ar spectrum in future calibrations.

Another irradiation took place in October 2024 at CNA in Sevilla, and the $^{37}$Ar source was injected into the GEM side of TREX-DM upon receiving it at LSC. Several calibration runs were succesfully taken, despite the challenges of operating the GEM side. The much lower energy threshold, $O(10)$~eV, allowed to see the entire energy spectrum, with both $^{37}$Ar peaks (270~eV and 3.2~keV) clearly visible (see Section~\ref{Chapter8_Calibration_TREX-DM}).

\section{Drift Velocity Measurements \texorpdfstring{Ar-1\%-iC$_{4}$H$_{10}$}{Ar-1\%iC4H10}}
\label{Chapter5_Drift_Velocity_Measurements}

Drift velocity of electrons is a very important parameter in the analysis routine, especially in the track reconstruction of particles such as alphas, because its value will determine the $\Delta Z$ length of the track, allowing for energy discrimination based on the expected track length (from sources such as NIST ASTAR~\cite{ESTAR_PSTAR_ASTAR}). Up to now, the drift velocity values had been extracted from simulations, but some experimental confirmation was lacking. To this end, prior to the March 2024 intervention, several 3-hour calibrations were taken to try to determine the drift velocity of Ar-1\%-iC$_{4}$H$_{10}$ at different drift fields. 

\vspace{2mm}
\textbf{\normalsize Experimental Approach}
\vspace{0mm}

The method relies on capturing events that produce correlated signals (such as Comptons) at both ends of the TPC: near the Micromegas readout and near the cathode. With a sufficiently long time window, these events are recorded by the DAQ within a single event. Given that the drift length is known ($d=160$~mm), the time difference $\Delta t$ between the earliest and the latest peaks for these events provides an estimate of the maximum drift time. This time difference is calculated multiplying the bin difference $k$ and the electronics' clock $c$ ($c=1/$sampling rate). All in all, the drift velocity can be estimated using the relation:

\begin{equation}
    v_{\mathrm{drift}}=\frac{d}{\Delta t}=\frac{d}{k\times c}
    \label{eq:chapter5_drift_vel_calculation}
\end{equation}

\begin{figure}[htb]
\centering
\includegraphics[width=1.0\textwidth]{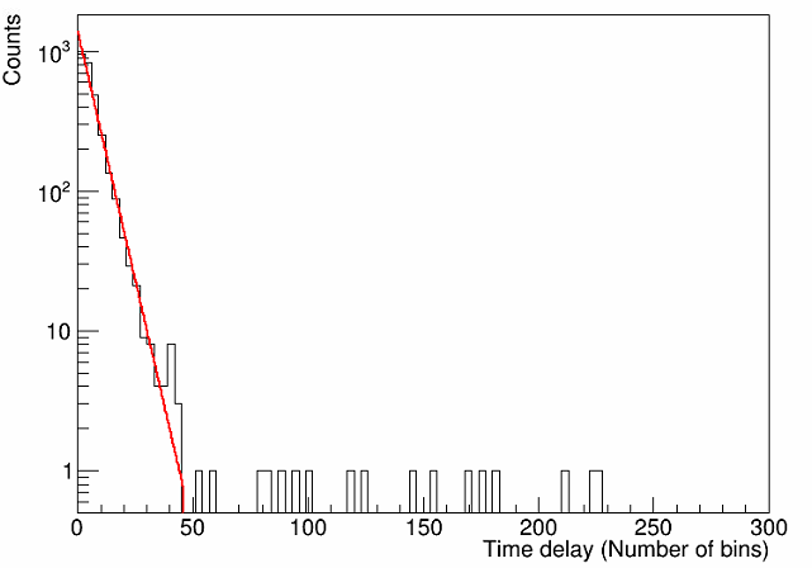}
\caption{Example of the cut-off determination for one of the calibration runs. \textit{MaxPeakTimeDelay} for 2-track events with a low-energy cut in \textit{SecondMaxTrackEnergy} (black) and exponential + cut-off fit overlaid (red). Note the y axis is in logarithmic scale, hence the fit appears linear. The parameters for this fitted run are: $E=100$~V/cm/bar, $d=160$~mm, $c=160$~ns/bin, $k=45.9\pm2.5$~bin.}
\label{fig:chapter5_max_peak_time_delay_fit}
\end{figure}

To identify these events, we plot the time between the earliest peak and the latest peak of all pulses for all events of a given run (observable \textit{MaxPeakTimeDelay} in REST-for-Physics), and we look for the time cut-off in the spectrum, which gives the maximum drift time (see Figure~\ref{fig:chapter5_max_peak_time_delay_fit} for an example).

For a fixed suitable clock setting (160~ns/bin, corresponding to window size $\sim 82$~$\upmu$s), the drift field was varied, with values of 60, 80, 100, 150, 200, 250, and 300~V/cm/bar. The position of the cut-off was expected to shift as the drift velocity changes.

\vspace{2mm}
\textbf{\normalsize Analysis}
\vspace{0mm}

To isolate the events of interest from pile-up (uncorrelated, independent events that fall within the same time window), an interesting observation is made based on a $^{109}$Cd calibration simulation using REST-for-Physics: correlated events exhibiting more than one track typically consist of a primary energy deposit of either approximately 19~keV or 5~keV, combined with a secondary deposit around 3 keV. This is interpreted as 22~keV/8~keV events from $^{109}$Cd/copper fluorescence, respectively, that produce a K-shell photoionisation in argon, after which the de-excitation is produced via 3~keV fluorescence X-ray, reabsorbed within the same event. 

\begin{figure}[htb]
\centering
\includegraphics[width=1.0\textwidth]{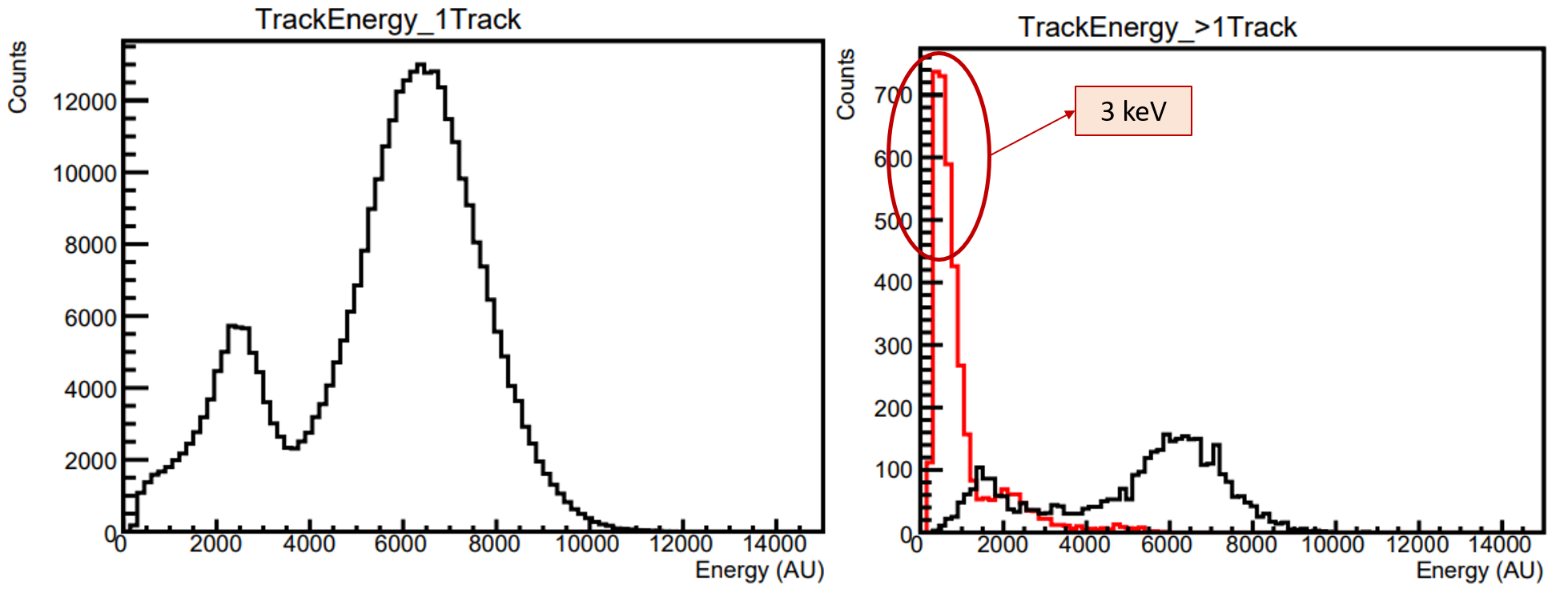}
\caption{$^{109}$Cd calibration spectra in arbitrary units of energy. Left: only 1-track events, peaks at 8~keV (copper fluorescence) and 22~keV (X-ray from $^{109}$Cd) are visible. Right: only events with more than one track, with the energy of the most energetic track in black, and the energy of the second most energetic track in red. Most events in the secondary track are 3-keV (argon K-shell fluorescence) sub-events.}
\label{fig:chapter5_one_track_two_tracks_spectra}
\end{figure}

\begin{figure}[htb]
\centering
\includegraphics[width=1.0\textwidth]{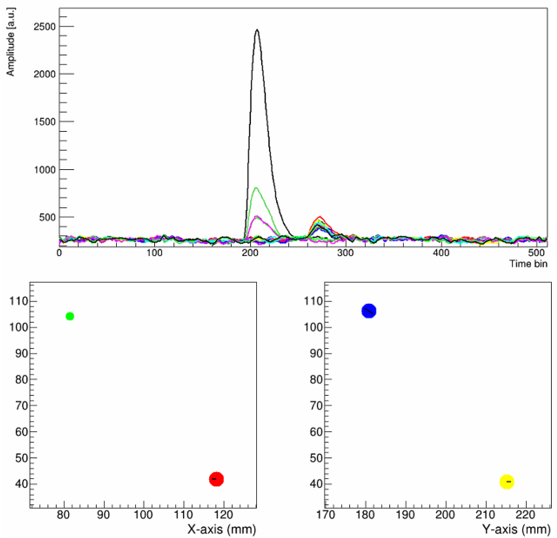}
\caption{Two-track event that passes the low-energy selection cut in \textit{SecondMaxTrackEnergy}. Top: raw view of the event, featuring two distinct sub-events. Amplitude is in ADC units and time is in bins. Bottom: track reconstruction of the event, highlighting the 2-track structure both in X and Y.}
\label{fig:chapter5_two_track_event}
\end{figure}

In the experimental data, a dedicated track analysis (implemented via the \textit{TRestTrackAnalysisProcess}) is used to extract detailed observables from each event. This module calculates key parameters such as the number of tracks, the energy of the most energetic track (\textit{MaxTrackEnergy}), and the energy of the second most energetic track (\textit{SecondMaxTrackEnergy}). In events where more than one track is present, a low-energy cut is applied based on the \textit{SecondMaxTrackEnergy}, thereby isolating events of interest. Figure~\ref{fig:chapter5_one_track_two_tracks_spectra} shows the energy spectrum for events with one track (left), displaying the full 8~keV and 22~keV peaks, and the energy spectrum for events with more than one track (right), where the spectrum for the most energetic track is plotted in black, and the spectrum for the second most energetic track is plotted in red. It is clear that the intuition gained from the simulation translates to the experimental data, as the second most energetic track spectrum is basically 3~keV sub-events. Figure~\ref{fig:chapter5_two_track_event} shows an example of a two-track event that passes the selection cut. Both the raw event and the track representation are shown.

The resulting \textit{MaxPeakTimeDelay} distribution for those selected events was then fitted to an exponential function with a defined cut-off, corresponding to the maximum drift time: 

\begin{equation}
    f(t) = 
    \begin{cases}
        A \exp\left(-Bt\right) & \mathrm{if~} t<k \\
        0 & \mathrm{if~} t\geq k
    \end{cases}
\end{equation}

where $t$ is the bin difference, $k$ is the cut-off and $A$ and $B$ are constants. Figure~\ref{fig:chapter5_max_peak_time_delay_fit} shows an example of a fitted run ($E=100$~V/cm/bar) with $k=45.9\pm2.5$~bin, which yields $v_{\mathrm{drift}}=2.18\pm 0.12$~cm/$\upmu$s according to Equation~\ref{eq:chapter5_drift_vel_calculation}. 

\vspace{2mm}
\textbf{\normalsize Observations and Discussion}
\vspace{0mm}

\begin{figure}[htb]
\centering
\includegraphics[width=1.0\textwidth]{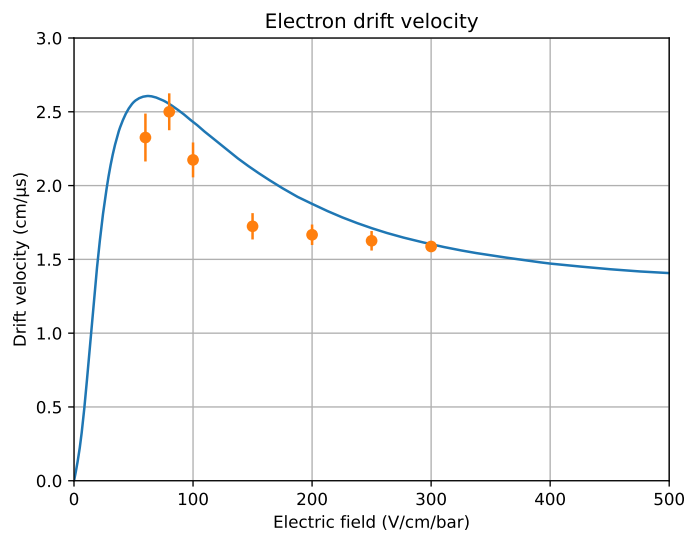}
\caption{Electron drift velocity for Ar-1\%iC$_{4}$H$_{10}$. Comparison between simulation (blue, values extracted from~\cite{gas_graphs}) and
experimental values (orange) derived from the method explained in the text. Error bars in experimental data only include error propagation of the fitted cut-off bin.}
\label{fig:chapter5_drift_velocity_simu_exp}
\end{figure}

Comparing the experimental values extracted from this method with the simulation (see Figure~\ref{fig:chapter5_drift_velocity_simu_exp}), the agreement is quite good. There are some discrepancies on the order of 10\%, probably due to the combination of uncertainties in the simulation and experimental uncertainties that have not been quantified, such as the effective drift distance value. However, these results validate the values used in the analysis routines. It is important to note that these tests were performed at a rate of approximately 35~Hz per detector side, ensuring enough statistics to reliably determine the cut-off in the time distribution.

While the current method has served as an essential cross-check, it is also worth mentioning that a full-fledged set-up to accurately measure drift velocities of different mixtures is being developed in Zaragoza.

%% file: CHAPTERS/Chapter6.tex
\chapter{Radon and the Background Problem} \label{Chapter6_Radon_problem}

{
\lettrine[loversize=0.15]{R}{adon} is a naturally occurring radioactive gas and one of the primary background sources in rare event experiments. Its decay chain produces a series of short-lived isotopes that can contaminate both the detector’s gas volume and its internal surfaces, if exposed to it long enough. In the TREX-DM experiment, initial measurements revealed background levels significantly higher than predicted. Detailed investigations showed that active $^{222}$Rn in the gas was responsible for most of the background events, while the presence of radon progeny (mainly $^{210}$Pb) on detector components led to additional low-energy background.

%
\section*{}
\parshape=0
\vspace{-20.5mm}
}

This chapter focuses on the progress of TREX-DM in understanding and mitigating its background problem. It covers the evolution of the issue, from the early detection of unexpected levels to the identification of both active radon and its long-lived progeny in the form of surface contamination. The chapter also reviews the strategies that have been implemented to reduce these background contributions, and outlines the diagnostic methods employed, including the development of the AlphaCAMM detector, designed to measure and characterise surface alpha contamination.

A brief description of this problem, together with some of the results discussed in this chapter, was published in~\cite{MicromegasPerspectives_2024}.

\section{Identification of the Problem} \label{Chapter6_Identification}

\begin{figure}[htbp]
\centering
\includegraphics[width=0.98\textwidth]{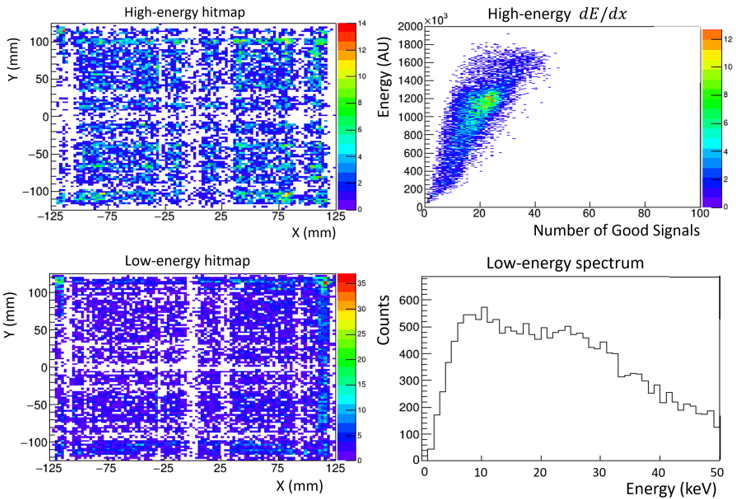}
\caption{Top: low-gain run ($V_{\mathrm{mesh}}=230$~V) to study the high-energy region. Left: hitmap of the North detector, where each point represents the mean position of an alpha event. Right: energy as a function of the number of good signals per event. The energy per channel is approximately constant, which is indicative of alpha events. Bottom: nominal-gain run ($V_{\mathrm{mesh}}=365$~V) to study the low-energy background. Left: hitmap of the North detector after some quality cuts to remove noise and unwanted signals. Right: background spectrum in the $(0,50)$~keV region, levels $\approx 1000$~dru.}
\label{fig:chapter6_2020_high_low_energy}
\end{figure}

As indicated in Section~\ref{Chapter2_Direct_Searches_Sensitivity}, the detection of potential Dark Matter interactions needs exceptionally low background levels. In the case of TREX-DM, the background model initially predicted levels in the range of 1-10 dru, as detailed in Section~\ref{Chapter5_Description_Background_Model}. However, the first experimental data collected with Ne-2\%iC$_{4}$H$_{10}$ at 4~bar during the commissioning phase in early 2020 revealed a significantly higher background level (approximately 500-1000 dru), orders of magnitude above the expectations from the background model. A typical background spectrum at nominal gain ($V_{\mathrm{mesh}}=365$~V) can be seen in the bottom right plot of Figure~\ref{fig:chapter6_2020_high_low_energy}.

These early runs contained an unusually high population of high-energy events that saturated the acquisition electronics. It was decided to investigate this region further, so in order not to saturate the DAQ, special low-gain runs ($ V_{\mathrm{mesh}}=210-235$~V, a gain approximately $100-500$ times lower than for $V_{\mathrm{mesh}}=365$~V) were taken, revealing clear evidence of alpha particles (see top right plot of Figure~\ref{fig:chapter6_2020_high_low_energy}). Initially, it was thought $^{210}$Pb surface contamination resulting from $^{222}$Rn exposure of internal detector surfaces was responsible for this contamination. This would manifest as 5.3~MeV alphas from $^{210}$Po, the final radioactive isotope in the decay chain (see Figure~\ref{fig:chapter6_radon_decay_chain}). Simulations suggested $^{210}$Pb from the cathode could be compatible with the low-energy spectrum from nominal-gain runs, and the proportion between high-energy and low-energy rates (roughly $1:1$, discussed in Section~\ref{Chapter6_Surface_Contamination}) also aligned with simulation predictions. However, the possibility that this surface contamination came from the PCB walls of the field cage was not ruled out.

Based on this evidence, in October 2020, the mylar cathode was replaced with a new one to test the hypothesis that $^{210}$Pb surface contamination on the cathode was the primary contributor to the elevated background. Also, a PTFE piece covering the field cage was introduced (see Figure~\ref{fig:chapter5_cathode_field_cage_last_ring}). However, data showed that both high-energy and low-energy rates remained unchanged, indicating that the problem was elsewhere.

\begin{figure}[htbp]
\centering
\includegraphics[width=1.0\textwidth]{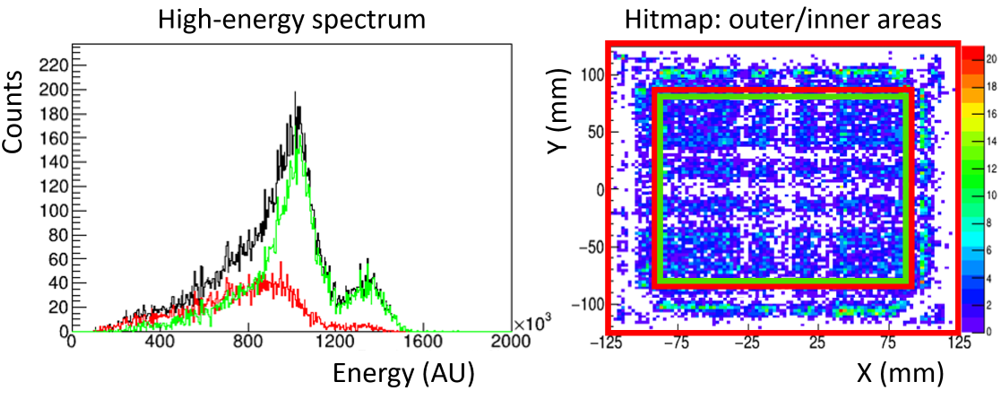}
\caption{Left: high-energy spectrum (black) separated into inner (green) and outer (red) regions of the detector. The full spectrum is compatible with $^{222}$Rn (5.5~MeV + 6.0~MeV + 7.7~MeV alphas) + $^{210}$Pb surface contamination (5.3~MeV alphas), see Figure~\ref{fig:chapter6_radon_decay_chain} for the full decay chain. Right: hitmap showing in squares (same colour legend) the areas considered for the spectrum.}
\label{fig:chapter6_radon_evidence}
\end{figure}

In November 2020, it was decided to collect data in sealed mode (with the chamber closed and no gas recirculation). This approach revealed a gradual decrease in both high-energy and low-energy rates, converging to a constant component. This observation led to the hypothesis that the background was primarily comprised of two components: active $^{222}$Rn in the detector volume would be the main source, while $^{210}$Pb surface contamination on inner components in contact with the gas would add a smaller, secondary contribution. Evidence in favour of this idea came from carefully examining the high-energy spectra: dividing the sensitive volume into an inner and outer part showed a radon-compatible spectrum (5.5~MeV + 6.0~MeV + 7.7~MeV alphas) for the inner part and a spectrum dominated by surface contamination (5.3~MeV alphas) near the borders of the hitmap (see Figure~\ref{fig:chapter6_radon_evidence}).

\begin{figure}[htbp]
\centering
\includegraphics[width=0.85\textwidth]{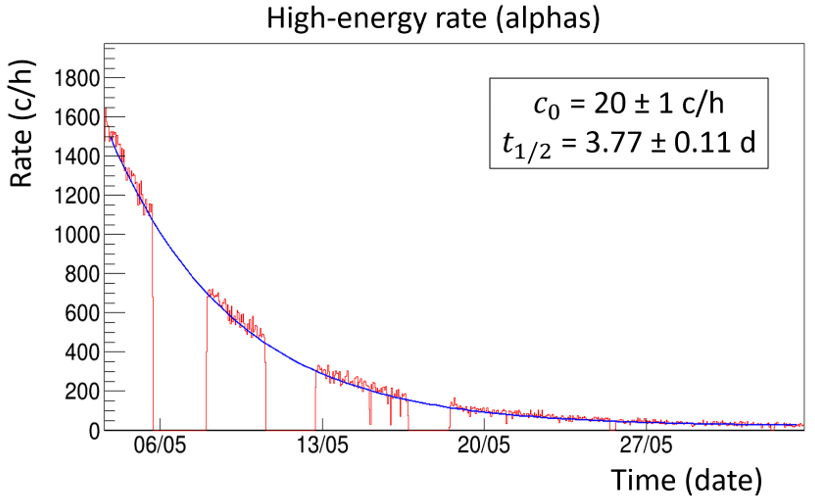}
\includegraphics[width=0.85\textwidth]{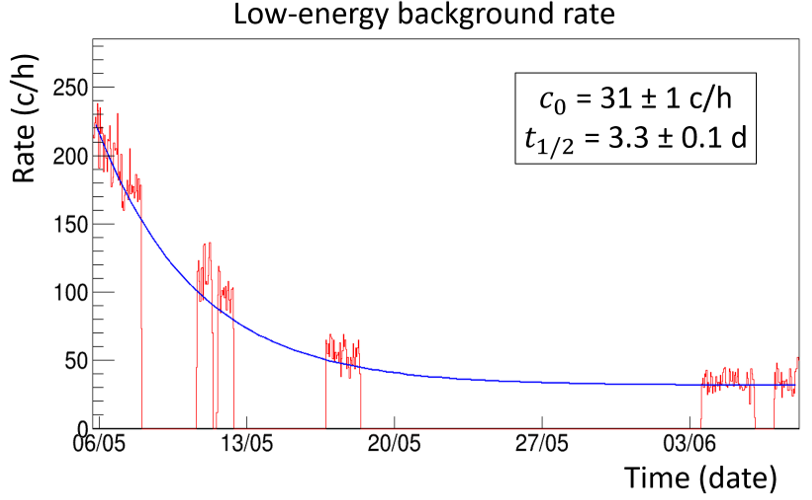}
\caption{Evolution of the high-energy (230~V, top) and low-energy (365~V, bottom) rates during one month of sealed-mode operation. Experimental data are shown in red, while the fit to Equation~\ref{eq:chapter6_exponential_decay} is plotted in blue. The relevant parameters of the fit, together with the statistical error, are also included.}
\label{fig:chapter6_2021_closed_vessel}
\end{figure}

However, this hypothesis was indisputably confirmed after operating in sealed mode for one month in May 2021, a long enough period to follow the decay. Alternating high-energy and low-energy runs revealed, in both regions, an exponential decay component consistent with the half-life of $^{222}$Rn ($t_{1/2}=3.82$~d)~\cite{lnhb_table_radionucleides}, superimposed on a constant component attributed to long-lived ($t_{1/2}=22.3$~y) $^{210}$Pb surface contamination. Figure~\ref{fig:chapter6_2021_closed_vessel} shows the evolution of the rate along with the exponential fit:

\begin{equation}
    R(t)=c_0 + c_1\exp\left(-\frac{\ln{2}}{t_{1/2}}t \right)
    \label{eq:chapter6_exponential_decay}
\end{equation}

where $c_0$ is the constant component and $t_{1/2}$ the half-life of the decay. The agreement with $^{222}$Rn's half-life is excellent, even though $t_{1/2}$ differs a bit for the low-energy region, probably due to having a smaller dataset than the high-energy region.

The clear identification of these two distinct background sources, active $^{222}$Rn and $^{210}$Pb surface contamination, set the direction for subsequent efforts to mitigate the background problem in TREX-DM, as will be explored in the following sections.

\section{Active \texorpdfstring{$^{222}$Rn}{222Rn}} \label{Chapter6_Active_Rn}

\begin{figure}[htbp]
\centering
\includegraphics[width=0.8\textwidth]{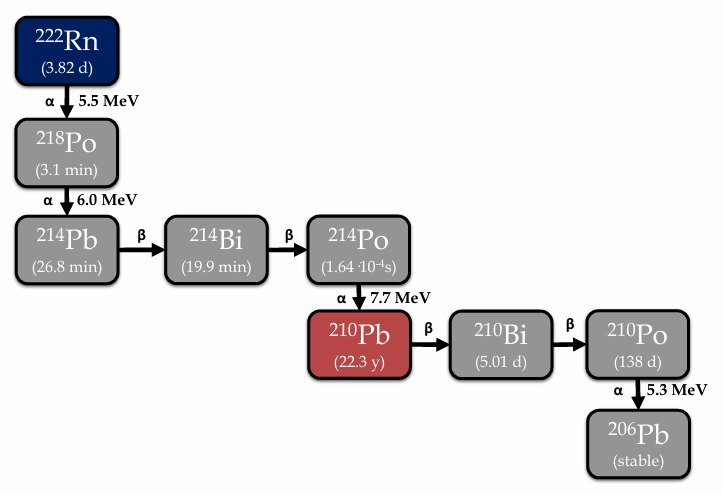}
\caption{$^{222}$Rn decay chain. Taken from~\cite{AlphaCAMM_2022}.}
\label{fig:chapter6_radon_decay_chain}
\end{figure}

As already mentioned, radon is a significant background source in many rare event search experiments, including Dark Matter and neutrinoless double beta decay searches. $^{222}$Rn comes from the primordial $^{238}$U decay chain, making it ubiquitous in most materials. What makes $^{222}$Rn particularly problematic is its gaseous nature combined with its relatively long half-life ($t_{1/2}=3.82$~d, the longest of any radon isotope). These properties allow it to emanate from the internal surfaces of materials and diffuse into the active volume of detectors. Once there, the beta decays of its progeny ($^{214}$Pb and $^{214}$Bi, see Figure~\ref{fig:chapter6_radon_decay_chain}) and the subsequent low-energy electrons and X-ray emissions induce signal-like events that can be difficult to eliminate through fiducialisation or offline techniques.

Efforts to control radon backgrounds can generally be classified into two broad categories: use of radon-clean materials and radon removal. The first approach focuses on using materials and components with low intrinsic radon emanation, which requires careful material screening. As an example, the XENON1T collaboration conducted an extensive campaign of radon emanation measurements across their detector components, including metals (e.g. titanium, stainless steel), gas purification systems (such as purifiers operating in both hot and cold conditions and recirculation pumps), and auxiliary items (PMTs, cables, epoxy, copper, and PTFE components). Their studies ultimately reduced the radon activity in the target volume to $4.5 \pm 0.1$~$\upmu$Bq/kg~\cite{XENON_2020_radon}, the lowest ever achieved in a xenon-based Dark Matter experiment. 

The second approach, radon removal, focuses on actively purifying the detector environment either by trapping radon before it enters the detector or by continuously removing it from the gas volume. Several methods have been developed to achieve this, including cryogenic radon distillation or radon adsorption on different materials. This is the strategy followed by underground laboratories such as LSM to produce radon-purified air~\cite{Hodak_2019}. 

In the context of TREX-DM, it was soon discovered (after the sealed mode runs in November 2020) that the moisture purifier used to maintain gas quality in the recirculation system was the main emanator, with a minor (10-20\%) contribution coming from the recirculation pump and the oxygen filter. This conclusion was reached by taking data isolating the different elements of the gas system. Different strategies were explored to mitigate radon contamination and reduce background levels to within the expected range. However, efforts shifted when a definitive solution based on a semi-sealed open-loop operation was found, which reduced the background levels from several hundred dru down to around 80-100 dru, bringing active radon contamination down to negligible levels. The following sections will discuss the various approaches attempted in TREX-DM, including material selection efforts based on testing different purifiers, and radon removal techniques using two different filtering methods. Finally, the solution adopted is described.

\subsection{Emanation Studies of Filters} \label{Chapter6_Emanation_Studies}

In gas-based rare event experiments operating in recirculation mode such as TREX-DM, oxygen and humidity filters are often necessary to maintain gas quality (check Section~\ref{Chapter3_Charge_Drift} to see how this affects the performance of the detector). However, while studying the background problem in TREX-DM, it was found that they emanate $^{222}$Rn in unacceptable amounts for the background goals of the experiment. The fact that these filters emanate varying amounts of $^{222}$Rn is not new, and it has been documented by different experiments. For example, in the NEXT-White detector, they observed a three-order-of-magnitude increase in $^{222}$Rn activity after using an ambient temperature purifier~\cite{NEXT_2018}, with the resulting electron background dominated by $^{214}$Bi.

\begin{figure}[htbp]
\centering
\includegraphics[width=1.0\textwidth]{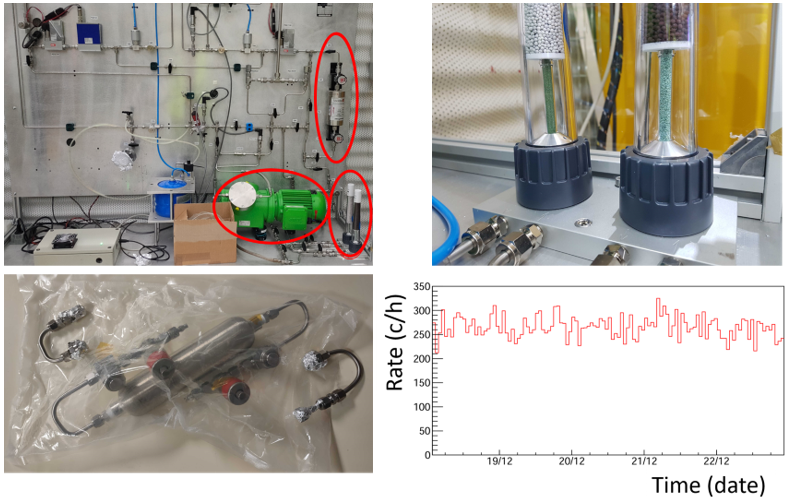}
\caption{Top left: gas panel of TREX-DM with recirculation pump, SAES FT400-902 filter and Agilent filters installed (all of them encircled in red). Top right: zoomed view of moisture (left) and oxygen (right) Agilent filters. Bottom left: custom-made filter sent from The University of Birmingham. Bottom right: stable HE rate achieved with both Agilent filters engaged.}
\label{fig:chapter6_filters_studies}
\end{figure}

In TREX-DM, a systematic investigation of different filters was carried out, with the idea of trying to find one with emanation levels suitable to our needs. To study this, two basic parameters were tracked, interspersing low-gain and nominal-gain runs: high-energy (HE) and low-energy (LE) rates. We defined the HE rate as all the alpha events detected by TREX-DM North\footnote{These studies were carried out before installing TREXDM.v2 detectors, so only one detector was in operation at the time.}, and LE rate as all the background events after cuts that fall within the 0-50~keV range. These definitions are, of course, a first approximation: on one hand, one would have to subtract the constant component from both rates to be totally correct; on the other hand, a smaller energy range should be considered. However, given how large the active radon contribution was, this "big numbers" approach was enough for practical purposes, and the range 0-50~keV was useful to track how the contamination was affecting the background. As already evidenced in Figure~\ref{fig:chapter6_2021_closed_vessel}, HE and LE rates are correlated as they both follow the $^{222}$Rn decay. Different filters were tested in TREX-DM over the course of $\sim$ 1 year:

\begin{itemize}
    \item \textbf{Agilent Gas Clean Filters}: these were the baseline commercial filters used in closed-loop operation. Of the two simultaneously used filters (O$_2$ and H$_2$O, see Figure~\ref{fig:chapter6_filters_studies}), the O$_2$ filter was found to be relatively clean, while the H$_2$O filter was identified as the primary emanator. With these two filters combined, the stable HE rate was around 250-280~c/h, as can be seen in Figure~\ref{fig:chapter6_filters_studies}, and the LE rate was around 110-120~c/h. As a reference, the HE spectrum corresponding to these filters is the one presented in Figure~\ref{fig:chapter6_radon_evidence}), while the LE one can be seen in Figure~\ref{fig:chapter6_2020_high_low_energy}.
    \item \textbf{SAES MicroTorr FT400-902 Filter}: another commercial O$_2$ + H$_2$O filter. This proved to be the worst of all tested filters, with high-energy rates 6-7 times higher than Agilent filters and low-energy rates 3 times higher. Due to its really high rates, this filter was selected for use immediately before closing the vessel to monitor the radon decay (see Figure~\ref{fig:chapter6_2021_closed_vessel}, initial rates correspond to this filter).
    \item \textbf{Custom-made O$_2$ + H$_2$O Filter}: developed by the University of Birmingham for the NEWS-G collaboration using low-emanation materials~\cite{filters_ieee}, this filter yielded the best results among all tested filters, with rates approximately half those of Agilent filters. However, the background level was still $O(100)$~dru.
\end{itemize}

The images of all these filters can be seen in Figure~\ref{fig:chapter6_filters_studies}, and a summary of the HE and LE rates achieved with them, together with some other relevant parameters, are listed in Table~\ref{table:chapter6_radon_studies}. The lowest HE rate achieved with these filters was around 150~c/h, which corresponds to an activity of $^{222}$Rn of roughly 1~mBq/L in the system\footnote{This number comes from subtracting 20~c/h for the constant component, considering 1 Bq of $^{222}$Rn has 3 alpha events (see Figure~\ref{fig:chapter6_radon_decay_chain}) until reaching the long-lived $^{210}$Pb, and taking into account the active volume for just one detector is around 10~L.}, and the associated LE rate is equivalent to background levels $O(100)$~dru. As a reference point, we mentioned above $4.5 \pm 0.1$~$\upmu$Bq/kg in 3.2 t of liquid xenon for the XENON1T experiment, which corresponds to around 10~$\upmu$Bq/L ($\rho_{\mathrm{Xe}}\approx 3$~kg/L at 2 bar and 175 K, point at which XENON1T operates~\cite{XENON1T}). Although it is a totally different experiment and the background may be influenced by radon in distinct ways, this comparison shows that we were still far from desirable levels. Therefore, the main conclusion from these studies is that none of the filters yielded good enough results, but the custom-made filter option was definitely the most promising, potentially with a lot of room for improvement through optimisation of base filtering materials and quantities.

\begin{table}[htbp]
\centering
\begin{center}
\begin{tabular}{c|c|c|c}  
\hline\hline
\multirow{2}{*}{\textbf{Test}}
 & $V_{\mathrm{mesh}}$ & \textbf{Energy} & \textbf{Stable Rate} \\
 & \textbf{(V)} & \textbf{Range} & \textbf{(c/h)} \\
\hline
\multirow{2}{*}{Agilent filters}  
  & 235 & HE & 250--280 \\  
  & 365 & LE & 110--120 \\    
\hline
\multirow{2}{*}{SAES filter}  
  & 235 & HE & 1500--1600 \\  
  & 365 & LE & 340--360 \\  
\hline
\multirow{2}{*}{Custom-made filter}  
  & 210 & HE & 150--170 \\  
  & 365 & LE & 60--70 \\  
\hline
\multirow{2}{*}{Molecular sieves}    
  & 210 & HE & 380--420 (200~g) \\  
  & 210 & HE & 300--340 (20~g) \\  
\hline
\multirow{1}{*}{Activated carbon}    
  & 210 & HE & 90--110 \\   
\hline
\multirow{2}{*}{Semi-sealed open-loop}  
  & 210 & HE & 20 \\  
  & 365 & LE & 30 \\  
\hline\hline
\end{tabular}    
\end{center}
\caption{Summary of TREX-DM background runs under various conditions.}
\label{table:chapter6_radon_studies}
\end{table}


\subsection{Radon Removal Tests}
\label{Chapter6_Radon_Removal}

Due to the insufficient results from the first approach testing filters, a second strategy was followed in parallel. This involves removing $^{222}$Rn from the system using specialised materials such as radon traps. Since $^{222}$Rn is a chemically inert noble gas, it can only be trapped via physical adsorption through van der Waals forces. The principle is to delay the passing of $^{222}$Rn long enough through adsorption/desorption processes for it to decay within the trapping material. The resulting daughters, being electrically charged, can remain indefinitely trapped in the material.
Effective radon traps must be highly porous, with pore sizes large enough to capture $^{222}$Rn atoms. Two different materials were tested in TREX-DM: molecular sieves and activated carbon.

\subsubsection{Molecular Sieves}
\label{Chapter6_Molecular_Sieves}

Molecular sieves are crystalline materials that possess pores with precisely defined sizes, which enable them to selectively adsorb small molecules based on their critical diameters: molecules smaller than the pore size can enter and be adsorbed within the internal structure, while larger molecules are excluded because they cannot physically fit through the openings. The name "sieve" alludes to a mechanical sieve or filter because they function similarly, but at the molecular level. Commercial molecular sieves are typically zeolites, which mainly consist of aluminosilicates.

In TREX-DM, molecular sieves with pore sizes of 5~\r{A} were evaluated for radon mitigation. These particular sieves were selected because their pore dimensions are well-suited to capture radon atoms, which have effective diameters reported between approximately 2.5 and 4.9~\r{A}~\cite{Ezeribe_2017}. For our experimental tests, commercially available 5~\r{A} sieves from Sigma-Aldrich were purchased in bead form with sizes ranging from 8 to 12~mm.

It is worth noting that previous studies~\cite{Ezeribe_2017, Gregorio_2021} demonstrating successful radon mitigation with molecular sieves were conducted using SF$_6$ as a carrier gas. SF$_6$ molecules, with their larger effective diameter, are excluded from 5~\r{A} pores, allowing the sieves to selectively trap radon without capturing the carrier gas. In contrast, the TREX-DM experiment utilised a neon-based gas mixture during these tests, where neon atoms (with diameters smaller than the pore size) can potentially compete with radon for adsorption sites. This fundamental difference may significantly impact the trapping efficiency, as the sieves could become partially or fully saturated with neon, leaving limited capacity for radon capture.

\begin{figure}[htb]
\centering
\includegraphics[width=0.38\textwidth]{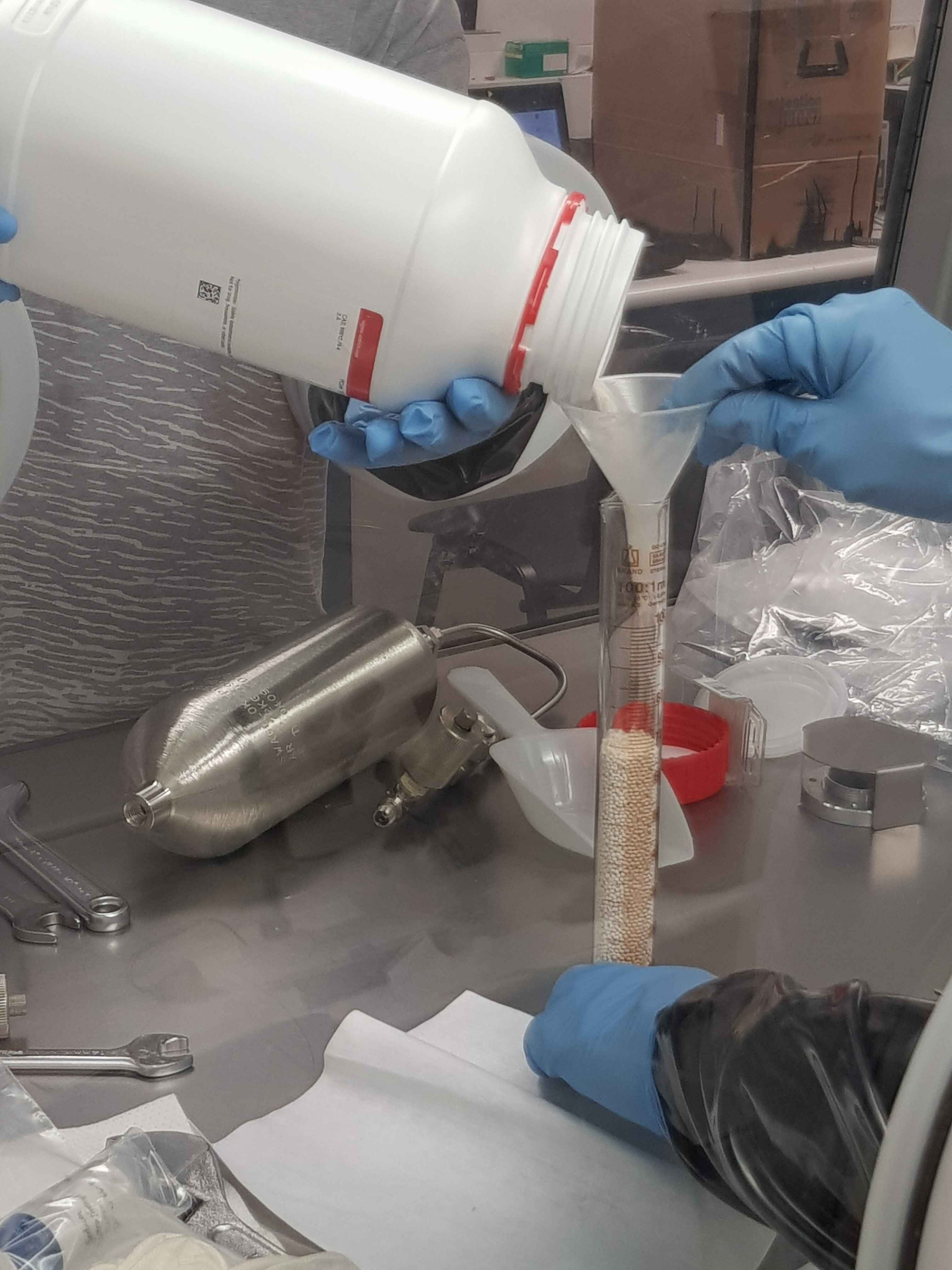}
\includegraphics[width=0.273\textwidth]{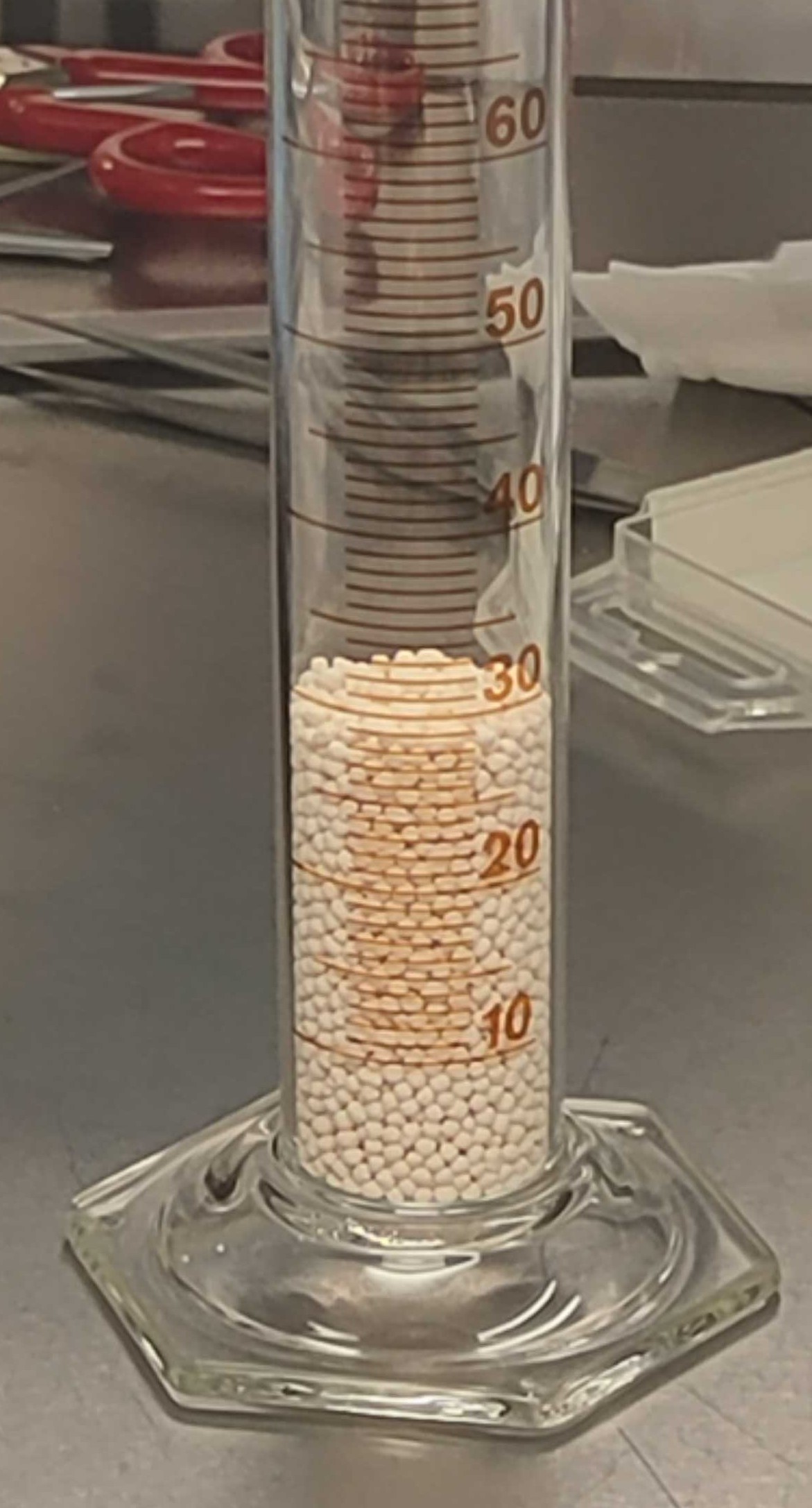}
\includegraphics[width=0.328\textwidth]{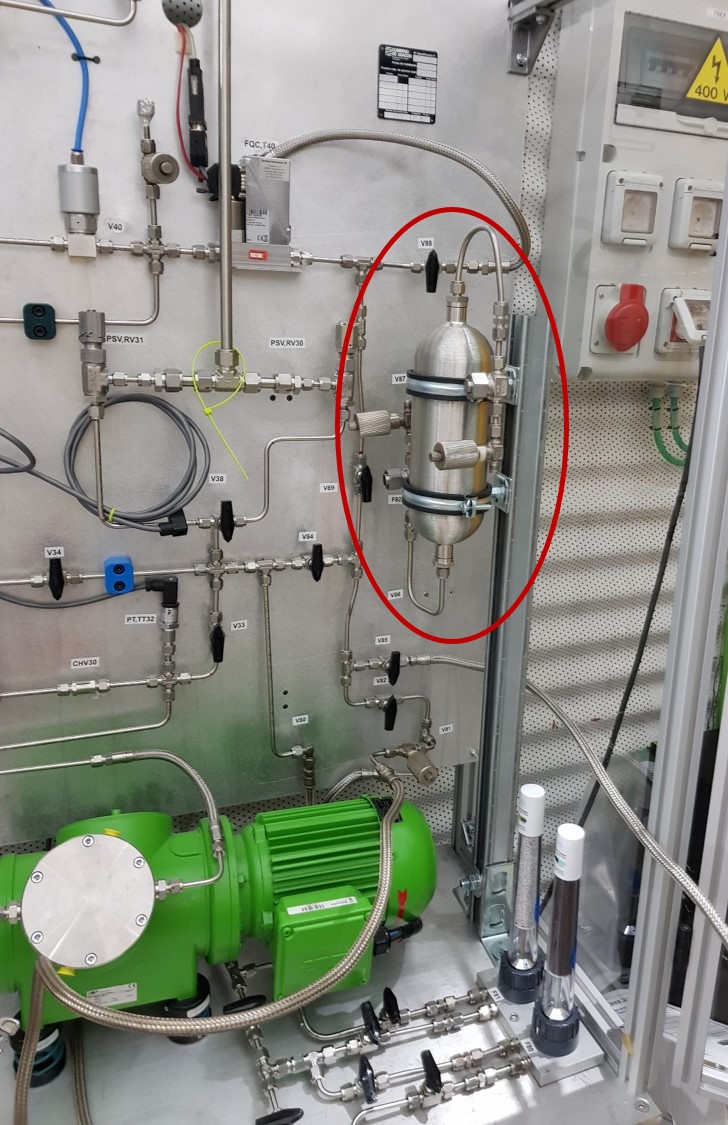}
\caption{Left: process of filling the bottle with molecular sieves in the glove box. Centre: calibrated container with 20~g of molecular sieves. Right: bottle with molecular sieves (encircled in red) installed in the gas panel of TREX-DM.}
\label{fig:chapter6_molecular_sieves}
\end{figure}

Our initial testing strategy involved a substantial quantity of material to ensure maximum trapping efficiency. As illustrated in Figure~\ref{fig:chapter6_molecular_sieves}, a dedicated 1-L vessel was prepared in a glove box with a controlled N$_2$ environment and filled with 200~g (approximately 290~ml, occupying 30\% of the vessel volume) of molecular sieves. This vessel was then integrated into the gas panel of the TREX-DM system. Calculations based on published adsorption capacities~\cite{Gregorio_2021} suggested that even a few grams of the material should theoretically suffice to trap the amount of $^{222}$Rn present in our system. However, uncertainties regarding the interaction with neon gas and potential variations in sieve properties prompted the conservative decision to use a larger quantity.

Prior to integration with the gas system, the vessel containing the molecular sieves was pumped down to a vacuum level of $10^{-3}$~mbar to minimise initial contamination. 

Contrary to expectations, the incorporation of 200~g of molecular sieves into the gas circulation system resulted in an increase in HE event rates to 380-420c/h (see Table~\ref{table:chapter6_radon_studies}), exceeding the baseline measurements observed with standard Agilent filters. This elevated rate was slowly reached from a baseline of 250-280 c/h, and stabilised within 3-4 hours of activating the sieves and persisted throughout the testing period. This observation deviated significantly from the anticipated behaviour: an initial sharp reduction in event rates followed by a gradual increase over several days as radon emanation from the sieves themselves\footnote{Even as these materials adsorb radon from the circulating gas, they simultaneously release radon generated from trace amounts of $^{238}$U present within their composition.} reached equilibrium.

The unexpected results were attributed to self-emanation from the molecular sieves that was not properly evacuated during pumping: both the macroscopic size of the beads and the fact that sieves can adsorb part of the self-emanated radon make it difficult to reach a vacuum level competitive in terms of radon with the quantities we are dealing with here ($O(\mathrm{mBq})$). Furthermore, the potential competition between neon and radon for adsorption sites may have significantly reduced the effective trapping capacity for radon compared to results observed in SF$_6$-based systems.

To explore whether a reduced quantity of molecular sieves might yield better results by diminishing the self-emanation contribution while maintaining good trapping capacity, a second test was conducted with only 20~g (29~ml) of material (one tenth of the original amount). In this second attempt, HE event rates stabilised at 300-340 c/h, representing a modest improvement over the 380-420 c/h observed with the 200~g configuration. This outcome suggested that self-emanation does indeed scale with mass, but the relationship may not be strictly linear because of self-absorption effects within the sieve matrix. In any case, the result was a net increase in background.

\subsubsection{Activated Carbon}
\label{Chapter6_Activated_Carbon}

Activated carbon (or charcoal) has been extensively studied as an effective radon adsorbent in the context of rare event experiments~\cite{Pushkin_2018} due to its high surface area. Unlike molecular sieves with their uniform pore structure, activated carbon possesses a wider distribution of pore sizes. This material undergoes an activation process that significantly enhances its porosity, thereby dramatically increasing the available surface area for adsorption. It is well-established that maintaining $^{222}$Rn traps at reduced temperatures (typically at dry ice temperature) substantially improves the trapping efficiency~\cite{Hodak_2019}. As with molecular sieves, self-emanation of $^{222}$Rn from trace $^{238}$U content is a significant concern.

In TREX-DM, an investigation of activated carbon materials was conducted through a series of tests initially performed in laboratory conditions in Zaragoza and subsequently \textit{in situ} at the LSC. A primary concern for our application was the potential absorption of iC$_4$H$_{10}$, as its molecular diameter is smaller than the pore size of activated carbon, which would alter the gain and performance of the detector.

\begin{figure}[htbp]
\centering
\includegraphics[width=1.0\textwidth]{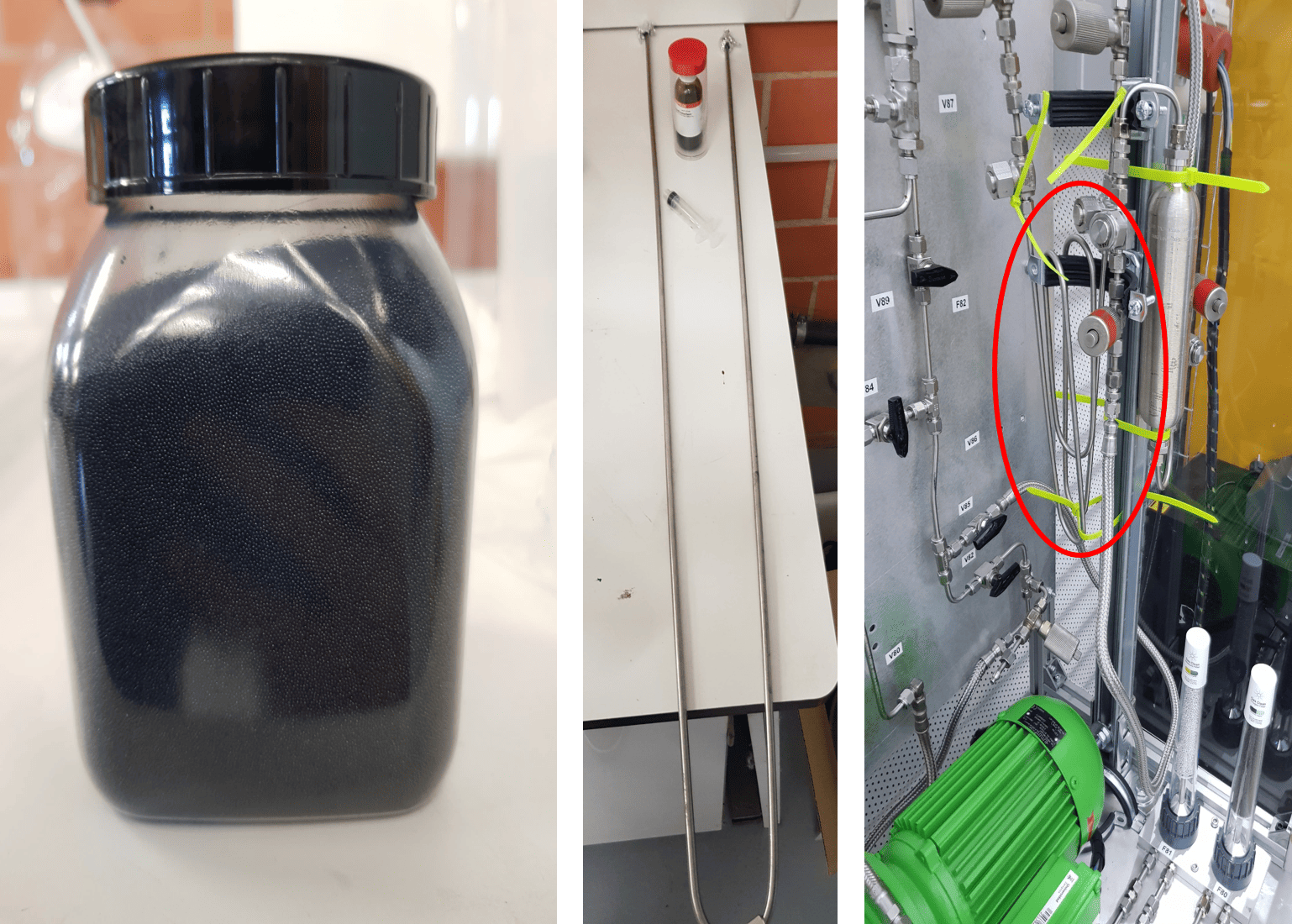}
\caption{Left: close-up view of activated charcoal. Centre: 1-m column used to contain the activated charcoal and small 10-g bottle of the product. Right: activated charcoal column installed in the gas panel of TREX-DM, right after the custom-made filter. The tube was bent several times to make it easier to handle.}
\label{fig:chapter6_activated_charcoal_tests}
\end{figure}

The initial tests began in November 2021 with a 250-g sample from Saratech, shown in the left image of Figure~\ref{fig:chapter6_activated_charcoal_tests}. It was placed in a 1-litre vessel (the same bottle previously used for molecular sieve tests). Open-loop measurements using a Binary Gas Analyser (BGA) revealed that this quantity of activated carbon completely absorbed the iC$_4$H$_{10}$ component from an Ar-1\%iC$_4$H$_{10}$ test mixture at room temperature.

Further tests were conducted with 10 g of Carboxen 572 from Sigma-Aldrich. This commercial activated carbon has pore sizes in the 10-12~\r{A} range and particle sizes between 0.2 and 0.8~mm. The idea was to see if a reduced quantity of activated carbon might exhibit less iC$_4$H$_{10}$ absorption. Here, rather than employing a vessel configuration, we adopted a column approach as it is usual in charcoal filters~\cite{Hodak_2019}. We used a stainless steel tube with a 4 mm internal diameter, formed into a U-shape with a total length of 1.7 m, as illustrated in Figure~\ref{fig:chapter6_activated_charcoal_tests}. This geometry was specifically chosen to maximise the path length through which the gas would travel, thereby increasing the time the gas spends within the filter and enhancing the probability of radon capture through adsorption/desorption processes.

Despite the significant reduction in activated carbon mass, open-loop tests with the Carboxen 572 sample yielded similar results to those observed with the larger Saratech sample: complete absorption of iC$_4$H$_{10}$ from the Ar-1\%iC$_4$H$_{10}$ mixture at room temperature. More tests with an Ar-2\%iC$_4$H$_{10}$ mixture confirmed that the carbon filter continued to absorb all available iC$_4$H$_{10}$, even at this higher quencher concentration.

To characterise the dynamic behaviour of activated carbon under more realistic operational conditions, closed-loop tests were performed. In these experiments, the system underwent multiple evacuation and refilling cycles while monitoring the gas composition with the BGA. The objective was to approach a saturation condition where the activated carbon reaches a stable iC$_4$H$_{10}$ concentration. This testing protocol involved a sequence of isolating the filter, pumping down the system, introducing fresh gas, and then re-exposing the filter to the fresh gas again. These tests yielded promising results, demonstrating that after several evacuation-filling iterations, the iC$_4$H$_{10}$ concentration in the test system approached the nominal value of the pre-mixed Ar-2\%iC$_4$H$_{10}$ bottle. Based on these findings, it was decided to test the charcoal filter at room temperature at LSC.

In January 2022, the activated carbon filter was installed in the TREX-DM gas system, positioned after the (engaged) custom-made O$_2$ + H$_2$O filter, see Figure~\ref{fig:chapter6_activated_charcoal_tests}. After approximately 14 hours of gas recirculation at maximum flow rate ($\sim$ 60 ln/h), gas composition measurements were performed. As anticipated from the laboratory saturation tests, the activated carbon trap did not completely remove the iC$_4$H$_{10}$ component. Instead, only a minor fraction (approximately 0.15\% of the total 2\%) was absorbed, leaving a 1.85\%iC$_4$H$_{10}$ concentration in the Ne-2\%iC$_4$H$_{10}$ mixture, as measured with a calibrated BGA. This value remained stable throughout the testing period, confirming that equilibrium conditions had been achieved with respect to iC$_4$H$_{10}$ absorption.

HE rate measurements conducted over several hours demonstrated that the rate stabilised in the range of 90-110~c/h (see Table~\ref{table:chapter6_radon_studies}). This represented approximately a 50\% reduction compared to the best previous configuration using only the custom-made filter. While this improvement was significant, the achieved radon levels remained several times higher than the constant background component identified in earlier studies (see Figure~\ref{fig:chapter6_2021_closed_vessel}), indicating that this approach should be refined, or other avenues to mitigate radon contamination should be pursued.

It should be noted that while lowering the temperature of the activated carbon would potentially enhance its radon adsorption efficiency, this was impractical in TREX-DM system due to constraints imposed by the quencher in our gas mixture. Specifically, the boiling point of iC$_4$H$_{10}$ (-11.7 $^\circ$C) would lead to condensation at the reduced temperatures typically reached for optimal radon trapping. This represents a fundamental limitation compared to experiments that use alternative quenchers with lower boiling points, such as methane (CH$_4$, boiling point -161.5 $^\circ$C), which can operate with radon traps at significantly lower temperatures.


\subsection{Semi-Sealed Open-Loop Approach}
\label{Chapter6_Semi-Sealed_Open-Loop}

After exploring active removal and material selection strategies for $^{222}$Rn mitigation, neither approach yielded sufficient background reduction in TREX-DM. The most successful approach proved to be operating the detector in a semi-sealed open-loop mode. 

The semi-sealed open-loop operation avoids the use of filters and radon traps altogether. Instead, the system is configured to allow only a minimal continuous flow of gas from the bottle through the detector and out to the exhaust. Implementation of this strategy requires excellent leak-tightness and low outgassing rates to make a minimal flow rate viable that results in an acceptable gas loss rate for the experiment. For TREX-DM, flow values below 1~l/h were sufficient, resulting in a complete gas renovation every two or three weeks, resembling a sealed-mode scenario but with continuous gas flow. In practice, this approach involves isolating the recirculation components that significantly contribute to radon emanation (the O$_2$ and H$_2$O filters and the recirculation pump). Long-term operation in this mode has demonstrated that the system maintains stable gas composition and quality as verified by BGA measurements, with detector gain and resolution remaining consistent.

\begin{figure}[htb]
\centering
\includegraphics[width=0.95\textwidth]{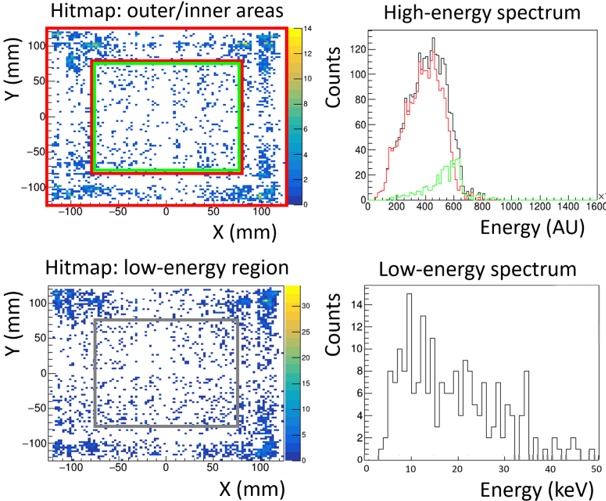}
\caption{Semi-sealed open-loop runs. Top: Low-gain run (210~V, 4.5 days, 0.9~l/h flow rate) showing the inner (green) and outer (red) detector areas in the hitmap, with corresponding high-energy spectrum (right) compatible with monoenergetic alphas from surface contamination ($^{210}$Pb chain). Bottom: Nominal-gain run (365~V, 4.5 days, 0.9~l/h flow rate). The background spectrum in the $(0,50)$~keV region (right) corresponds to the fiducial area of the hitmap in grey (left). The background level in this $15\times 15$~cm$^2$ area is $\approx 80$~dru. In both runs, the surface contamination near the borders due to the PTFE piece is visible.}
\label{fig:chapter6_2022_semi-sealed_open-loop}
\end{figure}

The implementation of this strategy yielded the most significant reduction in background levels: both HE and LE rates dropped to levels consistent with the constant component obtained from sealed-mode runs from May 2021 (compare Figure~\ref{fig:chapter6_2021_closed_vessel} with Table~\ref{table:chapter6_radon_studies}). This translated into low-energy background in the 0-50 keV range decreasing from approximately 600 dru to around 80 dru in the inner region of the detector, as shown in Figure~\ref{fig:chapter6_2022_semi-sealed_open-loop}. This inner region selection removes the surface contamination component from the borders, which now dominates the hitmap.

More evidence of the absence of $^{222}$Rn in the system came from the HE spectrum in Figure~\ref{fig:chapter6_2022_semi-sealed_open-loop}, where the 7.7 MeV peak characteristic of $^{222}$Rn alpha decays disappeared, leaving a spectrum compatible with monoenergetic alphas from the $^{210}$Pb chain. This spectrum confirms that $^{222}$Rn levels in the system are now below the statistical fluctuations of the constant contribution component associated with alphas from surface contamination. The width of the spectrum is likely due to incomplete alpha events, either because of the high number of dead strips in TREXDM.v1 North detector, or because the events are not fully contained in the sensitive volume. This is especially relevant for alphas originating from the borders, as they could deposit part of their energy in the PTFE piece surrounding the field cage.

The success of this approach demonstrates that, for TREX-DM, minimising the introduction of $^{222}$Rn into the system was finally more effective than attempting to trap radon once it was already present. The semi-sealed open-loop approach may be particularly suitable for other experiments with similar gas systems and background requirements where conventional radon mitigation techniques prove difficult.

\section{Surface Contamination} \label{Chapter6_Surface_Contamination}

Once active radon was removed from the system, radon-induced surface contamination was the main background component. This section begins by examining the nature of this issue, followed by a discussion of targeted interventions implemented to address it. We also present a refined analysis method developed to identify the origin of this contamination. Additionally, the section describes the development of the AlphaCAMM detector, a diagnostic tool conceived within the TREX-DM project to measure and characterise surface alpha contamination. Finally, we briefly explain how recent simulation studies have validated our experimental observations, particularly the direct proportional relationship between high-energy alpha events and low-energy background rates.

\subsection{Overview of the Problem}
\label{Chapter6_Overview}

Beyond active $^{222}$Rn, surface contamination from long-lived $^{222}$Rn progeny can be a potential background source for rare event physics experiments. When materials are exposed to environments containing $^{222}$Rn, its progeny (particularly $^{210}$Pb with its 22.3-year half-life) can accumulate on surfaces through a process known as plate-out. This surface contamination becomes a persistent source of background events that cannot be eliminated by simply removing active $^{222}$Rn from the system. This is especially critical on the inner surfaces facing the detector volume, where it can produce low-energy background due to the emission of X-rays and low-energy electrons.

Several studies have quantified and characterised this phenomenon across different materials. At SNOLAB, average $^{210}$Pb plate-out rates of 249 and 423 atoms/day/cm$^2$ were determined for polyethylene and copper, respectively, after exposing samples for three months to radon activity of 135 Bq/m$^3$~\cite{Stein_2018}. Material properties have been shown to influence the plate-out rate. For example, the LUX-ZEPLIN experiment found that the rate of radon daughter plate-out onto Teflon (PTFE) can be orders of magnitude higher than onto other materials~\cite{Morrison_2018}, making PTFE components (a widely used material in low-background searches) particularly susceptible to surface contamination in radon-rich environments.

There are two primary approaches to tackle surface contamination:

\begin{itemize}
    \item \textbf{Active removal}: this refers to the elimination of contamination from the surfaces by different means. For example, the removal of Rn-generated $^{210}$Po surface contamination from copper using methods such as chemical etching or electropolishing has been explored~\cite{Guiseppe_2017}. Other studies have examined different cleaning procedures to remove radon daughters from PTFE surfaces, concluding that nitric acid baths are the most effective~\cite{Bruenner_2021}.
    \item \textbf{Careful material selection and protection}: when contamination directly affects elements where active removal is not feasible (e.g., the surface of the detectors themselves, or very thin ($\sim \upmu$m) surfaces such as the mylar cathode used in TREX-DM), one must resort to using materials with low intrinsic $^{210}$Pb content and isolating all materials as carefully as possible.
\end{itemize}

The impact of surface contamination varies depending on detector technology. The high density and full 3D fiducialisation capabilities of noble liquid/dual-phase detectors enable self-shielding to mitigate surface contamination effects, an advantage not available in lighter gaseous media. Consequently, lighter gas mixtures (Ar- or, especially, Ne-based mixtures), such as those used in TREX-DM, are more affected by the low-energy emissions from the decay chain.

\subsection{First Reduction Attempt in TREX-DM}
\label{Chapter6_Reduction_Attempt}

In TREX-DM, after reducing the background contribution from active $^{222}$Rn through semi-sealed open-loop operation, the remaining background of approximately 80-100 dru was attributed to surface contamination from $^{222}$Rn progeny on inner components of the detector. This contamination manifested as alpha events in the high-energy spectrum and associated low-energy events that contributed to the background in the region of interest.

To test this hypothesis, an intervention was performed in February 2021 to machine a few millimetres off the walls of the PTFE piece inside the field cage. Due to mechanical constraints, only approximately 50-60\% of the piece was machined; the corners were left untouched because they needed to be bent, and machining would have made them too fragile. PTFE was specifically targeted given evidence from other experiments of its high sensitivity to radon daughter plate-out, as well as evidence from TREX-DM hitmaps obtained after radon had decayed.

\begin{figure}[htb]
\centering
\includegraphics[width=1.0\textwidth]{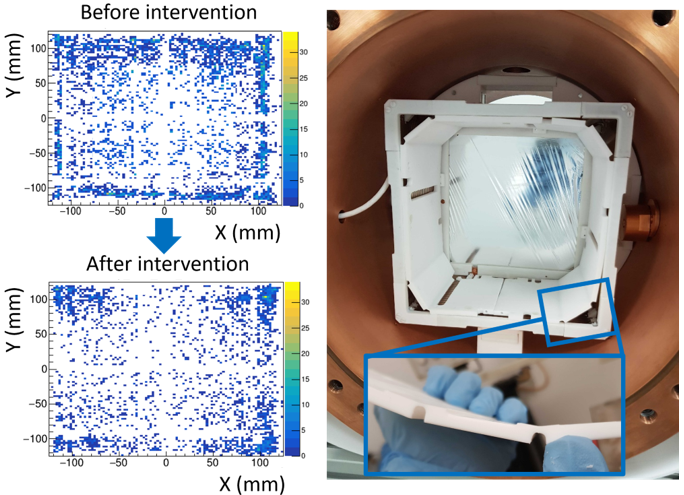}
\caption{Left: hitmaps in the low-energy region before (top) and after (bottom) the intervention. Run duration is roughly the same in both, so number of events is similar. A clear reduction is visible except near the unmachined parts. The corresponding hitmaps for the high-energy region display the same reduction pattern. Right: PTFE piece inside the field cage with a detailed view of the machined surface.}
\label{fig:chapter6_2021_surface_contamination_reduction}
\end{figure}

This intervention had a significant impact on both high-energy and low-energy rates near the borders of the detector, with both experiencing a reduction of approximately 50\%. In particular, there was a reduction in the number of events near the areas where the machining had taken place, in both energy regions. The remaining events were clustered around the unmachined corners of the piece, see Figure~\ref{fig:chapter6_2021_surface_contamination_reduction}. This confirmed that a substantial portion of the remaining background was indeed coming from surface contamination on inner components, particularly the PTFE parts in contact with the active volume.

Most importantly, this indicated a $1:1$ relation between alphas and low-energy events, as the reduction was roughly 50\% in both cases. This relation is further confirmed by examining the rates after implementing the semi-sealed open-loop approach: the HE rate was around 20 c/h and LE rate around 30 c/h in the whole detector area. Assuming the background model was correct except for the missing contribution from surface contamination, the HE region would be solely from $^{210}$Po alphas, while the LE region would include $^{210}$Pb chain emissions plus 1-10 dru from other sources in the background model predictions. This puts the relation in the range $1:1$ to $1.5:1$ for $\mathrm{LE}:\mathrm{HE}$, with the actual value likely in the middle. This proportionality was further verified by simulation studies, as explained in Section~\ref{Chapter6_Simulation_Studies}.

While the PTFE machining intervention shed light on the issue of surface contamination, it did not fully solve the background problem. As expected, background levels in the fiducial inner $15\times 15$ cm$^2$ region of the detector remained at around 80-100 dru, and there was still an alpha spectrum present in that inner region of the detector (see Figure~\ref{fig:chapter6_2022_semi-sealed_open-loop}). This indicated that surface contamination was present on multiple inner components beyond just the PTFE piece, potentially including the cathode and the Micromegas readout planes, as these were the only components that could affect the fiducial area.

A follow-up intervention in June 2022 involved replacing the cathode with another aluminised mylar and installing new, cleaner detectors. The new cathode had been stored and handled with care to prevent exposure to high-radon environments that the previous cathode (installed in October 2020) had experienced before the semi-sealed open-loop approach was implemented. The new detectors had been stored in an oven with controlled humidity, with their surfaces covered with plastic to prevent radon deposition. Additionally, the Micro-Pattern Technologies (MPT) workshop at CERN had been instructed to be careful with detector exposure to air.

Initial results indicated that background levels remained similar in the inner region of the detector. This was hypothesised to be due to the TREXDM.v1 North detector's dead strip problem. The installation of TREXDM.v2 detectors increased the active area of the central part by approximately 25\%, suggesting that previously some events may have been lost, and the actual background might have been higher. To properly analyse these changes and assess whether the cathode or the detectors contributed most to the background, an alpha directionality algorithm was implemented, as explained in the following section.

\subsection{Alpha Directionality Studies} \label{Chapter6_Alpha_Directionality_Studies}

To further analyse surface contamination, an alpha directionality algorithm to determine if alphas come from the cathode or the surface of the detectors was developed and implemented, which became feasible only after the installation of TREXDM.v2 detectors in June 2022. Previous attempts with TREXDM.v1 were limited by numerous dead strips in the readout plane, which significantly compromised track reconstruction capabilities. These analyses were primarily conducted after the experiment's relocation to Lab2500, coinciding with a transition from Ne-2\%iC$_{4}$H$_{10}$ to Ar-1\%iC$_4$H$_{10}$ as the gas mixture, accompanied by a reduction in pressure from 4 bar to 1 bar.

\begin{figure}[htb]
\centering
\includegraphics[width=1.0\textwidth]{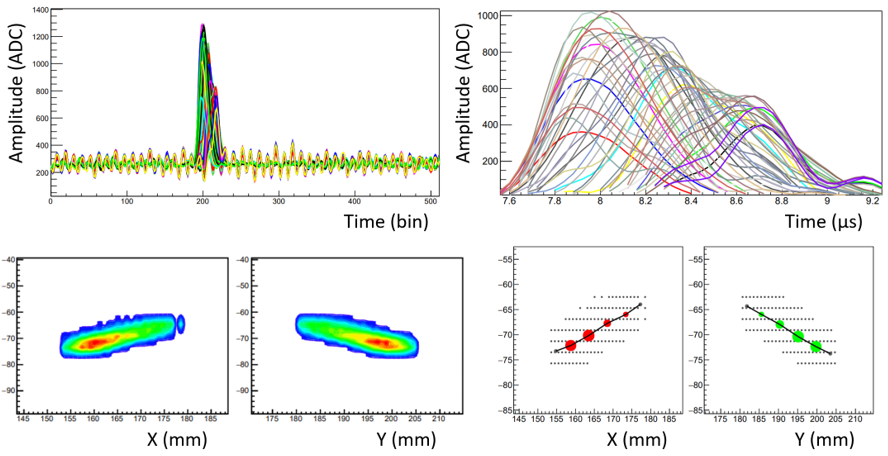}
\caption{Event originating on the cathode recorded with the North detector (angle is 1.28 rad) in a $15\times 15$~cm$^2$ fiducial region. The different plots show the whole analysis chain starting with the \textit{RawSignal} (top left, time in bins), following with the \textit{DetectorSignal} (top right, time in $\upmu$s), then the \textit{DetectorHits} (bottom left), and finally the track reconstruction (bottom right).}
\label{fig:chapter6_alpha_event_cathode_North}
\end{figure}

The alpha particle directionality algorithm is integrated within the REST-for-Physics framework (described in Section~\ref{Chapter5_Description_REST}). This algorithm was originally developed for the AlphaCAMM detector (see Section~\ref{Chapter6_AlphaCAMM}), where it was extensively tested before being used in TREX-DM. It works by clustering detector hits based on spatial proximity, then applying track reduction to merge nearby hits, followed by path minimisation to arrange hits in their optimal spatial sequence. This allows for the determination of the particle's trajectory, extracting key parameters such as track origin, endpoint, length, and orientation angle. The algorithm exploits the characteristic energy deposition pattern of alpha particles (specifically the Bragg peak discussed in Section~\ref{Chapter3_Interactions_Charged_Particles}) where maximum energy deposition occurs near the end of the track. This approach was confirmed by comparing reconstructed track lengths with theoretical predictions from NIST ASTAR~\cite{ESTAR_PSTAR_ASTAR}: for 5.3 MeV alphas in Ar-1\%iC$_4$H$_{10}$ at 1 bar, expected track lengths are approximately 4 cm, and reconstructed tracks agree with this value. Figure~\ref{fig:chapter6_alpha_event_cathode_North} shows the reconstruction of an alpha event coming from the cathode.

\begin{figure}[htb]
\centering
\includegraphics[width=1.0\textwidth]{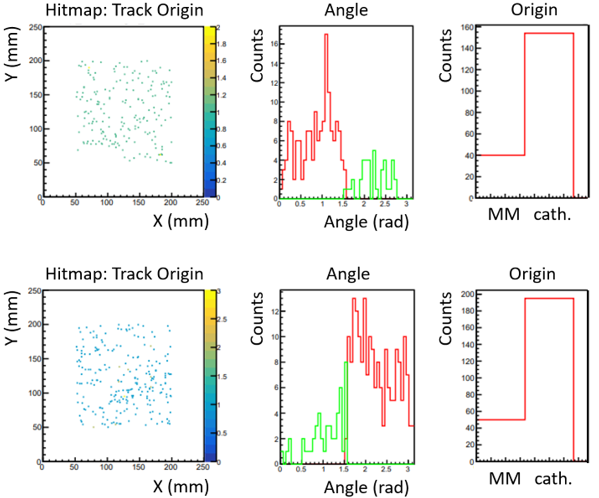}
\caption{Alpha directionality study for an 11.5-day low-gain run ($V_\mathrm{mesh}=230$~V), showing the proportion of alphas originating from the cathode is 80\% both for the North detector (top) and for the South detector (bottom). The hitmap represents the origin of each alpha event, as determined by the track reconstruction algorithm.}
\label{fig:chapter6_2023_alphas_directionality}
\end{figure}

Application of this directionality analysis in low-gain runs provided insights into the distribution of alpha-emitting contaminants within the detector. The analysis revealed that approximately 70-80\% of detected alpha particles originated from the cathode rather than from the Micromegas readout plane. This determination was made by analysing the track angles: angles near $\pi/2$ indicate horizontal tracks with undefined direction, while angles below $\pi/2$ (for North detector) or above $\pi/2$ (for South detector) mean tracks originating on the cathode. The hitmap with the origin of the tracks and the angle distribution of tracks for both detectors is shown in Figure~\ref{fig:chapter6_2023_alphas_directionality}, together with the direction histogram of those tracks, extracted from that angle distribution. These findings indicated that while careful handling protocols had successfully limited contamination on detector surfaces, the mylar cathode remained a significant source of background.

\begin{figure}[htb]
\centering
\includegraphics[width=0.83\textwidth]{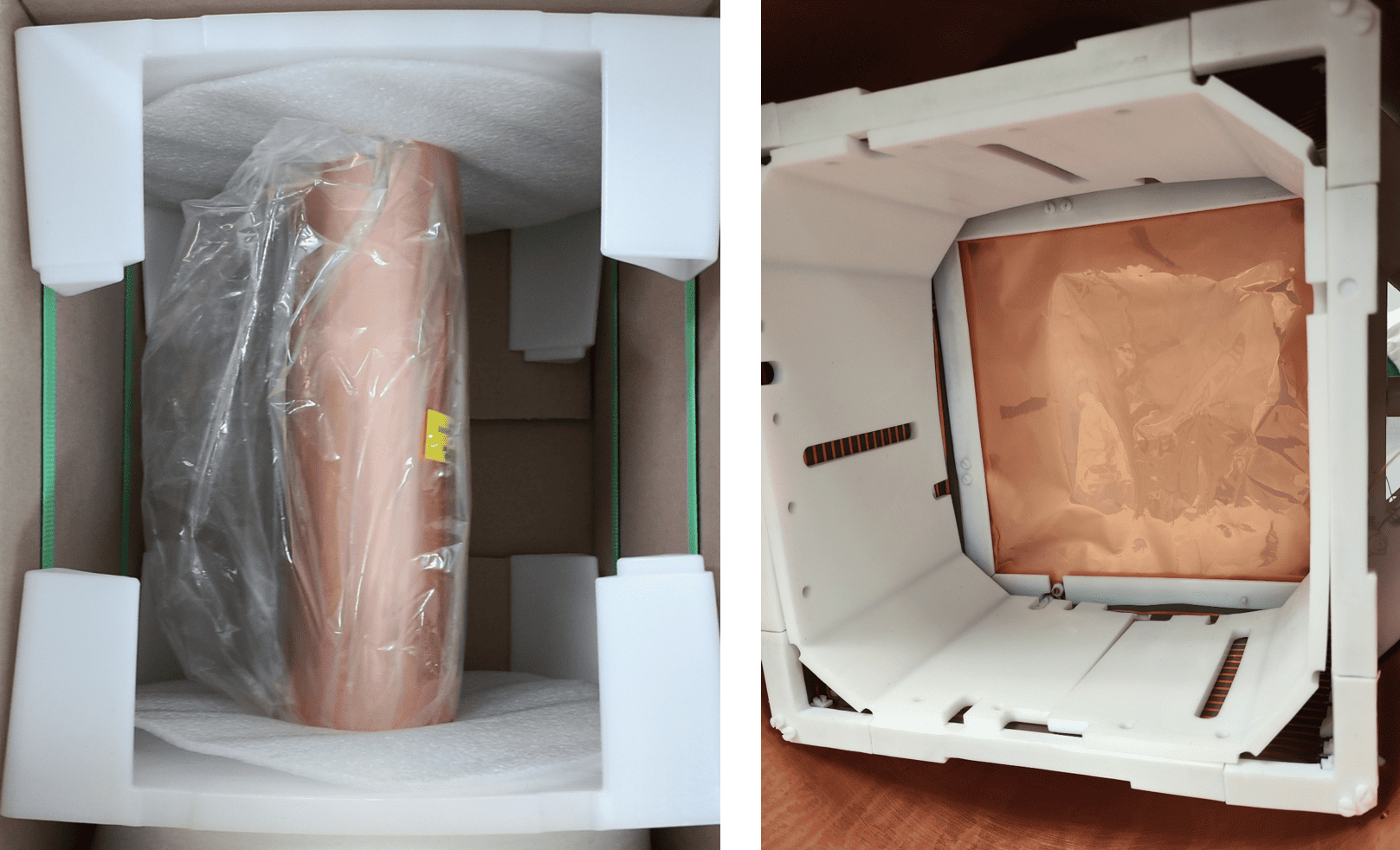}
\caption{Left: copper-kapton-copper roll stored. Right: new cathode installed.}
\label{fig:chapter6_2024_copper_cathode}
\end{figure}

Based on the results of this analysis, a targeted mitigation strategy was implemented during the March 2024 intervention. Given that the detector chamber needed to be opened for the installation of the GEM stage (described in Section~\ref{Chapter7_Installation}), it was decided to take advantage of this opportunity by implementing several changes simultaneously. Among these was the replacement of the original mylar cathode with a thin ($O(50)$~$\upmu$m) copper-kapton-copper foil structure, known for its radiopurity (same base materials as Microbulk detectors). The installation of this new cathode is shown in Figure~\ref{fig:chapter6_2024_copper_cathode}. However, runs after this intervention indicated that this modification did not achieve the expected reduction in alpha backgrounds, suggesting either intrinsic surface contamination of the cathode or, more likely, that the installation process is critical and requires even more careful protocols: despite storing the copper-kapton-copper roll sealed and untouched (see Figure~\ref{fig:chapter6_2024_copper_cathode}), additional precautions may be necessary, including minimising exposure to air during interventions, avoiding direct contact with the surface, and preventing wiping of nearby surfaces (especially PTFE) that might create electrostatic potential attracting radon progeny deposits.

At present, surface contamination remains the limiting background component in TREX-DM, with current background level in the inner area of the detector at around 100 dru. Ongoing efforts are focused on its reduction, with one promising approach under consideration involving redesigning the cathode with the goal of minimising the total material surface susceptible to surface contamination. One promising idea is the use of a cathode made of copper wires. The effectiveness of these mitigation strategies will directly impact the sensitivity of future data-taking campaigns, as discussed in Section~\ref{Chapter9_Sensitivity_Projections}.

\subsection{AlphaCAMM} \label{Chapter6_AlphaCAMM}

While standard screening techniques like HPGe spectroscopy, GDMS and ICPMS provide comprehensive material radioassays, they exhibit limited sensitivity to $^{210}$Pb surface contamination. To address this measurement gap, AlphaCAMM~\cite{AlphaCAMM_2022} (Alpha CAMera Micromegas) was conceived as a dedicated detector for direct screening of $^{210}$Pb surface contamination on flat samples.

\begin{figure}[htb]
\centering
\includegraphics[width=1.0\textwidth]{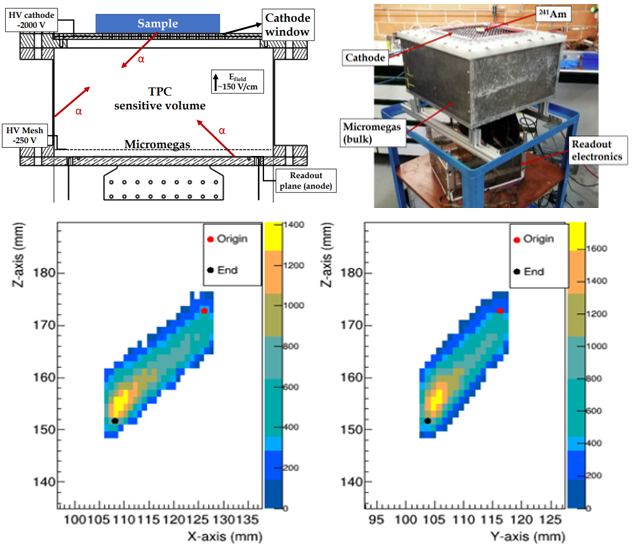}
\caption{Top: conceptual design of AlphaCAMM (left) together with the first prototype used to test the concept (right). Bottom: track reconstruction of a $^{241}$Am alpha event. Images taken from~\cite{AlphaCAMM_2022}.}
\label{fig:chapter6_alphacamm_prototype}
\end{figure}

AlphaCAMM is a Micromegas-read gaseous TPC developed to reconstruct alpha tracks originating from $^{210}$Po decay (the progeny of $^{210}$Pb, Figure~\ref{fig:chapter6_radon_decay_chain}). As illustrated in the schematic design of Figure~\ref{fig:chapter6_alphacamm_prototype}, the sample under test is placed on top of a thin cathode window designed to minimise energy loss of alpha particles. The alpha particles ionise a suitable gas mixture (like Ar-1\%iC$_4$H$_{10}$), and the resulting ionisation is then amplified and read on a Micromegas plane. 

Background modelling studies for TREX-DM indicate that $^{210}$Pb surface contamination below 100 nBq/cm$^2$ contributes negligibly to the overall experiment background, $O(1)$ dru, thus establishing AlphaCAMM's sensitivity target. Therefore, the detector aims for minimum detectable $^{210}$Pb activities of approximately 100 nBq/cm$^2$, with sensitivity upper limits of about 60 nBq/cm$^2$ at 95\% confidence level for a standard one-week measurement. Achieving these targets requires an intrinsic background level below $5\times10^{-8}$ alphas/cm$^2$/s. For a comprehensive view of the background model of AlphaCAMM, see~\cite{tesis_hector_2024}.

AlphaCAMM's key innovation lies in its signal discrimination capabilities. Using the algorithm described in Section~\ref{Chapter6_Alpha_Directionality_Studies}, AlphaCAMM provides precise topological information for each alpha track, allowing determination of both track origin and direction and, thus, discrimination between alpha particles coming from the sample and those from intrinsic detector contamination. 
The detector's initial prototype demonstrated the viability of the concept by successfully reconstructing alpha tracks from a $^{241}$Am calibration source placed on the cathode, as can be seen in Figure~\ref{fig:chapter6_alphacamm_prototype}.

\begin{figure}[htb]
\centering
\includegraphics[width=0.9\textwidth]{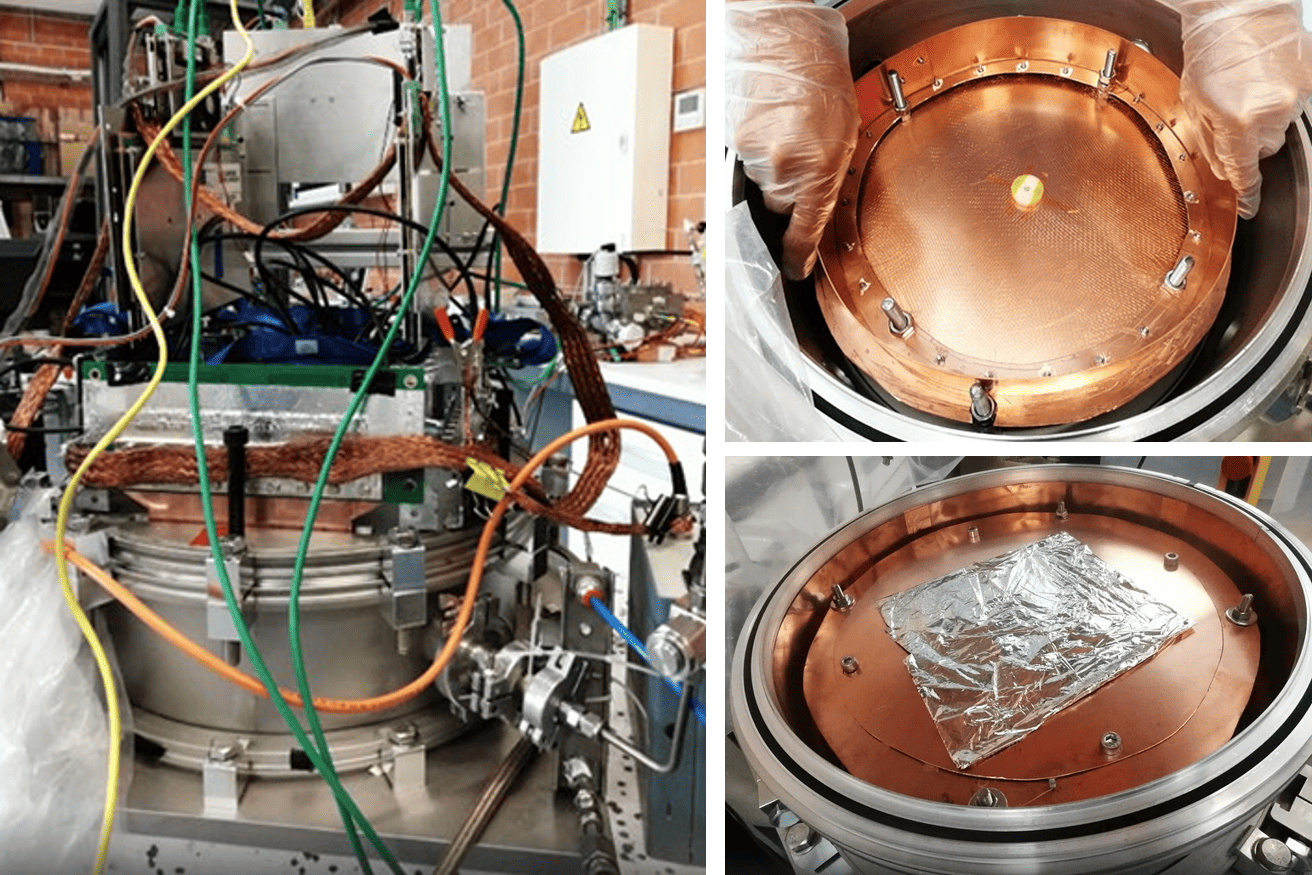}
\caption{Radiopure version of AlphaCAMM. Left: set-up with the closed stainless steel chamber, readout electronics and noise reduction arrangement. Right: chamber open with $^{241}$Am source on top of the cathode (top) and mylar cathode sample (bottom). Note that the drift length differs between pictures.}
\label{fig:chapter6_alphacamm_radiopure}
\end{figure}

Following the successful proof-of-concept, a radiopure version of AlphaCAMM was designed and constructed in 2022. This new iteration used radiopure materials including copper, stainless steel and PTFE, with a $25.6\times 25.6$~cm$^2$ microbulk Micromegas readout from the TREXDM.v2 batch. In this version, there is a slight design change with respect to the original concept: the sample under test is now positioned inside the sensitive volume instead of outside, sitting directly on top of a copper cathode at the bottom of a stainless steel chamber (see Figure~\ref{fig:chapter6_alphacamm_radiopure}). The Micromegas detector is installed on the chamber's endcap, facing downwards towards the sample.

Commissioning measurements conducted in Zaragoza demonstrated excellent spatial and energy resolutions using a $^{241}$Am source. However, background characterisation revealed intrinsic alpha levels of approximately $6\times10^{-6}$ alphas/cm$^2$/s, roughly two orders of magnitude above the target background level required to achieve the desired sensitivity of 100 nBq/cm$^2$. Investigations pointed to $^{220}$Rn and $^{222}$Rn emanation from the chamber walls as the likely source, despite the stainless steel having previously shown acceptable radiopurity in GDMS screening. Various mitigation strategies were implemented, like covering the chamber walls with copper and operating in sealed mode. These actions, combined with an improved offline analysis, reduced the background by a factor of 10. Though still above target sensitivity (current level approximately 1 $\upmu$Bq/cm$^2$ versus the goal of 100 nBq/cm$^2$), AlphaCAMM has begun measuring samples of interest.

The first measurement campaign focused on the aluminised mylar cathode used in TREX-DM for two years (June 2022 to May 2024), as shown in Figure~\ref{fig:chapter6_alphacamm_radiopure}, where this mylar serves as a sample on top of the AlphaCAMM cathode. As previously discussed, this particular component was identified as a potential source of alpha background inside TREX-DM. Initial results show contamination levels compatible with AlphaCAMM's current background, indicating that further background reduction is necessary. This is expected, as the mylar cathode contributed $O(100)$ dru in TREX-DM, which approximately matches the current sensitivity of 1 $\upmu$Bq/cm$^2$.

Planned measurements include the first GEM installed and subsequently replaced in TREX-DM, and samples from the copper-kapton-copper sheet batch currently used as the cathode in TREX-DM.

Despite not yet reaching its sensitivity goals, AlphaCAMM represents a significant milestone in efforts to measure and control surface contamination in rare-event experiments like TREX-DM, providing complementary data to traditional radiopurity assessments.

\subsection{Simulation Studies} \label{Chapter6_Simulation_Studies}

To investigate the contribution of surface contamination to the low-energy background spectrum in TREX-DM, dedicated simulation studies were performed in the context of an End of Degree Dissertation. This section is a brief summary of this work, so we refer to~\cite{tfg_alfas} for the original source. 

\begin{figure}[htb]
\centering
\includegraphics[width=0.7\textwidth]{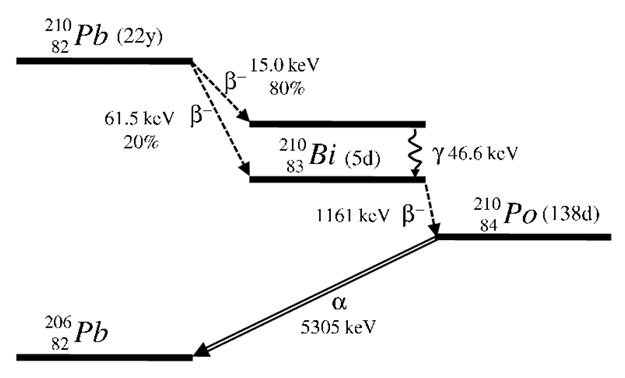}
\caption{Decay scheme of $^{210}$Pb. Extracted from~\cite{tfg_alfas}.}
\label{fig:chapter6_210Pb_decay}
\end{figure}

The simulations were designed to evaluate the relation between high-energy alpha events (5.3 MeV) from $^{210}$Po and low-energy events that could impact the detection sensitivity for low-mass WIMPs. They were carried out with the REST-for-Physics framework together with Geant4, using the full TREX-DM geometry. The simulations placed $^{210}$Pb atoms uniformly distributed on the cathode surface, where they would naturally accumulate due to $^{218}$Po and $^{214}$Pb being positively charged. The full decay chain from $^{210}$Pb to $^{206}$Pb was simulated, including all associated decay products and their energies. The $^{210}$Pb decay scheme (Figure~\ref{fig:chapter6_210Pb_decay}) is particularly relevant for low-energy background studies as it produces several low-energy components: a 15 keV beta decay (occurring 80\% of the time), a 61.5 keV beta decay (20\% of the time), and electronic de-excitation processes that yield X-rays and conversion electrons.

Initial simulations produced approximately 2.5 registered events per $^{210}$Pb decay, with a total of 625k events recorded from 250k $^{210}$Pb decays. Energy depositions were tracked both in terms of total energy and energy specifically deposited in the sensitive volume (the gas). The energy spectrum clearly showed the 5.3 MeV alpha peak from $^{210}$Po decay, as well as the continuous beta spectrum from $^{210}$Bi decay with a maximum energy of 1161 keV. At lower energies, the spectrum revealed structures around 46 keV from the electronic de-excitation of $^{210}$Bi* to its ground state.

Critically, the simulations indicated that the low-energy region (below 25 keV) recorded nearly 193k events, whereas the high-energy part registered $\sim$ 163k 5.3-MeV alpha events. This yields a ratio of 1.18 low-energy events per alpha event. This ratio aligns with experimental observations from TREX-DM, which suggested a relationship of approximately $1:1$ between low-energy events and alpha particles.

To reproduce experimental conditions more accurately, the simulated data underwent processing to account for detector resolution, charge diffusion, and signal conversion effects.
After processing, approximately 30\% of the data was lost due to quality cuts, but the 1.18 ratio remained consistent, which strengthened the results.

The simulation also provided insights into the origin of low-energy events. More than half of them originated directly from $^{210}$Pb decays, while a significant portion came from the $^{210}$Bi beta decay (1161 keV maximum energy). This contribution from $^{210}$Bi was attributed to edge effects, where electrons deposit only a small portion of their energy in the sensitive volume.

To summarise, these simulation studies provide strong evidence that radon progeny, particularly $^{210}$Pb surface contamination, contributes significantly to the low-energy background spectrum in TREX-DM. They also reinforce the approximate $1:1$ relationship between high-energy alphas and low-energy events discovered experimentally.

\section{Summary} \label{Chapter6_Summary}

TREX-DM has made substantial progress in understanding and mitigating its background problem. Through dedicated studies, two primary background sources were identified: active $^{222}$Rn in the detector volume and surface contamination from $^{210}$Pb. The implementation of a semi-sealed open-loop operation successfully reduced the contribution from active radon, bringing background levels from $\sim$ 1000 dru down to approximately 80-100 dru in the inner $15\times 15$ cm$^2$ fiducial area. Surface contamination, particularly on the cathode (responsible for 70-80\% of alpha emissions as confirmed by directionality studies), now constitutes the dominant background source. There is an approximate $1:1$ correspondence between alpha events and low-energy background, determined experimentally and later validated through simulation studies.

AlphaCAMM was developed to measure $^{210}$Pb surface contamination with a target sensitivity of 100 nBq/cm$^2$. Although it shows excellent track reconstruction capabilities, its current background level of approximately 1 $\upmu$Bq/cm$^2$ remains one order of magnitude above the target. Initial measurements of the TREX-DM mylar cathode yielded contamination levels comparable to AlphaCAMM's background, highlighting the need for further sensitivity improvements. Currently, the background comprises a combination of volumetric $^{222}$Rn and surface $^{210}$Pb from the cathode, though determining the dominant component remains challenging as volumetric contamination produces some alpha particles with trajectories towards the detector that are indistinguishable from those coming from the cathode. To address this limitation, a copper-wire cathode design is being developed with two goals: to minimise material surface area susceptible to contamination and to enable discrimination between volumetric and surface contamination. With this configuration, only events with track origins in the vicinity of the wires would indicate surface contamination, whereas track origins in the middle of the grid would be identified as volumetric contamination. This refinement in the offline analysis presents a promising avenue for further background reduction.

Future efforts for background reduction in TREX-DM will focus on implementing stricter protocols for detector component handling and installation, alongside redesigning the cathode using the same idea of a wire-based model to minimise material surface area. AlphaCAMM will serve as a testbench for this concept, allowing for feasibility assessment and optimisation before implementation in TREX-DM. These developments are critical to achieving the sub-10 dru background levels necessary for TREX-DM to reach its projected sensitivity to low-mass WIMPs.

%% file: CHAPTERS/Chapter7.tex
\chapter{GEM Preamplification Stage for Energy Threshold Reduction} \label{Chapter7_GEM-MM}

\lettrine[loversize=0.15]{A}{s} explained in Section~\ref{Chapter2_Direct_Searches_Sensitivity}, the energy threshold is one of the main factors to optimise in the search for low-mass WIMPs. The impact this parameter has on the sensitivity of TREX-DM in parameter space $(m_\chi ,\sigma_{n})$ will be discussed in Chapter~\ref{Chapter9_Sensitivity}, but for now suffice it to say that the lower the detection threshold, the better the sensitivity to the low-mass region. In this chapter, the experimental studies carried out with the aim of lowering the energy threshold of TREX-DM are discussed. These show promise to lower it to values comparable to the single-electron ionisation energy. This is possible thanks to the inclusion of an extra preamplification stage above the Micromegas, by means of the use of a GEM foil. The results of this chapter have been published in~\cite{GEM_2025}.

%
\section*{}
\parshape=0
\vspace{-13mm}
\vspace{-2mm}
\section{Motivation} \label{Chapter7_Motivation}

Despite the microbulk Micromegas readout's capability to attain high gains (even exceeding $10^6$ in controlled settings~\cite{Derre_2000}), the operational constraints of a full-scale experiment (the need for prolonged stable operation, robustness, absence of destructive discharges, extensive area coverage, fine readout segmentation, controlled electronic noise levels, and specific gas compositions and pressures) render an energy threshold slightly below 1 keV more realistic. Initially, TREX-DM aimed for a 0.4 keV threshold; however, challenges like electronic noise and leakage currents have limited the best achieved threshold to approximately 0.8-0.9 keV following the 2022 TREX-DM.v2 detector upgrade (see Figure~\ref{fig:chapter5_performance_comparison_detectors}).

Lowering the detection threshold to the single-electron ionisation energy ($\sim 20$~eV) could substantially enhance sensitivity to low-mass WIMPs. This potential has driven the exploration of incorporating a GEM foil as a preamplification stage above the microbulk Micromegas: the GEM amplifies the primary electron cloud before it enters the Micromegas gap, thereby increasing the overall gain of the readout system.

The idea of stacking several amplification structures is well-established, with configurations such as multi-GEM detectors and GEM-Micromegas hybrids documented in the literature~\cite{Sauli_2016,Kane_2002,Kane_2003,Zhang_2014,Aiola_2016}. These arrangements have so far been explored in the context of high-energy physics and/or high ionisation rates. Indeed, for these applications, qualities such as detector stability over time (absence of electric discharges) and detector ageing (degradation of the microstructures by sparks or charge accumulation effects) are crucial, and the division of the detector into several amplifying phases allows the overall amplification factor to be distributed, reducing the risk of discharges by being able to lower the amplification voltages. Notably, the upgrade of the COMPASS experiment in 2014-2015 successfully implemented a combination of a GEM and a bulk Micromegas~\cite{Neyret_2009,Neyret_2024}.

In contrast, the application of successive amplification stages in rare-event physics, particularly within low-background experiments like TREX-DM, is a novel approach. The objective here is to bring each stage to its maximum gain in order to improve the energy threshold. The GEM + Micromegas combination under investigation involves positioning a GEM foil above a microbulk Micromegas detector. In order to study it, tests have been conducted using both small-scale microbulk detectors and a full-scale detector from the batch evaluated for the TREX-DM.v2 upgrade. 

\vspace{-2mm}
\section{Definition of the General Set-Up and Method} \label{Chapter7_Definition_Set-Up_Method}

This section describes the elements and the operating mechanism of the joint GEM + Micromegas system, as well as the method used to study the effectiveness of said system. The gas used in these studies is Ar-1\%iC$_{4}$H$_{10}$, due to its importance for TREX-DM and its immediate availability, but with plans to extend the research to other Ar- and Ne-based such as Ar-10\%iC$_{4}$H$_{10}$ (see Section~\ref{Chapter9_Sensitivity_Projections}).

\vspace{2mm}
\textbf{\normalsize Description}
\vspace{0mm}

As described in Section~\ref{Chapter4_MPGDs}, in a TPC system with GEM/Micromegas, the gaseous volume is divided into drift and amplification regions. In this composite case, we define the drift region as the volume between the cathode and the top face of the GEM. It is in this region that the primary ionisations occur. The first amplification region is defined by the gap between the top and bottom face of the GEM, and with this manufacturing technology it is 50~$\upmu$m. Between the bottom face of the GEM and the mesh of the Micromegas, we have the so-called transfer region, which is the area in which the primary charge pre-amplified by the GEM drifts towards the Micromegas. Finally, the gap between the mesh and the anode, also 50~$\upmu$m for microbulk technology, defines the second amplification region. The charge is collected at the anode, which can be single-channel or pixelised (as in the case of TREX-DM), thus allowing 2D spatial reconstruction of the event.

In this way, with two amplification regions instead of one, we can achieve lower detection thresholds, although the operation of the detector becomes more complicated, as we will see later in Section~\ref{Chapter7_Challenges}. 

In the tests to be described, radioactive X-ray sources were used to check the operation of the system. For this purpose, a metal grid has been used as a cathode and the source has been placed on it, pointing towards the active volume of the detector. Most of the photons will ionise the gas in the drift volume, thus undergoing a double amplification. However, a percentage of the photons will travel through the drift region and pass through the GEM without interacting, ionising the transfer volume. This percentage depends on the mean free path of the photons in a given gas mixture at a given pressure. In any case, the important thing to note is that these photons will only receive the amplification provided by the Micromegas.

\begin{figure}[htbp]
\centering
\includegraphics[width=0.70\textwidth]{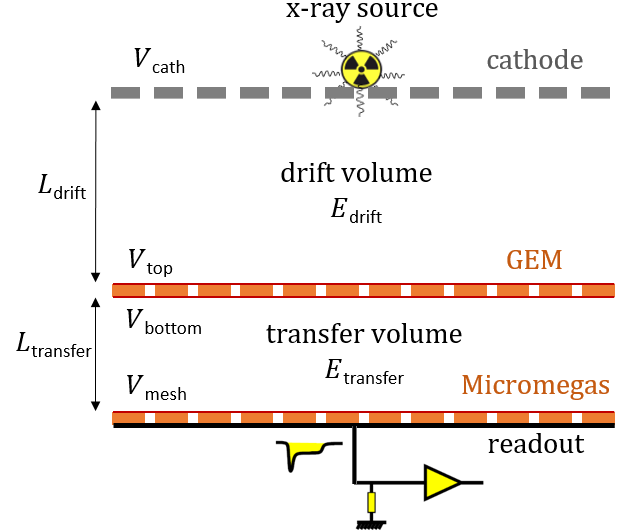}
\caption{Diagram of the different elements of the set-up. The readout can be unsegmented (Section~\ref{Chapter7_Small_Set-Up}) or segmented (Section~\ref{Chapter7_TREXDM_Set-Up}). The dimensions are not to scale. Source: own elaboration.}
\label{fig:chapter7_schematic_set-up}
\end{figure}

Regarding the voltages, the anode is kept grounded, the mesh is at $ V_{\mathrm{mesh}} $, and the bottom and top layers of the GEM are at $ V_{\mathrm{bottom}} $ and $ V_{\mathrm{top}} $, with the GEM preamplification voltage defined as $ V_{\mathrm{GEM}} = V_{\mathrm{top}}-V_{\mathrm{bottom}} $. The cathode is at $ V_{\mathrm{cath}} $. To avoid potential distortions in the transfer field $ E_{\mathrm{transfer}} $, the metallic support plate is kept at $ V_{\mathrm{plate}}=V_{\mathrm{mesh}} $. Two CAEN HV power supply modules (a 4-channel N1471H and a 2-channel N471A) are used to supply these voltages. A schematic view of the set-up can be seen in~\ref{fig:chapter7_schematic_set-up}.

\vspace{2mm}
\textbf{\normalsize Method}
\vspace{0mm}

To evaluate the performance of the preamplification system, calibration runs were performed using both a standard only-Micromegas configuration and a combined GEM~+ Micromegas setup. In the only-Micromegas configuration, the Micromegas mesh was powered while the GEM voltage was set to zero, thereby isolating and recording solely the events of interest. In contrast, for the GEM + Micromegas runs, both the mesh and the GEM were activated. In the former case, photons deposit their energy in the transfer region, whereas in the latter they deposit energy in the drift region. Comparing the energy spectra from these two configurations allows the determination of the relative amplification provided by the extra GEM stage.

We define the preamplification factor as the additional gain contributed by the GEM relative to a fixed Micromegas-induced gain. However, under realistic experimental conditions, the maximum voltage attainable with a standalone Micromegas is typically higher than that achievable with the GEM stage: in the composite system, a cloud of electrons already amplified by the GEM reaches the Micromegas, which means that maintaining the same voltage as if only the primary electron cloud arrived can result in detector instability due to excessive amplification. Consequently, we introduce the GEM effective extra gain factor (hereafter referred to as the GEM extra factor) as the amplification provided by the GEM in the optimised GEM + Micromegas set-up relative to the optimised only-Micromegas configuration. To put it more clearly, we define the gains:

\begin{equation}
    \begin{split}
    &\mathrm{\underline{\textbf{Only~MM:}}} \quad   
    \begin{cases}
        G^{\mathrm{ref}}_{\mathrm{MM}} \equiv G( V^{\mathrm{ref}}_{\mathrm{mesh}},0) \quad \mathrm{(reference~voltage)} \\
        G^{\mathrm{opt}}_{\mathrm{MM}} \equiv G( V^{\mathrm{opt}}_{\mathrm{mesh}},0) \quad \mathrm{(optimised~voltage)}
    \end{cases}
    \\ \\
    &\mathrm{\underline{\textbf{GEM + MM:}}} \quad 
    G_{\mathrm{GEM+MM}} \equiv G( V^{\mathrm{ref.}}_{\mathrm{mesh}},V_{\mathrm{GEM}} )
    \end{split}
    \label{eq:chapter7_gain_definitions}
\end{equation}

where $G( V^{\mathrm{opt}}_{\mathrm{mesh}},0)>G( V^{\mathrm{ref}}_{\mathrm{mesh}},0)$ because $V^{\mathrm{opt}}_{\mathrm{mesh}}>V^{\mathrm{ref}}_{\mathrm{mesh}}$. Therefore, the preamplification and extra gain factors can be calculated as:

\begin{equation}
    \mathrm{Preamp.~factor} = \frac{G_{\mathrm{GEM+MM}}}{G^{\mathrm{ref.}}_{\mathrm{MM}}} \quad \quad  \quad 
    \mathrm{GEM~extra~factor} = \frac{G_{\mathrm{GEM+MM}}}{G^{\mathrm{opt.}}_{\mathrm{MM}}}
    \label{eq:chapter7_preamp_factors_definitions}
\end{equation}

From these definitions, it is expected that Preamp. factor > GEM extra factor.

\section{Small Set-Up} \label{Chapter7_Small_Set-Up}
\subsection{Description} \label{Chapter7_Small_Set-Up_Description}

\begin{figure}[htbp]
\centering
\includegraphics[width=0.35\textwidth]{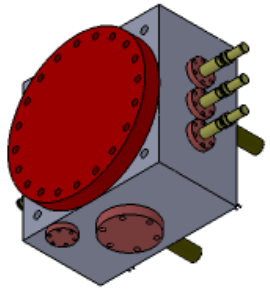}
\caption{3D view of the small high-pressure TPC used for the GEM + Micromegas set-up. The large CF160 blind flange in red is used to open and close the vessel.}
\label{fig:chapter7_small_chamber}
\end{figure}

A small (2.4~L) stainless-steel chamber certified for high pressure (up to 12~bar) is used to house the small microbulk Micromegas detector equipped with a GEM preamplification stage. This vessel (see Figure~\ref{fig:chapter7_small_chamber}) has been used extensively by the research group since its manufacture in 2004, and a detailed description is available at~\cite{tesis_paco}. After approximately one hour of using a Pfeiffer Vacuum HiCube 80 Classic Turbo Pump, the vacuum level achieved in this chamber is about $ 10^{-5} $~mbar. 

\begin{figure}[htbp]
\centering
\includegraphics[width=0.3\textwidth]{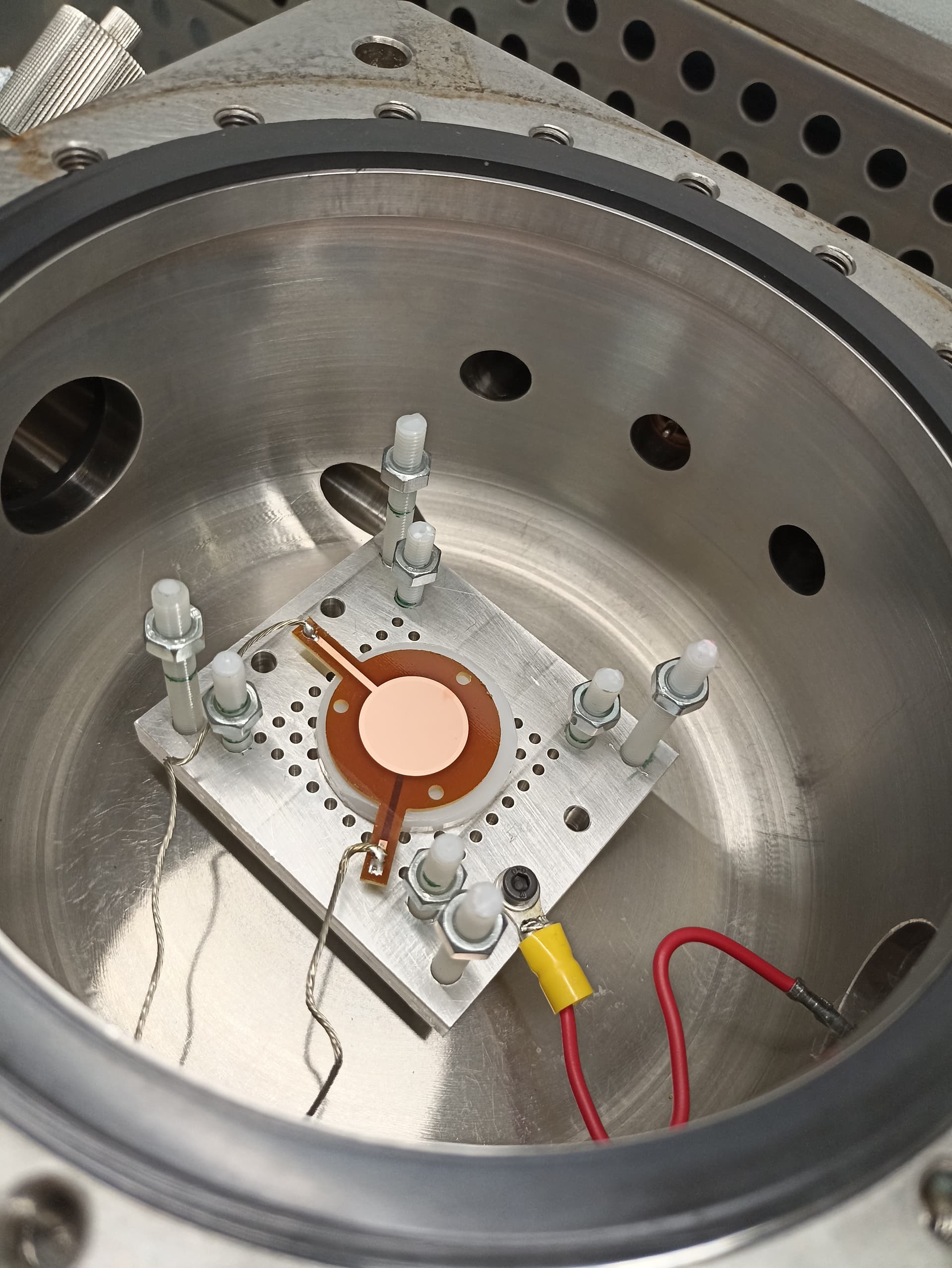}
\includegraphics[width=0.3\textwidth]{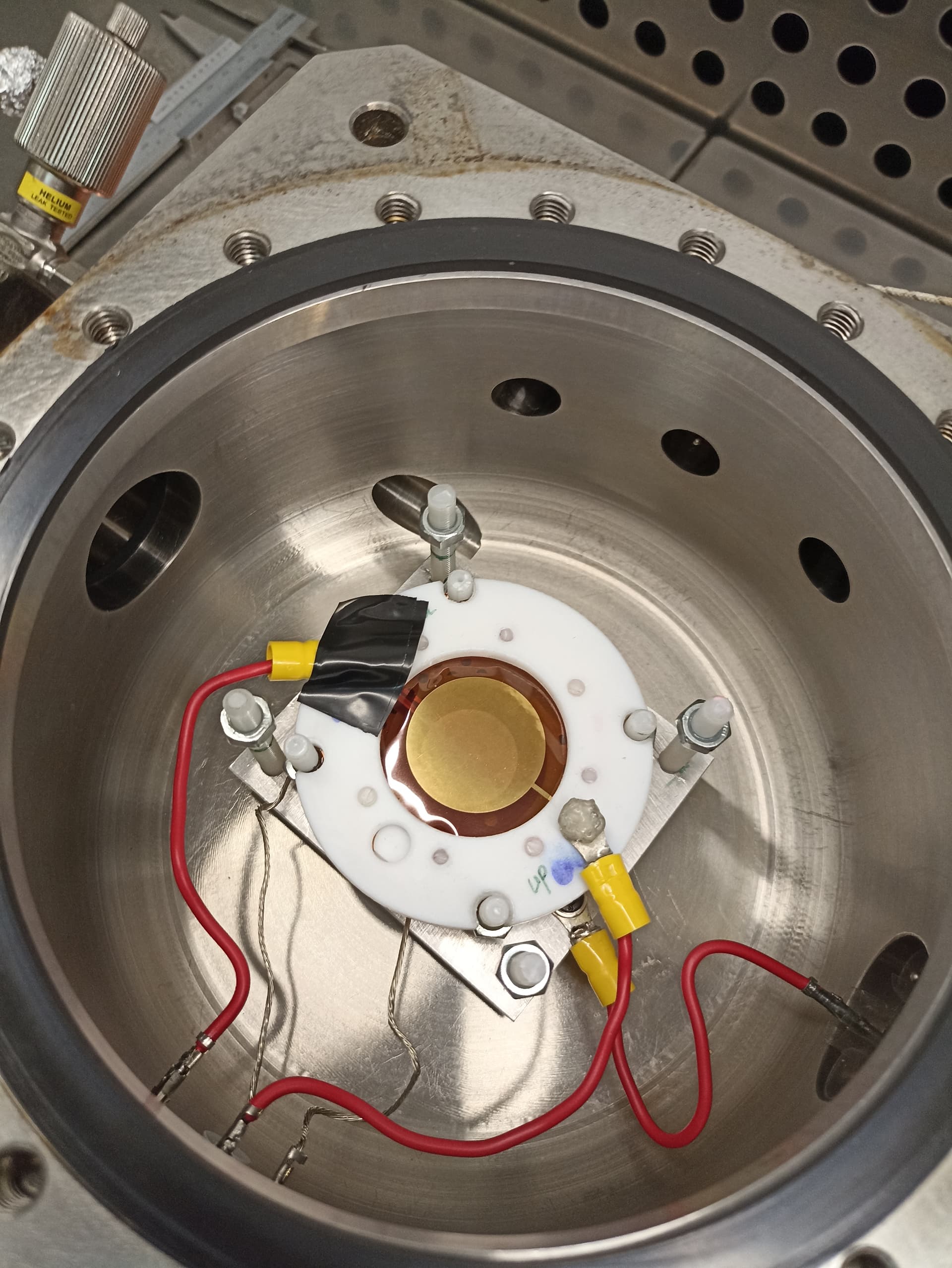}
\includegraphics[width=0.3\textwidth]{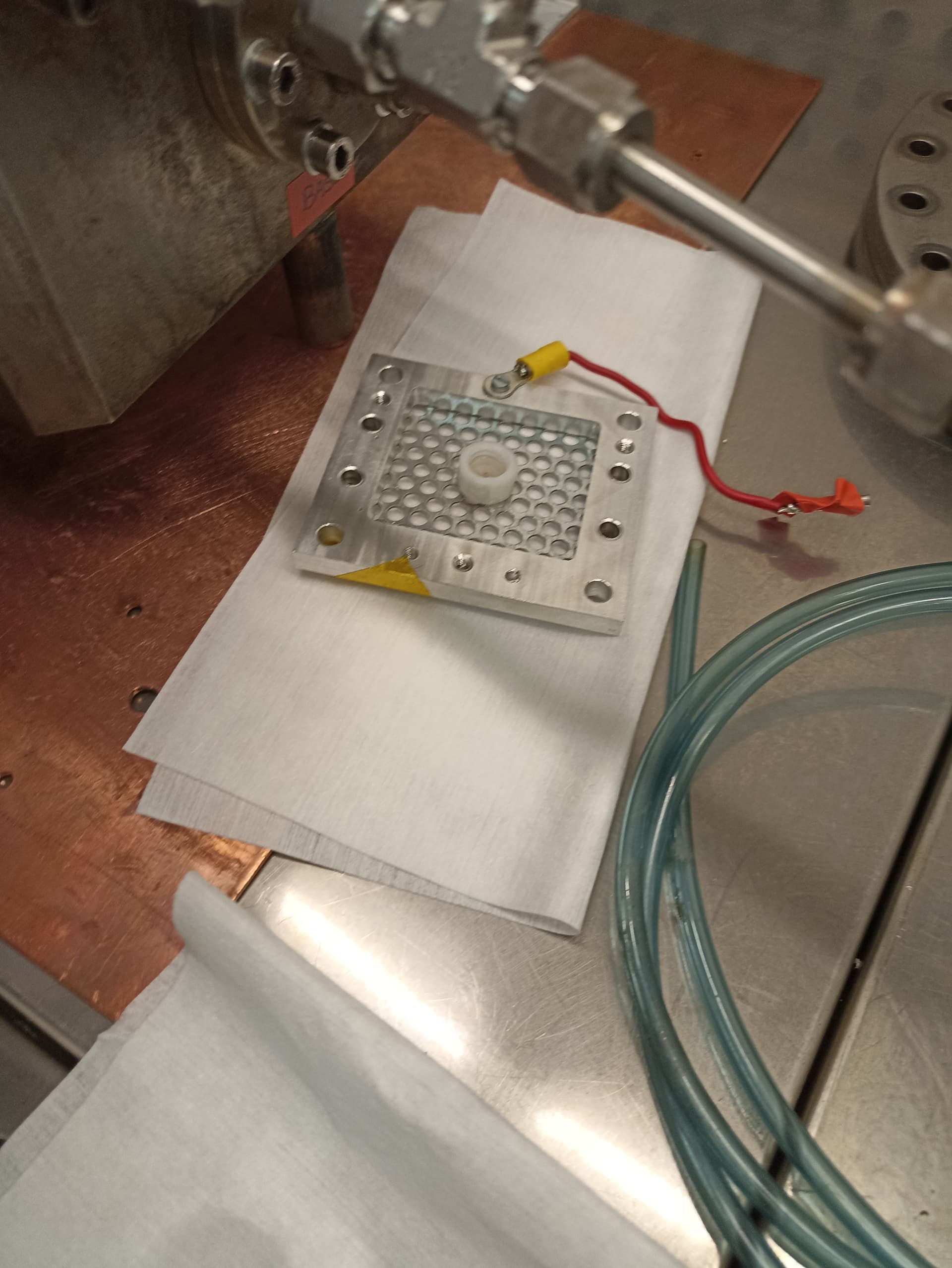}
\caption{Left: the metallic structure inside the vessel with the Micromegas on top. The inner PTFE pillars support the GEM at the correct distance, while the outer PTFE pillars hold the cathode in place. The connections for the anode, mesh, and plate to the feedthroughs are shown. Centre: a GEM foil is mounted on top of the Micromegas, along with the high-voltage connections. Right: the cathode, with an attached $^{55}$Fe source facing downwards.}
\label{fig:chapter7_small_chamber_set-up}
\end{figure}

\vspace{-1mm}
The microbulk Micromegas detector is positioned on a metallic support plate, separated by a PTFE piece. This Micromegas features a non-segmented, disc-shaped anode with a small, 2-cm-diameter active area, and a 50~$ \upmu $m gap between the mesh and anode. In these tests, a series of Micromegas with varying hole diameters (50-60~$\upmu$m) and pitches (100-110~$\upmu$m) were employed. A GEM stage of similar dimensions is mounted above the mesh, at a distance $ L_{\mathrm{transfer}} = $ 10~mm. The GEM has a total thickness of 60~$\upmu$m, comprising a 50-$\upmu$m kapton layer and two 5-$\upmu$m copper layers. It is characterised by a hole pitch of 140~$\upmu$m, with the copper holes having a diameter of 70~$\upmu$m and those in the kapton measuring 60~$\upmu$m. Above the GEM, a cathode (a stainless-steel grid) is positioned at a distance $ L_{\mathrm{drift}} = $ 13~mm. The cathode has an attached $ ^{55} $Fe source (X-ray at 5.9 keV) facing the ionisation volume. See Figure~\ref{fig:chapter7_small_chamber_set-up} for a view of the different elements.

As for the DAQ, the signal registered at the anode is initially sent to a preamplifier (Canberra Model 2005), and subsequently passes through an amplifier module (Canberra Model 2022 NIM module). Both the preamplified and amplified signals are read using an oscilloscope (Tektronix TDS5054). A tailor-made software for data acquisition and analysis is employed to operate the oscilloscope and process the data.

\subsection{Analysis and Results} \label{Chapter7_Small_Set-Up_Analysis_Results}

Both the preamplification factor and the GEM extra factor were determined at 1, 4 and 10~bar (the target pressure in TREX-DM).

\begin{figure}[htbp]
	\centering
	\includegraphics[width=0.85\textwidth]{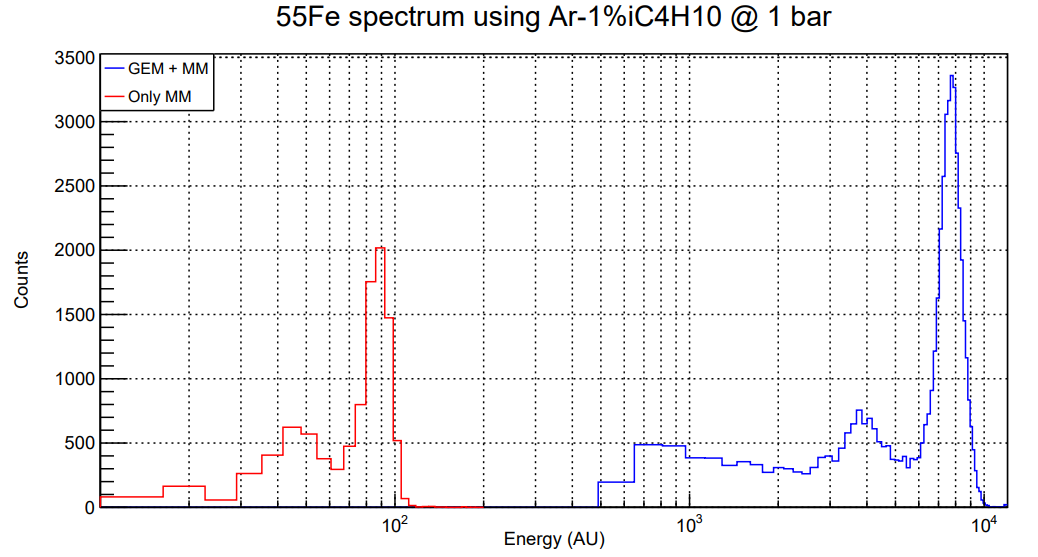}
    \includegraphics[width=0.85\textwidth]{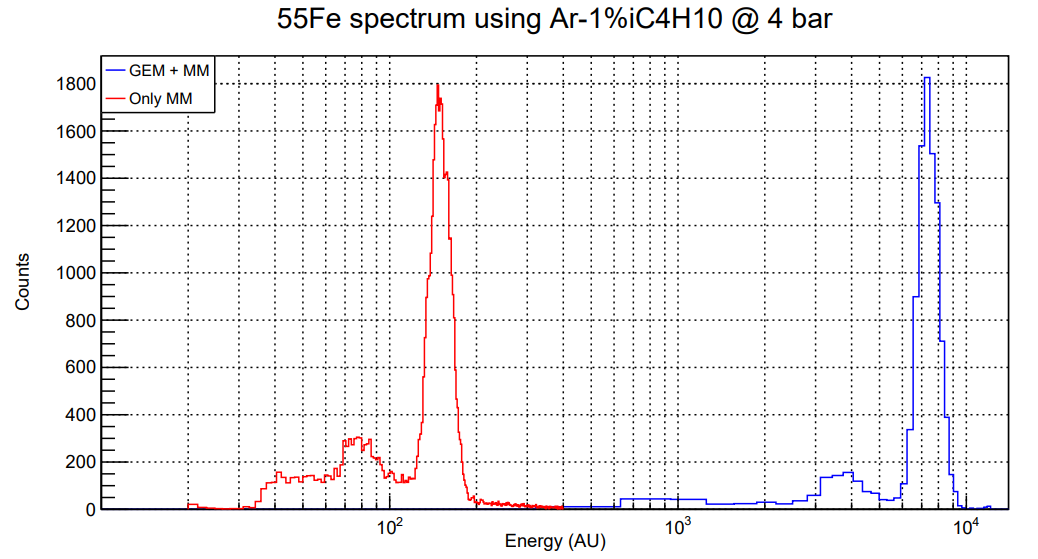}
    \includegraphics[width=0.85\textwidth]{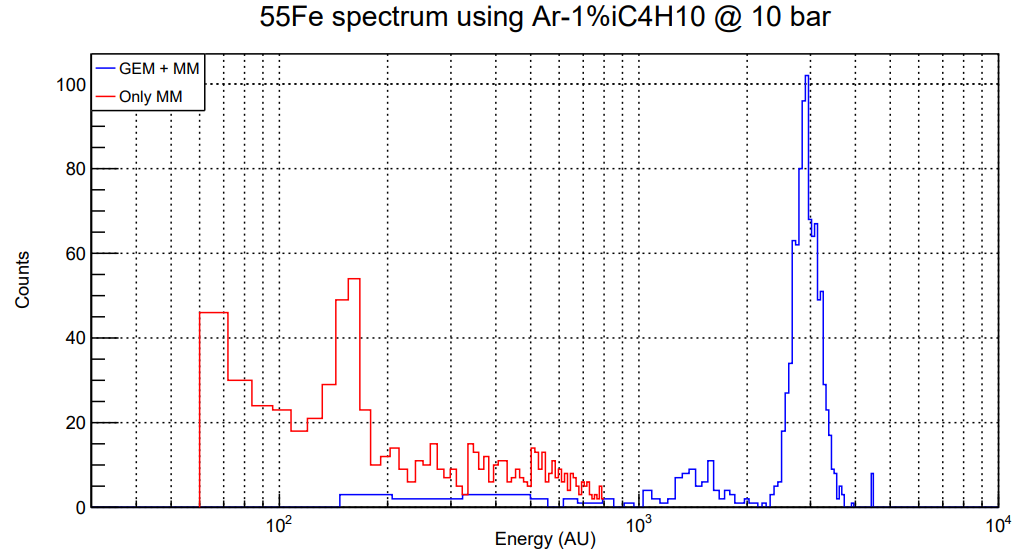}
    \vspace{-1mm}
	\caption{Energy spectra comparing calibrations using only Micromegas (red) and the combined GEM + Micromegas system (blue), acquired with a $ ^{55} $Fe source in the test set-up. The gas mixture employed is Ar-1\%iC$_{4}$H$_{10}$, and the horizontal axis is plotted on a logarithmic scale. From top to bottom: at 1~bar the GEM extra factor is approximately 90; at 4~bar, approximately 50; and at 10~bar, approximately 20. The voltages applied in these runs correspond to those listed in the fifth column (red lines) and the second and third columns (blue lines) of Table~\ref{table:chapter7_gem_mm_data}.}
    \label{fig:chapter7_spectra_gem_mm_small}
\end{figure}

Initial runs with the combined system show that it is quite easy to saturate the output of the amplifier module, given the high gains that are achieved by having the GEM on. Therefore, the first step in order to make the cleanest comparison possible between GEM + Micromegas runs versus only-Micromegas runs is to take test runs to select a DAQ dynamic range that is valid for both situations. Specifically, a compromise is sought between amplifying enough to trigger events in the Micromegas-only runs, but not so much as to saturate the electronics in the GEM-Micromegas runs. This approach permits a direct comparison of energy spectra, as illustrated in Figure~\ref{fig:chapter7_spectra_gem_mm_small}.

The highest stable operating voltages, as well as the achieved preamplification and GEM extra factors achieved for the small set-up are listed in Table~\ref{table:chapter7_gem_mm_data}. The only missing value is the preamplification factor at 1~bar, which could not be measured due to noise issues on the day of data acquisition. However, since the primary parameter of interest, the GEM extra factor, had already been established, the remaining measurements were carried out. 

\begin{table}[htb]\centering
\begin{center}
\begin{tabular}{c|c|c|c|c|c}  
\hline\hline
\textbf{Pressure} & $ V^{\mathrm{ref}}_{\mathrm{mesh}} $ & $ V_{\mathrm{GEM}} $ & \textbf{Preamp.} & $ V^{\mathrm{opt}}_{\mathrm{mesh}} $ & \textbf{GEM} \\
(bar) & (V) & (V) & \textbf{factor} & (V) & \textbf{extra factor} \\
\hline
1  & 305 & 310 & -  & 315 & 90 \\  
\hline
4  & 390 & 410 & 70 & 400 & 50 \\  
\hline
10 & 535 & 550 & 21 & 540 & 19 \\  
\hline
1  & 290 & 285 & 85 & 293 & 80 \\  
\hline\hline
\end{tabular}    
\end{center}
\caption{Measured preamplification and GEM extra gain factors for the two set-ups at different pressures. These factors are defined as the gain ratio between GEM + Micromegas runs ($V_{\mathrm{GEM}} \neq 0$~V) and only Micromegas runs ($V_{\mathrm{GEM}} = 0$~V). The first three rows correspond to the test set-up, while the last row represents the full-scale TREX-DM set-up.}
\label{table:chapter7_gem_mm_data}
\end{table}

A reference value for the maximum voltage in only-Micromegas runs was obtained from~\cite{TREXDM_Bckg_Assessment_2018}. For the combined runs, the starting point for both $V_{\mathrm{mesh}}$ and $ V_{\mathrm{GEM}} $ was set to a safe value, approximately 30-40~V below the reference voltage. From this point, the voltage was incremented in steps of 5~V (first in $V_{\mathrm{mesh}}$ and then in $ V_{\mathrm{GEM}} $) until unstable behaviour (typically sparks) was observed. Notably, it was found that even at the last stable $V_{\mathrm{mesh}}$, $ V_{\mathrm{GEM}} $ could be increased slightly further. 

Given that the objective of this set-up was merely to demonstrate the feasibility and potential of the combined GEM + Micromegas system under realistic experimental conditions in TREX-DM, a detailed study of the electron transmission and gain curves of the GEM + Micromegas system was deferred to the full-scale set-up (see Section~\ref{Chapter7_TREXDM_Set-Up}). In all cases, the drift and transfer fields were maintained at $ E_{\mathrm{drift}}= 100 $~V/cm/bar and $ E_{\mathrm{transfer}}= 100 $~V/cm/bar, values that typically lie within the electron transmission plateau of the Micromegas. Nonetheless, border effects cannot be entirely excluded, given the small active area and the absence of a field shaper.

Comparison of the 5.9~keV peak position in $^{55}$Fe calibration runs indicates maximum GEM extra factors of 90 (1~bar), 50 (4~bar) and 20 (10~bar). These values have been reproduced over several days and for different Micromegas detectors with the same micropattern, with variations contained within $\pm 20$\%. This variation is primarily attributable to slight differences in the maximum stable voltages achieved (typically $\pm 5$~V in either $V_{\mathrm{mesh}}$, $ V_{\mathrm{GEM}} $, or both). The decrease in gain with pressure is generally expected for MPGDs, although microbulk Micromegas do not exhibit such pronounced performance degradation at high pressures~\cite{Iguaz_2022}.

Several examples of these comparisons are shown in Figure~\ref{fig:chapter7_spectra_gem_mm_small}. At 10~bar, only-Micromegas runs are more challenging, as the mean free path of 5.9~keV photons in Ar-1\%iC$_{4}$H$_{10}$ at that pressure is approximately 2.3~mm~\cite{NIST_XCOM}. Photons that are not absorbed within the drift volume must go through the GEM foil, resulting in a more prominent exponential background and a less intense calibration peak (although the peak remains clearly visible). The resolutions (expressed as \%FWHM) for both preamplified and only-Micromegas runs are around 20\%, with the exception of the only-Micromegas run at 10~bar, which is approximately 30\% (largely due to the small number of events). Importantly, these findings indicate that the addition of a preamplification stage does not significantly degrade the resolution of the microbulk detectors.

\section{Full-Scale TREX-DM Set-Up} \label{Chapter7_TREXDM_Set-Up}
\subsection{Description} \label{Chapter7_TREXDM_Set-Up_Description}

In order to verify whether the promising results obtained with a small-scale test set-up can be reproduced under realistic experimental conditions (substantially larger readout area and drift distance), a dedicated test bench was prepared. The primary aim of this study is to evaluate the combined GEM + Micromegas detector configuration for reducing the energy threshold in the low-mass WIMP search undertaken by TREX-DM.

The test bench consists of a stainless-steel 50-L chamber that houses a spare TREXDM.v2 Micromegas detector (from the same batch as the ones deployed in TREX-DM) with an overlying GEM foil. The chamber is initially pumped down to approximately 10~mbar using the same Pfeiffer Vacuum HiCube 80 Classic Turbo Pump over the course of about one hour. It is then filled with an Ar-1\%iC$_{4}$H$_{10}$ gas mixture, and a continuous gas flow of 8~l/h is maintained for 72~h to ensure optimal gas quality. During these tests, the chamber pressure is kept at 1~bar in accordance with the design specifications.

The microbulk detector is mounted on the chamber's endcap, and a GEM foil of the same dimensions and amplification gap is positioned above the mesh at a transfer distance of $ L_{\mathrm{transfer}} = $ 10~mm. Above the GEM, a stainless-steel grid serves as the cathode and is located at a drift distance of $ L_{\mathrm{drift}} = $ 100~mm. Two $^{109}$Cd radioactive sources are attached to the cathode.

\begin{figure}[htb]
	\centering
    \includegraphics[width=0.45\textwidth]{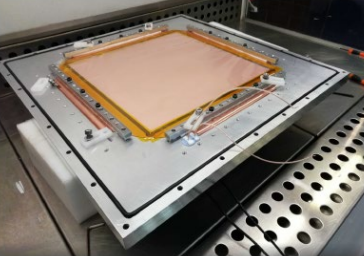}
    \includegraphics[width=0.45\textwidth]{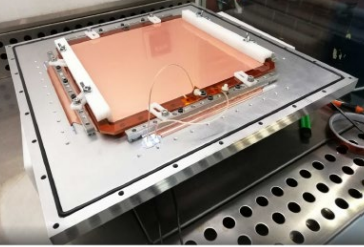}\\
	\includegraphics[width=0.45\textwidth]{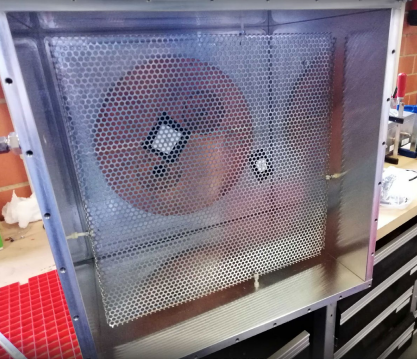}
    \includegraphics[width=0.45\textwidth]{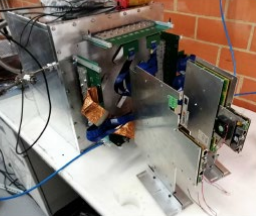}
	\caption{Top left: TREXDM.v2 Micromegas detector placed on the endcap of the chamber. Top right: GEM foil secured on top of the mesh. Bottom left: cathode grid with two $ ^{109} $Cd radioactive sources attached with black tape. Bottom right: closed vessel with DAQ.}
    \label{fig:chapter7_big_chamber_set-up}
\end{figure}

For signal readout from the TPC, a combination of a FEC equipped with AGET and a Feminos card (same as the ones used in TREX-DM, see Section~\ref{Chapter5_Description_DAQ}) is used. Data processing and analysis are performed using a routine based on REST-for-Physics (see Section~\ref{Chapter5_Description_REST}). Images detailing the assembly of the various components are presented in Figure~\ref{fig:chapter7_big_chamber_set-up}.

\subsection{Analysis and Results} \label{Chapter7_TREXDM_Set-Up_Analysis_Results}

\begin{figure}[htbp]
	\centering
	\includegraphics[width=0.495\textwidth]{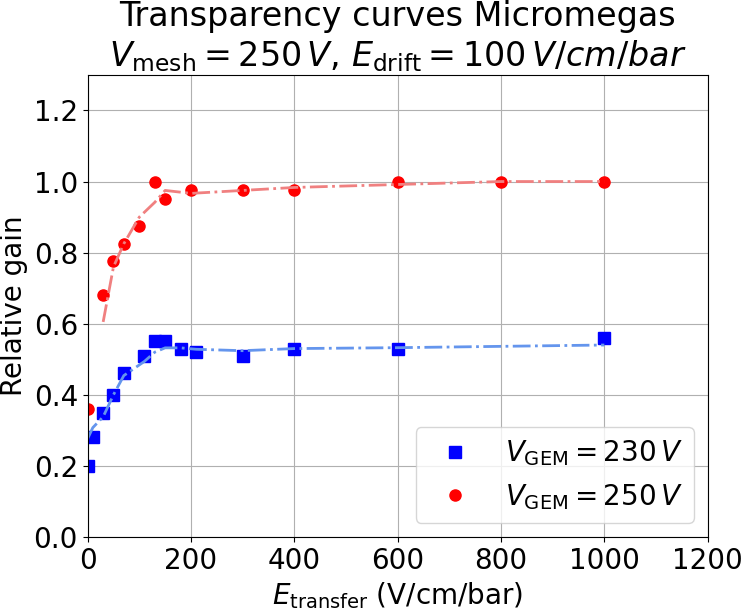}
    \includegraphics[width=0.495\textwidth]{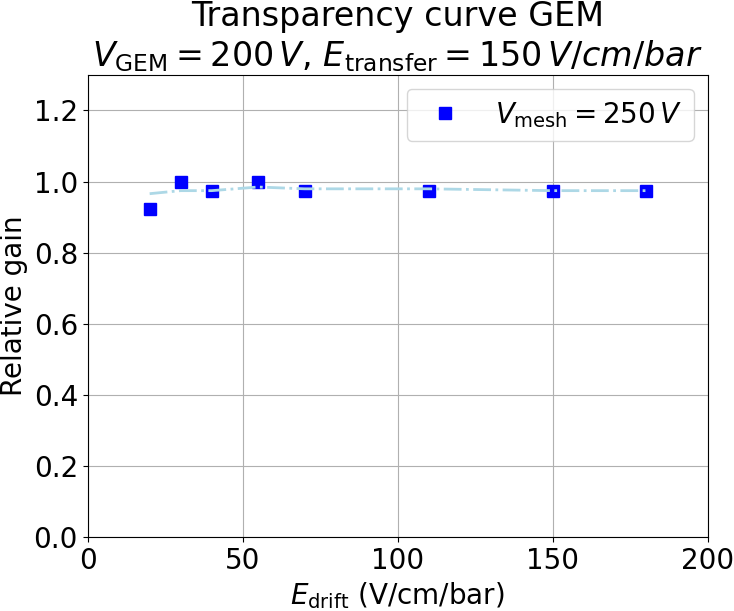}
	\caption{Electron transmission curves. The vertical axis represents the mean peak position, normalised to the maximum value, with statistical uncertainties below 1\% in both plots. Left: Micromegas transmission measured at a fixed mesh voltage for two different GEM voltages, with $E_{\mathrm{drift}}$ = 100~V/cm/bar ensuring complete GEM transparency. Right: GEM transmission recorded at fixed mesh and GEM voltages, with $E_{\mathrm{transfer}}$ = 150~V/cm/bar to operate within the plateau region of the Micromegas transmission curve.}
    \label{fig:chapter7_transparency_curves_full_scale}
\end{figure}

Although the primary objective was to replicate the small set‐up results at 1~bar (see Section~\ref{Chapter7_Small_Set-Up_Analysis_Results}), further investigations were carried out using the full-scale set-up. In this extended study, we examined the electron transmission (i.e. the transparency curves) of both the GEM foil and the mesh, as shown in Figure~\ref{fig:chapter7_transparency_curves_full_scale}. Notably, the GEM transmission reaches a plateau very rapidly, even at very low drift field values. This behaviour is consistent with earlier studies of electron transmission in GEMs~\cite{Sauli_2002}, which demonstrated that the collection efficiency increases with $V_{\mathrm{GEM}}$, so that full transparency is achieved at lower $E_{\mathrm{drift}}$ when $V_{\mathrm{GEM}}$ is raised.

In contrast, the Micromegas curves exhibit an intriguing phenomenon at $E_{\mathrm{transfer}} = 0$~V/cm/bar, where the relative gain remains non-zero. Here, photons converted in the drift volume are first amplified by the GEM, and due to the close proximity of the mesh, diffusion allows a fraction of these events to reach the Micromegas for further amplification. This effect is likely related to the small transfer gap, $L_{\mathrm{transfer}} \sim 1$cm, although additional tests at varying transfer distances would be needed for a more definitive explanation. Beyond $E_{\mathrm{transfer}} \approx 150$~V/cm/bar, the relative gain remains essentially constant. This is somewhat unexpected, as one would normally anticipate that the transparency curve depends on both the extraction efficiency of the GEM’s bottom layer and the collection efficiency of the Micromegas. While a plateau is typical for the Micromegas, it is generally expected that the GEM extraction efficiency should continue to increase in this region, potentially up to a few kV/cm/bar~\cite{GEM_2003}. We are considering several hypotheses: first, the observed plateau might actually be part of a slowly rising curve that could extend beyond the approximately 1~kV/cm/bar limit imposed by our set-up; second, the gas mixture may play a significant role in shaping the curves. Typically, GEM detectors employ noble gases combined with quenchers such as CH$_{4}$ or CF$_{4}$, owing to their high drift velocities and low diffusion coefficients compared to others such as iC$_{4}$H$_{10}$~\cite{diffusion_1984}. The Ar-1\% iC$_{4}$H$_{10}$ mixture, however, has not been characterised in the context of GEMs as far as we know, and its relatively higher diffusion coefficients may reduce the extraction efficiency from the GEM’s bottom layer.

\begin{figure}[htbp]
	\centering
	\includegraphics[width=0.495\textwidth]{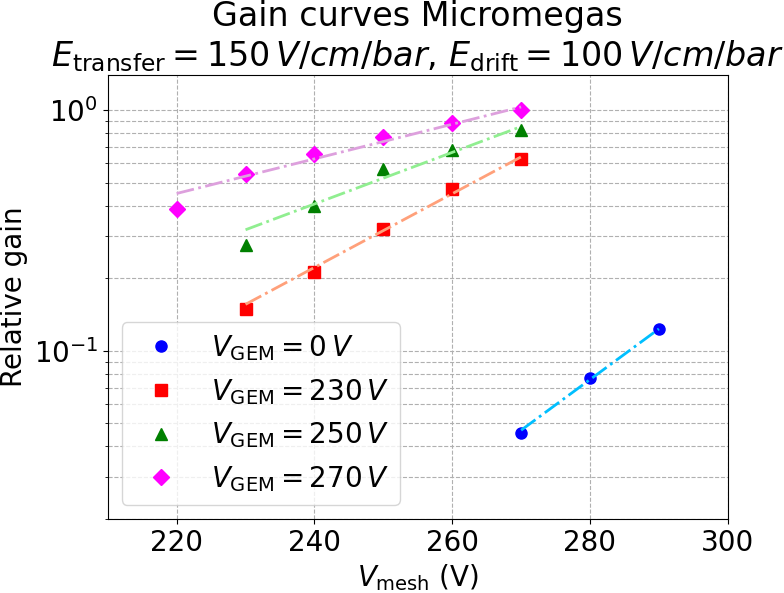}
    \includegraphics[width=0.495\textwidth]{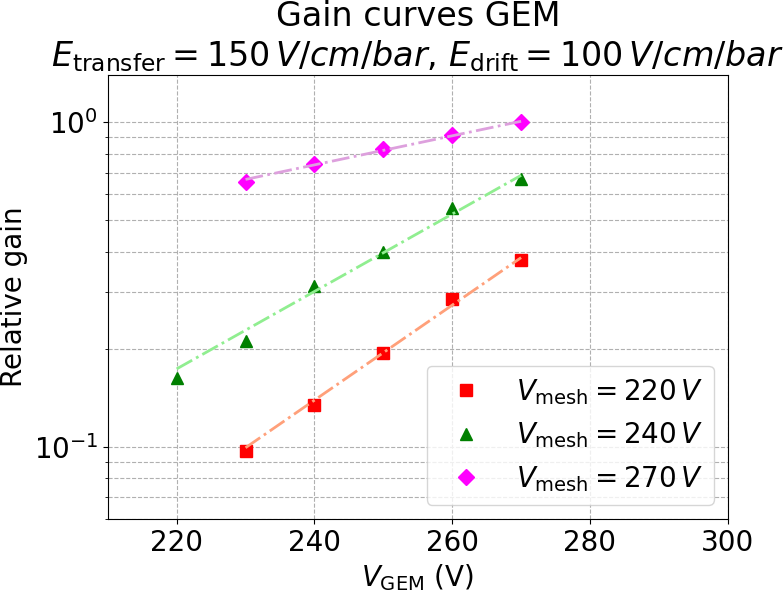}
	\caption{Gain curves. The vertical axis represents the mean peak position, normalised to the maximum value, with statistical errors below 1\% in both plots. Left: Micromegas curves recorded with a fixed $ V_{\mathrm{GEM}} $. Right: GEM curves obtained with a fixed $V_{\mathrm{mesh}} $. In both cases, $E_{\mathrm{drift}}$ and $E_{\mathrm{transfer}}$ are held constant. Note that the maximum gain is reached at the same data point in both plots ($V_{\mathrm{mesh}} $=270 V, $ V_{\mathrm{GEM}} $=270 V), which allows a direct comparison of the relative gain between them.}
    \label{fig:chapter7_gain_curves_full_scale}
\end{figure}

In addition to these transparency curves, we also studied gain curves. As illustrated in Figure~\ref{fig:chapter7_gain_curves_full_scale}, two types of gain curves were obtained: one set by varying $V_{\mathrm{mesh}}$ for a fixed $V_{\mathrm{GEM}}$ (Micromegas gain curves) and another by varying $V_{\mathrm{GEM}}$ for a fixed $V_{\mathrm{mesh}}$ (GEM gain curves). In both cases, the expected exponential increase with amplification voltage is observed. In the Micromegas curves, the case $V_{\mathrm{GEM}} = 0$~V (corresponding to the baseline only-Micromegas detector) is included, and comparison with the GEM + Micromegas curves already suggests extra gain factors on the order of 10 due to the GEM addition. Moreover, the fact that the curves are not perfectly parallel on a logarithmic scale indicates that the two amplification stages may not be completely independent, possibly as a result of some backflow; however, this does not compromise the main objectives of the study.

\begin{figure}[htbp]
	\centering
	\includegraphics[width=1.0\textwidth]{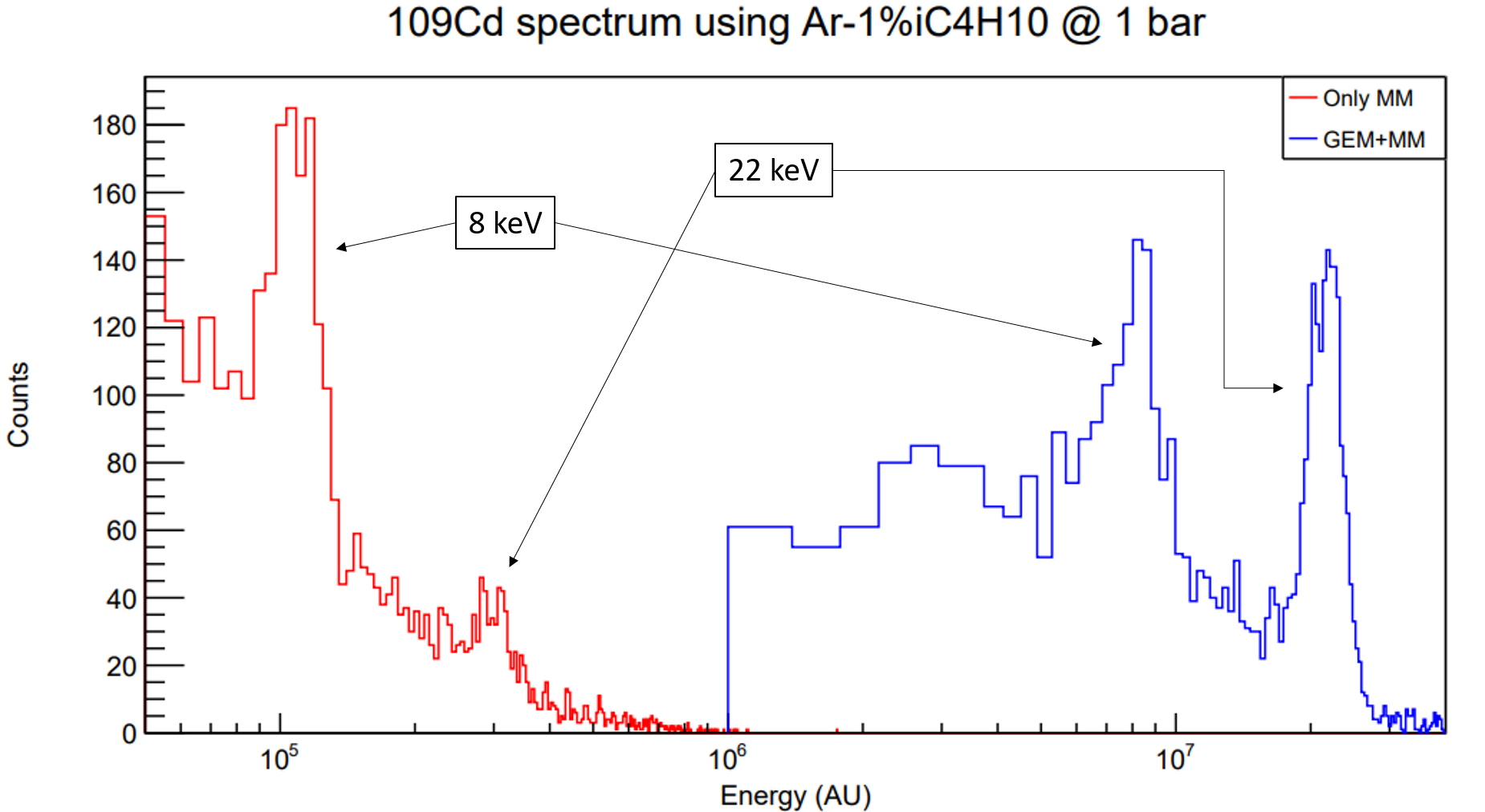}
	\caption{Energy spectrum comparison between $ ^{109} $Cd calibrations in the full‐scale set‐up. Red: only-Micromegas calibration. Blue: GEM + Micromegas calibration. The 8~keV peak corresponds to copper fluorescence at the Micromegas surface. The gas mixture employed is Ar-1\%iC$_{4}$H$_{10}$ at 1~bar. Note that the horizontal axis is plotted on a logarithmic scale and that an energy cut has been applied to the GEM + Micromegas run to remove background, thereby keeping the left side of the canvas clear.}
    \label{fig:chapter7_1bar_preamp_factor_trex_gem}
\end{figure}

Additional measurements were then performed to determine the maximum achievable GEM extra factor and preamplification factor. The voltages applied in these tests are summarised in the fourth row of Table~\ref{table:chapter7_gem_mm_data}. By comparing the position of the 8~keV copper fluorescence peak in $^{109}$Cd calibration runs (see Figure~\ref{fig:chapter7_1bar_preamp_factor_trex_gem}), a GEM extra factor of 80 was obtained, which is consistent with the findings from the small set‐up (Section~\ref{Chapter7_Small_Set-Up_Analysis_Results}). In these runs, the fields were maintained at $E_{\mathrm{drift}} = 100$~V/cm/bar and $E_{\mathrm{transfer}} = 150$~V/cm/bar, as indicated by the transparency curves.

It should be noted that the operating voltages in the full-scale set-up are lower than those in the small set‐up. This is due to the inherent difficulties of operating larger-area Micromegas (with 1 versus 1024 channels, the likelihood of leakage currents between the mesh and some channels is higher). Nonetheless, the full-scale set-up still demonstrates the potential to lower the energy threshold, closely replicating the real experimental conditions of TREX-DM.

\section{Installation in TREX-DM} \label{Chapter7_Installation}

In light of the successful laboratory tests, arrangements were made to install the GEM stage in TREX-DM. This initiative represents the first serious attempt to deploy a combined GEM + Micromegas readout in an operational rare-event physics experiment.

\begin{figure}[htbp]
\centering
\includegraphics[width=0.95\textwidth]{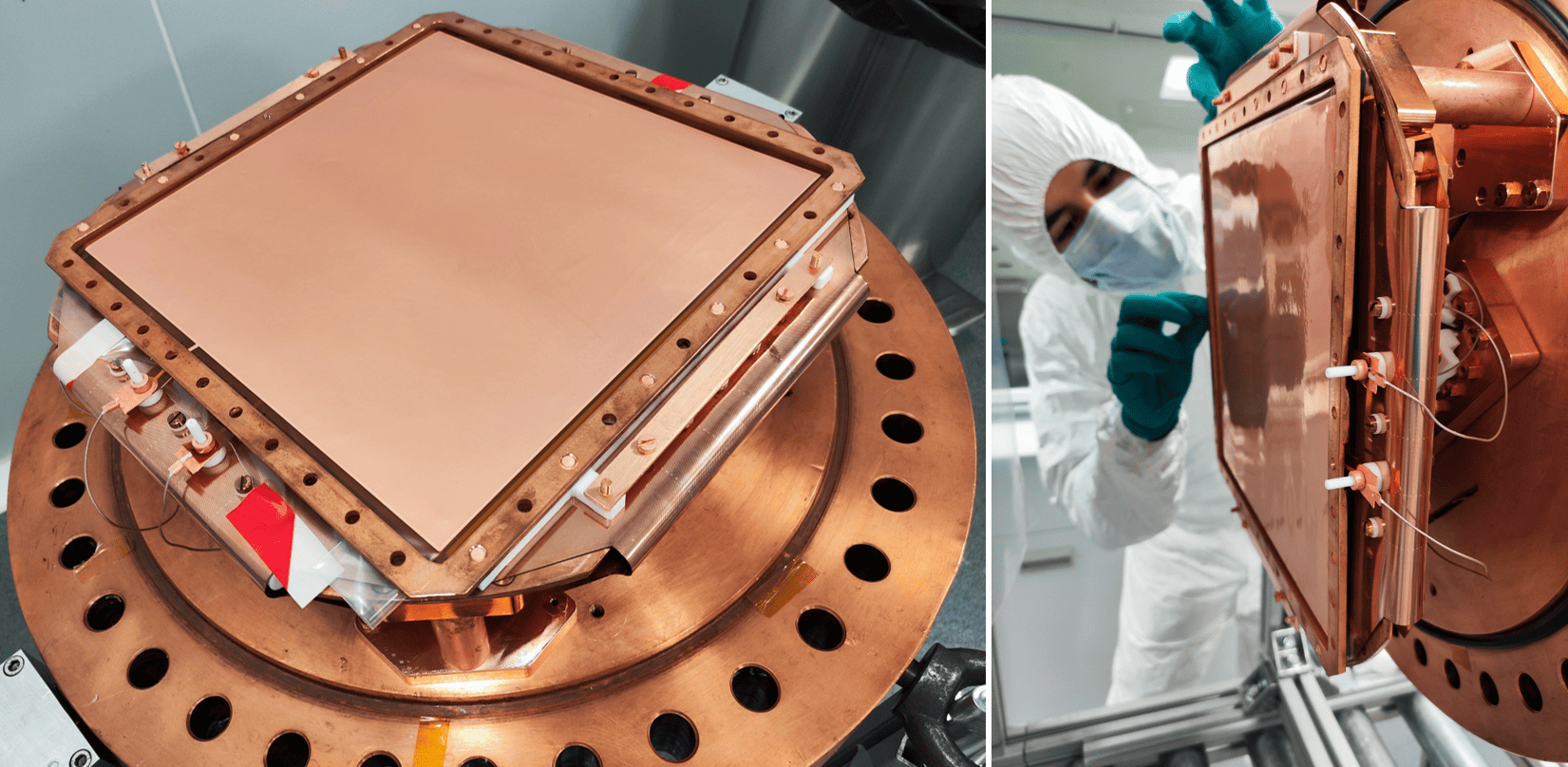}
\caption{GEM + Micromegas system in TREX-DM during installation in the clean room. The high-voltage connections to both the top and bottom layers of the GEM are visible. Left: the GEM is mounted atop the Micromegas (with a plastic cover protecting its surface). Right: final adjustments before closing the endcap.}
\label{fig:chapter7_gem_installation_trexdm}
\end{figure}

In March 2024, the TREX-DM detector was brought into the clean room at Lab2400 for a series of interventions, one of which was the installation of a GEM foil atop one of the microbulk Micromegas (with $ L_{\mathrm{transfer}} =1$~cm, same as in Section~\ref{Chapter7_TREXDM_Set-Up_Description}). The procedure was successfully completed (see Figure~\ref{fig:chapter7_gem_installation_trexdm}), and standard tests (such as leak-tightness, verification of electrical connections, and voltage ramp-up and ramp-down to confirm that the GEM layers charge and discharge as expected) yielded satisfactory results. However, after the chamber was relocated to Lab2500 and installed within its shielding, subsequent tests revealed a short-circuit between the GEM foil electrodes, rendering that half of the detector inoperative. Although the cause is not entirely clear, a plausible hypothesis is that the fault occurred during transport. This problem prompted another intervention to substitute the GEM foil.

In June 2024, the establishment of a new clean room in Lab2500 (facilitated by the collaboration with LSC) enabled the first in-situ intervention, during which the damaged GEM was replaced with a new unit. The new GEM was installed and electrical tests confirmed proper operation. Detailed work was undertaken to reinstall the electronics and re-optimise the electronic noise, with the aim of ensuring that the GEM side operates under optimal conditions (see Section~\ref{Chapter7_Challenges} for more information about the challenges of this system). A semi-stable operating point at 1~bar was reached, which allowed to take several low-energy calibrations using a $^{37}$Ar source (see Section~\ref{Chapter8_Calibration_TREX-DM}). These calibrations were a huge success and a milestone for the experiment, as they confirmed the enhanced energy threshold achieved in laboratory tests, and showed the realistic potential to bring it down to single-electron levels. The discussion of the implications of these results for TREX-DM's sensitivity is provided in Chapter~\ref{Chapter9_Sensitivity}.

\section{Challenges of the System} \label{Chapter7_Challenges}

This section outlines the challenges encountered in operating the combined GEM + Micromegas system, both during testing on the full-scale laboratory set-up and, more notably, following its installation in TREX-DM.

\vspace{2mm}
\textbf{\normalsize Mechanical Integration}
\vspace{0mm}

Combining both detectors requires very precise alignment and mechanical robustness, which is more complex than using a single detector separately. In particular, it is crucial to get the GEM at a homogeneous distance from the Micromegas along the whole $x-y$ plane. For this purpose, in both set-ups, pillars had to be added to hold the GEM at a fixed distance. In the case of the TREX-DM assembly, four pieces (one on each side of the GEM foil) have also been added to tighten it: being so thin (60~$ \upmu $m counting the copper foils), inhomogeneities easily appear which can be smoothed out by pulling on the four sides at the same time.

\vspace{2mm}
\textbf{\normalsize Electrical Connections}
\vspace{0mm}

Adding the GEM foil implies the use of two extra high-voltage feedthroughs to power the upper and lower foil, which adds complexity to the assembly as well as to the control system. On the other hand, this also implies the use of higher absolute voltages, on the order of kV, both for the GEM and the cathode. As a reference, if we used Ar-1\%iC4H10 in TREX-DM at 10 bar, keeping $ (V_{\mathrm{mesh}}, V_{\mathrm{GEM}} ) \sim (500, 500)$~V as in the small set-up (see Table~\ref{table:chapter7_gem_mm_data}), and requiring $ E_{\mathrm{drift}}=100$~V/cm/bar, $ E_{\mathrm{transfer}}=150$~V/cm/bar, we would need $( V_{\mathrm{mesh}}, V_{\mathrm{bottom}}, V_{\mathrm{top}}, V_{\mathrm{cath}} ) = (500,2000,2500,18500)$~V, taking into account that $ L_{\mathrm{transfer}} =1$~cm and $ L_{\mathrm{drift}} =16$~cm.

\vspace{2mm}
\textbf{\normalsize Detector Instability and Maintenance}
\vspace{0mm}

Operating with two amplification stages instead of one significantly increases the probability of overcurrents that can trip the high-voltage power supply, thereby compromising detector operation and data acquisition stability. Such overcurrents may be induced by excessively high gains in the GEM or Micromegas (which can trigger discharges during the avalanche process), by high-energy events (for example, alpha particles), or by discharges resulting from spatial inhomogeneities in the GEM/Micromegas, such as accumulated dust or manufacturing defects in the microholes. Depending on their nature, these discharges may cause irreparable damage, including the creation of a continuous conduction path between the top and bottom GEM layers, or between the mesh and a readout channel. In a segmented Micromegas, only a limited number of strips might be affected (see Section~\ref{Chapter5_Technical_Challenges_Leakage_Currents}). However, a short-circuit in the GEM is particularly detrimental since, being unsegmented, the entire foil is lost. This can lead to costly maintenance in terms of both resources and downtime, as the detector must be opened and reassembled to replace the GEM, as already described in Section~\ref{Chapter7_Installation}.

\vspace{2mm}
\textbf{\normalsize Gain}
\vspace{0mm}

The gain of a Micromegas/GEM detector is not always stable and may depend on factors such as fluctuations in pressure, temperature, and gas quality. By having two consecutive stages, it is to be expected that the variations in gain could be more pronounced, as the effect of the two stages is multiplicative.

On the other hand, the maximum voltages reached in the tests presented in Sections~\ref{Chapter7_Small_Set-Up} and~\ref{Chapter7_TREXDM_Set-Up} do not correspond to the optimal operating point under long-term conditions, as extended stability (over days) may be compromised by overcurrents in the GEM, the Micromegas, or both. Consequently, the optimal operating point has to be carefully determined, with an acceptable spark rate established to ensure detector recovery and minimal detector downtime.

\vspace{2mm}
\textbf{\normalsize Noise}
\vspace{0mm}

The inclusion of additional connections and components can lead to higher levels of electronic noise, so although the overall signal-to-noise ratio improves thanks to the addition of the GEM, the noise reduction has to be even more thorough than with only a Micromegas plane (see Section~\ref{Chapter5_Technical_Challenges_Noise}).

\vspace{2mm}
\textbf{\normalsize Ramping-Up and Ramping-Down Protocols}
\vspace{0mm}

Raising voltages on the Micromegas is a delicate process in itself, but the challenge is compounded when operating the GEM stage. This is primarily because it is essential to increase both $V_{\mathrm{top}}$ and $V_{\mathrm{bottom}}$ simultaneously in order to avoid creating an excessive potential difference, $V_{\mathrm{GEM}}$, which could damage the GEM foil.

Initial stability tests were performed in TREX-DM by applying 50~V to the bottom electrode and 250~V to the top, thereby establishing a 200~V difference. However, when attempting to achieve a similar differential at higher absolute voltages (with $V_{\mathrm{mesh}} = 250$~V, $V_{\mathrm{bottom}} = 400$~V, and $V_{\mathrm{top}} = 650$~V), the GEM top electrode tripped. It was suspected that sparking occurred between the GEM top and the last ring of the field cage, which is separated by only 1~cm.

To mitigate this risk, a specific voltage ramping protocol was developed: 

\begin{enumerate} 
    \item Increase $V_{\mathrm{mesh}}$, $V_{\mathrm{bottom}}$, $V_{\mathrm{top}}$, and the last ring voltage (controlled via $V_{\mathrm{cath}}$) simultaneously until the $V_{\mathrm{mesh}}$ value is reached. 
    \item Then, raise $V_{\mathrm{bottom}}$, $V_{\mathrm{top}}$, and the last ring voltage together until reaching the $V_{\mathrm{bottom}}$ value. 
    \item Next, increase $V_{\mathrm{top}}$ and the last ring voltage simultaneously until the $V_{\mathrm{top}}$ value is attained. 
    \item Finally, raise $V_{\mathrm{cath}}$ to its final value. 
\end{enumerate} 

This procedure is designed to prevent sparking between the GEM layers, the Micromegas, and the last ring.

Additionally, to avoid a significant voltage imbalance in the event of a trip on any channel (which could potentially lead to a fatal spark), a hardware-based interlock system was introduced to disable all CAEN channels if one channel trips. It was also deemed necessary to include the Spellman power supply in this system, thereby preventing the last ring of the field cage from retaining several hundred volts while the GEM top is at ground potential.

\vspace{2mm}
\textbf{\normalsize Radiopurity and Background Levels}
\vspace{0mm}

Although the base materials of the GEM foil are the same as those of the Micromegas (kapton and copper), an extra element is being added to the sensitive volume that is susceptible to contamination, especially surface contamination, as we have seen in Section~\ref{Chapter6_Surface_Contamination}. On the other hand, the improvement of the energy threshold means opening the sub-keV window, together with the possibility of having new background sources not foreseen in this energy range. The study of background runs in this combined system is currently underway.

%% file: CHAPTERS/Chapter8.tex
\chapter{\texorpdfstring{$^{37}$Ar}{37Ar} as a Low-Energy Calibration Source} \label{Chapter8_Ar37}

{
\lettrine[loversize=0.15]{A}{s} explained in Chapter~\ref{Chapter2_WIMPs}, Dark Matter searches focused on WIMPs are gradually shifting towards very low energy thresholds (sub keV). Consequently, understanding detector behaviour and energy reconstruction at these energies becomes critical. Specifically, given the developments presented in Chapter~\ref{Chapter7_GEM-MM}, there is a need in TREX-DM for high-statistics, low-energy calibration.
%
\section*{}
\parshape=0
\vspace{-20.5mm}
}

Traditional external gamma sources often struggle to provide reliable calibration at these energies due to detector material self-shielding and insufficient penetration, creating non-uniform event distributions. Additionally, many conventional calibration sources lack mono-energetic emissions in this energy range, making things more complicated. In this context, $^{37}$Ar has emerged as a particularly interesting calibration source for low-energy applications.

This chapter describes the method we followed to produce $^{37}$Ar for calibrating the TREX-DM detector. We begin by presenting the motivation to select $^{37}$Ar and briefly reviewing available production methods. We then discuss the initial tests conducted during a research stay at CEA Saclay to obtain a calibration source. Subsequently, we detail the adapted set-up designed to implement this technique specifically for TREX-DM, followed by a description of the irradiation process used to generate the source at Centro Nacional de Aceleradores (CNA) in Sevilla. Finally, we show and analyse the low-energy calibration spectrum obtained with TREX-DM, where an exceptional energy threshold of $O(10)$~eV was achieved.

\section{Introduction} \label{Chapter8_Introduction}

$^{37}$Ar ($t_{1/2}=35.04$~d) is a gas radioisotope that decays by electron capture ($Q=813.9$ keV) to $^{37}$Cl. The electron capture process creates a vacancy in either the K or L electron shell (or, with marginal probability, on the M shell), followed by electron rearrangement. This generates a cascade of characteristic X-rays and Auger electrons as outer electrons transition to fill the initial vacancy. The total energy of these deposits corresponds to the binding energies of the captured electrons: 2.82 keV from K-shell capture, occurring with 90\% probability, and 0.27 keV peak from L-shell capture, with approximately 9\% probability~\cite{lnhb_table_radionucleides}. These discrete, low-energy emissions align with the energy region of interest for WIMP searches.

As a noble gas, $^{37}$Ar can be homogeneously distributed within the active volume, ensuring uniform calibration both for dual-phase and gas TPCs. This eliminates the spatial biases associated with external sources and allows for full detector characterisation.

\begin{figure}[htb]
\centering
\includegraphics[width=1.0\textwidth]{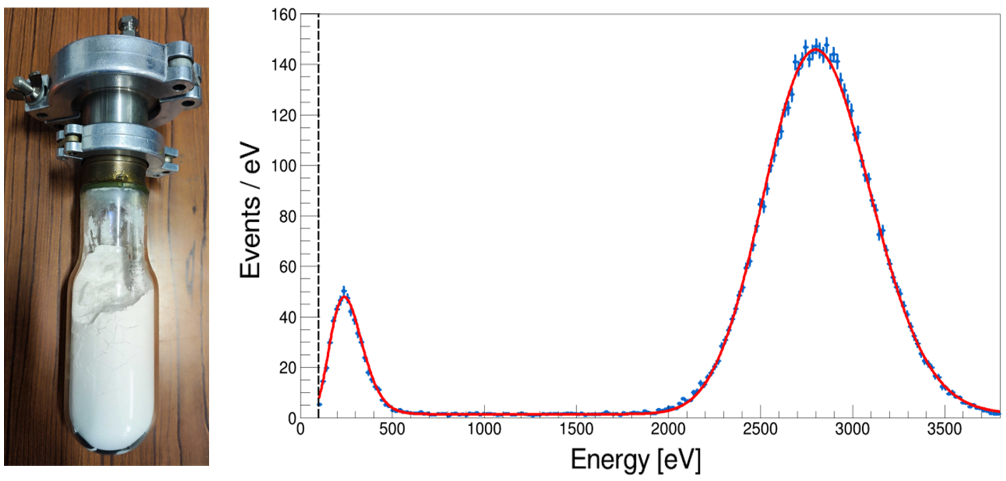}
\caption{Left: container with CaO powder used for $^{37}$Ar production. Image taken at CEA Saclay. Right: $^{37}$Ar calibration spectrum obtained by the NEWS-G collaboration, showing experimental data (blue) and fitted spectrum with flat background (red). The grey line indicates the energy threshold at approximately 100 eV. Plot taken from~\cite{NEWS-G_2019}.}
\label{fig:chapter8_powder_and_spectrum}
\end{figure}

The production of $^{37}$Ar for experimental purposes has been achieved through several methods, each with particular advantages in specific contexts:

\begin{itemize}
    \item Irradiation of $^{36}$Ar with thermal neutrons via the reaction $^{36}\mathrm{Ar}(n,\gamma)^{37}\mathrm{Ar}$. This can be performed using atmospheric argon~\cite{Sangiorgio_2013} or, as implemented by the XENON collaboration, using argon enriched in $^{36}$Ar~\cite{XENON1T_Ar37_2022}. The enrichment enhances the production of $^{37}$Ar while reducing the production of undesired isotopes such as $^{39}$Ar ($t_{1/2}=269$ years) and $^{41}$Ar ($t_{1/2}=110$ minutes).
    \item Proton irradiation via the reaction $^{37}\mathrm{Cl}(p,n)^{37}\mathrm{Ar}$, as investigated in~\cite{Kishore_1975,Weber_1985}, though this approach needs access to a proton beam facility.
    \item Neutron activation of calcium oxide (CaO) targets via the reaction $^{40}\mathrm{Ca}(n,\alpha)^{37}\mathrm{Ar}$. This is the most common production route, typically utilising fast neutrons from nuclear reactors or dedicated neutron generators. This technique has been successfully applied in dual-phase TPC contexts~\cite{Boulton_2017}, and has been used in gaseous TPCs by collaborations such as NEWS-G  for more than a decade~\cite{NEWS_Ar37_2014}. It is also the one selected for application in TREX-DM. Figure~\ref{fig:chapter8_powder_and_spectrum} shows a container with CaO in powder form and an $^{37}$Ar spectrum.
\end{itemize}

Recent research has demonstrated improvements in $^{37}$Ar production using nano-dimensioned CaO particles ($\sim 20$-nm diameter) compared to conventional micron-sized particles ($1-10$~$\upmu$m)~\cite{Kelly_2018}. The nanomaterial exhibits better liberation of $^{37}$Ar from the CaO matrix without chemical treatment, with measurements of the by-product $^{41}$Ar by gamma spectroscopy suggesting $(96 \pm 10)$\% liberation efficiency. This approach circumvents the need for radiochemical separation procedures that have been required in larger-scale productions.

The irradiation of natural calcium also produces several other radioisotopes, most notably $^{42}$K through the reaction $^{42}\mathrm{Ca}(n,p)^{42}\mathrm{K}$. $^{42}$K is a gamma-emmiting isotope ($E_\gamma \approx 1525$~keV) with a half-life $t_{1/2}=12.4$~h~\cite{lnhb_table_radionucleides} that can serve as a useful monitor for $^{37}$Ar production. However, it should be noted that the natural abundance of $^{40}$Ca (96.941\%) is significantly higher than that of $^{42}$Ca (0.647\%)~\cite{Kelly_2018}, resulting in greater $^{37}$Ar production compared to $^{42}$K (see Section~\ref{Chapter8_Irradiation_CNA_Estimation}).

\section{Tests at CEA Saclay} \label{Chapter8_Tests_Saclay}

While irradiation of CaO powder in a neutron reactor or specialised neutron facility offers faster production and higher activities of $^{37}$Ar, the use of a neutron source (such as an AmBe source) represents a viable alternative. This section details the experiments conducted at the end of 2022 during a three-month research stay at CEA Saclay, which aimed to test the $^{37}$Ar production technique and transfer the expertise to Zaragoza for its application in TREX-DM.

CEA Saclay provided an ideal environment, as some of the first successful tests to obtain an $^{37}$Ar source using this method were conducted there, initially by Ioannis Giomataris and colleagues~\cite{Giomataris_2012}. Our objective was to replicate their procedure and observe $^{37}$Ar events in a Spherical Proportional Counter (SPC, described in Section~\ref{Chapter4_SPC}).

\vspace{2mm}
\textbf{\normalsize Experimental Set-Up and First Attempt}
\vspace{0mm}

\begin{figure}[htb]
\centering
\includegraphics[width=1.0\textwidth]{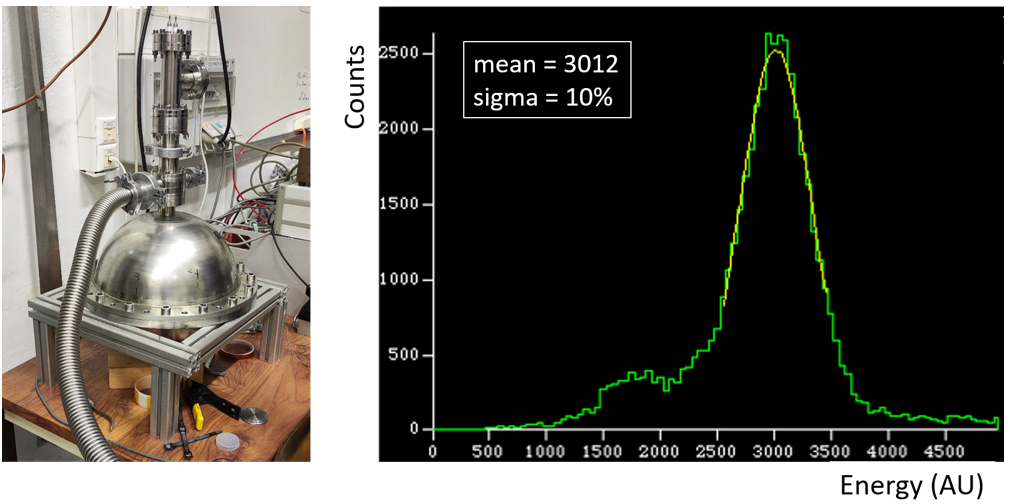}
\vspace{-8mm}
\caption{Left: 30-cm-diameter stainless-steel SPC used for the $^{37}$Ar tests. Right: $^{55}$Fe calibration spectrum (green) and Gaussian fit (orange) obtained prior to the first $^{37}$Ar test. Operating voltage was 1600 V with an acquisition time of approximately 20 minutes. The energy resolution at 5.9 keV is $\approx 25$\% FWHM.}
\label{fig:chapter8_CEA_sphere_and_spectrum}
\end{figure}

\begin{figure}[htb]
\centering
\includegraphics[width=1.0\textwidth]{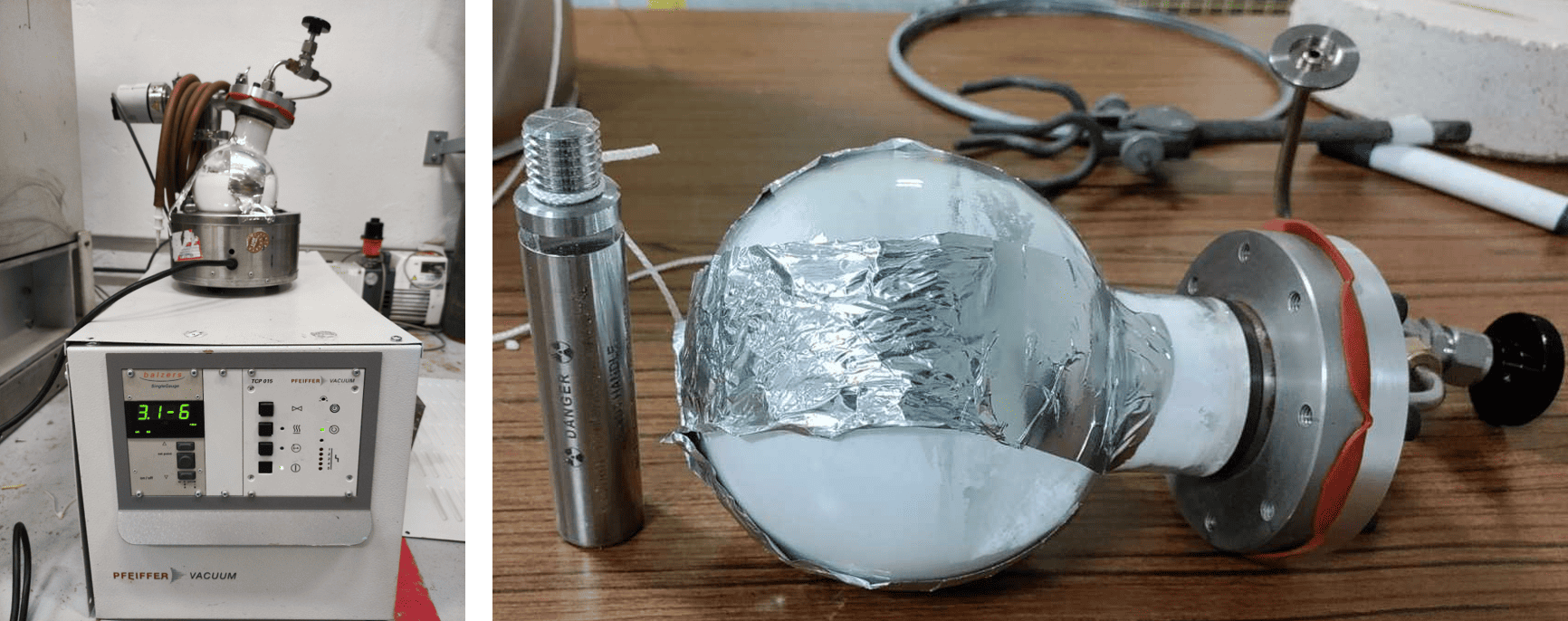}
\caption{Container with CaO powder being pumped down to vacuum (left) previous to its irradiation with an AmBe neutron source (right).}
\label{fig:chapter8_CEA_pumping_and_irradiation}
\end{figure}

For our tests, we used a 30-cm-diameter stainless-steel SPC (see Figure~\ref{fig:chapter8_CEA_sphere_and_spectrum}) equipped with an ACHINOS anode. The detector was filled with Ar-2\%CH$_4$ at a pressure of 600 mbar. The container with the CaO powder was pumped down to vacuum ($10^{-5}$-$10^{-6}$ mbar recommended) over 2-3 days (Figure~\ref{fig:chapter8_CEA_pumping_and_irradiation}). One potential issue encountered was that the pressure sensor was positioned near the pump, connected to the vessel via a small-diameter tube, suggesting that the actual pressure at the vessel might have differed from the measured value of approximately $3\times10^{-6}$ mbar, potentially by a factor of 10-100. The evacuated CaO container was then sent for irradiation to SPRE, where it was exposed to an AmBe neutron source (Figure~\ref{fig:chapter8_CEA_pumping_and_irradiation}). The irradiation process lasted approximately two weeks in total, although this was non-continuous as the neutron source was also required for other experiments.

Before injecting the gas from the irradiated container, we calibrated the SPC with a $^{55}$Fe source placed inside the sphere (at the bottom). The spectrum is shown in Figure~\ref{fig:chapter8_CEA_sphere_and_spectrum}. The sphere was then pumped down to vacuum ($10^{-5}$-$10^{-6}$ mbar), with the idea of letting the $^{37}$Ar gas diffuse into the system. When opening the valve to the $^{37}$Ar source, the pressure in the sphere rose to $10^{-1}$ mbar. Given that $V_\mathrm{sphere}/V_\mathrm{vessel}\sim O(10)$, this implied a pressure in the vessel after irradiation of $P_\mathrm{vessel}\sim 1$ mbar, six orders of magnitude higher than before irradiation. This increase was attributed to potential leaks in the vessel and outgassing from the CaO powder and the vessel materials.

Although the $^{55}$Fe signal was clearly observed, we found no clear indication of $^{37}$Ar events. The most reasonable explanation was that the activity was not enough: the neutron source used was less intense than in previous attempts (by a factor of 2-3), and the intermittent irradiation schedule likely reduced efficiency. Our estimates suggested that the produced $^{37}$Ar activity might have been in the order of 1 Hz, whilst the $^{55}$Fe spectrum shown in Figure~\ref{fig:chapter8_CEA_sphere_and_spectrum} had a significantly higher event rate of $\sim 30$ Hz. Consequently, the $^{37}$Ar peak at 2.82 keV may have been covered by the argon escape peak from the $^{55}$Fe source, which occurs at a similar energy ($\approx 5.9 - 3 = 2.9$ keV).

\vspace{2mm}
\textbf{\normalsize Second Attempt and Successful Detection}
\vspace{0mm}

For our second attempt, we modified the experimental protocol by removing the $^{55}$Fe source to eliminate the potential interference with the argon escape peak. We repeated the irradiation procedure, this time with approximately two weeks of effective irradiation time (spread out across three weeks of intermittent use of the source). To monitor detector stability, we recorded cosmic ray spectra before and after the injection of $^{37}$Ar to track any significant changes in gain.

\begin{figure}[htb]
\centering
\includegraphics[width=1.0\textwidth]{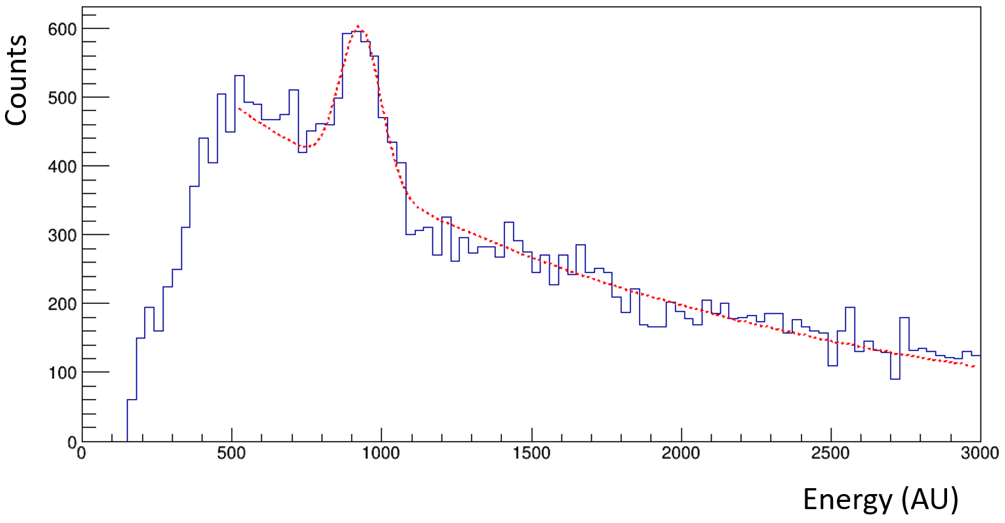}
\vspace{-8mm}
\caption{Low-energy spectrum obtained after injecting $^{37}$Ar during the second attempt ($V=1600$ V, run duration $\approx 100$ min). The blue histogram shows the experimental spectrum, with the dashed red line representing a fit to a Gaussian distribution over an exponential background. Because of limited statistics ($< 1$ Hz total $^{37}$Ar rate) and suboptimal noise and energy threshold conditions, the $^{37}$Ar signal appears only as an excess over the background.}
\label{fig:chapter8_CEA_Ar37_spectrum_fit}
\end{figure}

These adjustments proved successful, and we clearly observed $^{37}$Ar events corresponding to the 2.82 keV peak. Figure~\ref{fig:chapter8_CEA_Ar37_spectrum_fit} shows the low-energy spectrum after cuts obtained from the $^{37}$Ar run (at 1600 V for approximately 100 minutes, limited by sensor stability issues). The data reveals a distinct excess over the background, indicating the presence of $^{37}$Ar events. The data were converted and analysed off-line with ROOT, and the spectrum was fitted to a Gaussian distribution superimposed on an exponential background, showing excellent agreement.

As anticipated, the activity was relatively low ($< 1$ Hz), but sufficient to confirm the viability of the method. Although time constraints prevented further tests, this result validated the technique and provided a starting point for implementing it under more optimal conditions in Zaragoza. We concluded that with appropriate improvements, including a higher-activity source, extended irradiation times, and optimisation of the solid angle during irradiation, the results could be substantially improved for future calibrations in TREX-DM.

\section{Set-Up for TREX-DM} \label{Chapter8_Set-Up_TREX-DM}

Following the successful validation of the $^{37}$Ar production technique at CEA Saclay, the next step involved developing an optimised set-up for implementation in TREX-DM. 

\vspace{2mm}
\textbf{\normalsize Material Selection and Considerations}
\vspace{0mm}

The production of $^{37}$Ar requires CaO powder for neutron irradiation. An initial design decision involved selecting between micron-sized and nano-sized CaO powder. Given that the CaO previously tested at CEA Saclay was not specifically optimised (using standard powder rather than nano-CaO), we chose micron-sized CaO powder as the baseline, while maintaining the option to progress to nano-CaO if required for enhanced extraction efficiency~\cite{Kelly_2018}.

The quantity of CaO powder was determined with reference to the vessel used at CEA Saclay, which we estimated contained approximately $0.5-1$ kg of CaO (density 3.34 kg/L). However, the absolute quantity is less critical than optimising the surface area exposed to neutrons, which was addressed in the vessel design.

Purity represented a critical parameter to avoid unwanted activation of impurities during neutron irradiation. We selected CaO from Sigma Aldrich with a certified purity of 99.9\%. The Certificate of Analysis confirmed an actual purity of 99.985\%, with magnesium identified as the primary impurity, which presented minimal concern for our application.

\vspace{2mm}
\textbf{\normalsize Vessel Design and Construction}
\vspace{0mm}

\begin{figure}[htb]
\centering
\includegraphics[width=1.0\textwidth]{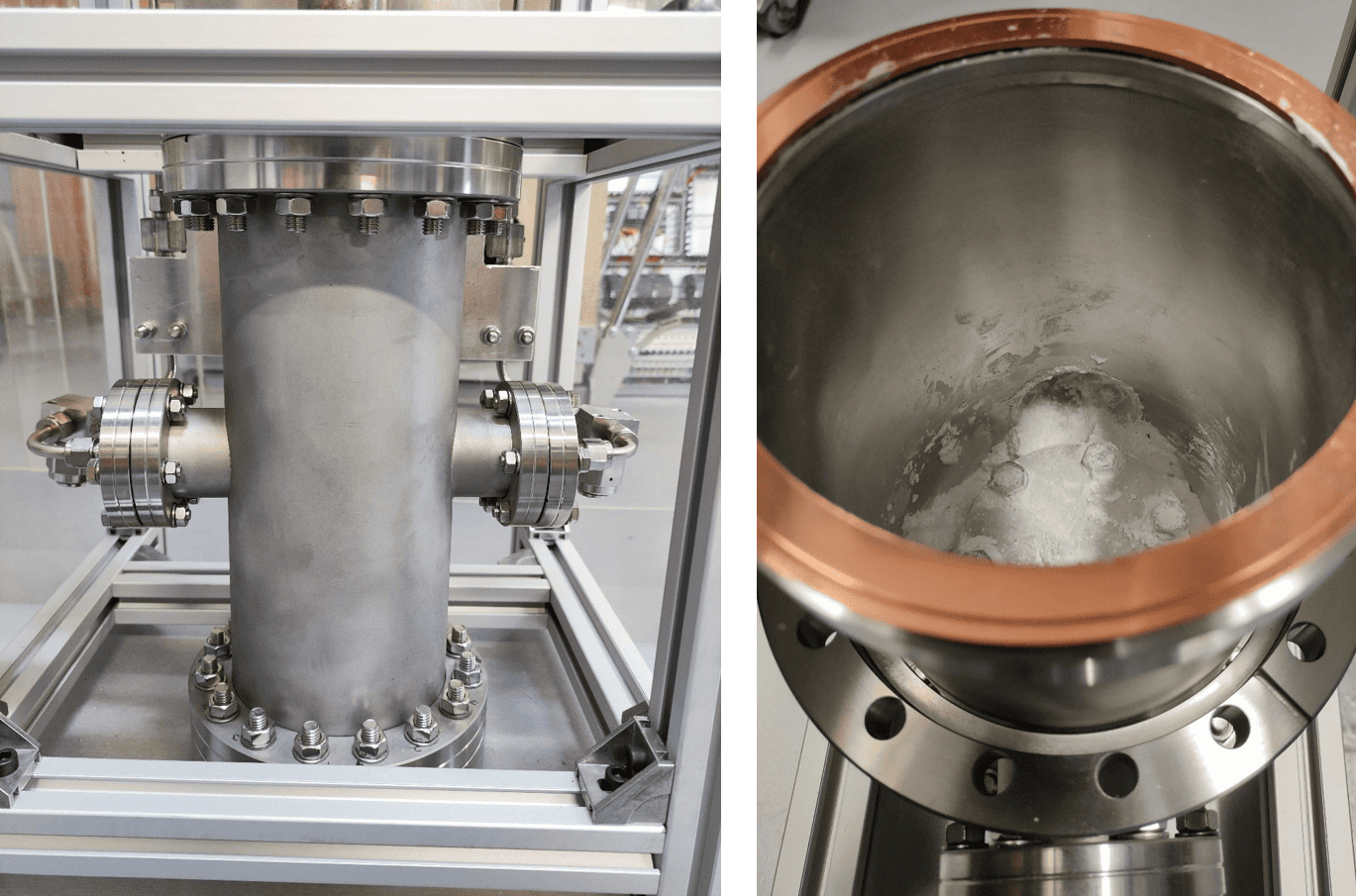}
\caption{Left: cross used as a container for the CaO powder. Right: vessel filled with 0.5 kg of powder. The idea is to leave the upper part and the side filters free so that the $^{37}$Ar can be liberated from the CaO and it can easily flow into the gas system.}
\label{fig:chapter8_TREX-DM_set-up_vessel}
\end{figure}

\begin{figure}[htb]
\centering
\includegraphics[width=1.0\textwidth]{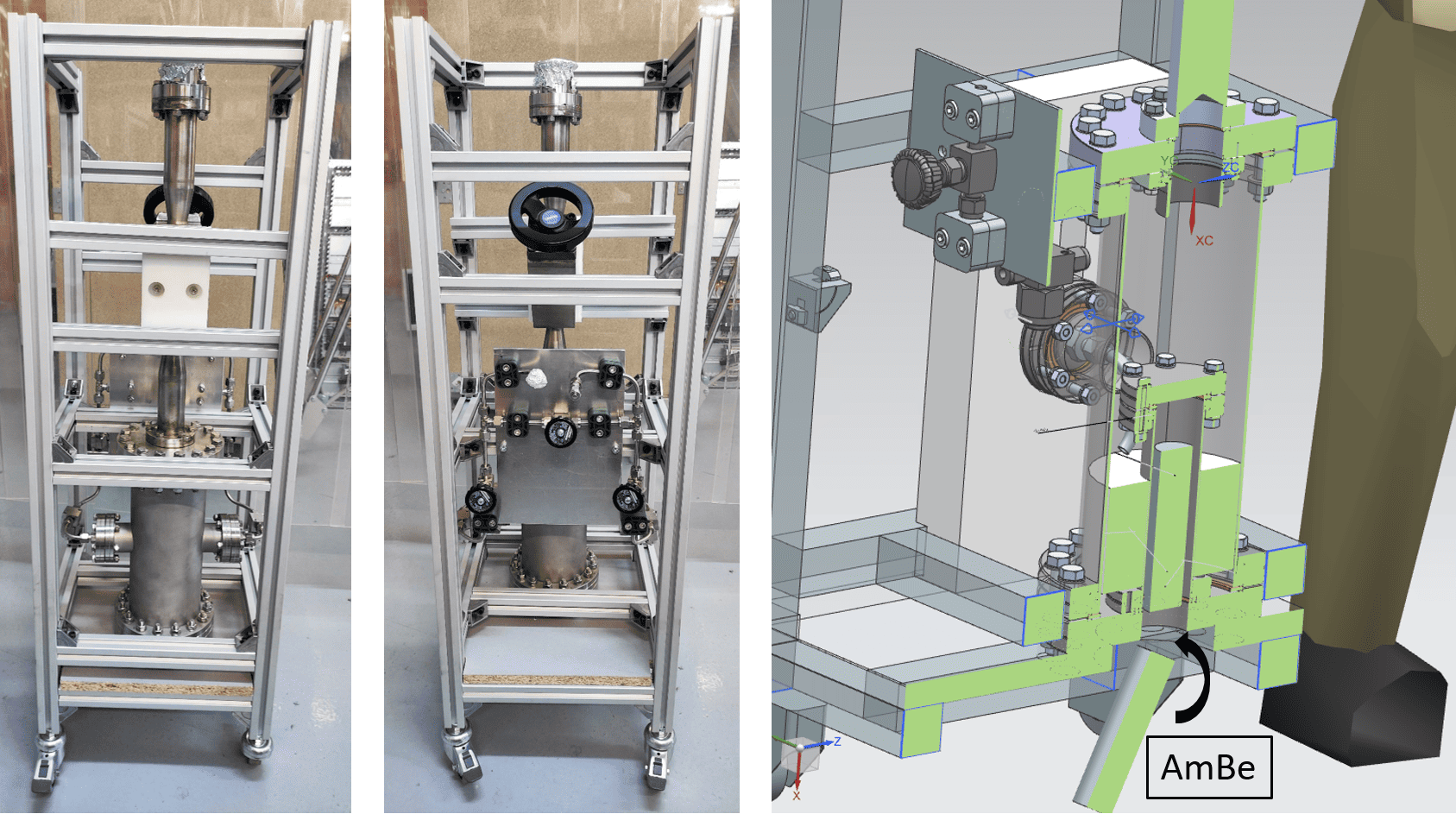}
\caption{Front (left) and back (centre) views of the set-up composed of the vessel, the particulate filters on the sides, the UHV valve on top, the support structure and a small gas panel to perform different operations on the system. Right: 3D view of the set-up with a person for scale. The view shows the vessel filled with CaO, highlighting the hole machined at the bottom of the cross to insert the AmBe source from Figure~\ref{fig:chapter8_CEA_pumping_and_irradiation}.}
\label{fig:chapter8_TREX-DM_set-up_structure}
\end{figure}

The irradiation vessel consisted of a cross CF DN40-CF DN100 with a 40-mm hole machined at the bottom (see Figure~\ref{fig:chapter8_TREX-DM_set-up_vessel} for the vessel and Figure~\ref{fig:chapter8_TREX-DM_set-up_structure} for the 3D showing the machined hole). This design was originally conceived for irradiation at CEA Saclay's SPRE facility, with the machined hole specifically intended to allow insertion of the AmBe source from below. This configuration would position the source along the central axis of the cylindrical distribution, with CaO powder completely surrounding it, thus optimising the solid angle for neutron exposure. However, due to logistical considerations, alternative irradiation facilities were explored, with the Centro Nacional de Aceleradores (CNA) in Sevilla ultimately selected (see Section~\ref{Chapter8_Irradiation_CNA}). Consequently, although the machined hole was incorporated into the design to accommodate the AmBe source, this feature has not been used as originally intended.

The vessel was leak and pressure tested (up to 10 bar) before using it. To prevent powder from entering into the gas system when connected to TREX-DM, two particulate filters (Swagelok 0.5 $\upmu$m filters) were installed on the sides of the cross. An Ultra High Vacuum (UHV) valve was mounted on the top section of the cross to achieve and maintain the required vacuum levels and pressure specifications during operation. In order to pump down to vacuum without the risk of powder entering the system, a 4 $\upmu$m filter was incorporated into the vacuum line.

An external support structure (Figure~\ref{fig:chapter8_TREX-DM_set-up_structure}) was designed and constructed to provide stability to the system, minimising the risk of the CaO buffer volume tipping or rolling over during handling and transportation. This structure also facilitated easier movement of the assembled system.

\vspace{2mm}
\textbf{\normalsize Assembled System}
\vspace{0mm}

The assembled system houses 0.5 kg of CaO powder within the vessel, with the upper section and side filters unobstructed. The goal of this configuration is to ensure that, once produced, the $^{37}$Ar can be efficiently liberated from the CaO matrix and flow freely into the gas system when required.

As shown in Figure~\ref{fig:chapter8_TREX-DM_set-up_structure}, the complete set-up includes the vessel, particulate filters, UHV valve, support structure, and a dedicated small gas panel designed to facilitate various operations. The 3D view illustrates the scale of the set-up and highlights the machined hole at the bottom of the cross, specifically designed to accommodate the AmBe source (previously shown in Figure~\ref{fig:chapter8_CEA_pumping_and_irradiation}).


\section{Irradiation at CNA Sevilla} \label{Chapter8_Irradiation_CNA}

As explained above, the initial idea was to conduct irradiation at CEA Saclay. However, due to the tight schedule of the SPRE service and the logistical challenges associated with transporting the source back and forth, an alternative option was sought, preferably within Spain. Given that irradiation with fast neutrons offers a higher cross-section~\cite{Kelly_2018} and suitable accelerator facilities can provide a high neutron flux (enabling $^{37}$Ar source production within hours rather than days), the Centro Nacional de Aceleradores (CNA) in Sevilla was selected, specifically their HiSPANoS irradiation facility~\cite{HiSPANoS_2024}.

The neutron generation process involves accelerated deuteron beams bombarding various targets. For the initial attempt in April 2024, we were assigned the beam with a thick beryllium target, which produces neutrons through the reaction $^{9}\mathrm{Be}(d,n)^{10}\mathrm{B}$. The resulting neutron spectrum is a continuum with a mean energy of approximately 5 MeV. The irradiation was planned for approximately 6 hours (representing a full beam day), with potential duration adjustments for future iterations. 

The objective was twofold: to produce $^{37}$Ar via neutron irradiation of $^{40}$Ca and to generate the by-product $^{42}$K from $^{42}$Ca. The latter serves as a monitor for reaction success, particularly through its 1525 keV gamma emission. This monitoring approach is essential because $^{37}$Ar itself cannot be directly traced, as its low-energy emissions are absorbed by the stainless-steel vessel.

\subsection{Estimation of \texorpdfstring{$^{37}$Ar}{37Ar} Activity through \texorpdfstring{$^{42}$K}{42K} Production}
\label{Chapter8_Irradiation_CNA_Estimation}

Prior to irradiation, establishing a method to estimate the produced $^{37}$Ar quantity based on $^{42}$K production was crucial. The following numerical approach provides a relation between the activities of these two isotopes.

For any radioactive isotope, the change in the number of atoms per unit time, $\frac{dN}{dt}$, is described by the fundamental equation:

\begin{equation}
    \frac{dN}{dt} = R - \lambda N 
    \label{eq:chapter8_disintegration_rate}
\end{equation}

where $R$ is the production rate of the isotope (assumed to be constant), and $\lambda=\ln2/t_{1/2}$ the decay constant, with $t_{1/2}$ the half-life. Assuming an initial isotope amount of zero ($N(0)=0$), the solution, expressed in terms of the isotope's activity $A$ (defined as $A=\lambda N$), is given by:

\begin{equation}
    A = \lambda N = R \left( 1 - \exp\left( -\frac{\ln 2}{t_{1/2}} t \right) \right)
    \label{eq:chapter8_activity_decay}
\end{equation}

In our specific case, both reactions involve neutron capture, so $R$ is defined by:

\begin{equation}
    R=\int (\phi_n^i-\phi_n^f)\ d\Omega dE
    \label{eq:chapter8_production_rate}
\end{equation}

Here, $\phi_n^i$ represents the initial neutron flux (in n/s/sr/MeV), $\phi_n^f$ is the final neutron flux in the same units, $\Omega$ is the solid angle and $E$ is the energy. This formula indicates that the number of lost neutrons per second is equal to the number of isotope atoms produced per second, given the neutron-to-isotope ratio $1:1$ in the reactions. Considering an interaction cross-section $\sigma$,  target number density $n$, and material thickness $z$, $\phi_n^f$ can be expressed as:

\begin{equation}
    \phi_n^f = \phi_n^i\exp\left( -n\sigma z \right)
    \label{eq:chapter8_final_flux_neutrons}
\end{equation}

For simplicity, the neutron spectrum is approximated as mono-energetic at its mean energy, acknowledging that this provides only an order-of-magnitude estimation. In addition, it is reasonable to assume that $\phi_n^i$ is isotropic, as neutrons are produced in all directions when the deuteron beam hits the beryllium target. With these considerations, Equation~\ref{eq:chapter8_production_rate} can be rearranged as:

\begin{equation}
    A=\phi_n^i \ \Omega \left( 1- \exp\left( -n\sigma z \right)\right) \left( 1 - \exp\left( -\frac{\ln 2}{t_{1/2}} t \right) \right)
    \label{eq:chapter8_activity_final_form}
\end{equation}

This equation remains valid during production time (with the beam on), but after that, activity follows a simple exponential decay, with the initial activity equal to the activity reached at the end of irradiation time. We can now make two approximations:

\begin{itemize}
    \item Since $n\sim N_A\sim 10^{23}$, $z\sim 1$ cm (CaO thickness traversed by neutrons), and $\sigma(^{37}\mathrm{Ar}) \sim 10^{-25}\mathrm{~cm}^2$, $\sigma(^{42}\mathrm{K}) \sim 10^{-26}\mathrm{~cm}^2$ for $^{37}$Ar and $^{42}$K production, respectively\footnote{These cross-sections are taken at 5 Mev from~\cite{nndc_ENDF_2011}.}, then $n\sigma z \sim 10^{-2}-10^{-3}$, and we can approximate $1- \exp\left( -n\sigma z \right)\approx n\sigma z$.
    \item Given $t_{\mathrm{beam}}\approx 6$ h, we will approximate $1 - \exp\left( -\frac{\ln 2}{t_{1/2}} t \right)\approx \frac{\ln 2}{t_{1/2}} t$. This is valid for $^{37}$Ar, because $t_{1/2}\approx 35 \mathrm{~d}\gg t_{\mathrm{beam}}$. For $^{42}$K, $t_{1/2}\approx 12.5\mathrm{~h}\sim t_{\mathrm{beam}}$, but the relative error introduced with respect to keeping the full exponential is $\approx 15$\%, so it is still reasonable given the approximate nature of our approach.
\end{itemize}

With these approximations, the activity for both reactions is:

\begin{equation}
    A \approx \phi_n^i \ \Omega \ n\sigma z \frac{\ln 2}{t_{1/2}} t
    \label{eq:chapter8_activity_final_form_approximated}
\end{equation}

Finally, $n$ can be expressed as $n=N_A \varepsilon\rho /m$, with $\rho$ the target density, $m$ the molecular mass and $\varepsilon$ the abundance of the target in the base material. It is reasonable to assume $\rho$ and $m$ are the same for $^{40}$CaO and $^{42}$CaO. However, as noted in Section~\ref{Chapter8_Introduction}, their natural abundances differ substantially. Consequently, the ratio of activities is:

\begin{equation}
    \frac{A( ^{37}\mathrm{Ar} )}{A( ^{42}\mathrm{K} )}\approx \frac{\varepsilon(^{40}\mathrm{Ca})}{\varepsilon(^{42}\mathrm{Ca})}\times \frac{\sigma(^{37}\mathrm{Ar})}{\sigma(^{42}\mathrm{K})} \times \frac{t_{1/2}(^{42}\mathrm{K})}{  t_{1/2}(^{37}\mathrm{Ar})} \approx \frac{0.9694}{0.0065} \times \frac{0.1\mathrm{~b}}{0.01\mathrm{~b}} \times \frac{0.5\mathrm{~d}}{35\mathrm{~d}} \approx 20
\end{equation}

Although this calculation provides a rough order-of-magnitude estimate, a simulation of the irradiation is still pending. However, this analysis demonstrates that $^{42}$K serves as an effective monitor, because its detection guarantees the presence of $^{37}$Ar.

\subsection{Before Irradiation}
\label{Chapter8_Irradiation_CNA_Before}

As previously explained, the strategy for estimating the produced $^{37}$Ar activity involves monitoring the $^{42}$K gamma emissions. To facilitate this, CNA provided two scintillation detectors: a sodium iodide (NaI) and a lanthanum bromide (LaBr$_3$).

\begin{figure}[htb]
\centering
\includegraphics[width=1.0\textwidth]{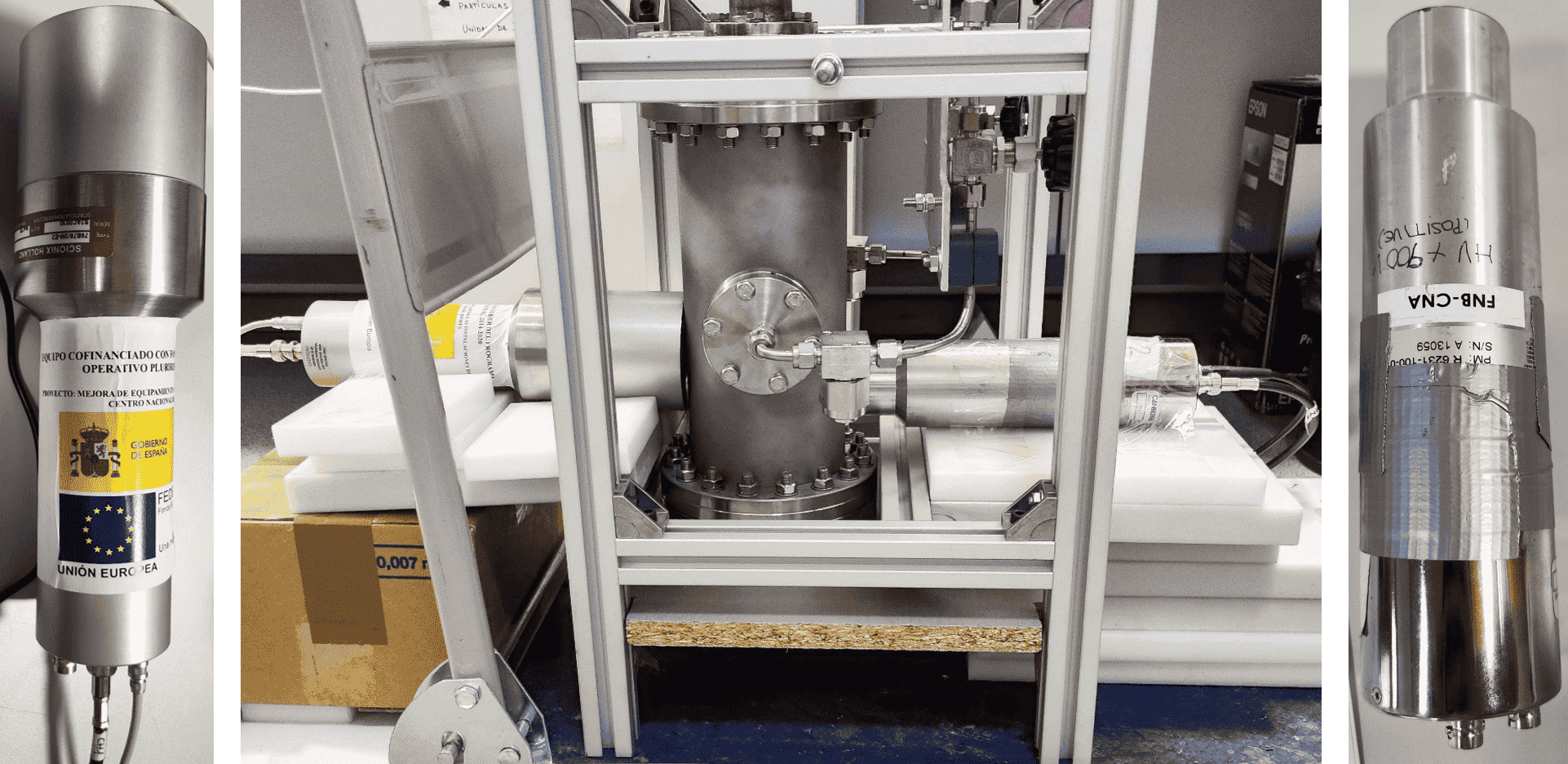}
\caption{Left: NaI detector. Centre: detector configuration for data-taking. Right: LaBr$_3$ detector.}
\label{fig:chapter8_CNA_before_irradiation_set-up}
\end{figure}

The day before irradiation was dedicated to detector calibration and background measurement. Calibration was performed using standard radioactive sources: $^{137}$Cs (662 keV gamma) and $^{60}$Co (gammas at 1173 keV and 1332 keV)\cite{lnhb_table_radionucleides}. A background run lasting 14 hours was taken to establish a baseline for post-irradiation spectrum comparison. The scintillation detectors and their configuration for background data acquisition are illustrated in Figure~\ref{fig:chapter8_CNA_before_irradiation_set-up}. Interestingly, this initial background run proved unexpectedly irrelevant, because during the first hours following irradiation, the recorded activity rates were several orders of magnitude higher than those observed in the background measurements.

\begin{figure}[htb]
\centering
\includegraphics[width=1.0\textwidth]{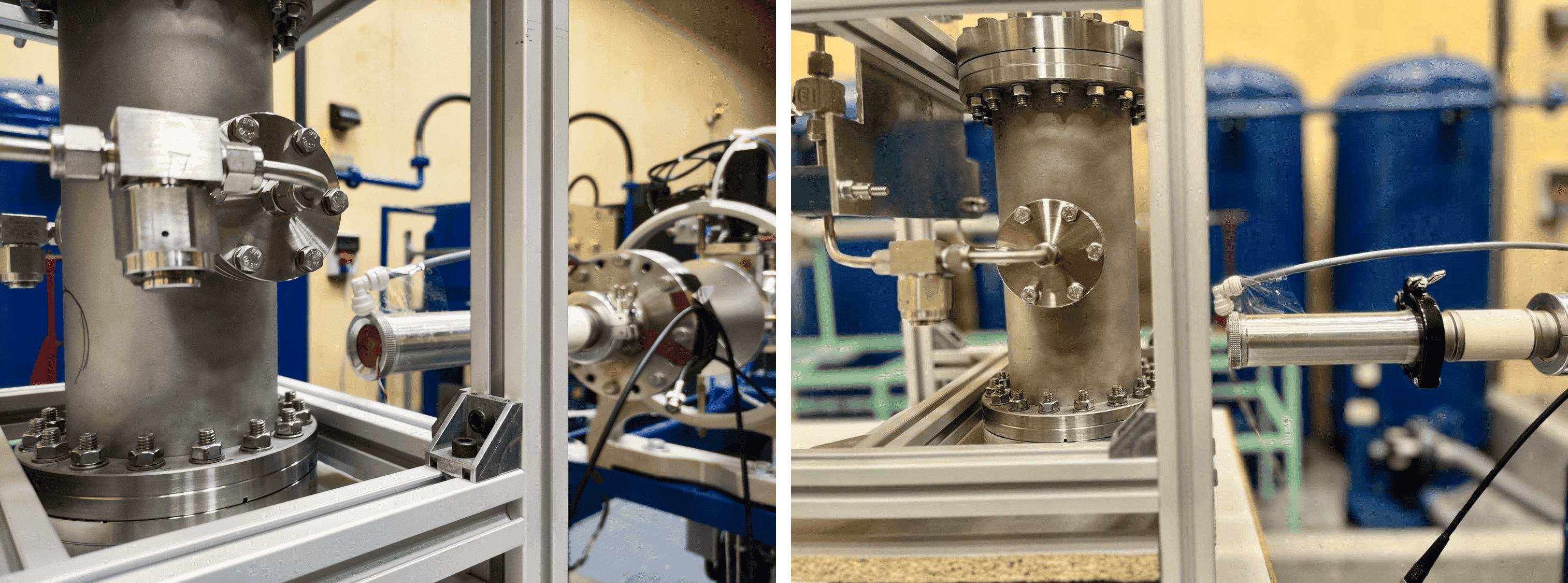}
\caption{Different views of the irradiation set-up at CNA irradiation hall. The distance beam-vessel is 75 mm.}
\label{fig:chapter8_CNA_vessel_beam}
\end{figure}

After these preparatory runs, the experimental set-up was placed directly in front of the beam, and was irradiated during $t_{\mathrm{beam}}\approx 6$ h. Figure~\ref{fig:chapter8_CNA_vessel_beam} provides a detailed view of the position of the vessel during the irradiation process.

\subsection{After Irradiation}
\label{Chapter8_Irradiation_CNA_After}

Following the irradiation process, the radioactivity of the vessel was monitored using a dosimeter. The vessel remained in the irradiation hall until the dose rate decreased below the environmental background radiation level ($<0.2\mathrm{~}\upmu$Sv/h).

\begin{figure}[htb]
\centering
\includegraphics[width=1.0\textwidth]{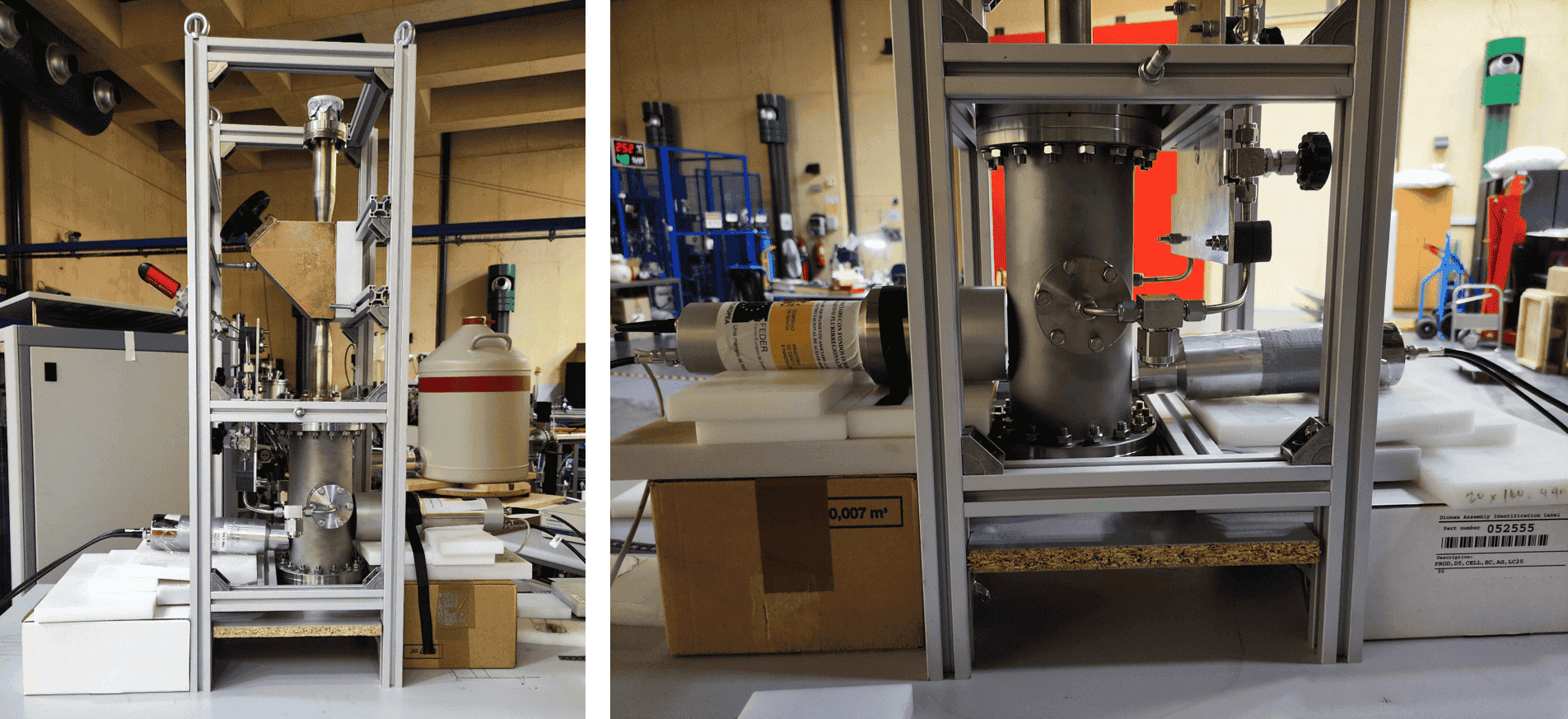}
\caption{Data-taking set-up with the two scintillation detectors after irradiation.}
\label{fig:chapter8_CNA_after_irradiation_set-up}
\end{figure}

To monitor the $^{42}$K gamma emissions, the scintillation detectors were repositioned to their pre-irradiation configuration, as shown in Figure~\ref{fig:chapter8_CNA_after_irradiation_set-up}. Initial calibration was performed using $^{137}$Cs and $^{60}$Co sources, followed by a 16-hour background run that started approximately one hour after irradiation was completed.

Identifying the $^{42}$K gamma proved challenging due to significant material activation. The region of interest at 1525 keV contained numerous unidentified spectral components, presumably arising from stainless-steel vessel activation~\cite{Gregoire_2001}. By tracking the time evolution of the potential $^{42}$K gamma peak location, a faster exponentially decaying component with an approximate half-life of 2.35 h was observed, which could not be definitively attributed to any known activated element. To address this complexity, a double exponential fit was applied to the histogram, including a fast exponential, a slower exponential for the $^{42}$K decay, and a flat background derived from pre-irradiation measurements.

\begin{figure}[htb]
\centering
\includegraphics[width=1.0\textwidth]{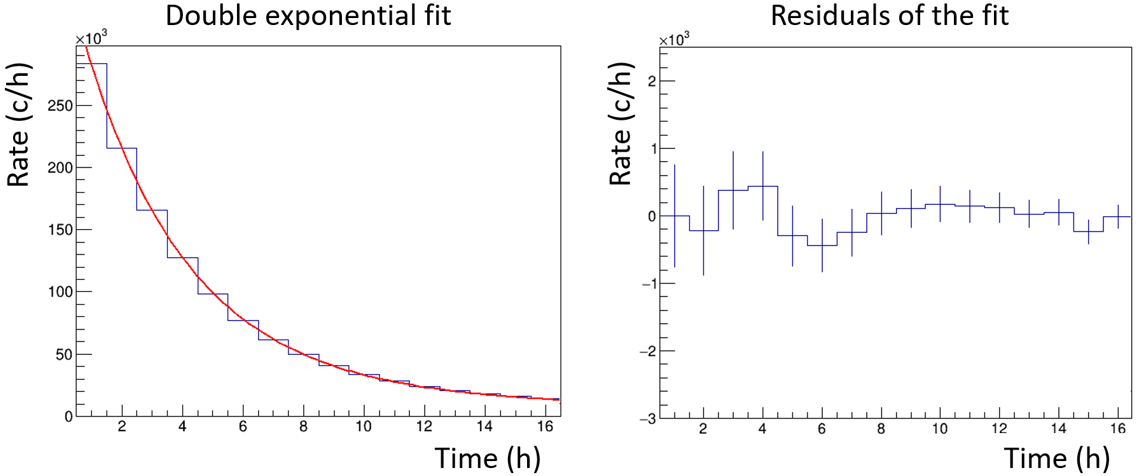}
\caption{Rate evolution after irradiation for the LaBr$_3$ detector in the region of interest around 1525 keV. Left: double exponential fit, with experimental data shown in blue and the fitted curve in red. Right: residuals plot showing the difference between histogram bins and the fit. This enables visualisation of statistical uncertainties, which are otherwise imperceptible in the main plot.}
\label{fig:chapter8_CNA_after_irradiation_fit}
\end{figure}

Figure~\ref{fig:chapter8_CNA_after_irradiation_fit} presents the results of the analysis for the LaBr$_3$ detector. The residuals of the fit are all centred around zero within statistical error margins, which validates the analysis. The statistical error was computed by combining the Poissonian error of each histogram bin with the statistical error of the fit. For context, the LaBr$_3$ detector had a flat background rate of 500 c/h in the region of interest (1525 keV) before irradiation, while the y-axis scale in Figure~\ref{fig:chapter8_CNA_after_irradiation_fit} is $10^3$ c/h. As noted in Section~\ref{Chapter8_Irradiation_CNA_Before}, this indicates that the post-irradiation rates remain orders of magnitude higher than the pre-irradiation background, even after 16 h of data-taking. Consequently, the flat background component in the fit can be considered negligible.

From the $^{42}$K exponential component of the fit, the initial activity of $^{42}$K immediately after irradiation was estimated to be in the range of $100-1000$ Bq. According to the estimation detailed in Section~\ref{Chapter8_Irradiation_CNA_Estimation}, this translates to an approximate $^{37}$Ar activity of $O(1-10)$~kBq. The same analysis applied to the NaI detector corroborated these numbers, thus validating the result.

\section{Low-Energy Calibration in TREX-DM} \label{Chapter8_Calibration_TREX-DM}

Following the initial irradiation of the CaO powder at CNA (see Section~\ref{Chapter8_Irradiation_CNA}), the container was installed in the TREX-DM gas system. The first detector calibration took place at the end of May 2024, approximately 1.5 months after the early April irradiation. This delay, caused by administrative challenges associated with introducing the source to LSC, led to a loss of about 65\% of the initially produced activity.

The injection of $^{37}$Ar gas into the TREX-DM chamber was performed in cycles as follows: 

\begin{enumerate} 
    \item The irradiated container was first filled with Ar-1\%iC$_{4}$H$_{10}$ gas at 2 bar. 
    \item The container was then connected to the chamber gas system, maintained at 0.9 bar. 
    \item After allowing the pressures to equilibrate, the chamber is closed and the container is filled again. Each injection cycle extracted approximately 50\% of the $^{37}$Ar, causing the chamber pressure to increase by about 0.04 bar per cycle. 
\end{enumerate} 

After five injection cycles, the chamber pressure reached 1.1 bar, with an estimated 95\% of the $^{37}$Ar transferred from the container. At that time, the GEM-MM readout plane was not yet operational (see Section~\ref{Chapter7_Installation}), and only the 2.82 keV emission peak could be observed. Nevertheless, this initial calibration validated the procedures developed for transferring the $^{37}$Ar. By analysing the rate of the 2.82 keV peak and considering the 1.5-month delay since irradiation, the initial container activity was inferred to be around 3 kBq, in agreement with the results presented in Section~\ref{Chapter8_Irradiation_CNA_After}.

In the first week of October 2024, a second irradiation of the CaO powder was conducted at CNA, following the same procedure described in Section~\ref{Chapter8_Irradiation_CNA}. Fifteen days after this second irradiation, $^{37}$Ar was injected into TREX-DM using the previously validated method. Due to the shorter time since irradiation, only two injections were performed this time, resulting in the transfer of an estimated 75\% of the container's activity.

\begin{figure}[htb]
\centering
\includegraphics[width=1.0\textwidth]{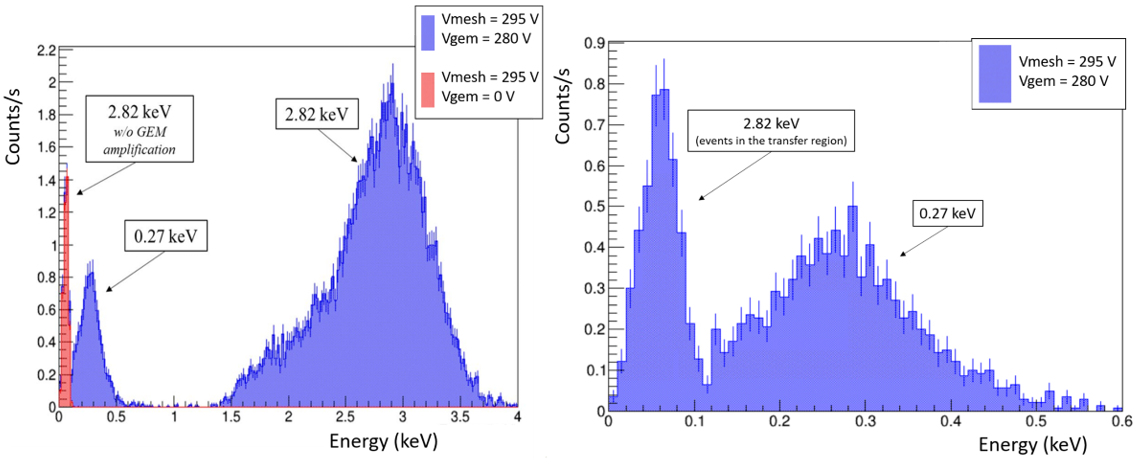}
\caption{Left: measured $^{37}$Ar spectrum in TREX-DM with the GEM-MM readout plane operational. The blue curve corresponds to a run with both $V_{\mathrm{mesh}}$ and $V_{\mathrm{GEM}}$ on, while the red curve shows the spectrum with only $V_{\mathrm{mesh}}$ active. Right: zoomed-in view of the low-energy region, showing an $O(10)$ eV energy threshold. Credit to Álvaro for the plot.}
\label{fig:chapter8_Ar37_spectrum_TREX-DM_and_zoomed_LE_region}
\end{figure}

During this second test, the GEM-MM readout plane installed on the North side was fully operational, enabling the acquisition of a full $^{37}$Ar spectrum. The left part of Figure~\ref{fig:chapter8_Ar37_spectrum_TREX-DM_and_zoomed_LE_region} shows the measured spectrum, where the blue curve corresponds to both $V_{\mathrm{mesh}}$ and $V_{\mathrm{GEM}}$ switched on, while the red curve was taken at the same $V_{\mathrm{mesh}}$ but with $V_{\mathrm{GEM}}$ off. The low-energy peak at 270 eV is clearly visible, marking a significant achievement for the experiment. The leftmost portion of the blue spectrum coincides with the red spectrum because those events deposit their energy in the transfer region between the GEM foil and the micromesh, rather than in the drift region. As expected, a run with only the Micromegas active (red) is equivalent to the portion of the GEM + Micromegas run (blue) in which ionisation occurs in the transfer region. 

The zoomed view of the low-energy region (right part of Figure~\ref{fig:chapter8_Ar37_spectrum_TREX-DM_and_zoomed_LE_region}) reveals that the detector threshold is around the single-electron ionisation energy (26 eV for argon, see Table~\ref{table:chapter3_ionisation_energy_fano_factors}). We leave the detailed analysis of the energy threshold for Section~\ref{Chapter8_Energy_Threshold_Determination}. 

The mean position of the 2.82 keV in the only-Micromegas run, $ (V_{\mathrm{mesh}}, V_{\mathrm{GEM}} ) = (295, 0)$~V, lies at around an equivalent energy of 50 eV. Comparing this with the calibrated 2.82 keV from the GEM + Micromegas run, $ (V_{\mathrm{mesh}}, V_{\mathrm{GEM}} ) = (295, 280)$~V, yields a preamplification factor of $50-60$. This is in line with the preamplification factor measured in the test bench from Section~\ref{Chapter7_TREXDM_Set-Up} (see Table~\ref{table:chapter7_gem_mm_data}), though slightly lower. This discrepancy is most likely because the 2.82 keV peak amplified by the GEM contains numerous saturating events, suggesting the actual position of the peak at those energies should be even higher.

\begin{figure}[htb]
\centering
\includegraphics[width=1.0\textwidth]{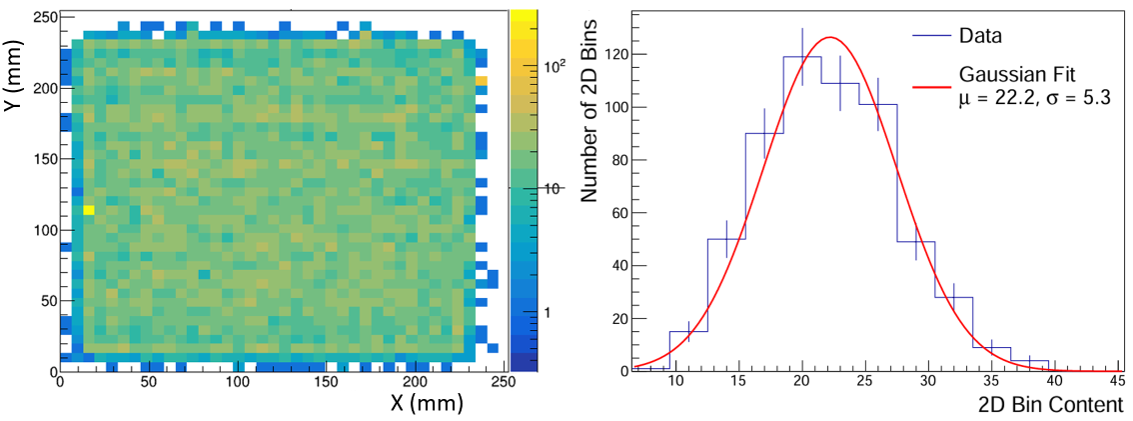}
\caption{Left: 2D hitmap of the $^{37}$Ar calibration spectrum acquired with the GEM-MM system, illustrating the spatial uniformity of the response. Right: 1D histogram of bin counts from the hitmap, fitted with a Gaussian function. The good agreement demonstrates the spatial homogeneity and proper functioning of the TREX-DM v2 detector.}
\label{fig:chapter8_Ar37_hitmap_TREX-DM_spatial_homogeneity}
\end{figure}

Furthermore, the corresponding hitmap in Figure~\ref{fig:chapter8_Ar37_hitmap_TREX-DM_spatial_homogeneity} provides a visual confirmation of the spatial homogeneity achieved during the $^{37}$Ar calibration. To assess this homogeneity more quantitatively, we consider the statistical nature of the bin contents in the 2D hitmap. Each bin records a number of counts, $N$, corresponding to the radioactive decays of $^{37}$Ar, which follow a Poisson distribution with mean $\mu$ and standard deviation $\sqrt{\mu}$. For large $\mu$, this distribution approaches a Gaussian with the same parameters.

By projecting a $20\times 20$ cm$^2$ fiducial cut\footnote{The goal is to remove border effects, such as events absorbed in the PTFE piece surrounding the field cage.} of the 2D hitmap into a 1D histogram (where the x-axis represents the number of counts per bin and the y-axis the number of bins with that count), if the calibration is homogeneous, then we would expect to observe a Gaussian distribution. This is shown in Figure~\ref{fig:chapter8_Ar37_hitmap_TREX-DM_spatial_homogeneity}, together with a Gaussian fit. The good agreement between the data and the fit confirms the expected behaviour. Moreover, the standard deviation extracted from the fit approximately matches $\sqrt{\mu}$, as expected.

This result not only supports the spatial uniformity of the calibration but also provides a fine-grained sensitivity map, confirming the absence of visible dead strips in the TREX-DM.v2 detectors, whereas this issue was recurring in the first TREX-DM.v1 iteration.

\begin{figure}[htb]
\centering
\includegraphics[width=0.75\textwidth]{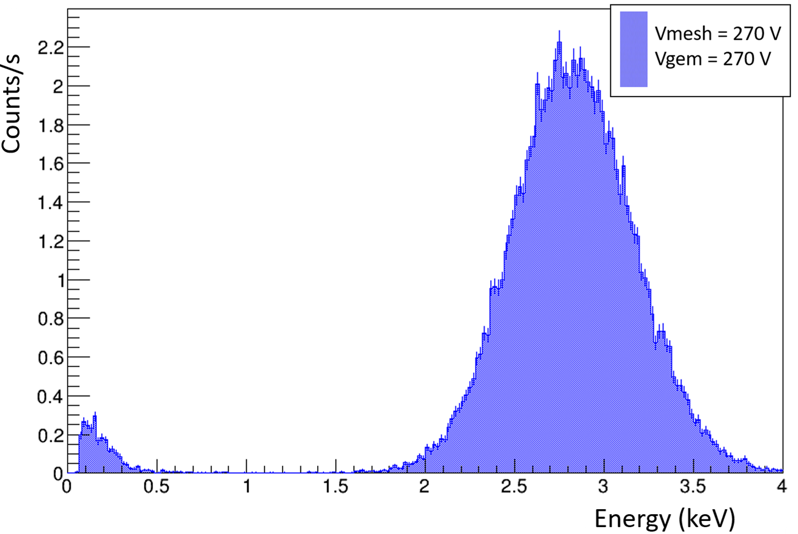}
\caption{$^{37}$Ar spectrum in TREX-DM acquired at conservative operating voltages ($V_{\mathrm{mesh}}=270$ V, $V_{\mathrm{GEM}}=270$ V). The 270 eV peak is still partially visible, demonstrating that practical energy thresholds of $200-300$ eV are achievable even under standard operating conditions. Note that the energy scale, calibrated with the 2.82 keV peak, shows nonlinearities at low energies, highlighting the importance of multi-point low-energy calibration.}
\label{fig:chapter8_Ar37_spectrum_TREX-DM_conservative}
\end{figure}

It should be noted that the run in Figure~\ref{fig:chapter8_Ar37_spectrum_TREX-DM_and_zoomed_LE_region} was taken at higher gain settings ($V_{\mathrm{mesh}}=295$ V, $V_{\mathrm{GEM}}=280$ V) than those used during operation in the three months following the successful replacement of the GEM foil ($V_{\mathrm{mesh}}=270$ V, $V_{\mathrm{GEM}}=270$ V). Achieving these higher gains involved painstaking detector operation (credit to Álvaro and Héctor for all their work), given the numerous technical challenges introduced by the GEM foil (see Section~\ref{Chapter7_Challenges}). Nevertheless, even at the safer operating voltage, the 270 eV peak was partly visible (see Figure~\ref{fig:chapter8_Ar37_spectrum_TREX-DM_conservative}), meaning an energy threshold of $200-300$ eV is realistic even in conservative scenarios. Note that the scale of Figure~\ref{fig:chapter8_Ar37_spectrum_TREX-DM_conservative} fails around the 270 eV peak, as this axis has been calibrated with the 2.82 keV peak, and a higher-energy calibration may not accurately characterise the detector response at $O(10-100)$ eV. This shows the importance of having multiple calibration energies to achieve proper detector characterisation.

Encouraged by the success of this calibration and the productive collaboration with the CNA team, our intention is to repeat this procedure regularly. While the frequency will vary depending on experimental needs, an average of two to three $^{37}$Ar calibrations per year is anticipated.

\subsection{Energy Threshold Determination}
\label{Chapter8_Energy_Threshold_Determination}

\begin{figure}[htb]
\centering
\includegraphics[width=1.0\textwidth]{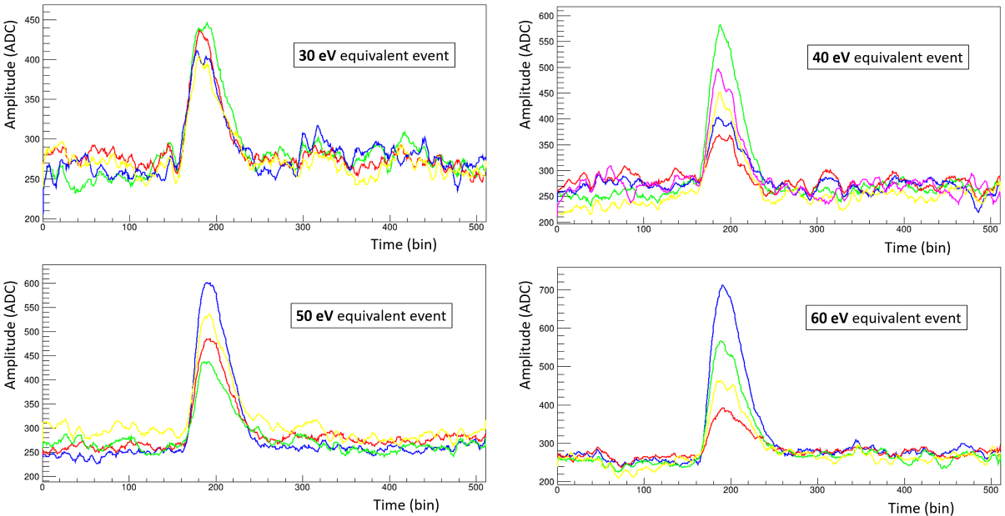}
\caption{Examples of low-energy events detected in TREX-DM with the GEM-MM readout during $^{37}$Ar calibration at high gains. The plots clearly show physical signals that are well distinguishable from the baseline noise, even for events with equivalent energies as low as 30 eV. These pulses demonstrate the exceptional sensitivity achieved with this detector configuration.}
\label{fig:chapter8_Ar37_TREX-DM_low-energy_events}
\end{figure}

Due to the importance of this successful low-energy calibration for this thesis and, in general, for the prospects of TREX-DM, in this section we discuss more carefully the lowest energy threshold achieved in the detector. First, we note in Figure~\ref{fig:chapter8_Ar37_spectrum_TREX-DM_and_zoomed_LE_region} that the 2.82 keV events that deposit their energy in the transfer region have energies equivalent to sub-100 eV. This indicates that even if we did not have a population of events with these energies in the drift region, we would be able to detect them, as the electronics threshold is clearly lower than 100 eV. Examining some of these low-energy events (see Figure~\ref{fig:chapter8_Ar37_TREX-DM_low-energy_events}), we observe that they are indeed physical events well above the baseline that can be confidently identified, even for events with an equivalent energy as low as 30 eV.

To determine the energy threshold rigorously, we analyse the 50 eV peak in the low-energy region from different runs with increasing mesh voltage, tracking the evolution of the energy threshold. In principle, one expects the energy threshold to decrease as voltage increases, with the peak appearing cropped at lower voltages and becoming progressively uncovered as voltage increases. To quantify this effect more rigorously, we fit the low-energy part near the threshold to a cropped-peak model, represented by a Gaussian multiplied by a smoothed step function. Therefore, to determine the energy threshold analytically, we fit the different spectra to the following function:

\begin{equation}
    f(E) = A \ \exp\left( -\frac{(E - \mu)^2}{2\sigma^2} \right) \times \frac{1}{2} \left[ 1 + \mathrm{erf} \left( \frac{E - E_{\text{thr}}}{\sqrt{2} \, \delta} \right) \right]
    \label{eq:chapter8_fit_energy_threshold}
\end{equation}

where $A$, $\mu$ and $\sigma$ are the amplitude, mean position and standard deviation of the Gaussian, $E_\mathrm{thr}$ is the energy threshold, and $\delta$ is a parameter that regulates the smoothness of the transition in the step function, which has been implemented through the error function. All five parameters are left free in the fit. In particular, the energy threshold $E_\mathrm{thr}$ is the key parameter that we seek to estimate. 

\begin{figure}[htb]
\centering
\includegraphics[width=1.0\textwidth]{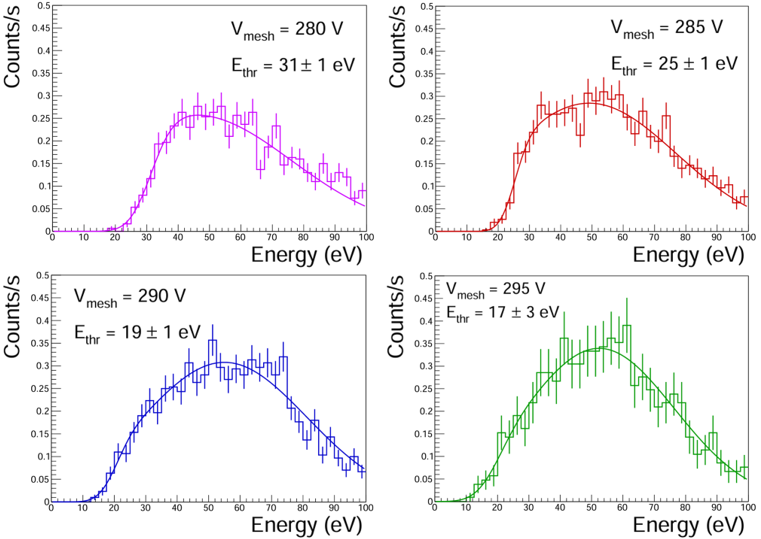}
\caption{Determination of the energy threshold in TREX-DM using a Gaussian multiplied by a smoothed step function (Equation~\ref{eq:chapter8_fit_energy_threshold}) for four $^{37}$Ar calibrations with different mesh voltages. The fitted energy threshold values and their statistical uncertainties are shown for each voltage. The threshold decreases systematically with increasing voltage, confirming the expected behaviour of the detector.}
\label{fig:chapter8_Ar37_TREX-DM_low-energy_threshold_fits}
\end{figure}

The fitted functions for four runs with mesh voltages of 280, 285, 290 and 295 V are shown in Figure~\ref{fig:chapter8_Ar37_TREX-DM_low-energy_threshold_fits}, along with the values of the energy thresholds derived from each fit. Several observations can be made:

\begin{enumerate}
    \item As expected, the energy threshold decreases with higher $V_{\mathrm{mesh}}$.
    \item The statistical error of $E_\mathrm{thr}$ is higher for 295 V. This is reasonable considering the shorter duration of this run (resulting in higher statistical errors in the bins) and the fact that the Gaussian peak is almost complete, leading to greater uncertainty in the step function cut.
    \item The fits at the highest gains (290 and 295 V) have limited sensitivity to $E_\mathrm{thr}$ due to the sparse statistics in the bins near the threshold (the Gaussian is barely cropped), leading to similar values for $E_\mathrm{thr}$.
    \item The method likely contains systematic errors from potential nonlinearities near the threshold, suggesting that the systematic uncertainty may exceed the statistical errors from the fit.
\end{enumerate}

\begin{figure}[htb]
\centering
\includegraphics[width=1.0\textwidth]{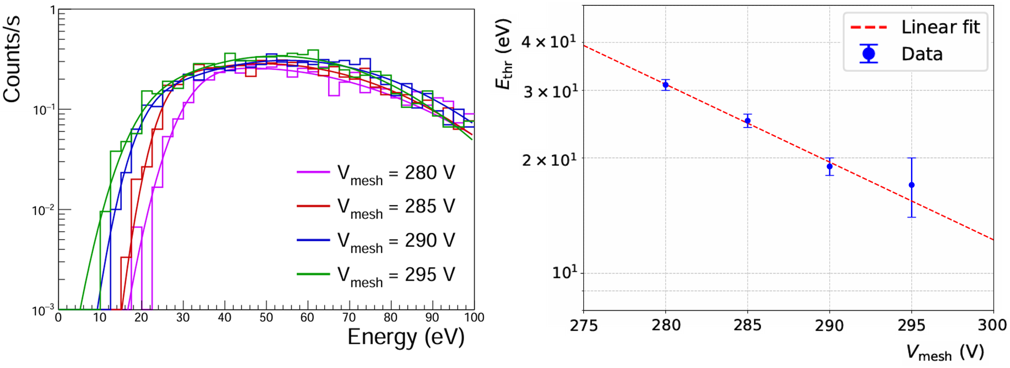}
\caption{Left: overlay of energy spectra from four runs at different mesh voltages (280, 285, 290 and 295 V) with their corresponding fits, shown on a logarithmic scale to highlight the threshold behaviour. Right: energy threshold in logarithmic scale versus mesh voltage, showing a fit describing the linear dependence between $\ln(E_\mathrm{thr})$ and $V_\mathrm{mesh}$. Note that the measurement at the highest gain (295 V) deviates slightly from this trend (but it is covered by the error bar) due to reduced fit sensitivity when the Gaussian peak appears nearly complete. The thresholds are near the single-electron ionisation energy of argon (26 eV), demonstrating the exceptional low-energy sensitivity achieved with TREX-DM.}
\label{fig:chapter8_Ar37_TREX-DM_low-energy_threshold_determination}
\end{figure}

The progression of the energy threshold with increasing voltage is clearly illustrated in Figure~\ref{fig:chapter8_Ar37_TREX-DM_low-energy_threshold_determination}, where we plot the four runs and their fits on the same graph, using a logarithmic y-axis to emphasise the step function. Bin errors are omitted to maintain clarity. Additionally, the figure shows the energy threshold (in logarithmic scale) with statistical errors derived from the fit as a function of applied voltage. Given the dependence $E_\mathrm{thr}\sim 1/G$, where $G$ represents the amplification gain of the system, and considering that $G\sim \exp(aV_\mathrm{mesh})$, with $a$ being some constant parameter, one would expect $E_\mathrm{thr}\sim \exp(-aV_\mathrm{mesh})$. Indeed, as demonstrated in Figure~\ref{fig:chapter8_Ar37_TREX-DM_low-energy_threshold_determination}, a linear fit on the coordinates $(V_\mathrm{mesh}, \ln(E_\mathrm{thr}))$ successfully describes the observed data, confirming the expected behaviour.

Even considering that the method likely contains systematic uncertainties (primarily due to nonlinearities near the threshold), this analysis allows us to assert that the energy threshold for the highest gain achieved (and even for lower gains) is situated around the single-electron ionisation energy. One potential systematic effect is that we are determining the threshold using 2.82 keV events in the transfer region, which may have a different event structure than true $20-50$ eV events. Consequently, while we may be sensitive to these 2.82 keV events, the efficiency could differ for genuine sub-100-eV events in the drift region.

All in all, this constitutes an outstanding result for TREX-DM. Future work will focus on obtaining more precise calibrations at these $O(10)$ eV energies, enabling a more accurate determination of the energy threshold. Some test set-ups are being developed, with a particularly promising approach involving the use of a UV laser for exact measurements at these ultra-low energies.

%% file: CHAPTERS/Chapter9.tex
\chapter{Sensitivity} \label{Chapter9_Sensitivity}

{
\lettrine[loversize=0.15]{I}{n} this chapter, we present how the results discussed in this thesis impact the sensitivity of TREX-DM and give an overview of different scenarios the experiment could achieve on various timescales. It also serves as a summary of the status of the experiment and future prospects.
%
\section*{}
\parshape=0
\vspace{-20.5mm}
}

\section{Method for Sensitivity Calculation}
\label{Chapter9_Sensitivity_Method}

We start by outlining how the sensitivity projections are obtained. We note that Equation~\ref{eq:chapter2_expected_signal_experiment} can be written in the form:

\begin{equation}
    S = \mathcal{E} \times \sigma_n \times \mathcal{I},
    \label{eq:chapter9_sensitivity}
\end{equation}

where $\mathcal{E} = M \times T$ is the exposure of the experiment (mass $M$ times live time $T$), $\sigma_n$ is the Spin-Independent WIMP-nucleon cross-section, and $\mathcal{I}$ is defined as the integrated interaction rate per unit mass and per unit cross-section. 

The quantity $\mathcal{I}$ is obtained by integrating the recoil energy spectra (previously converted to electron-equivalent energy by applying a quenching factor) between $E_\mathrm{thr}$ and $E_\mathrm{max}$\footnote{$E_\mathrm{max}$ is chosen as $E_\mathrm{thr}+ 1\mathrm{~keV}$, because this is the typical energy range relevant for $O(1)$ GeV WIMPs, as shown in Figure~\ref{fig:chapter2_WIMP_scattering_rate}. However, this value should be optimised for each WIMP mass and background rate.}, and summing over all nuclei in the gas mixture with their respective abundances, but leaving out the interaction cross-section.

Next, given a background rate $b$ in units of dru (c/keV/kg/day), the expected number of background events $B$ in a given energy interval is calculated by scaling $b$ with the exposure $\mathcal{E}$, assuming a flat background. For low-background conditions, the background-only hypothesis is modelled using a Poisson distribution with mean $B$, denoted $\mathcal{P}(N, B)$. In high-background regimes (i.e., large $B$), a Gaussian approximation with mean $B$ and standard deviation $\sqrt{B}$ may be employed.

In this work, the projections from Figure~\ref{fig:chapter9_exclusion_plot} are obtained by estimating the minimum number of signal events $S$ required to reject the background-only hypothesis at 90\% confidence level (corresponding to a significance of 10\%). This is done by solving:

\begin{equation}
    \sum_{N=0}^{S + B} \mathcal{P}(N, B) > 0.9
    \label{eq:chapter9_discovery_limit}
\end{equation}

which ensures that observing at least $S + B$ events due solely to background fluctuations would occur with less than 10\% probability. This procedure therefore defines a \textit{discovery limit}, as it identifies the smallest signal strength $S$ that would lead to the background-only hypothesis being rejected at the stated confidence level.

However, for setting \textit{exclusion limits}, the objective is reversed. Instead of rejecting the background-only hypothesis, one seeks to exclude signal hypotheses. That is, given an observed number of events $N$ (either taken as real data or drawn from Monte Carlo simulations assuming the background-only hypothesis), one determines the largest signal strength $S$ for which the background-plus-signal hypothesis is still consistent with the observed $N$ at the 90\% confidence level. Mathematically, this is expressed as:

\begin{equation}
    \sum_{n=0}^{N} \mathcal{P}(n, S + B) < 0.1
    \label{eq:chapter9_exclusion_limit}
\end{equation}

This yields a 90\% confidence level upper limit on the signal strength $S$ for a given $N$. When $N$ is not fixed by data, but simulated repeatedly under the background-only hypothesis, the distribution of such upper limits can be constructed. The sensitivity curve is then defined as the median upper limit over this ensemble. In practice, it has been shown that this median coincides with the upper limit obtained using the Asimov set~\cite{luca_lista_2023}, where $N = B$ is taken as a representative observation.

For large expected counts $B$, both the discovery and exclusion problems can be approximated using Gaussian statistics. Under this regime:

\begin{itemize}
    \item The background-only distribution is modelled as $\mathcal{N}(B, \sqrt{B})$
    \item The background-plus-signal distribution is approximated as $\mathcal{N}(B + S, \sqrt{B + S})$
\end{itemize}

where $\mathcal{N}(\mu,\sigma)$ denotes a normal distribution with mean $\mu$ and standard deviation $\sigma$.

At 90\% confidence (1.28$\sigma$ in the Gaussian limit, but roughly 1$\sigma$ for illustrative purposes), the discovery threshold corresponds to a signal satisfying:

\begin{equation}
    S \approx \sqrt{B} \label{eq:chapter9_discovery_limit_approximation}
\end{equation}

so that the observed count exceeds $B$ by approximately one standard deviation. Similarly, for the exclusion limit derived from the signal-plus-background hypothesis at $N = B$:

\begin{equation}
    S\approx \sqrt{S + B} \approx \sqrt{B} \label{eq:chapter9_exclusion_limit_approximation}
\end{equation}

which again yields $S \approx \sqrt{B}$ to leading order. Thus, in the large-$B$ regime, the discovery and exclusion limits approximately coincide. In all scenarios A$-$G presented in Figure~\ref{fig:chapter9_exclusion_plot}, the total number of background counts considered is $B\sim 100-5000$. For this reason, although the sensitivity curves are computed as discovery limits at 90\% confidence level, they closely approximate the corresponding exclusion limits.

Once $S$ is found, we can use it together with the estimated $\mathcal{I}$ and the known exposure $\mathcal{E}$ to calculate the corresponding cross-section $\sigma_n$ through Equation~\ref{eq:chapter9_sensitivity}. Repeating this process for different WIMP masses yields a set of $(m_\chi, \sigma_n)$ points that form the sensitivity curve.

\section{Sensitivity Projections}
\label{Chapter9_Sensitivity_Projections}

\begin{figure}[htb]
\centering
\includegraphics[width=1.0\textwidth]{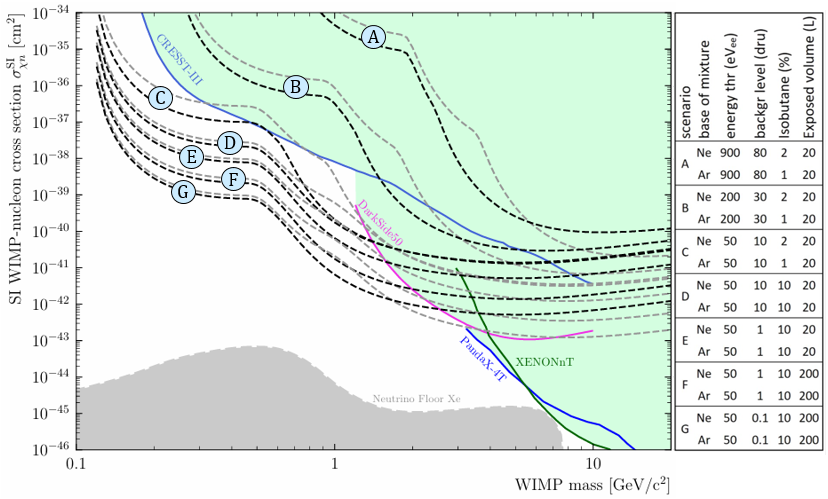}
\caption{Exclusion plot for the WIMP-nucleon SI elastic scattering cross-section vs. WIMP mass, with a focus on the low-mass region. Current bounds from leading experiments are included, together with different TREX-DM scenarios (each with 1~year exposure). Neon-based mixtures are shown in black and argon-based mixtures in grey. For clarity reasons, only the strongest bounds are included. The neutrino floor for a xenon target is also plotted. The references are the same as in Figure~\ref{fig:chapter2_WIMP_exclusion_plot}. Plots prepared using~\cite{wimp_limits_plotting_tool}.}
\label{fig:chapter9_exclusion_plot}
\end{figure}

Figure~\ref{fig:chapter9_exclusion_plot} shows the projected TREX-DM sensitivity curves for SI WIMP-nucleon scattering, assuming standard halo and astrophysical parameters. Several existing constraints from leading experiments are overlaid (see Section~\ref{Chapter2_Direct_Searches_Experiments}). The black and grey curves represent, respectively, neon- and argon-based mixtures under various experimental scenarios, each corresponding to different threshold, background, gas composition, and exposure assumptions. For a broader discussion of how these parameters affect direct WIMP searches, see Section~\ref{Chapter2_Direct_Searches_Sensitivity}.

\vspace{2mm}
\textbf{\normalsize Scenario A}
\vspace{0mm}

As of October 2022, before the experiment was dismantled and moved to Lab2500, TREX-DM achieved an energy threshold of about 900~eV (see Figure~\ref{fig:chapter5_performance_comparison_detectors}) and a background level of roughly 80~dru (see Figure~\ref{fig:chapter6_2022_semi-sealed_open-loop}), using a Ne-2\%iC$_4$H$_{10}$ mixture. In Figure~\ref{fig:chapter9_exclusion_plot}, this corresponds to scenario A. Although it does not enter unexplored regions of the parameter space, it demonstrates the current baseline performance.

\vspace{2mm}
\textbf{\normalsize Lowering the Threshold and Background (Scenarios B and C)}
\vspace{0mm}

As argued in Section~\ref{Chapter2_Motivation}, the sensitivity of Dark Matter experiments is shifting towards lower masses, which require a lower energy threshold. Chapter~\ref{Chapter7_GEM-MM} and the in-situ tests in TREX-DM described in Section~\ref{Chapter8_Calibration_TREX-DM} show that adding a GEM preamplification stage to the microbulk Micromegas can significantly reduce the energy threshold. Thresholds on the order of single-electron ionisation energy have been achieved in high-gain scenarios, and $O(100)$ eV in conservative operation. Improvements in stability are expected to bring routine detector operation close to the lowest threshold achieved.

At the same time, background in TREX-DM is currently dominated by surface contamination on the inner surfaces of the active volume. The PTFE pieces covering the field cage, the field cage itself, the cathode, and (to a lesser extent) the detector surfaces contribute. By applying a fiducial cut to select an inner region of the detector, the alphas directionality study described in Section~\ref{Chapter6_Alpha_Directionality_Studies} indicates that the primary background in the inner region originates from the cathode, with a secondary contribution from the detector surfaces. A clear strategy has emerged to reduce this background by replacing the existing cathode with one that has a smaller material budget, with copper wires being a natural candidate. In parallel, AlphaCAMM has started to screen material samples, including the mylar cathode previously used in TREX-DM, and is also serving as a test bench for surface contamination reduction. In fact, AlphaCAMM’s own background exceeds model expectations by a factor of 10, and is at the moment testing the idea of a wire-based cathode. Iterations of this design are expected to yield a polished version that can be implemented in TREX-DM. Such improvements are anticipated to lower TREX-DM's background to the order of 1 dru, especially considering that $^{40}$K, one of the main contributors in the background model, was reduced by at least a factor of 3 when transitioning from TREXDM.v1 to TREXDM.v2 detectors due to enhanced radiopurity (see Section~\ref{Chapter5_Description_Background_Model} and Section~\ref{Chapter5_Description_Detectors}).

Scenarios B and C combine these improvements in threshold and background. In particular, scenario B is considered a conservative projection, achievable in the coming months but still not reaching unexplored parameter space. By contrast, scenario C already extends into new regions and could also be realised in the near term. To facilitate these improvements, a new field cage design is being developed to enhance the detector's robustness and stability, while also reducing background through an optimised material budget. In addition, other avenues are under investigation to address detector stability, notably the implementation of resistive Micromegas technology. In resistive Micromegas, a resistive layer or pads form the anode, which are capacitively coupled to the readout pads via a ceramic substrate. The addition of resistive and ceramic layers between the amplification gap and the readout strips has shown improved stability against spark formation in similar applications~\cite{Attie_2013}. In TREX-DM, this approach aims to protect the readout channels, minimise the loss of strips, and enable operation at higher voltages. It should be noted that the realisation of these scenarios depends on the absence of unforeseen background sources at lower energies, a common challenge in ultralow-threshold WIMP experiments. While the GEM, constructed from the same radiopure materials as the microbulk Micromegas, is not expected to significantly contribute to the background, the overall background in the extended low-energy region remains to be characterised.

\vspace{2mm}
\textbf{\normalsize Gas Mixture Optimisation (Scenarios D and E)}
\vspace{0mm}

Optimising the gas mixture plays a crucial role in maximising sensitivity. TREX-DM was originally designed to search for WIMPs with masses below 10 GeV using neon as its primary gas, or argon depleted in $^{39}$Ar. Sensitivity estimates indicate that neon-based mixtures provide better performance compared to argon. However, during the commissioning phase, the ready availability of argon (not underground-sourced) has favoured its use during a significant portion of data-taking.

Equally important is the fraction of isobutane used. An increased isobutane content enhances the sensitivity to WIMPs below 1~GeV, as lighter target nuclei (such as hydrogen from the quencher) lead to more energetic recoils. In standard operation, as detailed in Section~\ref{Chapter5_Description_Detectors}, 1\% iC$_4$H$_{10}$ is used for argon and 2\% for neon to comply with flammability constraints.

The new isolated location at Lab2500 now offers the opportunity to work with mixtures containing a higher percentage of isobutane. Measures are underway to adapt both the gas system and the site to meet flammable gas regulations, with the aim of using mixtures with up to 10\%iC$_4$H$_{10}$ for either argon or neon. This increase in the quencher fraction does not significantly compromise the target mass, as it remains small compared to the base noble gas. Furthermore, a higher quencher concentration can improve the overall stability of the detector (see Chapter~\ref{Chapter3_Gaseous_Detectors} for a discussion on the effects of the quencher on the detection processes), potentially enabling operation at higher voltages and allowing for increased gains.

Scenarios D and E reflect these changes. Scenario D assumes the improvements of scenario C but with an increased isobutane content, while scenario E further incorporates an additional reduction in background levels. Notably, the sensitivity advantage that neon exhibited in scenarios A to C is diminished when the isobutane fraction is increased, resulting in similar performance for both neon- and argon-based mixtures. This outcome is realistic in the medium term, as tests to characterise 10\%iC$_4$H$_{10}$ mixtures are currently being conducted in Zaragoza, and the plans to adapt the TREX-DM site to handle flammable mixtures are advanced.

\vspace{2mm}
\textbf{\normalsize Long-Term Prospects (Scenarios F and G)}
\vspace{0mm}

Finally, scenarios F and G explore long-term goals. Both assume a factor of 10 increase in detector volume (to about 200~L total sensitive volume) and 1~year of exposure. Scenario G additionally envisions a background of 0.1~dru. Achieving these benchmarks would require building a larger vessel and new scaled up detectors, further refining radiopurity and background levels (with AlphaCAMM as a crucial screening tool), and consolidating the various stability upgrades mentioned above.


%% file: CHAPTERS/Chapter10.tex
\chapter*{Summary and Conclusions} \label{Chapter10_Summary_Conclusions}
\addcontentsline{toc}{chapter}{Summary and Conclusions}

In this thesis, we have discussed in depth TREX-DM, an experiment that aims to detect low-mass Weakly Interacting Massive Particles (WIMPs), and its promising developments regarding energy threshold reduction to be competitive in this region of the Dark Matter parameter space. The work presented here encompasses not only the theoretical foundations that motivate the search for Dark Matter particles but also the technological innovations implemented to enhance the sensitivity of the detector. Through improvements in background reduction, energy threshold optimisation, and low-energy calibration techniques, TREX-DM demonstrates significant potential as a competitive experiment in the search for sub-GeV WIMPs, a region that remains largely unexplored by conventional direct detection experiments.

The theoretical foundation of this work begins with substantial observational evidence for the existence of Dark Matter as a cold, non-baryonic component. As examined in Chapter~\ref{Chapter1_Dark_Matter}, the analysis of galactic rotation curves, galaxy cluster dynamics, gravitational lensing effects, and precision measurements of the CMB all converge towards the conclusion that Dark Matter constitutes a significant portion of the mass-energy budget of the Universe and plays a crucial role in structure formation. The chapter also reviews the primary theoretical candidates for Dark Matter, including WIMPs, axions, sterile neutrinos, and primordial black holes. The discussion briefly touches on modifications of gravity as an alternative explanation for the observed phenomena, though the focus remains on particle-based explanations.

Building upon this foundation, Chapter~\ref{Chapter2_WIMPs} provides a deeper examination of WIMPs as one of the best-motivated candidates for Dark Matter. While the canonical WIMP scenario, often referred to as the "WIMP miracle", provides an elegant connection between the weak scale and the observed Dark Matter abundance, there is substantial theoretical interest beyond this paradigm. By relaxing the condition $g_\chi \sim g_\mathrm{weak}$, a wider mass range becomes accessible, particularly extending towards the sub-GeV region. This broadening of the parameter space has renewed experimental interest in developing detection techniques sensitive to low-energy nuclear recoils, as conventional approaches face challenges in this region. Despite extensive experimental efforts across direct detection, indirect searches, and collider experiments, conclusive evidence for WIMPs remains elusive, highlighting the need for innovative detection strategies.

The technology used in this work is grounded in the fundamental processes governing gaseous detectors, discussed in Chapter~\ref{Chapter3_Gaseous_Detectors}. Here, we have analysed interactions involving photons, charged particles, and neutral particles, as well as the mechanisms leading to charge generation: primary ionisation, electron drift under the influence of electric fields, and charge amplification through avalanche multiplication. These factors, combined with the critical choice of gas mixture, collectively determine the detector gain and energy resolution. 

Among the variety of gaseous detectors available, Time Projection Chambers (TPCs) equipped with Micropattern Gaseous Detectors (MPGDs), particularly Micromegas, offer significant advantages for Dark Matter searches, as established in Chapter~\ref{Chapter4_TPCs_Micromegas}. These detectors provide three-dimensional event reconstruction capabilities with high spatial resolution, enabling effective background discrimination. Additionally, their intrinsic radiopurity makes them particularly suitable for rare event searches, as material selection and fabrication processes have been optimised to minimise background contributions.

The TREX-DM experiment leverages Micromegas technology in its search for low-mass WIMPs. Chapter~\ref{Chapter5_TREXDM} presents a comprehensive description of the experiment, detailing its design philosophy and operational principles. The experiment consists of a radiopure vessel housing a gaseous medium, surrounded by an optimised shielding system to reduce external backgrounds. The readout system employs the largest microbulk Micromegas ever built, developed through multiple iterations (TREXDM.v1 and TREXDM.v2) to enhance spatial and energy resolution and lower the energy threshold. The experimental infrastructure is supported by specialised subsystems, including a data acquisition system designed to handle high channel density, dedicated calibration sources to ensure stable and accurate energy measurements, a gas system to maintain a homogeneous detection medium, and a slow control system to monitor operational parameters. We have also discussed the technical challenges encountered during the development and commissioning phases, such as electronic noise, leakage currents, and gas leaks. The chapter concludes with drift velocity measurements in Ar-1\%iC$_{4}$H$_{10}$ mixture that serve as a benchmark for analysis validation.

A significant focus of this research has been the progressive reduction of background levels in TREX-DM, which have seen remarkable improvements from initial rates of approximately 1000 dru to current levels of 80-100 dru. As detailed in Chapter~\ref{Chapter6_Radon_problem}, this progress resulted from a comprehensive strategy targeting two primary sources: active radon and surface contamination. The initial high background was predominantly attributable to radon emanation from gas purifiers, prompting the development of specific mitigation techniques. After evaluating several strategies, the implementation of an open-loop operation with a low flow rate capable of compensating for leaks and outgassing (what we called the semi-sealed open-loop approach) proved most effective, reducing active radon to negligible levels compared to surface contamination.

Surface contamination from $^{222}$Rn progeny, particularly $^{210}$Pb and its decay products, was identified as the second major background source. Addressing this challenge led to the development of the AlphaCAMM detector, which uses the directional properties of alpha particles to study contamination on sample surfaces, and will be a key tool for material screening. Alpha directionality studies applied to TREX-DM revealed that most of the surface contamination originates from the cathode. On the other hand, simulation studies confirmed the experimentally observed $1:1$ correspondence between low-energy background and high-energy events. While background reduction remains challenging, there are promising solutions being developed, among which a wire-based copper cathode stands out. This is the most viable near-term approach to reduce the surface susceptible to contamination from radon progeny exposure.

One of the main results of this thesis is the enhancement of energy threshold through the addition of a GEM stage on top of the microbulk Micromegas detectors used in TREX-DM. The initial motivation for incorporating a GEM preamplification stage was to enhance the sensitivity of the detector to low-energy events, which is essential for detecting low-mass WIMPs. Chapter~\ref{Chapter7_GEM-MM} presents a systematic investigation of this combined detector configuration, beginning with small-scale prototype tests and progressing to a full-scale detector from the TREXDM.v2 batch. The small-scale tests validated this approach, demonstrating significant extra gain for different gas pressures using an Ar-1\%iC$_{4}$H$_{10}$ mixture. The successful implementation in the full-scale TREX-DM setup confirmed these results, achieving extra gain factors approaching 100 compared to only-Micromegas operation, and therefore showing potential for substantial energy threshold reduction. 

The result that solidifies this improvement in energy threshold is the successful calibration of the TREX-DM detector with a $^{37}$Ar source, as detailed in Chapter~\ref{Chapter8_Ar37}. This gaseous calibration source provides homogeneous, well-defined low-energy peaks at 0.27 keV and 2.82 keV that align with the region of interest for low-mass WIMP searches. The development process began with initial feasibility tests at CEA Saclay, validating the $^{37}$Ar production method through neutron irradiation of CaO powder. Building on these preliminary results, a dedicated set-up was developed for TREX-DM to house the powder and operate the calibration source. The production method was subsequently replicated at a suitable neutron facility such as CNA Sevilla. The reliability of this approach to generate a calibration source was confirmed by monitoring the by-product $^{42}$K as an indirect measure of $^{37}$Ar activity.

The integration of the $^{37}$Ar calibration source into TREX-DM was preceded by an intervention to install the GEM foil on one of the detector sides. Following successful commissioning of the combined GEM + Micromegas detector, the system was calibrated with $^{37}$Ar, registering a complete spectrum (0.27 and 2.82 keV peaks) thanks to the extra amplification provided by the GEM. Data analysis confirmed consistent and measurable signals within the target energy range, achieving remarkable energy thresholds at the level of single-ionisation energies (26 eV in argon) for the highest gains achieved, and thresholds of $O(200-300)$ eV at more conservative voltage combinations. This development represents a crucial milestone, demonstrating the experiment is capable of reliably detecting and interpreting low-energy events, an essential condition to probe the low-mass WIMPs parameter space.

Finally, Chapter~\ref{Chapter9_Sensitivity} collects all the results and projections related to background reduction and energy threshold improvement. Here, the sensitivity prospects of TREX-DM are discussed, presenting a comprehensive roadmap for the future of the experiment. With current background levels of 80-100 dru and conservative energy thresholds below 300 eV achieved through GEM preamplification, the experiment is well-positioned for further advances. Near-term improvements include background reduction through the implementation of a wire-based cathode (which will be first tested in AlphaCAMM before integrating it into TREX-DM) and enhanced detector stability through an improved field cage design and, eventually, through the use of resistive Micromegas technology. Working on detector stability will ensure that optimal performance is achieved in routine operation, reducing the energy threshold to the single-electron levels demonstrated during $^{37}$Ar calibrations. Medium-term strategies focus on increasing the isobutane content in the gas mixture to enhance sensitivity to low-energy nuclear recoils, while long-term plans include improved material selection for intrinsic background reduction and increased detector exposure through scaled-up vessel and detectors. Collectively, these modifications position TREX-DM to achieve competitive exclusion limits in the sub-GeV WIMP mass range, addressing a critical gap in the current experimental status of the $(m_\chi, \sigma_n)$ parameter space.

%% file: CHAPTERS/Chapter10_spa.tex
\chapter*{Resumen y Conclusiones} \label{Chapter10spa_Resumen_Conclusiones}
\addcontentsline{toc}{chapter}{Resumen y Conclusiones}
\vspace{-3mm}

En esta tesis, hemos discutido en profundidad TREX-DM, un experimento de búsqueda de Partículas Masivas Débilmente Interactuantes (WIMPs) de baja masa, y sus prometedores desarrollos en cuanto a la reducción del umbral de energía para ser competitivos en esta región del espacio de parámetros de la Materia Oscura. El trabajo que aquí se presenta abarca no sólo los fundamentos teóricos que motivan la búsqueda de partículas de Materia Oscura, sino también las innovaciones tecnológicas implementadas para mejorar la sensibilidad del detector. Gracias a las mejoras en la reducción de fondo, la optimización del umbral de energía y las técnicas de calibración de baja energía, TREX-DM demuestra un potencial significativo como experimento competitivo en la búsqueda de WIMPs sub-GeV, una región que permanece en gran medida inexplorada por los experimentos convencionales de detección directa.

La base teórica de este trabajo comienza con pruebas observacionales sustanciales de la existencia de Materia Oscura como componente frío no bariónico. Como se explica en el Capítulo~\ref{Chapter1_Dark_Matter}, el análisis de las curvas de rotación galácticas, la dinámica de los cúmulos de galaxias, los efectos de las lentes gravitacionales y las mediciones de precisión del CMB convergen en la conclusión de que la Materia Oscura constituye una parte significativa del balance masa-energía del Universo y desempeña un papel crucial en la formación de estructuras. En este capítulo también se examinan los principales candidatos teóricos de la Materia Oscura, como son los WIMPs, los axiones, los neutrinos estériles y los agujeros negros primordiales. La exposición incluye de forma breve las modificaciones de la gravedad como explicación alternativa de los fenómenos observados, aunque se centra en las explicaciones basadas en partículas.

Partiendo de esta base, el Capítulo~\ref{Chapter2_WIMPs} profundiza en el estudio de los WIMPs como uno de los candidatos mejor motivados para la Materia Oscura. Mientras que el escenario canónico de los WIMP, a menudo conocido como el "milagro WIMP", proporciona una conexión elegante entre la escala débil y la abundancia observada de Materia Oscura, existe un interés teórico sustancial más allá de este paradigma. Al relajar la condición $g_\chi \sim g_\mathrm{weak}$, se hace accesible un rango de masas más amplio, que se extiende particularmente hacia la región sub-GeV. Esta ampliación del espacio de parámetros ha renovado el interés experimental en el desarrollo de técnicas de detección sensibles a los retrocesos nucleares de baja energía, ya que los enfoques convencionales se enfrentan a retos en esta región. A pesar de los amplios esfuerzos experimentales realizados en detección directa, búsquedas indirectas y experimentos en colisionadores, siguen sin encontrarse pruebas concluyentes de la existencia de WIMPs, lo que subraya la necesidad de estrategias de detección innovadoras.

La tecnología empleada en este trabajo se basa en los procesos fundamentales que rigen los detectores gaseosos, analizados en el Capítulo~\ref{Chapter3_Gaseous_Detectors}. Aquí hemos estudiado las interacciones entre fotones, partículas cargadas y partículas neutras, así como los mecanismos que conducen a la generación de carga: ionización primaria, deriva de electrones bajo la influencia de campos eléctricos y amplificación de carga mediante multiplicación por avalancha. Estos factores, combinados con la elección adecuada de la mezcla de gases, determinan colectivamente la ganancia del detector y la resolución energética. 

Entre la variedad de detectores gaseosos disponibles, las Cámaras de Proyección Temporales (TPCs) equipadas con Detectores Gaseosos de Micropatrones (MPGDs), particularmente Micromegas, ofrecen ventajas significativas para las búsquedas de Materia Oscura, como se establece en el Capítulo~\ref{Chapter4_TPCs_Micromegas}. Estos detectores proporcionan ventajas como la reconstrucción tridimensional de eventos con alta resolución espacial, lo que permite una discriminación eficaz del fondo. Además, su radiopureza intrínseca los hace especialmente adecuados para la búsqueda de sucesos poco probables, ya que la selección de materiales y los procesos de fabricación se han optimizado para minimizar las contribuciones de fondo.

El experimento TREX-DM aprovecha la tecnología Micromegas en su búsqueda de WIMPs de baja masa. El Capítulo~\ref{Chapter5_TREXDM} presenta una descripción exhaustiva del mismo, detallando su filosofía de diseño y sus principios operativos. El experimento consiste en una vasija radiopura que alberga un medio gaseoso, rodeado por un sistema de blindaje optimizado para reducir el fondo externo. El sistema de lectura emplea el mayor detector \textit{microbulk} Micromegas jamás construido, desarrollado a través de múltiples iteraciones (TREXDM.v1 y TREXDM.v2) para mejorar la resolución espacial y energética y reducir el umbral de energía. La infraestructura experimental se apoya en subsistemas especializados, e incluye un sistema de adquisición de datos diseñado para manejar una alta densidad de canales, fuentes de calibración específicas para garantizar mediciones de energía estables y precisas, un sistema de gas para mantener un medio de detección homogéneo y un sistema de \textit{Slow Control} para supervisar los parámetros operativos. También se han tratado los retos técnicos encontrados durante las fases de desarrollo y puesta en marcha, como el ruido electrónico, las fugas de corriente y las fugas de gas. El capítulo concluye con mediciones de la velocidad de deriva en la mezcla Ar-1\%iC$_{4}$H$_{10}$ que sirven de referencia para la validación del análisis.

Un aspecto importante de esta investigación ha sido la reducción progresiva de los niveles de fondo en TREX-DM, que han experimentado mejoras notables desde unos niveles iniciales de aproximadamente 1000 dru hasta los niveles actuales de 80-100 dru. Como se detalla en el Capítulo~\ref{Chapter6_Radon_problem}, este progreso es el resultado de una estrategia exhaustiva que aborda dos fuentes principales: el radón volumétrico y la contaminación superficial. El elevado fondo inicial era atribuible predominantemente a la emanación de radón de los filtros de gas, lo que impulsó el desarrollo de técnicas específicas de mitigación. Tras evaluar varias estrategias, la operación en lazo abierto con un caudal bajo capaz de compensar las fugas y la desgasificación (lo que denominamos operación en lazo abierto semisellado) demostró ser la más eficaz, reduciendo el radón activo a niveles insignificantes en comparación con la contaminación superficial.

La contaminación superficial de la progenie de $^{222}$Rn, en particular $^{210}$Pb y sus productos de desintegración, se identificó como la segunda fuente de fondo más importante. Para hacer frente a este reto se desarrolló el detector AlphaCAMM, que utiliza las propiedades direccionales de las partículas alfa para estudiar la contaminación en las superficies de las muestras, y será una herramienta clave para la selección de materiales. Los estudios de direccionalidad de partículas alfa aplicados a TREX-DM revelaron que la mayor parte de la contaminación superficial procede del cátodo. Por otra parte, los estudios de simulación confirmaron la correspondencia $1:1$ observada experimentalmente entre el fondo de baja energía y los eventos de alta energía. Aunque la reducción del fondo sigue siendo un reto, se están desarrollando soluciones prometedoras, entre las que destaca un cátodo basado en hilos de cobre. Este es el enfoque más viable a corto plazo para reducir la superficie de material susceptible a contaminación por exposición a la progenie del radón.

Una de las principales conclusiones de esta tesis es la mejora del umbral de energía mediante la adición de una etapa GEM sobre los detectores \textit{microbulk} Micromegas utilizados en TREX-DM. La motivación inicial para incorporar una etapa de preamplificación GEM fue mejorar la sensibilidad del detector a eventos de baja energía, lo que es esencial para detectar WIMPs de baja masa. El Capítulo~\ref{Chapter7_GEM-MM} presenta una investigación sistemática de esta configuración combinada de detectores, comenzando con pruebas en prototipo a pequeña escala y progresando hasta un detector a gran escala del conjunto de TREXDM.v2. Las pruebas con detectores pequeños validaron este enfoque, demostrando una ganancia adicional significativa para diferentes presiones de gas utilizando una mezcla Ar-1\%iC$_{4}$H$_{10}$. La exitosa implementación en la configuración equivalente a TREX-DM confirmó estos resultados, logrando factores de ganancia extra cercanos a 100 en comparación con el funcionamiento con sólo Micromegas y, por lo tanto, mostrando potencial para una reducción sustancial del umbral de energía. 

El resultado que consolida esta mejora en el umbral de energía es la calibración del detector de TREX-DM con una fuente de $^{37}$Ar, como se detalla en el Capítulo~\ref{Chapter8_Ar37}. Esta fuente de calibración gaseosa proporciona picos de baja energía homogéneos y bien definidos a 0.27 keV y 2.82 keV que son relevantes para la región de interés en las búsquedas de WIMPs de baja masa. El proceso de desarrollo comenzó con pruebas iniciales de viabilidad en CEA Saclay, validando el método de producción de $^{37}$Ar mediante irradiación de polvo de CaO con neutrones. A partir de estos resultados preliminares, se desarrolló un montaje específico para TREX-DM para albergar el polvo y operar la fuente de calibración. El método de producción se reprodujo posteriormente en una instalación de neutrones adecuada como es el CNA en Sevilla. La fiabilidad de este método para generar una fuente de calibración se confirmó monitorizando el subproducto $^{42}$K como medida indirecta de la actividad $^{37}$Ar.

La integración de la fuente de calibración de $^{37}$Ar en TREX-DM fue precedida por una intervención para instalar la lámina GEM en uno de los lados del detector. Tras la exitosa puesta en marcha del detector combinado GEM + Micromegas, el sistema se calibró con $^{37}$Ar, registrando un espectro completo (picos de 0.27 y 2.82 keV) gracias a la amplificación adicional proporcionada por la GEM. El análisis de los datos confirmó señales consistentes y medibles dentro del rango de energía objetivo, alcanzando umbrales de energía notablemente bajos, del orden de la energía de ionización de electrones individuales (26 eV en argón) para las ganancias más altas, y umbrales de $O(200-300)$ eV en combinaciones de voltaje más conservadoras. Este avance representa un hito crucial, ya que demuestra que el experimento es capaz de detectar e interpretar de forma fiable sucesos de baja energía, una condición esencial para explorar el espacio de parámetros de WIMPs de baja masa.

Por último, el Capítulo~\ref{Chapter9_Sensitivity} recoge todas las conclusiones y proyecciones relacionados con la reducción del fondo y la mejora del umbral de energía. Aquí se discuten las perspectivas de sensibilidad de TREX-DM, presentando una hoja de ruta completa para el futuro del experimento. Con los actuales niveles de fondo de 80-100 dru y los umbrales de energía conservadores por debajo de 300 eV logrados gracias a la preamplificación GEM, el experimento está bien posicionado para nuevos avances. Las mejoras a corto plazo incluyen la reducción del fondo mediante la implementación de un cátodo de hilos (que será probado primero en AlphaCAMM antes de su integración en TREX-DM) y la mejora de la estabilidad del detector mediante un diseño mejorado de la jaula de campo y, a la larga, mediante el uso de la tecnología Micromegas resistiva. Trabajar en la estabilidad del detector garantizará que se logre un rendimiento óptimo en operación rutinaria, reduciendo el umbral de energía a los niveles de un solo electrón demostrados durante las calibraciones de $^{37}$Ar. Las estrategias a medio plazo se centran en aumentar el contenido de isobutano en la mezcla de gases para mejorar la sensibilidad a los retrocesos nucleares de baja energía, mientras que los planes a largo plazo incluyen una mejor selección de materiales para la reducción del fondo intrínseco y el aumento de la exposición mediante la construcción de una vasija y detectores más grandes. Como conclusión final, estas modificaciones permitirán a TREX-DM alcanzar límites de exclusión competitivos en el rango de masas de WIMP por debajo de los GeV, abordando este vacío en el estado experimental actual del espacio de parámetros $(m_\chi, \sigma_n)$.